\documentclass[11pt,twoside,a4paper,openright]{book}

\usepackage{floatflt} % allow floating figures
\usepackage{float}
\usepackage{pdfsync} % allow easy editing with pdf files
\usepackage[bookmarks=true, pdfpagelabels=true, colorlinks=true, linkcolor=black, urlcolor=black, citecolor=blue, anchorcolor=blue]{hyperref}

\usepackage[cyr]{aeguill}
\usepackage{booktabs}
\usepackage[utf8]{inputenc}
\usepackage[T1]{fontenc}
\usepackage{lmodern}
\usepackage[toc,page]{appendix}

\usepackage{amsmath,amssymb,bm}	  % math options
\usepackage{graphics,xcolor}      % graphic and colors
\usepackage{times}
\usepackage{graphicx}
\usepackage{emptypage}
\usepackage{cite}
\usepackage[export]{adjustbox}
\usepackage[compat=1.1.0]{tikz-feynman}   % Feynman Diagrams
\tikzfeynmanset{warn luatex=false}
\usepackage{contour}  
\usepackage{comment}              % easy commenting 
\usepackage{multirow}
\usepackage{subcaption}
\usepackage{slashed}
\usepackage{dsfont,bbold}  % allow blackboard bold math elements
\usepackage{arydshln}  % allow dashed lines in tables
\usepackage{listings}
\usepackage{lscape}
\usepackage{rotating}
\usepackage{setspace}
\usepackage{afterpage}
\usepackage{ulem}

\hypersetup{colorlinks=true}
\usepackage{etoolbox}% http://ctan.org/pkg/etoolbox

\usepackage[font=small,labelfont=bf]{caption}

\usepackage[a4paper,total={12cm,20cm},centering,headsep=.5cm]{geometry}
\usepackage{fancyhdr}
\usepackage{sectsty}
\sectionfont{\fontsize{16}{18}\usefont{OT1}{phv}{m}{n}\selectfont}
\subsectionfont{\fontsize{13}{15}\usefont{OT1}{phv}{m}{n}\selectfont}
\subsubsectionfont{\fontsize{11}{13}\usefont{OT1}{phv}{m}{n}\selectfont}

\usepackage[Lenny]{fncychap}
\ChNameVar{\fontsize{14}{16}\usefont{OT1}{phv}{m}{n}\selectfont}
\ChNumVar{\fontsize{60}{62}\usefont{OT1}{phv}{m}{n}\selectfont}
\ChTitleVar{\fontsize{24}{26}\usefont{OT1}{phv}{m}{n}\selectfont}

\makeatletter
\renewcommand\ps@plain{\let\@mkboth\@gobbletwo
     \let\@oddhead\@empty
     \def\@oddfoot{\reset@font\hfil}
     \let\@evenhead\@empty\let\@evenfoot\@oddfoot}

\patchcmd{\@makechapterhead}{\vspace*{50\p@}}{}{}{}% Removes space above \chapter head
\patchcmd{\@makeschapterhead}{\vspace*{50\p@}}{}{}{}% Removes space above \chapter* head

\g@addto@macro\normalsize{%
  \setlength\abovedisplayskip{10pt}
  \setlength\belowdisplayskip{10pt}
  \setlength\abovedisplayshortskip{10pt}
  \setlength\belowdisplayshortskip{10pt}
}

\makeatother

\usepackage{makeidx} 
\makeindex

\newcommand{\ket}[1]{\lvert#1\rangle} % Ket
\newcommand{\qprod}[2]{ \langle #1 | #2 \rangle} %Inner Product
\newcommand{\braopket}[3]{\langle #1 | #2 | #3\rangle} % Matrix Element
\newcommand{\MG}{{\sc MadGraph5}\_a{\sc MC@NLO} }
\newcommand{\fr}{{\sc Feyn\-Rules}}
\newcommand{\nloct}{{\sc NloCT}}
\newcommand{\mspin}{{\sc MadSpin}}
\newcommand{\MW}{{\sc MadWidth}}
\newcommand{\ma}{{\sc MadAnalysis\ 5}}
\newcommand{\pyt}{{\sc Pythia\ 8}}
\newcommand{\fj}{{\sc FastJet}}
\newcommand{\sfs}{{\sc SFS}}
\newcommand{\pyhf}{{\sc PyHF}}
\newcommand{\scr}[1]{\ensuremath{\mathcal{#1}}}

\newcommand{\ie} {{\it i.e.}\,}

\newcommand {\beq} {\begin{equation}}
\newcommand {\eeq} {\end{equation}}
\newcommand {\bea} {\begin{eqnarray}}
\newcommand {\eea} {\end{eqnarray}}

\newcommand{\bpm}{\begin{pmatrix}}      
\newcommand{\epm}{\end{pmatrix}}

\def\sing{S_1}
\def\oct{S_8}
\newcommand{\OWWW}{\mathcal{O}_{\widetilde{W}WW}}
\newcommand{\OBW}{\mathcal{O}_{\phi\widetilde{W}B}}

\newcommand{\JT}[1]{\color{red}{\bf #1 }\color{black}}

\newcommand{\obothi}{\hspace{-1mm}\begin{array}{c}\quad\\[-8mm]\scriptscriptstyle{\left({\wedge}\right)\,}\\[-1.7mm] \mathcal{O}_i\end{array}\hspace{-1mm}}

\newcommand{\stkout}[1]{\ifmmode\text{\sout{\ensuremath{#1}}}\else\sout{#1}\fi}

\begin{document}  

% Title, acknowledgment, table of contents...
%%%%%%%%%%%%%%%%%%%%%%%%%%%%%%%%%%%%%%%%%%%%%%%%%%%%
%
%      First pages ...
%
%
%%%%%%%%%%%%%%%%%%%%%%%%%%%%%%%%%%%%%%%%%%%%%%%%%%%

%% Titre du document
% \parbox[c][][c]{0.2\textwidth}{
% \includegraphics[height=1.0cm]{figures/logoUcl}
% }

\begin{figure}[t!]
\centering
\includegraphics[scale=0.3]{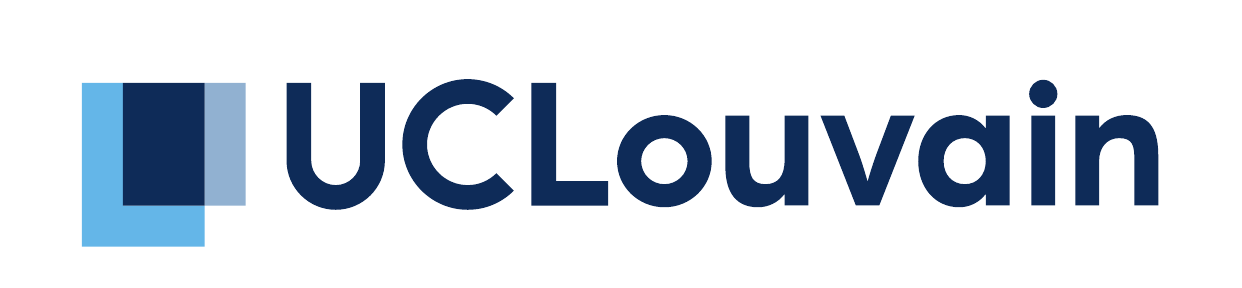}
\end{figure}

% \parbox[c][][c]{0.799\textwidth}{\vspace{0.05cm}
% \begin{flushright}
% \large Universit{\'e} catholique de Louvain\\[0.2\baselineskip]
% \normalsize Secteur des Sciences et Technologies\\[0.1\baselineskip] 
% Institut de Recherche en Math{\'e}matique et Physique\\[0.1\baselineskip]
% Center for Cosmology, Particle Physics and Phenomenology
% \end{flushright}
% }
\vspace{1.5cm}
\begin{center}
% \vspace{2cm}
\parbox{0.95\textwidth}{\fontsize{22}{30}\selectfont\centering{
			New Observables for Resonant and Non-Resonant New Physics Searches at Colliders
	}}
%\parbox{0.95\textwidth}{\fontsize{22}{30}\selectfont\centering{\sectionfont Search for New Physics in Double Higgs production with the CMS detector}}
% \vspace{0.3cm}
\end{center}
\vspace{0.6cm}
\begin{center}
Doctoral dissertation presented by \\
\vspace{2mm}
{\Large Julien Touchèque }\\
\vspace{2mm}
in fulfilment of the requirements for the degree of Doctor in Sciences.
\end{center}
\vspace{\fill}
\begin{center}
\begin{tabular*}{0.8\textwidth}{l @{\extracolsep{\fill}} r}
%{Thesis support committee :} & \\[3.5pt]
\multicolumn{2}{c}{\large Thesis support committee}                                  \\
                    &                                          \\
\toprule 
{Prof. Degrande Céline} (Supervisor) & UCLouvain, Belgium \\
{Prof. Lemaître Vincent}  & UCLouvain, Belgium \\
{Prof. Maltoni Fabio}  & UCLouvain, Belgium \\
\bottomrule 
\end{tabular*}

\vspace*{0.5cm}
\textsl{July, 2026}\\[1pt]
\end{center}
\thispagestyle{empty}

%%% acknowledgment

\hfill
\newenvironment{acknowledgements}%
    {\cleardoublepage\thispagestyle{empty}\null\vfill\begin{center}%
    \bfseries Acknowledgements\end{center}}%
    {\vfill\null}

\begin{acknowledgements}

It is only the beginning of the thesis, but this work marks the end of a long journey. It started, like a cliché straight out of a movie, in the storeroom at the back of my primary school library when I opened a book on the stars and their formation, and now ends with this manuscript on the search for New Physics. While I may have started out by myself, I would not have arrived at this point alone.

I want to thank first and foremost my partner, Esther, without whom this manuscript would genuinely not have seen the light of day. She not only supported me and believed in me when I did not believe in myself, but also kicked me in the backside when I needed it. I am very lucky (she will understand this pun) to have her in my life, and I love her with all my heart.

I will always be indebted to Céline, my supervisor, who accepted me as her PhD student and guided me through the challenges of a PhD. I am proud of the work we accomplished together. Without her, none of this would have been possible, and I would not be the researcher I am today.

I would like to extend my gratitude to all the members of my thesis committee: Eleni Vryonidou, Benjamin Fuks, Vincent Lemaître and Fabio Maltoni. It was an honour to have so many experts agree to review my manuscript. Their comments and questions have considerably improved both the clarity and the quality of this thesis.

From the bottom of my heart, I thank my Mum and Dad, who made it possible for me to achieve my dream of becoming a physicist. They always did the impossible to create the best conditions for me to succeed.

I owe a great deal to my Belgian and Australian families, whom I wish I could see more often, for their unwavering support. It was always a pleasure to explain to them what I was researching.

I thank my childhood friends Alexis, Laurent and Camille, who have been there through all the good times and the bad, my friends from university, the Maison des Sciences, the physics programme and Brussels, as well as my former flatmates.

I want to make special mention of Luc, Matteo and Haolin for the privilege of working with them on challenging and stimulating research. I truly cherished collaborating with them.

\newpage

Finally, I thank all the people at CP3 whom I had the chance and the pleasure to meet during my PhD years. I saw many generations arrive and leave, but I want to mention in particular Samip for all the football arguments, Andres for being my first office mate, Maxime for the sleepless nights playing Balatro, Olivier for never getting tired of my MadGraph questions, the lunch group, the football and badminton gangs, my colleagues from the pheno meetings, and several people I am sure I am forgetting.

\end{acknowledgements}

%%% abstract

\hfill
\newenvironment{Abstract}%
    {\cleardoublepage\thispagestyle{empty}\null\vfill\begin{center}%
    \bfseries Abstract\end{center}}%
    {\vfill\null}

\begin{Abstract}

The discovery of the Higgs boson cemented the form of the Standard Model of particle physics (SM) as we know it and observations have so far confirmed SM predictions about the couplings of the Higgs with other particles. If the particle physics community used to face too few models other than the Higgs before its observation to explain the origin of the particle masses, too many beyond-the-Standard-Model (BSM) models can be explored to overcome the remaining shortcomings of the SM. As a result, generic models present a useful intermediary step to point towards the potential extension of the SM probing multiple Ultraviolet (UV) models simultaneously. In this work, two examples of generic models are investigated.

Off-shell effects of potential heavy new particles are provided in the Standard Model Effective Field Theory (SMEFT) through higher dimensional operators. In particular, charge-parity ($CP$) symmetry violating contributions from Wilson coefficients are ranked by using several $U(1)$ global flavour symmetries on $CP$-odd operators of the Warsaw basis. Asymmetries of simple observables, such as triple products, are proven to be more sensitive to the SMEFT operators compared to the cross-section and more efficient than most of the usual observables used in experiments. We use the dileptonic decays of the diboson $W^{\pm} Z$ and $W^{\pm} \gamma$ processes to demonstrate the efficiency of these observables compared to the theoretical and measurable optimal observables.

On-shell effects of two heavy scalars are independently searched for in four-top production. These new scalars are singlet and octet of $SU_C(3)$ that couple to the top quark. The signal consists of the associated production of a scalar with a top quark pair and of pair production in the case of the octet where the fully hadronic, the leptonic and the same-sign-lepton dileptonic decay channels are considered. The decays of the massive scalars generate boosted top-quark pairs producing event topologies that are suppressed in the SM. The processes are generated at next-to-leading order (NLO) in quantum chromodynamics (QCD) with a careful renormalisation procedure and the correct selection of loop diagrams. Using improvements in top-jet tagging techniques and in the main background simulation as well as a new technique to select the events, a new particle search is performed to constrain the respective mass and coupling to top quarks of new resonances. The analysis yields improved limits by up to an order of magnitude compared to previous analyses.

\end{Abstract}

\clearpage

\begin{center}

\vspace{121pt}

{\LARGE Associated Publications:}
\vspace{22pt}
\begin{description}
\item[\cite{Degrande:2021zpv}] 
Degrande C{\'e}line and Touch{\`e}que Julien, “A reduced basis for CP violation in SMEFT at colliders and its application to diboson production”, \textit{JHEP} 04 (2022) 032.
\item[\cite{Darme:2025leu}]
Darm{\'e} Luc, Fuks Benjamin, Li Hao-Lin, Maltoni Matteo and Touch{\`e}que Julien, “Searching for top-philic heavy resonances in boosted four-top final states”,  \textit{JHEP} 11 (2025) 091.
\end{description}

\vspace{15pt}

{\Large Publications not included:}
\begin{description}
\item[\cite{Darme:2024epi}]
Darm{\'e} Luc, Fuks Benjamin, Li Hao-Lin, Maltoni Matteo, Mattelaer Olivier and Touch{\`e}que Julien, “Novel approach to probing top-philic resonances with boosted four-top tagging”, \textit{Phys. Rev. D}, vol. 111, no. 5, pp. 055037, 2025. 
\end{description}

\vfill
\begin{tabular}{llc} 
\multicolumn{3}{c}{\large Thesis Jury}     \\
                        &             &    \\
\toprule 
Pr. Degrande Céline     & Promoter    & Université Catholique de Louvain  \\
Pr. Maltoni Fabio       & President   & Université Catholique de Louvain  \\
Pr. Lemaître Vincent    & Secretary   & Université Catholique de Louvain  \\
Pr. Fuks Benjamin       & Lecturer    & Sorbonne Université  \\
Pr. Vryonidou Eleni     & Lecturer    & University of Cyprus \\
\bottomrule
\end{tabular}

\vspace{44pt}

\end{center}

%%% table of content
\pagestyle{empty}
\tableofcontents

% Introduction
%%%%%%%%%%%%%%%%%%%%%%%%%%%%%%%%%%%%%%%%%%%%%%%%%%%%
%
%     Introduction
%
%%%%%%%%%%%%%%%%%%%%%%%%%%%%%%%%%%%%%%%%%%%%%%%%%%%
\chapter*{Introduction}
\addcontentsline{toc}{chapter}{Introduction}
\setcounter{page}{1}

\hfill
\begin{minipage}{8cm}
{\it 
``Do not go gentle into that good night, \\
Old age should burn and rave at close of day; \\
Rage, rage against the dying of the light. [...]''}

\hfill poem by Dylan Thomas (1914–1953)
\end{minipage}

\vspace{0.5cm}

Science is a branch of philosophy of Nature dedicated to explaining the different phenomena observed in the Universe. Particle physics focuses on phenomena involving the fundamental components of matter. The scientific method to describe our environment has always relied on the modelling of its components and their interactions. The generated model must provide testable predictions, meaning that observations can be confronted with the predictions to build confidence in the model or disprove it. The back-and-forth between theory and experiment has challenged scientists in both fields to collaborate and improve the model. In particular, discoveries and predictions of new particles were milestones in establishing the SM as it is known today. Their observation have required the building of many particle colliders with different designs: the SLC, the LEP, the Tevatron, the LHC among others. The particle physics landscape would be very different without this fruitful joint effort.

After a few years of data taking at the LHC, the discovery of the Higgs boson looked at the time like the well-earned trophy following the efforts in the theoretical description of the symmetry breaking mechanism in the SM and in the precision of the experimental analysis to reveal the last elusive particle. However, it ended up putting the particle physics community in a sort of deadlock. Despite the SM offering a genuine understanding of the particle mass origin and the multiple, sometimes chaotic, particle interactions, the model is not without shortcomings. The comprehensive list of the SM issues is not strictly defined but some have been raised historically. The mere fact that our observable Universe appears almost entirely made of matter remains a puzzle to solve. After their prediction by Dirac and their discovery by Anderson, anti-particles are a standard component of matter. However, the abundance of matter derived from the SM interactions, and the $CP$ violation mechanism it contains leading to the matter-antimatter asymmetry, cannot accurately describe the current state of the Universe. Another more technical issue is the suppression of the asymmetry due to the nature of the phase transition dictated by the observed value of the Higgs mass. The transition from the symmetric to the broken phase is smooth so any excess of matter or anti-matter is compensated resulting in a too small matter density. That is why some mechanism must be added to the SM to make the transition more abrupt and stronger.

Solutions to this problem have been addressed by many BSM models that extend the SM with new symmetries or particles. For instance, models with right-handed neutrinos give rise to more $CP$ violating interactions which in turn could explain the present domination of matter over antimatter via leptogenesis. Supersymmetric models introduce more particles with more interactions, providing more $CP$ violating parameters. Models with extended Higgs particles or composite-Higgs models can have additional $CP$ violating parameters as well. Depending on the specific expression of the models, they could solve the issue with the phase transition of the Higgs. The landscape of BSM models is hard to explore without further indication.

Nevertheless, one should not feel lost when facing this challenge. Tools have been devised to scan the theoretical landscape with limited assumptions on the BSM model in the form of \textit{generic models}. Generic models describing off-shell and on-shell contributions of BSM particles can act as beacons to guide towards the next update of the SM. The great advantage of generic models is that they can always be reinterpreted to fit different UV-complete models where more particles and/or symmetries are properly introduced. Fitting multiple complete models with one simplified model offers greater flexibility and more opportunity to search for evidence of BSM.

In Chapter~\ref{chap:sm}, the importance of the cross-section at particle colliders is established since it connects the probability of final state observation in detectors to the Lagrangian density of the model. The derivation of the SM Lagrangian follows from its gauge symmetries and its degrees of freedom. The Brout-Englert-Higgs mechanism resulting from the Higgs potential induces the elementary particle masses. $CP$ violation in the SM via quark-mixing interactions is explained and its flavour symmetry dependence is emphasised. Chapter~\ref{chap:sm} ends with a short description of some remaining issues within the SM.

Then, Chapter~\ref{chap:bsm} contains the presentation of generic models. Generic models consist of two complementary descriptions : effective field theories and simplified models. First, the top-down approach to off-shell contributions by the construction of Effective Field Theories (EFT) via matching is discussed with a toy example and the 4-Fermi EFT. Then, the SMEFT is introduced in the bottom-up approach to off-shell contributions. The Warsaw basis is presented by listing the allowed operator classes and briefly reviewing how redundant operators are removed using the symmetries of the SM and the particle equations of motion. Simplified models reproducing the on-shell effects of added particles are also presented. The possible additions to the SM spectrum are sorted by their representations of the SM symmetries.

Chapter~\ref{chap:smeftcpv} presents a general $CP$ search in the SMEFT. It first defines that the leading $CP$ violating contributions to the amplitudes are from $CP$-odd dimension-six SMEFT operators and why the cross-section is not a good observable in this case. Two sets of $U(1)$ flavour symmetries reduce the $CP$-odd operators to either 10 or 17 operators respectively depending on whether the bottom quark mass is neglected or not. A review on observables sensitive to $CP$-odd operators obtained in the case where the bottom quark is massless, is given. The review takes into account observables directly measured in particle colliders but also indirect observables from low-energy measurements, optimal observables and machine learning (ML) approaches. Triple product asymmetries are proposed as simple, direct and efficient observables to target $CP$ violating contributions by two bosonic dimension-six $CP$-odd operators in diboson production. They are compared to asymmetries of existing angular observables from the review and the ones with the most sensitivity are highlighted. Differential cross-section results with respect to centre-of-mass energy are shown to assess the energy growth behaviour of the higher order $CP$-odd operators. The expected limits on the Wilson coefficients are then computed as a function of the luminosity.

Lastly, a search for new particles in four-top production at the LHC is presented in Chapter~\ref{chap:fourtops}. Two simplified models consisting of scalar singlet and octet of $SU_C(3)$ are investigated in the fully hadronic, leptonic and dileptonic decay channels of the four-top final state. After detailing the renormalisation of the two models and the selection of amplitudes to cancel the divergences, the NLO contributions to the cross-section are provided as K factors. The reconstruction of the four tops uses improvements in ML techniques to tag wide hadronic jets while the top decaying leptonically are reconstructed by assuming that $W$ bosons decay on-shell. A new technique using the invariant masses of top pairs allows us to gain sensitivity to signal and reduce the SM backgrounds. Limits on the respective parameter spaces of the scalars are greatly improved compared to a recast CMS analysis.

In Chapter~\ref{chap:conclusion}, we discuss the results of the two analyses mentioned above and conclude. 

\vspace{1cm}

Disclaimer : A conversation with the chatbot Claude (Opus 4.8) was used to check for spelling errors and grammar mistakes during the writing process of this manuscript. At times, it provided suggestions to improve the quality of the text which were taken into account when the changes were deemed relevant. Claude was also used to generate plots in Appendices but it did not generate the data it relies on.

% Chapters
%!TEX root = main.tex

%%%%%%%%%%%%%%%%%%%%%%%%%%%%%%%%%%%%%%%%%%%%%%%%%%%%
%
%      Chapter 1 :
%
%
%%%%%%%%%%%%%%%%%%%%%%%%%%%%%%%%%%%%%%%%%%%%%%%%%%%

\chapter{The Standard Model of Particle Physics}
\label{chap:sm}
\pagestyle{fancy}

\hfill
\begin{minipage}{7cm}
{\it  ``The most important is not what we know 
\\ but what we don’t yet know.''}

\hfill Citation from François Englert (1932 - 2026)
\end{minipage}

\vspace{0.5cm}

This chapter begins by presenting in Section~\ref{sec:QFT} how particle physicists relate observations of final states to operators of the Lagrangian density (and vice versa) through the cross-section. The definition of the cross-section of particle collisions in QFT will be developed from its expression in classical physics. Secondly, the SM Lagrangian is detailed in Section~\ref{sec:SMLag}. The gauge symmetries from the SM gauge group are presented with their respective vector bosons, then the degrees of freedom in the SM are gradually introduced. The Brout-Englert-Higgs (BEH) mechanism is explained by showing the Higgs potential and demonstrating how it is responsible for the mass of particles. Thirdly, it will be quickly proven in Section~\ref{sec:CPsymmetries} how the quark-mixing interactions violate the $\mathrm{CP}$ symmetry when the quarks are represented in the mass basis after flavour rotations. The Jarlskog invariant is derived by exploiting the flavour transformations of the quarks. Finally, the remaining issues in the SM are shown in Section~\ref{sec:SMIssues} emphasising on the particular problem of matter-antimatter asymmetry requiring $\mathrm{CP}$ violation.

This chapter was written using standard Quantum Field Theory textbooks such as Refs.~\cite{Schwartz:2014sze, Peskin:1995ev}. These textbooks were made possible thanks to historical works by Gell-Mann \cite{GELLMANN1964214}, Zweig \cite{Zweig:1964ruk}, Glashow \cite{Glashow:1961tr}, Weinberg \cite{PhysRevLett.19.1264}, Salam \cite{Salam:1959zz}, Higgs \cite{PhysRevLett.13.508}, Englert and Brout \cite{PhysRevLett.13.321} among many others.

\newpage

\section{From Observations to Particle Operators}
\label{sec:QFT}

Particle colliders are designed to measure the probability of transforming an initial state $\ket{i}$ into a targeted final state $\ket{f}$. The latter is observed in detectors and the former generally consists of two beams $\ket{i} = \ket{b_1 b_2}$ circulating in opposing directions. For instance, the beams at the LEP were $\ket{i}_{\text{LEP}} = \ket{e^- e^+}$ and at the LHC are $\ket{i}_{\text{LHC}} = \ket{pp}$. When the two beams are aligned, their particle content potentially collides. Most of the time, particles miss each other but sometimes particles meet and interact. These collisions of particles are called \textit{events}\footnote{It is important to keep in mind that some collisions are not recorded as events because particles in the final state are not detected due to the geometry of the collider, the trigger, etc. }. However, the probability of interaction is not directly measured by experiments. What is observed is the number of measured states $\ket{f}$ collected over numerous collisions of particles. The probability of interaction and the number of measured states are related through the \textit{cross-section}.

\begin{figure}[t]
    \centering
    \includegraphics[width=0.55\linewidth]{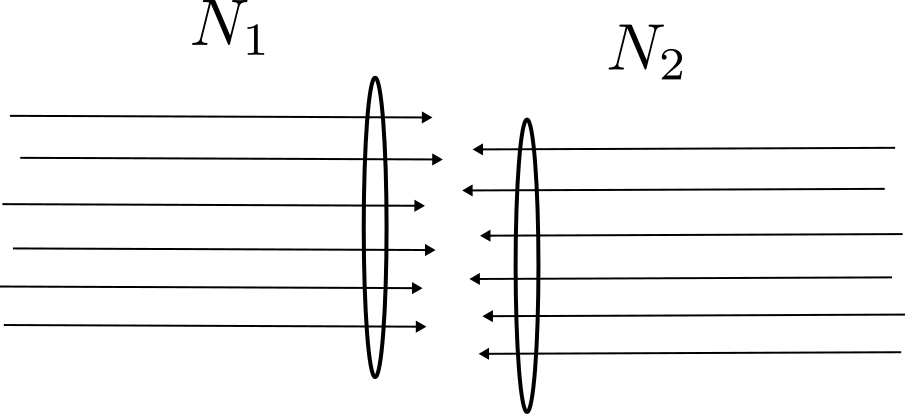}
    ~~~\vline~~~
    \includegraphics[width=0.35\linewidth]{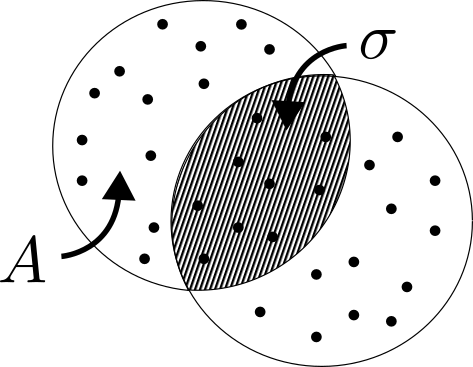}
    \caption{Representation of the classical cross-section between particle beams. \textit{Left}: The beam 1 is coming from the left and beam 2 from the right. \textit{Right}: Each circle represents a beam of area $A$ containing $N_i$ particles shown as black dots. The cross-section $\sigma$ is the crossed area where the two beams overlap. }
    \label{fig:classicalcrosssection}
\end{figure}

The cross-section is first defined in classical physics before being presented in the QFT picture. Figure~\ref{fig:classicalcrosssection} shows the two beams at the time of collision. The number of events $N_{ev}$ is determined by the number of incident particles $N_{i}$ in each beam, over a period of time $T$, uniformly distributed over the size of the beams $A$ and by $\sigma$ the surface over which the beams overlap. The latter quantity is what is called the cross-section and it is represented by the hatched region of Figure~\ref{fig:classicalcrosssection}. The two regions in white do not overlap so there are no events associated with their particle content\footnote{If the beams are not dense enough, particles in the hatched region from different beams can miss too.}. The number of events reads as
\begin{equation*}
    N_{ev} = \sigma \frac{N_1 N_2}{A} . 
\end{equation*}

In the frame where the particles in beam 2 are targeted at rest by those in beam 1, the probability of interaction per target is $P=N_{ev}/N_2$ and the incoming flux is $\Phi = N_1/AT$. The cross-section of a target hit by a beam is then given by  
\begin{equation}
\label{eq:TargetClassicalCrossSection}
	\sigma = P/(T\Phi).
\end{equation}

To get the cross-section of two beams facing each other, the flux in Eq.(\ref{eq:TargetClassicalCrossSection}) has to be adapted. In this configuration, it is given by the difference between the flux of the two beams $\Phi=|\vec{\Phi}_1 - \vec{\Phi}_2|$ where each flux is the density of particles in the beam times the beam velocity $\vec{v}_i$, $\vec{\Phi}_i = (N_i/V_i) \vec{v}_i$ where $V_i$ is the volume containing all interacting particles from beam $i$. Thus, if the density of particles is identical in the two beams $N_1/V_1 = N_2/V_2 = N/V$, 
\begin{equation}
    \Phi = \frac{|N_1 \vec{v}_1|}{V} - \frac{|N_2 \vec{v}_2|}{V} = \frac{N}{V} |\vec{v}_1 - \vec{v}_2 | .
\end{equation}
So, the classical cross-section is the area given by
\begin{equation}
\label{eq:crosssection}
    \sigma = \frac{V}{T} \frac{1}{|\vec{v}_1 - \vec{v}_2|} \frac{P}{N} =  \frac{V}{T} \frac{1}{|\vec{v}_1 - \vec{v}_2|} dP.
\end{equation}
dP corresponds to the probability of interaction per particle in the beams.

In QFT, the concept of the cross-section as a surface onto which the beams interact is replaced by the interaction probability of the beams. By analogy with its classical counterpart, the quantum cross-section is defined as the rate of interaction per unit of flux. So, $\sigma$ in QFT reads as
\begin{equation}
    d\sigma = \frac{V}{T} \frac{1}{|\vec{v}_1 - \vec{v}_2|} dP.
\end{equation}
where the probability $dP$ must be derived from the quantum state interactions.

The states $\ket{i}$ and $\ket{f}$ are supposed to be asymptotically free in space and time. We temporarily add $\pm\infty$ as an argument to emphasise the asymptotic free nature of the states, $\ket{i (-\infty)}$ and $\ket{f (\infty)}$. The interaction picture is adopted here : the states evolve only under the interaction Hamiltonian $H_{int.}$ and the operators carry the time dependence from the free Hamiltonian $H_0$. We use the time evolution operator in the interaction picture $U_{int.}$ to evaluate the states at some finite interaction time $t$, $\ket{i(t)}$ and $\ket{f(t)}$. The operator is  defined between two points in time $t_1$ and $t_2$ as $U_{int.}(t_1, t_2) \equiv e^{-i H_{int.} (t_1 - t_2)}$. We find that
\begin{equation}
\begin{split}
    & \ket{i(t)} = \lim_{T\to\infty} U_{int.}(t, -T/2)~ \ket{i(-T/2)} = \lim_{T\to\infty} e^{-i H_{int.} (t+T/2)}~ \ket{i(-T/2)}, \\
    & \ket{f(t)} = \lim_{T\to\infty} U_{int.}(t, T/2)~ \ket{f(T/2)}  =  \lim_{T\to\infty} e^{-i H_{int.} (t-T/2)}~ \ket{f(T/2)} .
\end{split}
\end{equation}
Therefore, we get
\begin{equation}
\label{eq:Smatrixelements}
    \qprod{f(\infty)}{i(-\infty)} = \lim_{T\to\infty} \braopket{f(t)}{e^{i H_{int.} T}}{i(t)} \equiv \braopket{f(t)}{ \mathcal{S} }{i(t)} .
\end{equation}
The whole unitary operator in-between the asymptotic states is called the \textit{scattering matrix}, or $\mathcal{S}$-matrix. Now that the $\mathcal{S}$-matrix has been defined, we can drop the indices of the states keeping implicit their asymptotic freedom. The $\mathcal{S}$-matrix elements $\mathcal{S}_{fi}$ are defined by Eq.~(\ref{eq:Smatrixelements}). The $\mathcal{S}$-matrix must be unitary to preserve the total probability  
\begin{equation}
\label{eq:SmatrixUnitary}
    \mathcal{S} \mathcal{S}^\dagger = \mathbb{1} = \mathcal{S} \mathcal{S}^{-1},
\end{equation}
with $\mathbb{1}$ being the identity matrix. The role of $\mathcal{S}$-matrix is to describe how the asymptotic initial state relates to the asymptotic final state accounting for all possible interactions. On the one hand, particles can just behave trivially without reacting to each other, that is no interactions occur and $\ket{f}=\ket{i}$. On the other hand, there can be interactions between the fields which will cause some particles to be destroyed and others to be created, $\ket{f} \neq \ket{i}$. Both possibilities are represented by the $\mathcal{S}$-matrix when expressed as
\begin{equation}
\label{eq:Smatrixdef}
    \mathcal{S} = \mathbb{1} + i \mathcal{T}.
\end{equation}
The first term is responsible for the trivial behaviour and the second term with the transfer matrix $\mathcal{T}$ deals with the interactions. The trivial term is not as interesting as the interacting term, so we focus on the latter. Even if particles interact, they must obey the conservation of the total 4-momentum. Thus, the transfer matrix is written as
\begin{equation}
\label{eq:Tmatrixdef}
    \mathcal{T} = (2 \pi)^4 \delta^{(4)}\left( \sum_f p^{\mu}_f - \sum_i p^{\mu}_i \right) \mathcal{M}.
\end{equation}
The momenta are summed over the particles of the incoming and outgoing states. Since we are interested in understanding how particles interact, we look at the matrix elements $\braopket{f}{\mathcal{M}}{i}$
\begin{equation*}
\begin{split}
    \braopket{f}{\mathcal{S} - \mathbb{1}}{i} & = i~(2 \pi)^4 \delta^{(4)}\left( \sum_f p^{\mu}_f - \sum_i p^{\mu}_i \right) \braopket{f}{\mathcal{M}}{i}, \\
    & = \braopket{f}{\mathcal{S}}{i},
\end{split}
\end{equation*}
where in the last line we considered different initial and final states so $\braopket{f}{\mathbb{1}}{i}$ reduces to $\qprod{f}{i}=0$.

Now, the differential probability $dP$ of observing the final state in an element of the phase space  $d\Pi$ is defined by
\begin{equation}
\label{eq:diffprob}
    dP  = \frac{|\braopket{f}{S}{i}|^2}{\qprod{f}{f}~ \qprod{i}{i}} d\Pi  = (2 \pi)^8 \delta^{(4)}(0) \delta^{(4)}\left( \sum_f p^{\mu}_f - \sum_i p^{\mu}_i \right) \frac{|\braopket{f}{\mathcal{M}}{i}|^2}{\qprod{f}{f}~ \qprod{i}{i}} d\Pi.
\end{equation}
Note that, in 3 dimensions, we have $ \delta^{(3)}(0) = \frac{1}{(2\pi)^3} \int_V d^3~\vec{x} = V/(2\pi)^3$. So, in 4 dimensions, we must get $ \delta^{(4)}(0) = TV/(2\pi)^4$. The normalisation of the states in the denominator of Eq.~(\ref{eq:diffprob}) is fixed by the creation operators $\ket{\phi} = \hat{a}^{\dagger}_p \ket{\Omega} = \frac{1}{\sqrt{2 E_p}} \ket{p} $, with $\ket{\Omega}$ being the vacuum state in the interacting theory, and the product that follows $\qprod{p}{p}=(2\pi)^3 (2 E_p) \delta^{(3)}(0)$. We obtain the normalisation of the states
\begin{equation}
\label{eq:NormalisationStates}
    \qprod{f}{f} = \prod_{f} (2E_f V), ~~
    %\text{  and  } 
    \qprod{i}{i} = \prod_{i} (2E_i V).
\end{equation}
The element of phase space $d\Pi$ is obtained by multiplying the differential momentum $d^3\vec{p}_f$ of the particles in the final state with a normalising factor. The latter follows from the normalisation of single-particle momentum states mentioned above. So,
\begin{equation}
\label{eq:PhaseSpace}
    d\Pi = \prod_{f} \frac{V}{(2\pi)^3}~ d^3\vec{p}_f.
\end{equation}

By injecting Eqs.~(\ref{eq:NormalisationStates}) and (\ref{eq:PhaseSpace}) in Eq.~(\ref{eq:diffprob}), the differential probability takes the following form\footnote{The total volume $V$ and the time of running $T$ naturally cancel in the final expression which thus becomes valid for $V, T \to \infty$. }
\begin{equation*}
    dP = (2 \pi)^4 TV \delta^{(4)}\left( \sum_f p^{\mu}_f - \sum_i p^{\mu}_i \right) \frac{|\braopket{f}{\mathcal{M}}{i}|^2}{\left( \prod_{i} (2 E_i V)  \prod_{f} (2 E_f) \right)}  \prod_{f} \frac{1}{(2\pi)^3}~ d^3\vec{p}_f.
\end{equation*}
The expression is shortened by defining the Lorentz-invariant phase space $d\Pi_{LIPS}$
\begin{equation}
\label{eq:LorentzInvariantPhaseSpace}
    d\Pi_{LIPS} \equiv (2 \pi)^4 \delta^{(4)}\left( \sum_f p^{\mu}_f - \sum_i p^{\mu}_i \right) \prod_{f} \frac{1}{2 E_f} \frac{1}{(2\pi)^3}~ d^3\vec{p}_f,
\end{equation}
such that
\begin{equation}
\label{eq:diffprobability}
    dP = \frac{T V}{\prod_{i} (2 E_i V) } |\braopket{f}{\mathcal{M}}{i}|^2 d\Pi_{LIPS}.
\end{equation}

 Gathering the results from Eqs.~(\ref{eq:crosssection}) and (\ref{eq:diffprobability}), we find the differential cross-section $d\sigma$ which stands for the probability to experimentally observe the final state $\ket{f}$ in an element of the phase space $d\Pi$ from two colliding beams of particles that form the initial state $\ket{i}$ is determined by
\begin{equation}
    d\sigma = \frac{1}{(2 E_1)~ (2 E_2)} \frac{1}{| \vec{v}_1 - \vec{v}_2|} |\braopket{f}{\mathcal{M}}{i}|^2 d\Pi_{LIPS}.
\end{equation}
The stronger the interaction is, expressed as a larger matrix element, the larger the cross-section is if everything else remains the same. A larger cross-section means that events are more likely and experimentalists are able to observe the targeted final state more often. The total number of events $N_{ev}$ is now defined as 
\begin{equation}
    N_{ev} = \int d\sigma \int L(t) dt, 
\end{equation}
where $L(t)$ is the instantaneous luminosity. After being provided with the different parameters of the experiment, the task of theorists is reduced to computing the $\mathcal{S}$-matrix elements.

The $\mathcal{S}$-matrix describes the inner product of the initial and final states at some point in time $t$ by the right-hand side of Eq.(\ref{eq:Smatrixelements}). On the left-hand side of Eq.(\ref{eq:Smatrixelements}), the asymptotically free states are excitations of the vacuum $\ket{\Omega}$ such that, in a $n$ particles process $2 \to (n-2)$,  
\begin{equation}
\begin{split}
    \ket{i} & = 2 \sqrt{E_1 E_2}~ \hat{a}^{\dagger}_{p_1}(-\infty)~ \hat{a}^{\dagger}_{p_2}(-\infty) \ket{\Omega}, \\
    \ket{f} & = 2^{(n-2)/2} \sqrt{E_3 \ldots E_n}~ \hat{a}^{\dagger}_{p_3}(\infty) ~...~ \hat{a}^{\dagger}_{p_n}(\infty) \ket{\Omega}.
\end{split}
\end{equation}
Using the relation of the field $\phi_I(x)$ describing the particle $I$, 
\begin{equation}
    \sqrt{2 E_{p_I} } [ \hat{a}_{p_I}(\infty) - \hat{a}_{p_I}(-\infty) ] = i \int d^4x~ e^{i p_I x} (\square_I + m_I^2) \phi_I(x), 
\end{equation}
with $I=\{1, \ldots, n\}$ and the time-ordering operation $\mathbb{T}\{...\}$ to ensure the chronological order of particle creation (this way $\hat{a}_{p_I}(-\infty)$ operators are pushed to the right and $\hat{a}_{p_I}(+\infty)$ operators are pushed to the left)\footnote{The fields must vanish at the boundaries located at infinite values of time and space.}, the Lehmann-Symanzik-Zimmermann reduction formula allows one to express the S-matrix elements as 
\begin{multline}
    \braopket{f}{\mathcal{S}}{i} = \left[ i \int d^4x_1~ e^{-i p_1 x_1} (\square_1 + m_1^2)  \right] ... \left[ i \int d^4x_n~ e^{i p_n x_n} (\square_n + m_n^2)  \right] \\
    \braopket{\Omega}{\mathbb{T}\{\phi_1(x_1)...\phi_n(x_n)\}}{\Omega}.
\end{multline}

In the classical field theory, the equations of motion of particles are derived from Euler-Lagrange equations of the action $S$. It is defined as 
\begin{equation*}
    S = \int d^4 x~ \mathcal{L}, 
\end{equation*}
and the Euler-Lagrange equations follow from the variation of the action  
\begin{equation*}
    \delta S = 0 \to \partial_{\mu} \left( \frac{\delta}{\delta (\partial^\mu \phi_I)}~ \mathcal{L} \right) - \frac{\delta}{\delta \phi_I }~ \mathcal{L} = 0.
\end{equation*}
The Lagrangian density $\mathcal{L}$ can be separated in a kinetic term $\mathcal{L}_{0}$ and an interacting term $\mathcal{L}_{int.}$ with $\mathcal{L} = \mathcal{L}_{0} + \mathcal{L}_{int.}$. If $\phi_I$ represents a scalar for instance, then the equations of motion result in
\begin{equation}
	\label{eq:GeneralEulerLagrangeEquation}
    (\square_I + m_I^2) \phi_I  = \mathcal{L}'_{int.} ,
\end{equation}
where $\mathcal{L}'_{int.} = \delta \mathcal{L}_{int.}/\delta \phi_I $\footnote{Eq.(\ref{eq:GeneralEulerLagrangeEquation}) is valid as long as there are no interaction involving derivatives. See Section \ref{sec:WarsawBasisDerivation} for examples of interaction operators with covariant derivatives. }. However, due to the commutation relations inherent in QFT, notably $[ \phi(t, \vec{x}), \partial_t \phi(t, \vec{x'})] = i \delta^{(3)}(\vec{x}-\vec{x'}) $, the presence of the time-ordering operator gives an additional term. For example, 
\begin{multline*}
     (\square_1 + m_1^2) \braopket{\Omega}{\mathbb{T}\{\phi_1(x_1) \phi_2(x_2)\}}{\Omega}  \\ 
     = \braopket{\Omega}{\mathbb{T}\{ (\square_1 + m_1^2) \phi_1(x_1) \phi_2(x_2)\}}{\Omega} - i \delta^{(4)}(x_1-x_2) .
\end{multline*}
Generalising for the $n$ particle process, we get
\begin{multline}
\label{eq:SchwingerDysonEq}
    (\square_I + m_I^2) \braopket{\Omega}{\mathbb{T}\{ \phi_1 ~...~ \phi_I ~...~ \phi_n \}}{\Omega} = \braopket{\Omega}{\mathbb{T}\{ \phi_1 ~...~ \mathcal{L}'_{int.}[\phi_I] ~...~ \phi_n \}}{\Omega}  \\
        - i  \sum_{J \neq I} \delta^{(4)}(x_I - x_J)  \braopket{\Omega}{\mathbb{T}\{ \phi_1 ~...~ \phi_{J-1} \phi_{J+1} ~...~ \phi_n \}}{\Omega}.
\end{multline}
The second term in Eq.(\ref{eq:SchwingerDysonEq}) does not exist in classical field theory\footnote{The $\delta^{(4)}(\ldots)$ term results from the combination of $\delta (t-t')$ coming from time derivatives acting on time step functions $\Theta(t-t')$ introduced by the time-ordering operator $\mathbb{T}$ and $\delta^{(3)} (\vec{x}-\vec{x}')$ from the anti-commutation relations mentioned above.} and consists of contact interactions. The latter generate quantum effects not present in classical field theory, e.g. closed loop interactions.

We see that combinations of operators in the Lagrangian density give the matrix elements between the states which, in turn, allow one to compute the cross-section and the number of collisions when the beam characteristics are taken into account. Thus, to explain all interactions between particles, one has to write the most complete Lagrangian possible with all possible degrees of freedom, i.e. all fundamental particles, whose terms respect the symmetries of Nature. Ensuring that symmetries are respected is far from trivial, but this was the tremendous task employed to write the SM Lagrangian density $\mathcal{L}_{SM}$. Ultimately, when the predictions on the cross-sections are computed, we compare them with the data gathered in the particle detectors.

\section{Building the Standard Model}
\label{sec:SMLag}

\subsection{Naive Dimensional Analysis}
\label{subsec:SMNDA
}

We now determine the different types of operators that can be built in the SM Lagrangian with Naive Dimensional Analysis (NDA) of the action. These terms are not yet well-defined as we have not listed the degrees of freedom, nor the symmetries they must respect. These terms only display the type of interactions that are allowed to exist.

The action $S$ is a dimensionless quantity $[S]=0$ in natural units ($\hbar = c = 1$). Within natural units, the dimension of the mass parameter is set to 1 leading to the name \textit{mass dimension} and the dimension of $[dx]$ is 1. Thus, the SM Lagrangian density has a dimension $[\mathcal{L}_{SM}]=4$. All generic Lagrangian terms with a dimension up to 4 can be constructed for all types of particles. Note that some Lagrangian terms will be forbidden by the symmetries of the SM.

First, we derive the dimension of the fields from the free theory containing only $\mathcal{L}_0$. For the real scalar field $\phi$, the kinetic terms are dimensionally consistent if
\begin{equation}
    4 = \left[ \frac{1}{2} (\partial_\mu \phi)^\dagger \partial^\mu \phi \right] = 2 + 2 [\phi] \to [\phi] = 1.
\end{equation}
The dimension of the scalar field is further verified by its mass term 
\begin{equation}
\label{eq:MassDimensionOfMass}
    4 = \left[ \frac{1}{2} m^2 \phi^2 \right] = 2[m] + 2 \to [m] = 1.
\end{equation}
The kinetic term of the vector field $A_\mu$ yields a mass dimension of 1 for $A_\mu$
\begin{equation}
    4 = \left[ F_{\mu\nu} F^{\mu\nu} \right] \to 2 =\left[ F_{\mu\nu} \right] = \left[ \partial_\mu A_\nu + \partial_\nu A_\mu \right] \to \left[ A_\mu \right] = 1.
\end{equation}
For the fermionic field $\psi$, the kinetic and the mass terms respectively imply
\begin{equation}
\begin{split}
    & 4 = \left[ \overline{\psi} \partial^\mu \psi \right] = 1 + 2 [\psi] \to [\psi] = 3/2, \\
    & \left[ m \overline{\psi} \psi \right] = 1 + 2.(3/2)= 4.
\end{split}
\end{equation}

Now for the interactions in $\mathcal{L}_{int.}$, the scalar bosons can interact with themselves through self-interactions 
\begin{equation*}
\begin{split}
    \left[ \phi \phi \phi \right] & = 3, \\
    \left[ \phi \phi \phi \phi \right] & = 4.
\end{split}
\end{equation*}
An interaction term with three scalars is dimensionally possible (but it would not respect gauge invariance as we will see). 
Similarly, vector fields allows for self-interactions
\begin{equation*}
	\begin{split}
		\left[ A_\mu A_\nu A_\rho \right] & = 3, \\
		\left[ A_\mu A_\nu A_\rho A_\sigma \right] & = 4.
	\end{split}
\end{equation*}
Contractions of Lorentz indices forbid self-interactions between three vector fields due to one free index and Lorentz invariance determines the indices of the four fields. However, indices with three vector fields can be correctly contracted with the addition of one derivative in the operator
\begin{equation*}
	\left[ \partial_\mu A_\nu A_\rho A_\sigma \right] = 4.
\end{equation*} 
Now the interactions between different types of particles are considered. We have the case of two vector bosons $A_\mu$ and two scalar bosons $\phi$
\begin{equation*}
    \left[ (A_\mu \phi)^\dagger (A^\mu \phi) \right] = 4,
\end{equation*}
or one vector boson and 2 scalars with a derivative to contract the Lorentz indices
\begin{equation*}
	\left[ (\phi^\dagger \partial_\mu  \phi) A^\mu \right] = 4.
\end{equation*}
There can be an interaction between a vector boson $A_\mu$ and two fermions $\psi$
\begin{equation*}
    \left[ \overline{\psi} A_\mu \psi \right] = 4.
\end{equation*}
The vector boson can be replaced by a scalar boson $\phi$
\begin{equation*}
    \left[ \overline{\psi} \phi \psi \right] = 4.
\end{equation*}

\subsection{Gauge Symmetries}
\label{subsec:SMGaugeSymmetries}

Nature has indicated so far that the SM elementary states transform under the gauge group $G=SU_{C}(3) \times SU_{L}(2) \times U_{Y}(1)$. Each subgroup $G'$ is described by its Lie algebra which is determined by the generators $\{t^{\alpha}(G')\}_{\alpha}$. One vector gauge boson $X_{\mu}^{\alpha}(G')$ is associated to each generator. The generators of non-Abelian Lie algebras are taken in the adjoint representation of their respective gauge group in the SM. The non-Abelian $SU_{C}(3)$ gauge group has eight generators $T^{A}=\frac{1}{2} \lambda^{A}$ with $\lambda^{A}$ being the eight traceless Gell-Mann matrices and $A=\{1, \ldots, 8\}$. There are thus eight gluons $X_{\mu}^{\alpha}(SU_C(3)) = G^{A}_{\mu}$ responsible for the strong interaction. The non-Abelian $SU_{L}(2)$ gauge group has three generators $\tau^{i} = \frac{1}{2} \sigma^{i}$ with $\sigma^{i}$ being the three traceless Pauli matrices and $i= \{1, 2, 3 \}$. The three weak bosons $X_{\mu}^{\alpha}(SU_L(2))  = W^{i}_{\mu}$ carry the weak force. Finally, the generator of the Abelian $U_Y(1)$ is the hypercharge number $Y$, so \sout{the Abelian} $U_{Y}(1)$ has one neutral vector boson $X_{\mu}^{\alpha}(U_{Y}(1)) = B_{\mu}$.  

The generators of the two non-Abelian groups satisfy the following commutation relations defining the structure constants,
\begin{equation}
\label{eq:SMstructureconstants}
    \left[ T^{A}, T^{B} \right] = i f^{ABC} T^{C}, ~~ \left[ \tau^{i}, \tau^{j} \right] = i \epsilon^{ijk} \tau^{k}.
\end{equation}
The numbers $f^{ABC}$ and $\epsilon^{ijk}$ are called the structure constants of the Lie algebra associated to the corresponding non-Abelian group. The normalisation of the generators defined above with their respective matrices forces them to satisfy the following trace conditions
\begin{equation}
    \text{Tr}\left[ T^{A} T^{B} \right] = \frac{1}{2} \delta^{AB}, ~~ \text{Tr}\left[ \tau^{i} \tau^{j} \right] = \frac{1}{2} \delta^{ij}.
\end{equation}
The general covariant derivative $D_{\mu}$ is formed by the combination of the partial derivative and the vector gauge bosons,
\begin{equation}
\label{eq:SMGeneralCovariantDerivative}
    D_{\mu} = \partial_\mu + i g_{s} T^{A} G^{A}_{\mu} + i g_{2} \tau^{i} W^{i}_{\mu} + i g_{1} \frac{Y}{2} B_{\mu}.
\end{equation}
The parameters $g_s$, $g_2$ and $g_1$ are the respective coupling constants of the gauge groups $SU_C(3)$, $SU_L(2)$ and $U_Y(1)$. We adopt the "slashed" convention contracting sums involving the four Dirac matrices $\gamma^\mu$, $\mu= \{0,1,2,3\}$, and 4-vectors such as $ \gamma^{\mu} D_{\mu} \equiv \slashed{D}, \gamma^{\mu} \partial_{\mu} \equiv \slashed{\partial}, ...$

The field strength tensors of the gauge bosons are defined by the commutator of covariant derivatives $\left[ D_{\mu}, D_{\nu} \right] = i g X^{\alpha}_{\mu \nu}(G')~ t^{\alpha}(G')$ where $D_\mu$ is restricted to one gauge term in Eq.(\ref{eq:SMGeneralCovariantDerivative}) and $t^{\alpha}(G')$ are the generators of the non-Abelian groups $SU_C(3)$ or $SU_L(2)$. The commutation relations of covariant derivatives are independent of the representation of the generators $t^{\alpha}(G')$. Therefore, the three field strength tensors are given by
\begin{eqnarray}
\label{eq:GluonFieldStrengthTensor}
    G^{A}_{\mu \nu} & = & \partial_\mu G^{A}_{\nu} - \partial_\nu G^{A}_{\mu} - g_s f^{ABC} G^{B}_{\mu} G^{C}_{\nu}, \\
    W^{i}_{\mu \nu} & = & \partial_\mu W^{i}_{\nu} - \partial_\nu W^{i}_{\mu} - g_2 \epsilon^{ijk} W^{j}_{\mu} W^{k}_{\nu}, \\
    B_{\mu \nu} & = & \partial_\mu B_{\nu} - \partial_\nu B_{\mu}.
\end{eqnarray} 
The dual tensor field $\tilde{X}_{\mu\nu}$ is the product of $X_{\mu \nu}$ with the totally antisymmetric Levi-Civita tensor $\epsilon^{\mu\nu \rho\sigma}$,
\begin{equation}
    \tilde{X}_{\mu\nu} = \frac{1}{2} \epsilon_{\mu\nu \rho\sigma} X^{\rho \sigma}.
\end{equation}
The Levi-Civita tensor values follow from its fully antisymmetric property and $\epsilon^{0123}=+1$.
The SM Lagrangian terms with only the vector gauge bosons are
\begin{equation}
\label{eq:SMLagGauge}
\mathcal{L}_{gauge} = G^{A}_{\mu \nu} G^{A~ \mu \nu} + W^{i}_{\mu \nu} W^{i~ \mu \nu} + B_{\mu \nu} B^{\mu \nu} .
\end{equation}

One possible pure-gauge term involves the dual tensor of the gluon tensor with a parameter $\theta$ 
\begin{equation*}
	\mathcal{L}_{\theta} = \theta \frac{g_s^2}{32 \pi^2} G^A_{\mu \nu} \widetilde{G}^{A~ \mu \nu}.
\end{equation*}
However, this term does not contribute to the dynamics of the gluon in a perturbative expansion since it is equivalent to a total derivative. It can be proven that $\mathcal{L}_{\theta}$ violates $CP$ and generates an electric dipole moment (EDM) for mesons and hadrons, in particular a neutron EDM $d_n$ whose experimental value is very small $|d_n| < 1.8~10^{-26} e$cm \cite{Abel:2020pzs}, which in turn gives $\overline{\theta}<10^{-10}$ \cite{Ai:2026ovp} with $\overline{\theta}= \theta + arg~ det~ M$ and $M$ the mass matrix of the three lightest quarks defined in the following Section~\ref{subsec:SMDoF}. As a result, we neglect $\mathcal{L}_{\theta}$ in $\mathcal{L}_{SM}$.

\subsection{Degrees of freedom}
\label{subsec:SMDoF}

Particles are the degrees of freedom in QFT. The SM degrees of freedom will be listed by three corresponding quantum numbers : their colour $C$, their weak isospin $\pm1/2$ and their hypercharge $Y$. Since the SM is a chiral theory, left-handed and right-handed particles are independent fields carrying different charges of the gauge subgroups. The SM particles are separated in two groups called leptons and quarks with three generations in each species. The quantum number associated to the generation is called the flavour.

For the leptons, the left-handed lepton field $L_L$ is associated to a single right-handed lepton field $l_R$. There are no right-handed neutrino $\nu_R$ implying they remain massless in the SM as we will see below. The leptonic fields transform as singlets under $SU_C(3)$, so they do not carry a colour charge. However, $L_L$ is a $SU_L(2)$ doublet so it has two components $\nu_L$ and $l_L$ with non-zero weak isospin, respectively $1/2$ and $-1/2$. Both components have a hypercharge of $-1$. The corresponding quantum numbers of the unique right-handed lepton field $l_R$ are $(0, -2)$. The flavours of the lepton generations are respectively called the electron, the muon and the tau. For the neutrinos, they are the electron neutrino, the muon neutrino and the tau neutrino. The lepton fields are thus written as 
\begin{equation}
\label{eq:LeptonFields}
    L_L = 
        \begin{pmatrix}
        \nu_L \\
        l_L
        \end{pmatrix}
        = \left\{
        \begin{pmatrix}
        \nu_{eL} \\
        e_L
        \end{pmatrix}
        ,
        \begin{pmatrix}
        \nu_{\mu L} \\
        \mu_L
        \end{pmatrix}
        ,
        \begin{pmatrix}
        \nu_{\tau L} \\
        \tau_L
        \end{pmatrix}
        \right\},
    ~~
    l_R = \left\{ e_R, \mu_R, \tau_R \right\}.
\end{equation}

Now for the quarks, the left-handed field $Q_L$ is associated to two right-handed fields $u_R$ and $d_R$. All three transform as triplets under $SU_C(3)$, so they carry a colour charge $(C=1)$ and are thus connected to the gluon field. The two components $U_L$ and $D_L$ of the $SU_L(2)$ doublet $Q_L$ have a non-zero weak isospin charge of $\pm 1/2$, like left-handed leptons. As a result, their quantum numbers are $(1/2, 1/3)$ and $(-1/2, 1/3)$ respectively. The right-handed up-quark field $u_R$ has $(0, 4/3)$ and the right-handed down-quark field $d_R$ has $(0, -2/3)$. The flavours of the up-type quark field components are the up, the charm and the top quarks. For the down-type quark components, we call them the down, the strange and the bottom quarks. We collectively write the quark fields as
\begin{equation}
\label{eq:QuarkFields}
\begin{split}
    Q_L &= 
        \begin{pmatrix}
        U_L \\
        D_L
        \end{pmatrix}
        = \left\{
        \begin{pmatrix}
        u_{L} \\
        d_L
        \end{pmatrix}
        ,
        \begin{pmatrix}
        c_{L} \\
        s_L
        \end{pmatrix}
        ,
        \begin{pmatrix}
        t_{L} \\
        b_L
        \end{pmatrix}
        \right\}, \\
    u_R &= \left\{ u_R, c_R, t_R \right\},
    ~~~~
    d_R = \left\{ d_R, s_R, b_R \right\}.
\end{split}
\end{equation}

\begin{table}
\centering
\begin{tabular}{|c||cccccccc|}
\hline
    Fields & $l_L$ & $\nu_L$ & $l_R$ & $U_L$ & $D_L$ & $u_R$ & $d_R$ & $h$ \\
\hline
    $SU_C(3)$ & $1$ & $1$ & $1$ & $3$ & $3$ & $3$ & $3$ & $1$ \\
    $SU_L(2)$ & $2$ & $1$ & $1$ & $2$ & $2$ & $1$ & $1$ & $1$ \\
\hline
\hline
    Fields & $l_L$ & $\nu_L$ & $l_R$ & $U_L$ & $D_L$ & $u_R$ & $d_R$ & $h$ \\
\hline
    %C & $0$ & $0$ & $0$ & $1$ & $1$ & $1$ & $1$ & $0$ \\
    $\tau^3$ & $-1/2$ & $1/2$ & $0$ & $1/2$ & $-1/2$ & $0$ & $0$ & $-1/2$ \\
    $Y$ & $-1$ & $-1$ & $-2$ & $1/3$ & $1/3$ & $4/3$ & $-2/3$ & $1$ \\
\hdashline
    $Q$ & $-1$ & $0$ & $-1$ & $2/3$ & $-1/3$ & $2/3$ & $-1/3$ & $0$ \\
\hline
\end{tabular}
\caption{Summary of the representations (above) and the quantum numbers (below) of the fermion and Higgs fields in the SM. The last line is the electric charge $Q = \tau^3 + \frac{1}{2} Y$.}
\label{tab:SMQuantumNumbers}
\end{table}

The free kinetic terms of the fermionic particles are merged with their interaction to the gauge bosons by minimal coupling presented in Eq.~(\ref{eq:SMGeneralCovariantDerivative}) such that they are invariant under gauge symmetries,
\begin{equation}
\label{eq:SMLagKinetic}
\begin{split}
    \mathcal{L}_{kin.} & = i \overline{\psi} \slashed{D} \psi, \\
        & = i \overline{L}_L \slashed{D} L_L + i \overline{l}_R \slashed{D} l_R +  i \overline{Q}_L \slashed{D} Q_L + i \overline{u}_R \slashed{D} u_R + i \overline{d}_R \slashed{D} d_R.
\end{split}
\end{equation}

We cannot write mass terms of the fermion fields like $m \overline{\psi}_L \psi_R$ since they would not respect $SU_L(2) \times U_Y(1)$ gauge invariance.

Elaborating on the subject of particle masses, the photon and the gluon are massless particles, as required by gauge invariance, so their mass term can simply be discarded. However, the weak bosons are massive while $m_{W}^2 W^I_{\mu} W^{I~\mu}$ does not respect the $SU_L(2)$ gauge symmetry either. At first glance, it seems there is an issue with the mass of particles but they will actually emerge from the ingenious BEH mechanism\footnote{A more correct name, more inclusive, should actually be the Anderson-Brout-Englert-Ginzburg-Guralnik-Hagen-Higgs-Kibble-Landau mechanism.}.

\subsection{The Higgs Scalar and the BEH Mechanism}
\label{subsec:SMHiggs}

Before explaining how particles get their mass through the BEH mechanism, let us present the Higgs field and its Lagrangian terms. The Higgs field $\varphi$ is a complex scalar field transforming as a doublet of $SU_L(2)$ with a hypercharge of 1. The Higgs Lagrangian $\mathcal{L}_{Higgs}$ is composed of its kinetic term and its potential $V(\varphi)$,
\begin{equation}
\label{eq:HiggsLag}
    \mathcal{L}_{Higgs} = (D_{\mu} \varphi)^{\dagger} (D^{\mu} \varphi) - V(\varphi).
\end{equation}
We have already generated possible Lagrangian terms in the form of a mass-type $m_\varphi^2 \phi^2$ and a $\phi^4$-type self-interaction terms. The $\phi^3$ self-interactions term was possible by NDA but breaks $SU_L(2)$ invariance in practice. So, the Higgs potential is made of two possible terms, 
\begin{equation}
    V(\varphi) = m_{\varphi}^2 \left( \varphi^\dagger \varphi \right) + \lambda \left( \varphi^\dagger \varphi \right)^2.
\end{equation}
The dimensionless positive-definite parameter $\lambda$ is the strength of the Higgs self-interactions.

\begin{figure}
    \centering
    \includegraphics[scale=1.05]{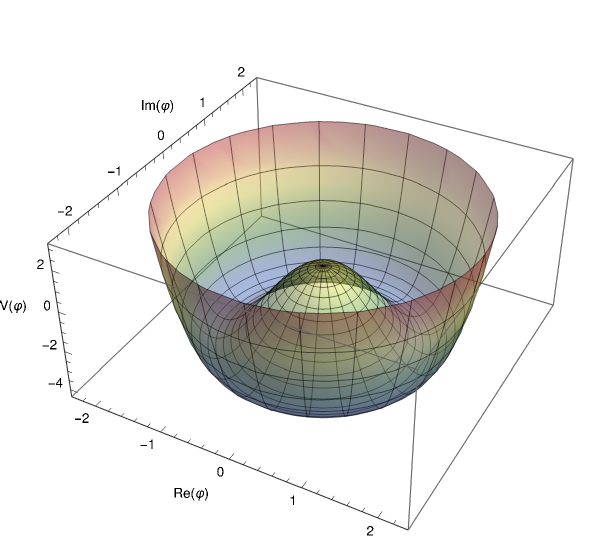}
    \includegraphics[scale=1.10]{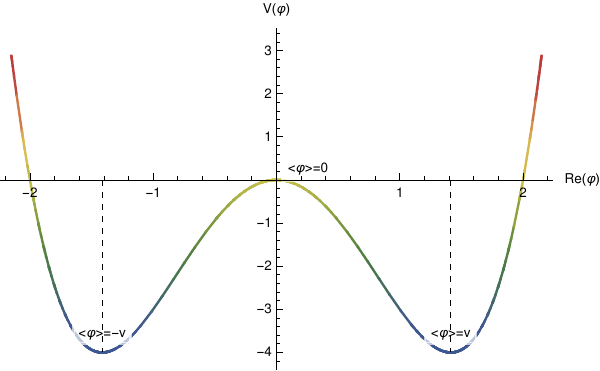}
    \caption{Plot of the Higgs potential. Left: 3D view of the Higgs potential $V(\varphi)$ with respect to the real and imaginary parts of the Higgs scalar boson. Right: 2D profile view of the Higgs potential $V(\varphi)$ with respect to the real part of the Higgs scalar boson for $Im(\varphi)=0$. The $SU_L(2)$ vacuum sits at the local maximum with $\langle \varphi \rangle = 0$. The true vacuum is the circle at the bottom of the 3D plot $\langle \varphi \rangle = v$, corresponding to the two minima on the profile view $\langle \varphi \rangle = \pm v$, that is $\theta_{\varphi}=\pm \pi$.}
    \label{fig:HiggsPotential}
\end{figure}

In order to find the stable minimum $\varphi_0$ from the potential $V(\varphi)$, we cancel the variation of the potential with respect to the Higgs field,
\begin{equation}
    0 = \left\langle \frac{\delta V(\varphi)}{\delta \varphi^{\dagger}} \right\rangle \Big|_{\varphi=\varphi_0} \to 0 = \langle\varphi_0\rangle \left( m_{\varphi}^2 + 2 \lambda   \langle \varphi_{0}^\dagger \varphi_0 \rangle \right). 
\end{equation}
If $m_{\varphi}^2>0$, then $\varphi_0 = (0~~ 0)^T$ is the global minimum of the potential. However, if $m_{\varphi}^2<0$, we note that the solution $\varphi_0 = (0~~ 0)^T$ becomes a local maximum. There is now a degenerate set of solutions for the local minima $\varphi_0 = e^{i\theta_{\varphi}} (0~~ v/\sqrt{2})^T$ with $\theta_{\varphi}$ the $SU_L(2) \times U_Y(1)$ phase and $v \equiv \sqrt{- m_{\varphi}^2 / \lambda}$. The parameter $v$ is called the expectation value of the Higgs field in the vacuum. Figure~\ref{fig:HiggsPotential} displays the Higgs potential as a function of the real and imaginary part of the Higgs boson. For clarity, a profile view of the potential along the plane $Im(\varphi)=0$, i.e. $\theta_{\varphi}=\pm \pi$, is also provided. The solution $\varphi_0 = (0~~ 0)^T$ is symmetric under $SU_L(2) \times U_Y(1)$ but is an unstable configuration of the Higgs field. As a result, the Higgs field is said to "fall" at the bottom of the potential in one of the degenerate solutions to stabilise (fixing the value of $\theta_{\varphi}$ in the process). Despite the Higgs vacuum expectation value not being symmetric under $SU_L(2) \times U_Y(1)$, the Lagrangian terms remain nonetheless symmetric at high energies. We label the $SU_L(2) \times U_Y(1)$ electroweak symmetry as \textit{spontaneously broken} and are left with the electromagnetic $U_{EM}(1)$ gauge symmetry,
\begin{equation}
    SU_L(2) \times U_Y(1) \xrightarrow{EWSSB} U_{EM}(1).
\end{equation}

We can write without loss of generality the vacuum solution of the Higgs field $\varphi_0$ as
\begin{equation}
    \varphi_0 = 
        \begin{pmatrix}
        0 \\
        v/\sqrt{2}
        \end{pmatrix},
\end{equation}
and choose a particular parametrisation of the Higgs field where the Higgs particle corresponds to a radial excitation from the minimum $\varphi_0$,
\begin{equation}
\label{eq:HiggsRepresentation}
    \varphi = e^{i \frac{\pi^i \tau^i}{v} }
        \begin{pmatrix}
        0 \\
        (v+h)/\sqrt{2}
        \end{pmatrix}.
\end{equation}
By further defining the Higgs mass $m_h$ with $m_{h}^2 \equiv -2 m_{\varphi}^2$, the potential is expressed only in terms of the real component $h$
\begin{equation}
    V(\varphi) = - \frac{m_h^4}{16 \lambda} + \frac{1}{2} m_h^2 h^2 + \lambda v h^3 + \frac{1}{4} \lambda h^4.
\end{equation}
The first term is a constant and does not affect the dynamics of the Higgs, so it is usually neglected.

The three $\pi^i$ fields are the Nambu-Goldstone bosons. After the EWSSB, the Nambu-Goldstone bosons are absorbed by the massless gauge fields $W^i$ to act as their longitudinal component on top of their natural transverse degrees of freedom. The weak bosons can now be written as massive bosons, as needed.

We see that our parametrisation of the Higgs field in Eq.~(\ref{eq:HiggsRepresentation}) is invariant under transformations related to the generator $T = \tau^3 + \frac{Y}{2} \mathbb{1}_2$ with $\mathbb{1}_2$ the $2\times2$ identity matrix. The gauge boson associated with the unbroken generator $T$ remains massless and corresponds to the photon. Thus, the charge $Q$ of the generator $T$ is the quantum number associated with the electric charge, listed on the last line of Table~\ref{tab:SMQuantumNumbers}.

A linear combination of the $\tau^{1,2}$ generators simplifies the commutation relation between them and $T$. By defining $\tau^{\pm} = \tau^{1} \pm i \tau^{2}$, then
\begin{equation}
    [T, \tau^{\pm}] = [\tau^3, \tau^{\pm}] = \pm \tau^\pm.
\end{equation}
The factor in front of $\tau^\pm$ suggests that the electric charge of the bosons associated to these generators are $\pm 1$. So, we define the charged massive weak bosons
\begin{equation}
\label{eq:ChangeOfChargedFields}
    W_\mu^{\pm} \equiv \frac{1}{\sqrt{2}} ( W^1_\mu \mp i W^{2}_\mu).
\end{equation}

With a $SU_L(2)$ phase rotation, amounting to fixing the gauge, the Nambu-Goldstone bosons are  gauged away and $\varphi = \frac{v+h}{\sqrt{2}} \begin{pmatrix} 0 \\ 1 \end{pmatrix}$. This particular choice is called the unitary gauge. We can look at the Lagrangian terms of Eq.(\ref{eq:HiggsLag}) free of $h$ but proportional to $v$
\begin{equation}
\label{eq:NonDiagBosonMass}
    (D_{\mu} \varphi)^{\dagger} (D^{\mu} \varphi) \supset \frac{g_2^2 v^2}{8} \left( (W_\mu^1)^2 + (W_\mu^2)^2 + \left( \frac{g_1}{g_2} B_\mu - W_\mu^3 \right)^2 \right).
\end{equation}
The last squared term mixes $B_\mu$ and $W^3_{\mu}$, so we apply a rotation of $B_\mu$ and $W^3_{\mu}$ to get canonical mass terms. Let us define the weak angle $\theta_W$ such that 
\begin{equation}
    \tan \theta_W =  \frac{g_1}{g_2}.
\end{equation}
Then, we find the expression of the fields associated to the neutral gauge bosons distinguished by their masses : the massive field $Z_\mu$ and the massless photon field $A_\mu$,
\begin{equation}
\label{eq:ChangeOfNeutralFields}
\begin{cases}
    Z_\mu \equiv \cos{\theta_W} W^3_\mu - \sin{\theta_W} B_\mu, \\
    A_\mu \equiv \sin{\theta_W} W^3_\mu + \cos{\theta_W} B_\mu.
\end{cases}
\end{equation}
The rotation of the fields means that the electromagnetic constant $e$, determined by the covariant derivative now expressed with $Z_\mu$ and $A_\mu$, is set to
\begin{equation*}
    e = g_1 \cos{\theta_W}  = g_2 \sin{\theta_W},
\end{equation*}
due to the second line of Eq.(\ref{eq:ChangeOfNeutralFields}). Now, the kinetic term of the Higgs Lagrangian becomes 
\begin{multline}
    (D_{\mu} \varphi)^{\dagger} (D^{\mu} \varphi) = \frac{1}{2} \partial_\mu h \partial^\mu h + m_W^2 W^+_\mu W^{-~ \mu} + \frac{1}{2} m_Z^2 Z_\mu Z^{\mu} \\ 
    + \frac{1}{8} (2vh + h^2)(2g_2^2 W^+_\mu W^{-~ \mu} + (g_1^2 + g_2^2) Z_\mu Z^{\mu}).
\end{multline}
The $Z$ boson mass $m_Z$ and the $W^{\pm}$ boson mass $m_W$ are respectively expressed by $\frac{1}{2} v\sqrt{g_1^2 + g_2^2}$ and $\frac{1}{2} g_2 v$.

\subsection{Yukawa Interactions and Fermion Masses}
\label{subsec:SMYukawa}

The interactions between the fermion fields and the Higgs field are determined by the Yukawa Lagrangian $\mathcal{L}_{Yuk.}$,
\begin{equation}
\label{eq:SMlagYukawa}
\begin{split}
    \mathcal{L}_{Yuk.} & = - (Y_\psi)_{ab} \overline{\psi}_a \varphi \psi'_b + h.c. \\ 
        & = -(Y_u)_{ab} \overline{Q}_{La} \tilde{\varphi} u_{Rb} -(Y_d)_{ab} \overline{Q}_{La} \varphi d_{Rb} - (Y_e)_{ab} \overline{L}_{La} \varphi e_{Rb} + h.c. 
\end{split}
\end{equation}
where $\tilde{\varphi} = i \sigma^2 \varphi^{*}$. The Yukawa matrices $Y_\psi$ are complex $3\times 3$ matrices carrying generation indices $a$ and $b$. Each term is invariant under the whole SM gauge symmetry group. The mass terms of the fermions arise when we insert the parametrisation of the Higgs field written in the unitary gauge in Eq~(\ref{eq:HiggsRepresentation}). Contributions proportional to $v$ define the complex mass matrices as $M_{u,d,e} \equiv vY_{u,d,e}/\sqrt{2}$ such that
\begin{equation}
\label{eq:SMMassMatrix}
\begin{split}
    \mathcal{L}_{Yuk.} & = -(M_u)_{ab} \overline{U}_{La} u_{Rb} -(M_d)_{ab} \overline{D}_{La} d_{Rb}  - (M_e)_{ab} \overline{l}_{La} e_{Rb} + h.c. \\
        & -(Y_u)_{ab} \ h \ \overline{U}_{La} u_{Rb} -(Y_d)_{ab} \ h \ \overline{D}_{La} d_{Rb}  - (Y_e)_{ab} \ h \ \overline{l}_{La} e_{Rb} + h.c.
\end{split}
\end{equation}

We want to replace the complex quark mass matrices $M_\psi$ with a diagonal mass matrix $M'_\psi = diag(m_{\psi_1}, m_{\psi_2}, m_{\psi_3}) \equiv v Y'_\psi/\sqrt{2} $ by using unitary matrices to rotate the fields such that, for the down-type quark fields, we have
\begin{equation}
	\label{eq:DownQuarksRotationForCP}
	\begin{cases}
		D_L \to D'_L = J_d D_L, \\
		d_R \to d'_R = K^\dagger_d d_R,
	\end{cases}
\end{equation}
while for the up-type quark fields,
\begin{equation}
	\label{eq:UpQuarksRotationForCP}
	\begin{cases}
		U_L \to U'_L = J_u U_L, \\
		u_R \to u'_R = K^\dagger_u u_R.
	\end{cases}
\end{equation}
Thus, the Yukawa matrices $Y_\psi$ can be written as
\begin{equation}
	\label{eq:YukawaMatricesDiag}
	Y_\psi = J_\psi Y'_\psi K^\dagger_\psi.
\end{equation}

Since $Y_\psi$ are complex, the product with their hermitian conjugate is hermitian and its eigenvalues real. These real eigenvalues are proportional to the squared masses of the fermions. Using the rotation matrices of the left-handed fields $J_\psi$, 
\begin{equation}
    Y_d Y_d^\dagger \propto J_d M'^2_d J^\dagger_d, ~~ Y_u Y_u^\dagger \propto J_u M'^2_u J^\dagger_u.
\end{equation}
and using the rotation matrices of the right-handed fields $K_\psi$, 
\begin{equation}
    Y_d^\dagger Y_d \propto K_d M'^2_d K^\dagger_d, ~~ Y_u^\dagger Y_u \propto K_u M'^2_u K^\dagger_u.
\end{equation}
The mass terms of the quarks in the Yukawa Lagrangian become
\begin{equation}
    \mathcal{L}_{Yuk.} \supset - \overline{D}_L J_d M'_d K^\dagger_d d_R - \overline{U}_L J_u M'_u K^\dagger_u u_R + h.c.
\end{equation}
The mass terms of the leptons are similarly derived. We say that we move from the interaction basis to the mass basis as parts of the Yukawa terms become canonical mass terms for the quarks,
\begin{equation}
\label{eq:DiagonalFermionMassTerm}
    \mathcal{L}_{Yuk.} = - M'_d \overline{D'}_L d'_R - M'_u \overline{U'}_L u'_R - M'_l \overline{l'}_L e'_R + h.c.
\end{equation}

\begin{comment}
There are still 6 $U(1)$ global symmetries associated to the rephasing of the fields leaving the mass terms invariant. We have 3 angles $\alpha_a$ to rephase the down-type quark fields, one for each generation, 
\begin{equation}
\begin{cases}
    d'_L \to e^{i\alpha_a} d'_L , \\
    d'_R \to e^{i\alpha_a} d'_R ,
\end{cases}
\end{equation}
while for the up-type quark fields, there are 3 angles $\beta_a$ such that
\begin{equation}
\begin{cases}
    u'_L \to e^{i\beta_a} u'_L , \\
    u'_R \to e^{i\beta_a} u'_R .
\end{cases}
\end{equation}
\end{comment}

Kinetic terms of the right-handed quark fields remain unaffected by the rotation from the interaction basis to the mass basis. However, that is not the case for the left-handed quark fields $Q_L$. As they are charged under $SU_L(2)$, the off-diagonal contributions in the covariant derivative proportional to $W^i_\mu \tau^i$ (corresponding to the charged bosons $W^\pm_\mu$) mix the fields. The mixing Lagrangian $\mathcal{L}_{mix.}$ reads as
\begin{multline}
    \mathcal{L}_{kin.} \supset \mathcal{L}_{mix} = \frac{e}{\sqrt{2}\sin{\theta_W}} \Big[ W^+_\mu \overline{u'}_{La} \gamma^\mu (V_{CKM})_{ab}~ d'_{L b} \\ + W^-_\mu \overline{d'}_{L a} \gamma^\mu (V^\dagger_{CKM})_{ab}~ u'_{L b}   \Big],
\end{multline}
where the matrix $V_{CKM} \equiv J^{\dagger}_u J_d$ is called the Cabibbo-Kobayashi-Maskawa (CKM) matrix\cite{PhysRevLett.10.531,Kobayashi:1973fv}. Since $V_{CKM}$ is a complex unitary matrix, it is described by 9 degrees of freedom (3 angles and 6 complex phases). Using $U(1)$ global flavour symmetries from the quark field rephasing, 5 of the 6 complex phases are absorbed and we reduce the degrees of freedom to 3 angles and 1 phase, respectively $\theta_{12}$, $\theta_{13}$, $\theta_{23}$ and $\delta_{CKM}$. The remaining complex phase $\delta_{CKM}$ cannot be absorbed in the quark rephasings because $V_{CKM}$ is unaffected when all transformations are identical, $d'_L \to e^{i\alpha} d'_L$ and $u'_L \to e^{i\alpha} u'_L$. We can write $V_{CKM}$ in the standard parameterisation as
\begin{multline}
\label{eq:StandardCKMmatrix}
    V_{CKM} = \\ \begin{pmatrix}
               c_{12}c_{13} & s_{12}c_{13} & s_{13}e^{-i\delta_{CKM}} \\
-s_{12}c_{23}-c_{12}s_{23}s_{13}e^{i\delta_{CKM}} & c_{12}c_{23}-s_{12}s_{23}s_{13}e^{i\delta_{CKM}} & s_{23}c_{13} \\
s_{12}s_{23}-c_{12}c_{23}s_{13}e^{i\delta_{CKM}} & -c_{12}s_{23}-s_{12}c_{23}s_{13}e^{i\delta_{CKM}} & c_{23}c_{13}
              \end{pmatrix}
\end{multline}
where $c_{ij}= \cos{\theta_{ij}}$ and $s_{ij}=\sin{\theta_{ij}}$. The angles are in the first quadrant by convention, that is $s_{ij},c_{ij} \geq 0$.

\subsection{The SM Lagrangian}
\label{subsec:SMLag}

Finally, the SM Lagrangian is the sum of the four Lagrangians introduced in Subsections~\ref{subsec:SMGaugeSymmetries} to \ref{subsec:SMYukawa}
\begin{equation}
\label{eq:SMLag}
    \mathcal{L}_{SM} = \mathcal{L}_{gauge} + \mathcal{L}_{kin.} + \mathcal{L}_{Higgs} + \mathcal{L}_{Yuk.}
\end{equation}
$\mathcal{L}_{SM}$ should include the Lagrangian of the Faddeev-Popov ghosts $\mathcal{L}_{gh.}$ to be complete. However, ghosts are not presented here because they are unphysical states and are not observable in colliders. Their purpose in reconciling Lorentz invariance of $\mathcal{L}_{SM}$ and unitarity in the presence of massless vector bosons is important albeit only mathematical. The reader is referred to standard QFT textbooks \cite{Schwartz:2014sze,Peskin:1995ev} for full explanations on their necessity, definition and derivation.

\section{The Fundamental Arrow of Time}
\label{sec:CPsymmetries}

In addition to local gauge symmetries reviewed in Subsection~\ref{subsec:SMGaugeSymmetries}, global symmetries of charge $C$, parity $P$ and time $T$ need to be addressed. In particular, the violation of $CP$ symmetry has fundamental implications for the Universe and its time evolution. Any local QFT must respect $CPT$ invariance. $CPT$ invariance is strongly supported by the equality of the respective masses of particles and their antiparticles, and their respective lifetimes if they decay. This means that observing $CP$ violation ultimately shows that fundamental interactions violate $T$ symmetry, i.e. Nature possesses a natural arrow of time. This natural arrow of time is manifested by the observed dominance of matter over antimatter which will be discussed in the next Section.

Classical relativistic mechanics dictates the first transformation rules under the $P$ and $T$ symmetries. The rest comes with the invariance of QED under the $C$, $P$ and $T$ symmetries. Thanks to a small subset of these rules, we can see that all terms of the SM are invariant except terms involving products of up- and down-type quarks. The relevant rules are the following transformation relations of fermionic bilinears under $CP$
\begin{equation}
\begin{split}
    \overline{\psi}_i \psi_j & \xrightarrow{CP} e^{i(\xi_j - \xi_i)} \overline{\psi}_j \psi_i, \\
    \overline{\psi}_i \gamma_5 \psi_j & \xrightarrow{CP} - e^{i(\xi_j - \xi_i)} \overline{\psi}_j \gamma_5 \psi_i, \\
    \overline{\psi}_i \slashed{X} \psi_j & \xrightarrow{CP} e^{i(\xi_X + \xi_j - \xi_i)} \overline{\psi}_j \slashed{X} \psi_i, \\
    \overline{\psi}_i \slashed{X} \gamma_5 \psi_j & \xrightarrow{CP}  e^{i(\xi_X + \xi_j - \xi_i)} \overline{\psi}_j \slashed{X} \gamma_5 \psi_i, 
\end{split}
\end{equation}
and the definitions of the projectors $P_{L/R} = (\mathbb{1} \pm \gamma_5)/2$. The spinors $\psi_i$ can be either up- or down-type quark fields. The phases $\xi$ are arbitrary, they can be set to 0 as done in Ref~\cite{Schwartz:2014sze}. The phases are opposite for anti-particles, so
\begin{equation*}
\begin{split}
	W^{+\mu} & \xrightarrow{CP} - e^{i\xi_W} W^{-}_{\mu}, \\
	W^{-\mu} & \xrightarrow{CP} - e^{-i\xi_W} W^{+}_{\mu}.
\end{split}
\end{equation*}
Physical quantities that change sign under $CP$ symmetry are called $CP$-odd.

When building the SM Lagrangian in the last section, we wrote the mixing Lagrangian in the mass basis. Under $CP$ symmetry, following the transformations listed above, it becomes
\begin{multline}
    \mathcal{L}_{mix.} \xrightarrow{CP} \mathcal{L}^{CP}_{mix.} = \frac{e}{\sqrt{2}s_W} \Big[ W^+_\mu \overline{U}'_L (V_{CKM})^* \gamma^\mu D'_L \\
        +  W^-_\mu \overline{D}'_L (V_{CKM})^T \gamma^\mu U'_L  \Big].
\end{multline}
The invariance of $\mathcal{L}_{mix.}$ under $CP$ seems to be related to the complex nature of $V_{CKM}$. If the mixing matrix is real, that is $\delta_{CKM}=0$, the SM Lagrangian is invariant under $CP$.

However, we could have just as easily worked in the interaction basis. Then, the condition under which $CP$ is preserved appears again to be related to the complex nature of the quark Yukawa matrices $Y_q$. Indeed, for the up-quark terms for instance, 
\begin{equation}
\begin{split}
    \mathcal{L}_{up \ Yuk.} & = - \frac{v}{\sqrt{2}} \Big[ \overline{U}_{Li} \left( Y_u \right)_{ij} u_{Rj} + \overline{u}_{Rj} \left( Y_u^\dagger \right)_{ji} U_{Li}  \Big] \\
    & \xrightarrow{CP} \mathcal{L}^{CP}_{up \ Yuk.} = - \frac{v}{\sqrt{2}} \Big[ \overline{U}_{Li} \left( Y_u^* \right)_{ij} u_{Rj} + \overline{u}_{Rj} \left( Y_u^T \right)_{ji} U_{Li}  \Big].
\end{split}
\end{equation}
Thus, $\mathcal{L}_{up \ Yuk.}$ is $CP$ invariant if $Y_u = Y_u^*$ and the cause of $CP$ violation in the SM is linked with the complex nature of the matrices $V_{CKM}$ and $Y$. However, these matrices are not flavour invariant because complex phases can be absorbed by flavour rotations of the quarks as shown in the previous Section.

When the Yukawa matrices were diagonalised in Eq.(\ref{eq:YukawaMatricesDiag}), the CKM matrix was obtained by the flavour matrices $J_{u/d}$. If $J_u = J_d$, $V_{CKM}$ would be $\mathbb{1}$ in the quark Yukawa Lagrangian with diagonal mass matrices for any flavour transformations. This means that the argument for $CP$ violation in the SM is actually related to the Yukawa matrices that cannot be diagonalised simultaneously instead of simply having complex Yukawa matrices.

To prove it, flavour transformations are used again. In Eqs.(\ref{eq:DownQuarksRotationForCP}) and (\ref{eq:UpQuarksRotationForCP}), both left-handed and right-handed quarks fields were rotated but we are free to apply flavour transformations on right-handed fields only. By taking  
\begin{equation}
\begin{split}
	d_R \to d''_R = K_d J^\dagger _d d_R, \\
	u_R \to u''_R = K_u J^\dagger _u u_R,
\end{split}
\end{equation} 
the Yukawa matrices are hermitian and are diagonalised thanks to the $J_\psi$ matrices 
\begin{equation*}
		Y_\psi = J_\psi Y'_\psi J^\dagger_\psi = \frac{\sqrt{2}}{v} J_\psi M'_\psi J^\dagger_\psi
\end{equation*}
with $\psi=u,d$. Then, we can look at the commutator of the Yukawa matrices since commuting matrices are simultaneously diagonalisable, %. We will not repeat the transformations provided in Eqs.~(\ref{eq:DownQuarksRotationForCP}) and (\ref{eq:UpQuarksRotationForCP}) here. Instead, to make the commutator more tractable, we only rotate the right-handed quark fields,
%\begin{equation}
%\begin{cases}
%    d_R \to d_R'' = K_d J^\dagger_d d_R, \\
%    u_R \to u_R'' = K_u J^\dagger_u u_R.
%\end{cases}
%\end{equation}
%The commutator now is 
\begin{equation}
    [Y_u, Y_d] = \frac{2}{v^2} [J_u M'_u J^\dagger_u, J_d M'_d J^\dagger_d] = \frac{2}{v^2} J_u [M'_u, V_{CKM} M'_d V^\dagger_{CKM} ] J^\dagger_u.
\end{equation}
If the two quark Yukawa matrices commute, then the determinant of their commutator has to cancel. The mass matrices are diagonal and we have introduced the standard parametrisation of the CKM matrix in Eq.(\ref{eq:StandardCKMmatrix}), so the determinant of the commutator is proportional to differences in quark masses and the CKM parameters. All factors containing a CKM angle are gathered in a single quantity $J$ such that
\begin{multline}
\label{eq:YukCommutator}
    \text{det} ([Y_u, Y_d]) = \\
    -i \frac{16}{v^6} (m_t - m_c) (m_t - m_u) (m_c - m_u) (m_b - m_s) (m_b - m_d) (m_s - m_d) J
\end{multline}
where
\begin{equation}
\label{eq:JarlskogInvariant}
\begin{split}
    J & = s_{12} s_{23} s_{13} c_{12} c_{23} c_{13}^2 \sin(\delta_{CKM}), \\
        & = (3.12^{+0.13}_{-0.12}) \times 10^{-5}.
\end{split}
\end{equation}
The flavour invariant quantity $J$ is called the Jarlskog invariant
\footnote{The Jarlskog invariant is related to what is called the unitarity triangle. By unitarity of $V_{CKM}$, $\sum_i V_{ij} V^*_{ik} = \delta_{jk}$, a triangle can be drawn and the Jarlskog invariant corresponds to twice the area of the triangle. See Ref.\cite{ParticleDataGroup:2024cfk} for more details and figures.} 
and its numerical value is provided by the Particle Data Group \cite{ParticleDataGroup:2024cfk} (smaller uncertainties were reported at the 2025 CKM workshop by the CKMfitter Collaboration
%https://indico.cern.ch/event/1440982/contributions/6583169/attachments/3136159/5566274/CKMFitter@CKM25_OD_light.pdf
but their results have not been published at the time of writing). In a general parametrisation, it is given by 
\begin{equation}
    Im\left( V_ {ij} V_{kl} V^*_{il} V^*_{kj} \right) = J \sum_{m,n} \epsilon_{ikm} \epsilon_{jln}.
\end{equation}
We see in Eq.~(\ref{eq:JarlskogInvariant}) that $J$ is proportional to the sine of the complex CKM phase $\delta_{CKM}$. Therefore, the Yukawa matrices would commute if that parameter were to vanish or if the quark masses were degenerate. Practically, degenerate masses would allow an extra $SU(2)$ flavour rotation to further reduce the number of degrees of freedom in $V_{CKM}$, allowing one to absorb $\delta_{CKM}$ in particular.

In conclusion, any flavour invariant $CP$ violating effect in the SM is going to be proportional to $\text{Im}~\text{det}[Y_u, Y_d]$ and very small. First, light quark masses in Eq.~(\ref{eq:YukCommutator}) are quite small making their differences even smaller. Secondly, processes sensitive to $CP$ violation generally require the involvement of the three generations. Such processes depend on off-diagonal elements of $V_{CKM}$ which have small values, and are loop-suppressed.

\section{The Limitations of the Standard Model}
\label{sec:SMIssues}

The absence of visible matter-antimatter annihilation in the Universe lets us assume that it is predominantly composed of matter. However, the Big Bang is theorised as a fully symmetric state. In order to explain the transition from a symmetric state to the current asymmetric state, Nature has to violate $C$ and $CP$. Nature also needs to violate the baryon number $B$ and to depart from its early thermal equilibrium. Otherwise, conjugate processes would erase any generated asymmetry. These are Sakharov's conditions presented in Ref.~\cite{Sakharov:1967dj}.

The violation of $B$ is not trivial in the SM. Non-perturbative effects in the electroweak sector, related to transitions between $SU_L(2)$ gauge field configurations in different vacua, do not preserve $B$. They also violate the lepton number $L$, but preserve their difference $B-L$. These effects are called sphaleron processes because they are transitions between spherically symmetric solutions. Quantum tunnelling through instanton processes is exponentially suppressed at low temperatures, so we cannot observe them at present times or in current particle colliders. However, at earlier stages of the Universe, the energy barrier could have been lowered enough to allow transitions by thermal fluctuations. See Ref. \cite{Matchev:2025ivr} for a recent detailed derivation.

The departure from thermal equilibrium could be explained by the phase transition of the SM during the EWSSB. However, the observed Higgs mass suggests that there is not a phase transition but actually a smooth crossover between the symmetric phase $\langle \varphi \rangle = 0$ and the broken phase $\langle \varphi \rangle \neq 0$. Unfortunately, this process does not satisfy Sakharov's conditions as predicted in Ref.\cite{Bochkarev:1987wf}.

We saw that the SM violates $CP$ due to the presence of a non-zero $J$ in the quark sector. Despite the SM respecting Sakharov's condition on the presence of $CP$ violation, theoretical estimations derived from the SM do not support the observations and a huge tension arises, see Ref.~\cite{Canetti:2012zc}. Matter and antimatter annihilate mostly through electromagnetic interactions. Thus, the density of photons in the present Universe $n_{\gamma}$ at $10^{-4}$ eV should give an estimate on the total density of particles initially present at a temperature of roughly 1GeV. Baryonic antimatter is not naturally present. We observe traces of it only in particle showers from cosmic rays. So, its density is basically null at present times $n_{\overline{B}} ~|_{T \sim 10^{-4} \text{eV}} \approx 0$. The asymmetry $\eta$ between baryonic matter density $n_{B}$ and baryonic antimatter density is then given by
\footnote{The approximation assumes entropy conservation between $T \sim 1\ GeV$ and $T \sim 10^{-4}$eV.}
\begin{equation}
    \eta = \frac{n_B - n_{\overline{B}}}{n_B + n_{\overline{B}}} \Big|_{T\sim 1 \text{GeV}} \approx \frac{n_B}{n_\gamma} \Big|_{T\sim 10^{-4} \text{eV}} \neq 0.
\end{equation}
Based on the cosmic microwave background (CMB) and baryon acoustic oscillations, the experimental value of $\eta$ is $\eta_{exp.} \approx 10^{-10}$. As mentioned, this is problematic since the estimation of $\eta$ from the SM in Ref.~\cite{Shaposhnikov:1986jp} gives $\eta_{SM} \approx 10^{-20}$. Therefore,
\begin{equation}
    \frac{\eta_{exp.}}{\eta_{SM}} \approx 10^{10}.
\end{equation}

This discrepancy is one of the fundamental flaws of the SM when compared to observations. On top of the issue with the baryon asymmetry, the SM fails to explain the strong $CP$ problem, the neutrino masses, the presence of dark matter, the stability of the Higgs mass, the hierarchy of the fermion masses and the magnitude of spacetime expansion attributed to the dark energy. On a more theoretical aspect, we can also mention that no coherent QFT formulation of gravity has been formalised yet.

All possible Lagrangian terms respecting the gauge symmetries of the SM group were obtained in the second section of this chapter except a term involving the gluon dual field tensor $\overline{\theta} G^A_{\mu \nu} \widetilde{G}^{A~\mu \nu}$. Under the $CP$ symmetry, it acquires a minus sign meaning that it violates $CP$. Strong $CP$ violating interactions could theoretically exist but there has not been any evidence supporting it. The limits on $\overline{\theta}$ obtained from electric dipole moments are strong. This could suggest the existence of a particle, called the axion, cancelling this effect as demonstrated in Refs.~\cite{PhysRevLett.37.8, PhysRevD.16.1791, CALLAN1976334, BELAVIN197585, Cheng:1987gp}.

In the SM, neutrinos are strictly massless. The absence of a right-handed neutrino field forbids Dirac masses, while gauge invariance under $SU_L(2)$ prevents Majorana masses for the left-handed neutrino field. However, observations of oscillations from solar and atmospheric neutrinos in Refs.~\cite{PhysRevLett.12.300, PhysRevLett.12.303, PhysRevLett.20.1205, SNO:2001kpb, SNO:2002tuh, Super-Kamiokande:1998kpq, Chen:2025ipd} suggest that their flavour and mass bases do not coincide. The respective eigenstates mix like the quark eigenstates and a mass matrix $M_{\nu}$ should exist.

Dark matter as a massive, stable, non-baryonic and electrically neutral matter state was introduced to solve the missing mass at various scales in the Universe. Rotation curves of galaxies cannot be reproduced with the mass of visible matter, in particular in the outer regions, see Refs.~\cite{Rubin:1970zza, Rubin:1980zd}. Gravitational lensing effects observed from the Bullet Cluster in Refs.~\cite{Clowe:2003tk, Clowe:2006eq} are too large compared to the expectation of the measured mass. Anisotropies of the CMB repeatedly measured by the Planck telescope in Refs.~\cite{Planck:2015fie, Planck:2015bue, refId0, Planck:2018vyg, Planck:2019evm} exceed the estimations from the visible matter in the Universe. However, there are no candidates for such a state of matter in the SM.

Quantum corrections to the Higgs mass from loop processes are quadratic in the energy cut-off scale $\Lambda$. No restrictions exist on $\Lambda$, so it can be as large as the Planck scale $M_P$ (see the introduction of the review on supersymmetry in Ref.~\cite{Martin:1997ns}). The Higgs will be extremely heavy but it would contradict the measured Higgs mass of $125~\text{GeV}$ by CMS in Ref.~\cite{CMS:2024eka} and ATLAS in Ref.~\cite{ATLAS:2023oaq}. Some mechanism should then protect the Higgs mass from these large contributions either via a (possibly broken) symmetry or new states in the loops.

The masses of the fermions are generated after the EWSSB due to the non-zero vacuum expectation value and the Yukawa couplings. The latter are free parameters of the SM. Nevertheless, they seem to follow a similar pattern across generations within each species. We see in the reported value of the Yukawa couplings from the Particle Data Group \cite{ParticleDataGroup:2024cfk} that every third generation fermion is heavier than the second, which is heavier than the first. This apparent hierarchy has no origin in the SM.

The redshift of galaxies indicates that the Universe is still expanding \cite{SupernovaSearchTeam:1998fmf, Perlmutter:1999rr}. The origin of this  energy driving the expansion is still unknown. It could phenomenologically be parametrised by the cosmological constant $\Lambda_{cc}$, a constant term in Einstein's formulation of general relativity. But, if that constant was related to the vacuum expectation value from the SM, the dark energy densities \cite{Planck:2015bue} derived from these two quantities would be separated by $\sim$ 60 or 120 orders of magnitude depending on the cut-off scale set at the electroweak or Planck scale \cite{RevModPhys.61.1, Martin:2012bt}. Comparing the orders of the energies themselves instead of the energy densities results in a smaller discrepancy of 13 orders of magnitude.

Many UV-complete theories have been constructed and proposed to solve one or more of the problems listed above. The SM is recovered from the UV theories at low (high) energy when heavier (lighter) particles are introduced, or at least approximated. However, no statistical evidence demonstrates which model should be favoured. Testing all possible model extensions would require an infinite amount of time and resources. We will see in the next chapter what are the theoretical tools at our disposal to potentially point to the next Standard Model of Particle Physics.

%!TEX root = main.tex

%%%%%%%%%%%%%%%%%%%%%%%%%%%%%%%%%%%%%%%%%%%%%%%%%%%%
%
%      Chapter 2 :
%
%
%%%%%%%%%%%%%%%%%%%%%%%%%%%%%%%%%%%%%%%%%%%%%%%%%%%

\chapter{Models Beyond the Standard Model}
\label{chap:bsm}
\pagestyle{fancy}

\begin{minipage}{10cm}

\hfill  {\small\it ``Necessity is the mother of innovation.''}

\hfill {\small \textit{Proverb}}
\end{minipage}

\vspace{0.5cm}

In this chapter, some informal derivations of EFT will be presented. Benefits from both top-down and bottom-up approaches will be shown, in particular, how we can match UV theories to the EFT and how the EFT can point to specific UV theories. We will use the 4-Fermi theory as a historic and illustrative example in both approaches. Then, the case of the SMEFT is explored, more specifically the derivation of the Warsaw basis and its leading CP-odd operators. The leading effects are obtained by using additional $U(1)$ symmetries present in massless approximations. Finally, if an EFT cannot be used, generic models take the form of simplified models. After shortly presenting the concept of simplified models, a non-exhaustive list of the particles appended to the SM spectrum is shown.

\newpage

\section{Top-down Approach by Matching}
\label{sec:BasicEFT}

As presented in Section \ref{sec:QFT}, experiments aim at observing some states in the detectors. These states are composed of particles observed by leaving energetic tracks in the detectors, but heavier particles decay before they can even leave any energy deposit. We can wonder whether it is worth keeping all fields in the theory if we cannot detect them. The answer is it depends on the energy scales involved, in particular the balance between the mass of the particles and the reach of the collider. The energy reach of a collider corresponds to the highest value that the centre-of-mass energy $\sqrt{\hat{s}}$ can take.

\subsection{Toy example}
\label{sec:ToyModelMatching}

As a back-of-the-envelope derivation, let us assume a toy theory formed by two massless fermions $\psi_{1,2}$ and a massive real scalar $\phi$ of mass $M_{UV}$. The scalar is heavy compared to the energy reach ($M_{UV} \geq E$) and is coupled only\footnote{$\phi$ could as easily be coupled to $\psi_2$, but this particular case is more interesting for the upcoming Section \ref{sec:BasicSMEFT}.} to $\psi_1$ with the coupling $\lambda$.

If $\phi$ is generated through a s-channel diagram then rapidly decays, it will be invisible in the detector. However, its coupling to the other particles has a visible effect in the cross section which will follow a Breit-Wigner distribution, 
\begin{equation}
\label{eq:ResonantCrossSection}
    \sigma \propto \left| \vcenter{\hbox{\includegraphics[scale=0.25]{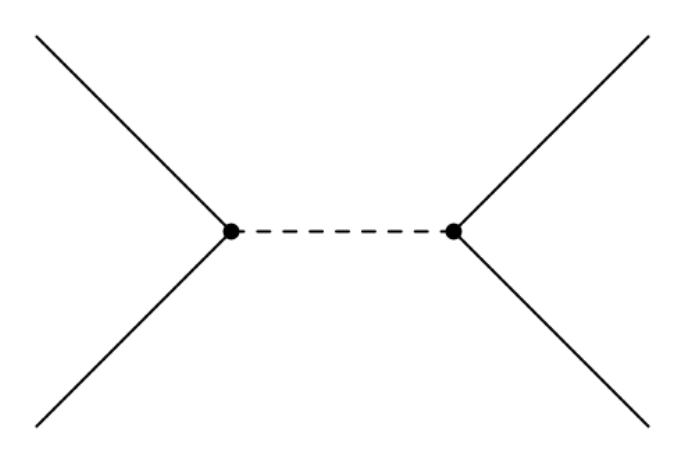}}} \right|^2 \propto \lambda^4 \frac{1}{(M_{UV}^2 - p^2)^2 + (\Gamma_{UV} M_{UV})^2}.
\end{equation}
We suppose here that the width $\Gamma_{UV}$ of $\phi$ is small compared to its mass, so the second term of the denominator is neglected. The peak of the Breit-Wigner distribution is observable only if $\sqrt{\hat{s}}$ is greater or of the same order as $M_{UV}$. Otherwise, if the momentum of $p^2 \sim \hat{s} \ll M_{UV}^2$, the coupling becomes effectively constant
\begin{equation}
\label{eq:EffectiveCrossSection}
    \sigma \propto \frac{\lambda^4}{M_{UV}^4} \frac{1}{\left( 1 - \frac{p^2}{M_{UV}^2} \right)^2 + \left( \frac{\Gamma_{UV}}{M_{UV}} \right)^2 } \sim \frac{\lambda^4}{M_{UV}^4} .
\end{equation}

\begin{figure}
    \centering
    \includegraphics[width=\linewidth]{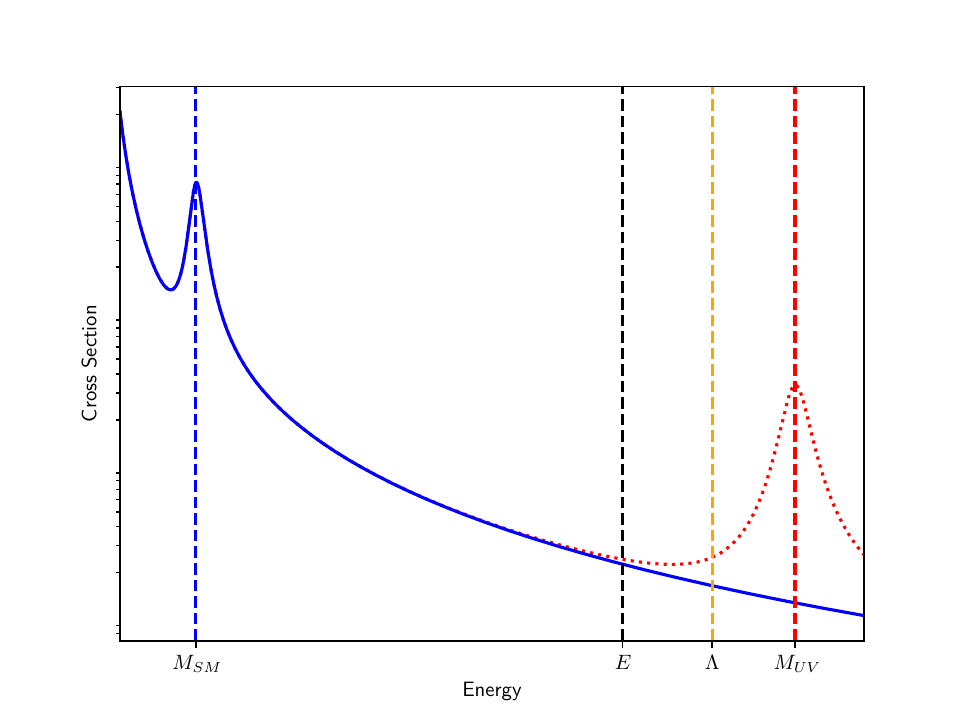}
    \caption{Representation of the cross section with respect to the partonic center-of-mass energy for the SM and some UV-complete theory (PDF effects are excluded). The former is shown in blue, the latter is shown by the red dotted line. Some SM resonance is highlighted by the vertical dashed blue line. We see that past the EFT energy scale $\Lambda$ in orange, we must use the UV-complete theory with the UV resonance in red.}
    \label{fig:EFTFig}
\end{figure}

Figure~\ref{fig:EFTFig} presents an example of the SM and some UV-complete theory where there is one massive particle with a mass $M_{UV}$ much larger than the energy reach $E$. We represent a Breit-Wigner distribution associated to a massive particle present in the SM decaying before reaching the detectors (such as the resonance of a $Z$ boson for instance). It is clear that we are forced to include the SM resonance in our predictions and an equation similar to Eq.(\ref{eq:ResonantCrossSection}) must be used. 

However, it is also clear that the experiment is blind to the presence of the massive particle from the UV-complete theory because the resonance sits at energies larger than the energy reach $E$ of the collider. The cross section of the UV-complete theory overlaps with the SM up to the energy reach $E$ and small deviations appear at higher energies. At the cut-off scale $\Lambda$, the deviations increase then dominate until the $\phi$ resonance is reached at $M_{UV}$. An expression such as Eq.(\ref{eq:EffectiveCrossSection}) closely reproduces the observations.

The heavy mass limit is now derived formally. The degree of freedom that represents the heavy particle $\phi$ is removed from the UV-complete theory by keeping identical values for the respective cross sections of the UV-complete theory and the resulting EFT. This amounts to \textit{match} the matrix elements from the UV action $S_{UV} = \int d^4x~ \mathcal{L}_{UV}[\phi, \psi_{1,2}]$ in the EFT action $S_{EFT}[\psi_{1,2}] = \int d^4x~ \mathcal{L}_{EFT}[\psi_{1,2}] \equiv \Gamma[\psi_{1,2}]$,
\begin{equation}
    \braopket{\Omega}{T\{ \psi_1 ~...~ \psi_n \}}{\Omega}_{S_{UV}[\psi_{1,2},\phi]} = \braopket{\Omega}{T\{ \psi_1 ~...~ \psi_n \}}{\Omega}_{\Gamma[\psi_{1,2}]}.
\end{equation}
The action of the UV-complete theory is derived from the UV Lagrangian density $\mathcal{L}_{UV}$,
\begin{equation}
\label{eq:BasicUVLagrangian}
    \mathcal{L}_{UV}[\psi_{1,2}, \phi] = i \overline{\psi}_1 \slashed{\partial} \psi_1 + i \overline{\psi}_2 \slashed{\partial} \psi_2 + \frac{1}{2} \partial_{\mu} \phi \partial^{\mu} \phi - \frac{1}{2} M_{UV}^2 \phi^2 + \lambda \phi \overline{\psi}_1 \psi_1 .
\end{equation}
and the Euler-Lagrange equation associated to $\phi$,
\begin{equation}
\label{eq:EOMPhi}
    (\square + M_{UV}^2 ) \phi = + \lambda \overline{\psi}_1 \psi_1 \to \phi = \frac{\lambda}{\square + M_{UV}^2} \overline{\psi}_1 \psi_1 ,
\end{equation}
where $\square \phi = - p^2 \phi$ in momentum space.

The heavy mass limit in momentum space imposes $p^2 \ll M_{UV}^2$. Fluctuations around its classical trajectory are suppressed and thus neglected. Therefore, $\phi$ does not deviate from its classical configuration and does not propagate over long distances due to its large mass. This allows the replacement of Eq.(\ref{eq:EOMPhi}) in Eq.(\ref{eq:BasicUVLagrangian}),
\begin{equation}
\label{eq:UVtoEFTexpansion}
\begin{split}
    \mathcal{L}_{EFT}[\psi_{1,2}] & = i \overline{\psi}_1 \slashed{\partial} \psi_1 +  i \overline{\psi}_2 \slashed{\partial} \psi_2 + \frac{\lambda^2}{2} \overline{\psi_1} \psi_1 \left(\frac{1}{\square + M_{UV}^2}\right) \overline{\psi}_1 \psi_1, \\
        & = \sum_{i=1,2} i \overline{\psi}_i \slashed{\partial} \psi_i + \frac{\lambda^2}{2 M_{UV}^2} \overline{\psi_1} \psi_1 \overline{\psi}_1 \psi_1 \\
        & ~~~~ - \frac{\lambda^2}{2 M_{UV}^4} \overline{\psi_1} \psi_1 \square \overline{\psi}_1 \psi_1 + \dots
\end{split}
\end{equation}
In the second equality, the expression in brackets has been expanded such that 
\begin{equation*}
    \frac{1}{\square + M_{UV}^2} = \frac{1}{M_{UV}^2} \left( 1 - \frac{\square}{M_{UV}^2} + \dots \right).
\end{equation*}

We say that $\phi$ has been integrated out
\footnote{
This statement is more explicit in the path integral formalism but strictly identical
\begin{equation*}
    \int \mathcal{D}\overline{\psi_i} \mathcal{D}\psi_i~ exp\left\{i \int d^4x~ \mathcal{L}_{EFT}[\psi_{1,2}] \right\} = \int \mathcal{D} \phi \mathcal{D}\overline{\psi_i} \mathcal{D}\psi_i~ exp\left\{i \int d^4x~ \mathcal{L}_{UV}[\phi, \psi_{1,2}] \right\}.
\end{equation*}
}
and the EFT has fewer degrees of freedom $\psi_{1,2}$. The removal of one degree of freedom has led to an infinite tower of operators made of combinations of $\psi_1$. There were no interactions between $\phi$ and $\psi_2$ in the UV-complete theory, so there are no additional terms involving $\psi_2$ in the EFT. The EFT energy scale $\Lambda$ roughly corresponds to UV mass $M_{UV}$ and ensures the perturbative behaviour of the EFT. The operators are sorted in a perturbative expansion where higher dimension operators are suppressed by powers of $1/M_{UV}$. In a more general theory, there will be many more operators. So, one can focus on the first operators that are relevant to track deviations compared to a theory involving only $\psi_{1,2}$, and neglect the higher order ones. The NDA introduced in section~\ref{sec:SMLag} sorts the operators with increasing mass dimension. 

\begin{figure}[t]
    \centering
    \begin{minipage}{0.43\textwidth}
        \centering
        \includegraphics[width=\textwidth]{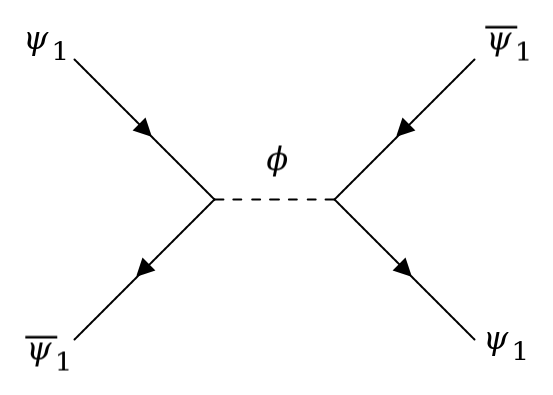}
        %\caption{Left Figure}
    \end{minipage}
    \hfill 
    \begin{minipage}{0.15\textwidth}
        \centering
        $\xrightarrow{ p^2 \ll M_{UV}^2 }$
    \end{minipage}
    \hfill 
    \begin{minipage}{0.40\textwidth}
        \centering
        \includegraphics[width=\textwidth]{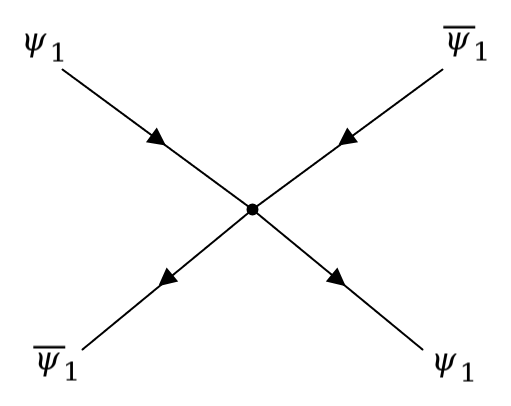}
        %\caption{Right Figure}
    \end{minipage}
\caption{Feynman diagrams representing the tree-level matching procedure of the s-channel via a $\phi$ resonance in the UV-complete theory to a 4-Fermi operator made of $\psi_1$ fields at lower energies. }
\label{fig:TreeLevelMatching}
\end{figure}

This is the \textit{top-down} approach where one starts from the UV-complete theory defined at high energy and generates the EFT valid at lower energies. The matching procedure described above is applied at tree-level, see Figure~\ref{fig:TreeLevelMatching}, by virtue of the classical equation of motion used for the heavy field. A loop-level matching could be considered by accounting for loop diagrams. Once $p^2 \sim M_{UV}^2$ (or $p^2 \sim \Lambda^2$), the perturbative expansion of the EFT breaks down and the UV-complete theory must be used. An early conclusion is the following : physics at different energy scales can be described by different theories that are related in some non-trivial way by the matching conditions.

\subsection{4-Fermi Theory}
\label{sec:4Fermi}

A great example of a top-down approach is the description of $\beta$-decays at low energy from weak interactions in the SM at high energies.

The muon decay $\mu^-(p_1) \to e^-(p_2) ~ \overline{\nu}_e(p_3) \nu_\mu(p_4)$ is a good example usually presented since it covers some interesting properties of EFTs. On the one hand, the process can be computed with Fermi's EFT made of four-fermion operators with $V-A$ Lorentz structures due to its low energy. On the other hand, the SM process is mediated via a $W^-$ boson with momentum $q$ such that $q=p_1-p_4=p_2+p_3$.

The degree of freedom represented by the $W$ boson is removed with a tree-level top-down approach and the EFT relies on leptons, 
\begin{equation}
    \braopket{\Omega}{T\{ \psi_1 ~...~ \psi_n \}}{\Omega}_{S_{SM}[l,W^\pm]} = \braopket{\Omega}{T\{ \psi_1 ~...~ \psi_n \}}{\Omega}_{\Gamma_{Fermi}[l]}.
\end{equation}
Here, the respective matrix elements are matched
\begin{equation}
     \braopket{e^- \overline{\nu}_e \nu_\mu}{\mathcal{M}_{SM}}{\mu^-} = \braopket{e^- \overline{\nu}_e \nu_\mu}{\mathcal{M}_{Fermi}}{\mu^-} .
\end{equation}

The SM interaction between leptons and $W$ bosons is written in the off-diagonal terms of the left-handed lepton term of Eq.(\ref{eq:SMLagKinetic}). Using the left projector $P_L$, the SM matrix element is
\begin{equation}
\label{eq:MuonDecayFullSM}
    \mathcal{M}_{SM} = \frac{g_2^2}{8(q^2 - m_W^2)} \big[ \overline{u}(p_2) \gamma^\sigma (1-\gamma_5) v(p_3) \big] \big[ \overline{u}(p_4) \gamma_\sigma (1-\gamma_5) u(p_1) \big]
\end{equation}
For centre-of-mass energies much smaller than the $W$ boson mass, $q^2 \sim \hat{s} \ll m_W^2$, the derivation from Eq.(\ref{eq:UVtoEFTexpansion}) can be applied. The leptons are assumed to be massless. This assumption is reasonable since $m_{e,\mu} \ll m_W$ and facilitates the derivations. The first generated operator is kept as the only relevant operator, so Eq.(\ref{eq:MuonDecayFullSM}) becomes
\begin{equation}
\label{eq:MuonDecaySimplifiedSM}
     \mathcal{M}_{SM} \approx - \frac{g_2^2}{8 m_W^2}  \big[ \overline{u}(p_2) \gamma^\sigma (1-\gamma_5) v(p_3) \big] \big[ \overline{u}(p_4) \gamma_\sigma (1-\gamma_5) u(p_1) \big] .
\end{equation}

A 4-lepton operator is written with a constant coefficient $C_{Fermi}$ and an EFT cut-off scale $\Lambda_{Fermi}$ that will match to the simplified SM matrix element. The leptonic operator is
\begin{equation}
\label{eq:FermiEFTLagrangian}
    \mathcal{L}_{Fermi} \supset -\frac{C_{Fermi}}{4\Lambda_{Fermi}^2}  (\overline{\mu} \gamma^\sigma (1 - \gamma_5) \nu_\mu) (\overline{e} \gamma_\sigma (1 - \gamma_5) \nu_e).
\end{equation}
Historically, the Lagrangian term was defined with only one constant, the Fermi constant $G_F$, instead of $C_{Fermi}$ and $\Lambda_{Fermi}$. One can verify that the two expressions of $\mathcal{L}_{Fermi}$ are identical as long as  
\begin{equation}
    G_F = C_{Fermi} / 2\sqrt{2} \Lambda_{Fermi}^2 .
\end{equation}
In practice, the value of $G_F$ is determined by the muon partial decay width $\Gamma(\mu \to e \overline{\nu}_e \nu_\mu)$, such that
\begin{equation}
    \frac{1}{\tau_{\mu}} \equiv \Gamma(\mu \to e \overline{\nu}_e \nu_\mu) = \frac{G_F^2}{2} \frac{m^5_\mu}{96 \pi^3},
\end{equation}
where $\tau_{\mu} = 2.197 \times 10^{-6}$s is the measured muon lifetime and $m_\mu = 105.66$ MeV its mass. So, $G_F = 1.166 \times 10^{-5}$ GeV$^{-2}$. The EFT matrix element derived from the 4-lepton operator is easily obtained as
\begin{equation}
\label{eq:Muondecay4Fermi}
    \mathcal{M}_{Fermi} = - \frac{C_{Fermi}}{4\Lambda_{Fermi}^2} \big[ \overline{u}(p_2) \gamma^\sigma (1-\gamma_5) v(p_3) \big] \big[ \overline{u}(p_4) \gamma_\sigma (1-\gamma_5) u(p_1) \big]
\end{equation}

Comparing Eqs.(\ref{eq:MuonDecaySimplifiedSM}) and (\ref{eq:Muondecay4Fermi}), the matching condition between the two theories is
\begin{equation}
\label{eq:MatchingConditionFermiTheory}
    \frac{C_{Fermi}}{\Lambda_{Fermi}^2} = \frac{g_2^2}{2 m_W^2}.
\end{equation}
This is a key result since computations are generally easier in EFTs than in UV theories at low energies $p^2 \ll m_W^2$. A small caveat remains in the degeneracy left in the ratio $C_{Fermi}/\Lambda_{Fermi}^2$. It will be experimentally impossible to disentangle the two constants, only the ratio can be extracted from measurements.  

For energies close to the $W$ boson mass, $m_W\approx 80~\text{GeV}$, the EFT description with four-fermion operators cannot be applied anymore. The complete SM theory has to be used to generate the Breit-Wigner distribution of the $W$ resonance.

\begin{figure}[t]
    \centering
    % First Figure
    \begin{minipage}{0.40\textwidth}
        \centering
        \includegraphics[width=\textwidth]{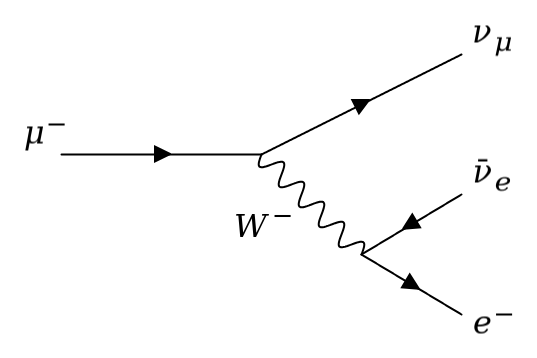}
        %\caption{Left Figure}
    \end{minipage}
    \hfill % Adds space between
    % Equation in the middle
    \begin{minipage}{0.15\textwidth}
        \centering
        $\xrightarrow{ p^2 \ll m_W^2 }$
    \end{minipage}
    \hfill % Adds space between
    % Second Figure
    \begin{minipage}{0.40\textwidth}
        \centering
        \includegraphics[width=\textwidth]{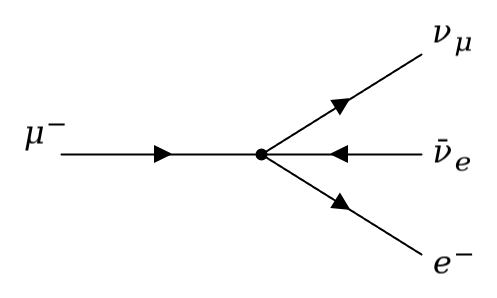}
        %\caption{Right Figure}
    \end{minipage}
\caption{Feynman diagrams representing the tree-level matching procedure of the muon decay by the $W$ emission to a 4-lepton operator. }
\end{figure}

\subsection{Running and Mixing}
\label{sec:RunningAndMixing}

Even if the matching is performed at tree-level, loop effects indirectly affect the value of parameters. Indeed, quantum fluctuations can give non-negligible logarithmic contributions to parameters.

In dimensional regularisation (DR) of loop amplitudes, the UV cut-off is $1/\epsilon$ where $\epsilon$ is the deviation from the four standard dimensions ($\epsilon \equiv 4-d$). Loop amplitudes can give divergent contributions when the limit $\epsilon \to 0$ is taken. A theory is called renormalisable when a finite number of counterterms $\delta_X$ are required to absorb the UV divergences from the field wavefunctions, particle masses, coupling parameters, $\ldots$  Renormalisable theories in four dimensions have Lagrangian terms with maximum mass dimension of 4. This further motivates why we listed the possible terms of the SM at the beginning of Section~\ref{sec:SMLag}.

The exact expression of the counterterms depends on the scheme, e.g. the on-shell, the minimal subtraction (MS) or the $\overline{MS}$ renormalisation schemes. In the latter, the divergent contributions to the loop amplitudes are cancelled by the counterterms with a universal constant $e^{\gamma_E}/4\pi$ ($\gamma_E$ being the Euler-Mascheroni constant). The minimal subtraction scheme does not include the constant term.

In any renormalisation scheme, renormalisation conditions fix the respective values of the renormalised quantities which are initially evaluated at some unphysical scale $\mu$. A physical scale $\mu_0$ must be set at which the conditions are considered. For instance, $\mu_0$ corresponds to the particle masses in the on-shell scheme. Using the expression of the counterterms of the renormalised theory, the parameters carry a dependence on both $\mu$ and $\mu_0$. Thus, $\mu$ is named the renormalisation scale. Since $\mu$ is arbitrary, it should not impact physical observables, such as the cross-section, when they are computed at all orders. At fixed order however, observables keep a residual dependence and the variation of the results when $\mu$ changes offers an estimation of the theory uncertainty associated with the perturbative treatment of the observable.

The UV divergences lead to scale-dependent contributions with factors of $\log( \mu/\mu_0 )$. As physical parameters are measured at different energy scales, logarithms of the ratio between the scales can be large if they are separated by orders of magnitude. These large logarithms are directly related to the existence of UV divergences in loop processes. The large logarithms are resummed thanks to the Renormalisation Group Equations (RGE).

The RGE follow from the fact that a bare parameter $g$ in the Lagrangian density does not depend on $\mu$ by definition, thus $d g/d\mu = 0$. Once the theory is renormalised, the bare coupling is substituted by a renormalised coupling $g_R(\mu)$ thanks to a renormalisation constant $Z_{g} \approx 1 + \delta_g$. The convention $g_R=g/Z_{g}$ is used here. The $\beta$-function describes how $g_R$ varies with the scale $\mu$ when $g_R$ is written as the dimensionless quantity $\alpha$, $\alpha = g_R^2/4\pi$, 
\begin{equation}
    \beta(\alpha) \equiv \mu \frac{d \alpha}{d \mu} .
\end{equation}
By defining the anomalous dimension $\gamma_{\alpha}$ of $\alpha$ as
\begin{equation*}
    \gamma_{\alpha} = \frac{\mu}{\alpha} \frac{d \alpha}{d \mu},
\end{equation*}
the $\beta$-function is
\begin{equation}
    \beta(\alpha) = \gamma_{\alpha}~ \alpha .
\end{equation}
We call this evolution with respect to the scale the running of the parameters.

In a more general case of a Lagrangian with multiple couplings, one loop amplitude may involve UV divergences from different couplings. This means that the counterterm of a coupling $C_i$ and the renormalisation condition needed to renormalise it depend on different couplings $C_j$. As couplings run between the scales, it introduces a mixing between them and the anomalous dimension takes the form of a matrix $\gamma_{ij}$, 
\begin{equation}
\label{eq:AnomalousDimensionMatrix}
     \mu \frac{d C_i(\mu)}{d \mu} = \gamma_{ij}(\mu) C_j(\mu).
\end{equation}
Diagonal components $\gamma_{ii}$ dictate the running of the respective couplings and the off-diagonal ones $\gamma_{ij}$ describe how $C_{j \neq i}$ mix with $C_i$. If $C_i$ is set to 0 at some high scale, then it can be generated due to its mixing with a non-zero $C_j$.

A convenient example to demonstrate the running of parameters would have been to compute the UV divergent part $\mathcal{M}_{\text{div.}}$ of the 1-loop matrix element $\mathcal{M}_{1-\text{loop}}$ of a photon exchange in muon decay. Then, the counterterm removing the UV divergence would have provided the anomalous dimension of $G_F$. Unfortunately, this contribution actually vanishes due to identical electric charges of the muon and the electron
\begin{equation*}
	\gamma_{G_F}^{\text{QED}} \propto Q_e - Q_\mu = 0.
\end{equation*}

However, we can imagine that the neutral scalar $\phi$ from the toy model is coupled to leptons with coupling $e$ such that it is exchanged with momentum $k$ between the negatively charged muon and electron as in QED. We further assume that this theory has been renormalised. The renormalised quantities relevant here are the bare coupling parameter $e_0$ and the bare lepton fields. The renormalised QED coupling is $e_R$ defined as
\begin{equation*}
	e_0 = Z e_R \text{  where } Z = 1 + \frac{e_r^2}{16 \pi^2} \frac{4}{3 \epsilon},
\end{equation*}  
and the renormalisation constants of the muon $Z_{\mu}$ and the electron field $Z_{e}$ are introduced such that the bare fields $e^0,\mu^0$ are
\begin{equation}
	e^0 = \sqrt{Z_e}~e~~ \text{  and  } ~~ \mu^0 = \sqrt{Z_\mu}~\mu.
\end{equation}
The two renormalisation constants are supposed to be equal since both particles carry identical electric charge
\begin{equation}
	Z_{l} \equiv Z_\mu = Z_e = 1 - \frac{e^2_R}{16\pi^2}\frac{2}{\epsilon}.
\end{equation}
The neutrinos are not charged under QED so they do not need to be renormalised and $Z_\nu = 1$ in our calculation\footnote{In the electroweak theory they will be renormalised but it is beyond the illustrative purpose of this example.}.

The matched scalar Lagrangian to consider instead of Eq.(\ref{eq:FermiEFTLagrangian}) is 
\begin{equation}
	\mathcal{L} = \frac{G}{\sqrt{2}} (\overline{\mu} e)(\overline{\nu}_e \nu_\mu).
\end{equation}
The coupling $G$ acts as a parameter equivalent to the Fermi constant $G_F$ in this case and is the parameter being renormalised. Now, $\mathcal{M}_{\text{div.}}$ reads in DR as 
\begin{equation}
    \mathcal{M}_{1-\text{loop}} \supset \mathcal{M}_{\text{div.}} = \mathcal{M}_0 \left( (-i) \mu^{4-d} \int \frac{d^d k}{(2\pi)^d} d \frac{e^2_R}{k^4}  \right) = \mathcal{M}_0 \left( \mu^{\epsilon} \frac{e^2_R}{2\pi^2} \frac{1}{\epsilon}  \right).
\end{equation}
The dimension $d$ arises from Dirac matrices contractions. $\mathcal{M}_0$ is the tree-level matrix element 
\begin{equation}
    \mathcal{M}_0 = \frac{G}{\sqrt{2}} \overline{u}(p_2) u(p_1) \overline{u}(p_3)  v(p_4) .
\end{equation}
By introducing the renormalisation constant $Z_G$ and the counterterm $\delta_G$ such that $Z_G \approx 1 + \delta_G$, the renormalised parameter $G_R \equiv G /Z_G$ gives an extra contribution
\begin{equation}
    \mathcal{M}_{count.} = \mathcal{M}_0 \delta_G.
\end{equation}
Absorbing the divergent contribution by means of the counterterm gives
\begin{equation}
    \delta_G = - \frac{e^2_R}{16\pi^2} \frac{8}{\epsilon}.
\end{equation}
Since the bare Lagrangian does not depend on the scale $\mu$, the RGE for $G_R$ is 
\begin{equation}
    0 = \mu\frac{d}{d\mu} \left( \frac{G_R Z_G}{Z_l} \right).
\end{equation}
The anomalous dimension of G $\gamma_{G}$ is given by
\begin{equation}
    \gamma_{G} \equiv \frac{\mu}{G_R}\frac{d}{d\mu} G_R = - \frac{3\alpha_e}{2\pi},
\end{equation}
where $\alpha_e = e^2_R/4\pi$. Then,  
\begin{equation}
    G_R(\mu) = G_R(\mu_0)~ exp \left[ \int_{\alpha_e(\mu_0)}^{\alpha_e(\mu)} \frac{\gamma_{G_F}}{\beta(\alpha_e)} \right] = G_R(\mu_0) \left( \frac{\alpha(\mu)}{\alpha_e(\mu_0)} \right)^{-\frac{9}{4}}, 
\end{equation}
when the $\beta$-function of the QED constant $\beta(\alpha_e)= 2\alpha_e^2/3\pi$ is used.

Since the matching of the electroweak theory to the 4-Fermi EFT is done when the $W$ boson is integrated out, the value of $G_F$ is set at $\mu_0=m_W$. At the scale of interest for the muon decay $\mu=m_\mu$, the value of $G_F$ increases by 5\% compared to its value at the higher scale. This increase of $G_F$ as the scale is lowered follows from the negative sign of the anomalous dimension $\gamma_{G_F}$.

\begin{figure}
    \centering
    \includegraphics[width=\linewidth]{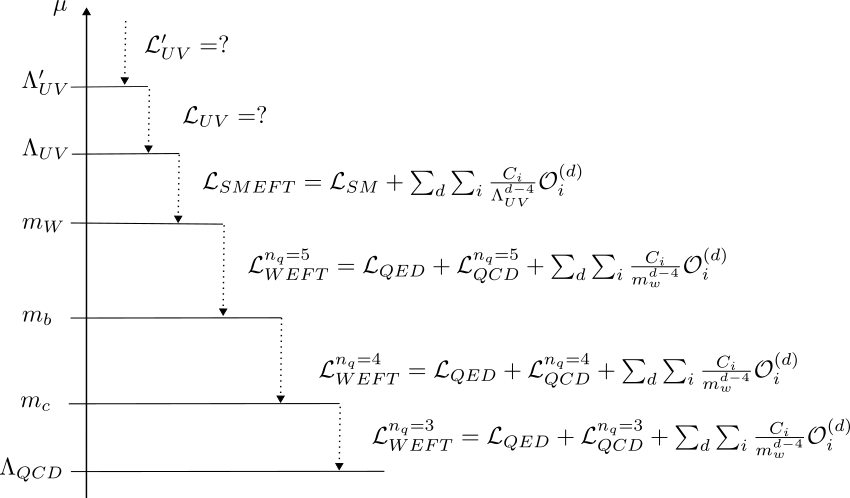}
    \caption{Example of a "stairway" of matching and running. Two different UV theories are matched at the scale $\Lambda_{UV}'$ then the couplings are run until the UV energy scale $\Lambda_{UV}$ with an effective field theory of the SM.  }
    \label{fig:MatchingAndRunningStairway}
\end{figure}

The combination of matching, running and mixing allows one to study the infrared (IR) behaviour of a UV-complete theory with simpler amplitudes. Note that multiple matchings between theories are permitted as long as their respective running is taken into account. For instance, to study the electric dipole moments with different UV theories in Ref.~\cite{Engel:2013lsa}, one can run the couplings down to the UV scale $\Lambda_{UV}$ and match them with an effective field theory of the SM. Then, the Wilson coefficients are run again down to the $W$ mass. They are matched with the Weak Effective Field Theory by integrating out the heavy fields (top, Higgs, $W$ and $Z$). The new Wilson coefficients are run until the mass of the bottom quark where it is integrated out, and so on until all degrees of freedom of decaying particles have been removed. At that point, the scale is the QCD scale $\Lambda_{\text{QCD}}$ where we can stop the matching and running. Figure~\ref{fig:MatchingAndRunningStairway} displays this sequence of running and matching, even considering the possibility of matching between two UV theories at a large scale $\Lambda_{UV}$.

\section{Bottom-up Approach by Basis Creation}
\label{sec:BasicSMEFT}

The tree-level matching procedure demonstrated above in the top-down approach, while powerful, is theoretically limited. The operators of the EFT cannot be constructed at the matching scale without first fully defining the UV-complete theory. If the goal is to find which model could explain deviations in cross-section distributions, the top-down approach is less efficient because we would need to parse all UV models and derive the possible mixing between operators down to the energy scale of colliders.

Nevertheless, the reverse procedure is possible by listing all possible higher order operators built from the degrees of freedom at low energies, reducing the number of operators with symmetry relations and obtaining an operator basis. The basis maps all interactions from any UV-complete theory as long as it involves the same degrees of freedom and respects the symmetries of the EFT. This reverse procedure is called the \textit{bottom-up} approach.

\subsection{Toy Example}

In section \ref{sec:ToyModelMatching}, a toy model with three particles was introduced. $\phi$ is a heavy scalar that decouples from the fermions $\psi_{1,2}$. In the bottom-up approach, it is not possible to know the existence of $\phi$ a priori. Thus, the kinetic terms of the fermions are supplemented with all higher-order operators in the Lagrangian $ \mathcal{L'}_{EFT}[\psi_{1,2}]$. The first ones are four-fermion operators so, all possible higher-order four-fermion operators made of $\psi_1$ and $\psi_2$ are considered
\begin{equation}
\begin{split}
    \mathcal{L'}_{EFT}[\psi_{1,2}] & = \sum_{i=1,2} i \overline{\psi}_i \slashed{\partial} \psi_i + \frac{C_{1}}{\Lambda^2} \overline{\psi_1} \psi_1 \overline{\psi}_1 \psi_1 + \frac{C_{2}}{\Lambda^2} \overline{\psi_2} \psi_2 \overline{\psi}_2 \psi_2 \\
        & ~~~~ + \frac{C_{3}}{\Lambda^2} \overline{\psi_1} \psi_1 \overline{\psi}_2 \psi_2 + \frac{C_{4}}{\Lambda^2} \overline{\psi_1} \psi_2 \overline{\psi}_2 \psi_2 + \dots
\end{split}
\end{equation}
It is only when an experiment, designed to measure cross-sections of collisions involving $\psi_1$ and $\psi_2$, is confronted with the predictions depending on $C_i$ that the value of the ratios $C_i/\Lambda^2$ can be constrained. In particular, the constraints on the ratios will be centred around 0 except for $C_1/\Lambda^2$ assuming that the experiment generates enough events for a dedicated analysis to be statistically significant. From there, the conclusion of a heavy particle coupling only to $\psi_1$ and not to $\psi_2$ could be made. Even if theoretical arguments further suggest the existence of $\phi$, its observation generally requires\footnote{This is true at $\mathcal{O}(\Lambda^{-2})$ but when dimension-eight operators at $\mathcal{O}(\Lambda^{-4})$ are included positivity constraints of the Wilson coefficients can exclude new particle states. } building a collider such that its energy reach is sufficiently large to produce $\phi$ on-shell and experiments see a Breit-Wigner distribution similar to the ones displayed in Figure~\ref{fig:EFTFig}.

\subsection{4-Fermi Theory}

An excellent example of the limitation of the top-down approach is again the 4-Fermi theory. Before the derivation of the electroweak theory, the $V-A$ Lorentz structure was one possibility among many, including the scalar $S$ $(\mathcal{O}_{S} = \mathbb{1})$, the pseudoscalar $P$ $(\mathcal{O}_P=\gamma_5)$, the vector $V$ $(\mathcal{O}_V=\gamma_\mu)$, the axial vector $A$ $(\mathcal{O}_A=\gamma_\mu \gamma_5)$, the tensor $T$ $(\mathcal{O}_T=\sigma_{\mu \nu})$ and any linear combinations between them.

\begin{table}[t]
    \centering
    \begin{tabular}{ccc}
    \hline
    \hline
        \multirow{ 2}{*}{Coupling} & \multirow{ 2}{*}{$\frac{\pi}{mE} \frac{d\sigma}{dy}(\nu_{e} e^{-} \to \nu_{e} e^{-})$} & \multirow{ 2}{*}{$\frac{\pi}{mE} \frac{d\sigma}{dy}(\overline{\nu}_{e} e^{-} \to \overline{\nu}_{e} e^{-})$} \\
            &   &   \\
    \hline
        $|C_s|^2 + |C_P|^2$ & $(1-y)^2/2$          & $1/2$  \\
        $|C_T|^2$           & $4(1+y)^2$           & $4(1-2y)^2$    \\
        $\text{Re}\{(C^*_S + C^*_P)C_T\}$ & $-2(1-y^2)$ & $-2(1-2y)$\\
        $|C_{V-A}|^2$       & $4$                  & $4(1-y)^2$     \\
        $|C_{V+A}|^2$       & $4y^2$               & $4y^2$         \\        
    \hline
    \end{tabular}
    \caption{Differential cross-section expressions of neutrino-electron scattering depending on the coefficient(s) of the four-fermion operators. $m$ is the electron mass, $E$ the initial energy of the (anti-)neutrino and $y$ is the rapidity of the outgoing electron.  }
    \label{tab:4FermiXsecExpressions}
\end{table}

In Table 6.1 of Ref.~\cite{Quigg:2013ufa} reproduced in Table~\ref{tab:4FermiXsecExpressions}, we see the different expressions of the differential cross-section for the scattering processes $\nu_e e \to \nu_e e$ and $\overline{\nu}_e e \to \overline{\nu}_e e$ after considering the matrix elements
\begin{equation}
    \mathcal{M} = \sum_i \mathcal{M}_i = \sum_i \widetilde{C}_i (\overline{\nu}_e \mathcal{O}_i e) (\overline{e} \mathcal{O}_i \nu_e).
\end{equation}
The Lorentz structures $\mathcal{O}_i$ have been listed above and $\widetilde{C}_i$ are their corresponding dimensionful Wilson coefficients ($[\widetilde{C}]=-2$). The comparison with data from experiments conducted in the 1950s ultimately led to the $V-A$ operator $\mathcal{O}_{V-A} = \gamma_\mu(1-\gamma_5)$ \cite{Wu:1957my, PhysRev.106.386, PhysRev.109.1015, PhysRev.109.193, PhysRev.109.1860}. The main message of this short section is that, without prior knowledge of the high energy theory, listing all possible interactions and subsequently comparing the predictions with experiments already bore fruit. It pointed towards the structure of the underlying UV-complete theory and was the foundation to establish the SM electroweak theory.

\subsection{The Standard Model Effective Field Theory}

At the time of writing, no statistically significant deviation from the SM has been observed at colliders and no BSM resonance has been detected at the LHC. On the contrary, bounds on new resonances and NP effects are getting stronger. This observation may hint that an energy gap exists between the LHC energy reach and the NP scale $\Lambda$. If $\Lambda$ sits sufficiently above the TeV range then NP naturally decouples from the SM \cite{Appelquist:1974tg}. Therefore, all possible NP effects can be parametrised with an EFT of the SM at energies $E \ll \Lambda$, just as was done with the 4-Fermi theory.

The symmetries of the SM are preserved by higher order operators, \textit{i.e.} the Lorentz symmetry and the $SU_C(3) \times SU_L(2) \times U_Y(1)$ gauge symmetry. This corresponds to the framework of the SMEFT whose Lagrangian $\mathcal{L}_{SMEFT}$ is defined as
\begin{equation}
\label{smeft expansion}
    \mathcal{L}_{SMEFT} = \mathcal{L}_{SM} + \sum_{d=5}^{\infty} \frac{1}{\Lambda^{d-4}} \mathcal{L}_{d}.
\end{equation}
The first term $\mathcal{L}_{SM}$ is the usual SM Lagrangian defined in Section~\ref{sec:SMLag} and the $\mathcal{L}_{d}$ terms contain operators of dimension $d$, $\{\mathcal{O}^{(d)}_i\}_{i,d>4}$, with their respective Wilson coefficients $\{C_i\}_i$. The operators parameterise all new interactions originating from an unknown UV-complete theory. The flavour symmetries are not mentioned here, but they will be discussed below.

We quickly mention the Higgs EFT (HEFT) which is a complementary theory to the SMEFT where the Higgs is considered as a singlet $h$ rather than being part of the $SU_L(2)$ doublet $\varphi$. The Goldstone bosons, also included in the $\varphi$ doublet in the SM(EFT), are then independent of $h$ in HEFT. Thus, Eq.(\ref{eq:HiggsRepresentation}) is separated into two independent pieces, one of them being the real scalar $h$ and the other the non-linear parametrisation of the Goldstone bosons 
\begin{equation*}
    U = e^{i \pi^i \tau^i}.
\end{equation*}
The HEFT operators are not as constrained by $SU_L(2)$ invariance as SMEFT operators. As a result, all SMEFT operators can be written in terms of HEFT operators after EWSSB, but not all HEFT operators can be written in terms of SMEFT operators. Positivity constraints set the boundaries in the parameter space of Wilson coefficients where HEFT operators are theoretically valid but not within SMEFT \cite{Remmen:2024hry, Chakraborty:2024ciu}. See Refs.~\cite{Brivio:2016fzo, Brivio:2017vri} for a complete presentation of HEFT and its comparison to the SMEFT, and Ref.~\cite{Barducci:2025ati} for the translation of Higgs couplings between SMEFT and HEFT.

The infinite number of coefficients makes a complete analysis impossible. Therefore, only the leading terms of the $1/\Lambda$ expansion are kept to focus on the largest NP contributions. At the Lagrangian level, we see that each $\mathcal{O}^{(d)}_i$ is suppressed by $d-4$ powers of $\Lambda$, so let us write the higher-order operators.

We proceed to write the operators as for the SM in Section~\ref{sec:SMLag} by NDA. The action is dimensionless, so the respective dimensions of the fields are
\begin{equation}
    [\varphi] = 1, ~~~ [A_\mu] = 1, ~~~ [\psi] = 3/2.
\end{equation}
In Section~\ref{sec:SMLag}, the SM operators were built with the fields, the covariant derivative $D_\mu$ and the field strength tensors $X_{\mu \nu}$. Their respective dimensions follow from their definitions and are
\begin{equation}
    [D_\mu] = 1, ~~~ [X_{\mu \nu}] = 2 .
\end{equation}
As a result, the higher order covariant operators are  
\begin{equation}
    \mathcal{O}^{(d)} = (\overline{\psi} \psi)^{N_{\overline{\psi} \psi}} (X_{\mu \nu})^{ N_X}  (D_\mu)^{N_D} (\varphi)^{N_{\varphi}},
\end{equation}
where $N_i$ are the respective counting numbers such that, for $d > 4$,
\begin{equation}
    d = 3N_{\overline{\psi} \psi} + 2 N_X + N_D + N_\varphi.
\end{equation}

The first NP contribution sits at the $\Lambda^{-1}$ order through operators with dimension $d=5$. No operators with $d=5$ can be constructed only out of bosonic fields due to $SU_L(2)$ gauge symmetry ($\varphi$ being a $SU_L(2)$ doublet) and contractions of Lorentz indices carried by covariant derivatives. This indicates that potential dimension-five operators involve a fermion pair $\overline{\psi}\psi$, that is $N_{\overline{\psi} \psi} = 1$. Considering the hypercharge invariance, the possible dimension-five operators are further reduced to the ones defined by $N_\varphi=2$ and left-handed lepton fields in the $\overline{\psi}\psi$ pair. Thus, the Lagrangian $\mathcal{L}_5$ consists of the single Weinberg operator $\mathcal{O}^{(5)}$ \cite{PhysRevLett.43.1566}
\begin{equation}
    \mathcal{L}_5 = \frac{C_i}{\Lambda} \left( \overline{L}_{L,i}^C \widetilde{\varphi} \right) \Big( \widetilde{\varphi} L_{L,i} \Big) + h.c.
\end{equation}
where $i$ stands for the flavour index and $l^C$ the charge conjugate of the left-handed lepton field. This term provides a neutrino Majorana mass term after EWSSB and violates the conservation of the lepton number $L$. However, $L$ is assumed to be conserved and therefore $\mathcal{L}_5$ is neglected. This leaves the $\Lambda^{-2}$ order as the leading source of NP effects through operators with dimension $d=6$.

The Lagrangian expansion is truncated at the $\Lambda^{-2}$ order of $\mathcal{L}_6$ corresponding to dimension-six operators,
\begin{equation}
\label{operator expansion}
    \mathcal{L}_{SMEFT} \sim \mathcal{L}_{SM} + \frac{1}{\Lambda^{2}} \mathcal{L}_{6} = \mathcal{L}_{SM} + \sum_{i} \frac{C_i}{\Lambda^{2}} \mathcal{O}_i + \sum_{i} \left( \frac{\hat{C}_i}{\Lambda^{2}} \hat{\mathcal{O}}_i + h.c. \right).
\end{equation}
The operators in $\mathcal{L}_6$ are classified in two categories. The $\mathcal{O}_i$ category represents hermitian operators, whereas the $\hat{\mathcal{O}}_i$ category represents non-hermitian operators. Ultimately, we want the set $\left\{ \mathcal{O}_i , \hat{\mathcal{O}}_{i'} \right\}$ to act as a basis in the dimension-six operator space with their respective Wilson coefficients $\left\{C_i , \hat{C}_i \right\}$. In the absence of absorptive phases\footnote{Absorptive complex phases can arise when new degrees of freedom at low energies create the dimension-six operators as discussed in Ref.\cite{Brehmer:2017lrt}. This case is not considered here. }, $C_i$'s must be real by hermiticity. On the contrary, $\hat{C}_i$'s can be complex.

As $CP$ effects will ultimately be investigated, the following conclusion can already be drawn: $CP$-odd operators originating from the $\mathcal{O}_i$ class immediately introduce $CP$ violating effects with real Wilson coefficients while operators from the $\hat{\mathcal{O}}_i$ class rely on the imaginary part of their coefficients to create such effects. Many global fits have included non-hermitian operators but have overlooked $CP$ violating effects in their analysis by imposing $CP$ symmetry, \textit{i.e.} $\hat{C}_i$'s are considered as real and the hermitian $CP$-odd operators are discarded. See Refs. \cite{Buckley:2015nca,Ellis:2018gqa,Hartland:2019bjb,Brivio:2019ius,Falkowski:2019hvp,Basan:2020btr,Bissmann:2020mfi,Bissmann:2019gfc} for examples of such analyses. Note that the constraints on Wilson coefficients are computed after setting the NP scale $\Lambda$ to some value (usually 1 TeV) since it is experimentally impossible to constrain $C_i$ and $\Lambda$ independently.

The Wilson coefficients from Eq.(\ref{operator expansion}) are implicitly evaluated at the matching scale such that $C_i(\mu)$ and $\hat{C}_i(\mu)$. They are allowed to run between the NP scale $\Lambda$ and the electroweak scale $v$ (or $m_W$) as dictated by the anomalous dimension matrix $\gamma_{ij}$ in Eq.(\ref{eq:AnomalousDimensionMatrix}). The constraints on Wilson coefficients, more precisely on their ratio with $\Lambda$, extracted from the comparison of predictions with data of sensitive processes have to be adjusted before translating them to constraints on some parameters of a UV-complete theory. Moreover, as mentioned in Section~\ref{sec:RunningAndMixing}, off-diagonal coefficients of $\gamma_{ij}$ mix the operators. A non-zero Wilson coefficient at the energy scale of the experiment can be absent at the matching scale, and come from another Wilson coefficient generated by the UV-complete theory.

After the truncation of $\mathcal{L}_{SMEFT}$, the total amplitude of a process sensitive to a $CP$-odd operator is
\begin{equation}
\mathcal{M}=\mathcal{M}_{SM} + \sum_i \mathcal{M}_{i} + \mathcal{O}\left(\Lambda^{-4}\right)\qquad
\end{equation} 
where $\mathcal{M}_{SM}$ is the SM amplitude and $\mathcal{M}_i$ the amplitude involving one vertex from one relevant $CP$-odd dimension-six operator either $\mathcal{O}_i$ or $\hat{\mathcal{O}}_i$. Squaring the total amplitude gives 
\begin{equation}\label{full amplitude expansion}
     \| \mathcal{M} \|^2  = \| \mathcal{M}_{SM} \|^2 + \sum_i \mathcal{M}_{int,i} + \sum_i \| \mathcal{M}_{i} \|^2 + \sum_{i\neq j} \mathcal{M}_{int,ij} + \ldots %\mathcal{O}\left(\Lambda^{-4}\right).
\end{equation}
where $\mathcal{M}_{int,i} \equiv 2 \Re e  \Big\{ \mathcal{M}_{SM}^* \mathcal{M}_{i} \Big\}$ and $\mathcal{M}_{int,ij} \equiv 2 \Re e  \Big\{ \mathcal{M}_{i}^* \mathcal{M}_{j} \Big\}$. Any $CP$ violation from $\mathcal{M}_{SM}$ in the first term of Eq.(\ref{full amplitude expansion}) is neglected as $CP$ violating effects from the CKM phase $\delta_{CKM}$ are negligible\footnote{This is supported by the small value of the Jarlskog invariant generated by $\delta_{CKM}$, as presented in Section~\ref{sec:CPsymmetries}.}. Just like with the operators, any $\Lambda^{-4}$ contributions are consistently neglected to focus on the leading contributions from the $CP$-odd operators, embodied by the $\Lambda^{-2}$ terms. In Eq.(\ref{full amplitude expansion}), they come from the interference between $\mathcal{M}_{SM}$ and $\mathcal{M}_{i}$. The third and fourth terms are respectively the interference amplitude between two dimension-six operators and the square of an amplitude with one dimension-six operator. They are $\Lambda^{-4}$-suppressed so they are discarded. In fact, obtaining the full $\Lambda^{-4}$ contributions would require including dimension-eight operators and considering diagrams with two vertices from dimension-six operators which, by interfering with the SM amplitude, contribute at that order. As a result, we can study independently each dimension-six operator in the amplitude of a sensitive process since the NP contribution is only linear in the Wilson coefficients,
\begin{equation}\label{amplitude expansion}
   \| \mathcal{M} \|^2 \sim  \| \mathcal{M}_{SM} \|^2 + \sum_i \mathcal{M}_{int,i}.
\end{equation}

\subsection{Derivation of the Warsaw basis}
\label{sec:WarsawBasisDerivation}

The complete derivation of the \textit{Warsaw basis} of dimension-six operators is not presented here, but the necessary relations are explained below. The reader is invited to see Refs.~\cite{Buchmuller:1985jz, Grzadkowski:2010es} for the comprehensive derivation.

In Ref.~\cite{Grzadkowski:2010es}, the construction of the Warsaw basis is divided into three parts depending on the number of fermion pairs. It begins with the bosonic operators, then the single-fermionic-current operators and ends with the four-fermion operators. We follow the same steps here.

\begin{table}[t]
    \centering
    \begin{tabular}{c|c|c|c|c}
        Classes & $N_{{\overline{\psi} \psi}}$ & $N_X$ & $N_D$ & $N_\varphi$ \\
        \hline
         $X^3$ & 0 & 3 & 0 & 0 \\
        \hline
         $X^2 D^2$ & 0 & 2 & 2 & 0 \\
        \hline
         $X^2 \varphi^2$ & 0 & 2 & 0 & 2 \\
        \hline
         $X D^4$ & 0 & 1 & 4 & 0 \\
        \hline
         $X \varphi^2 D^2$ & 0 & 1 & 2 & 2 \\
        \hline
         $X \varphi^4$ & 0 & 1 & 0 & 4 \\
        \hline
         $\varphi^2 D^4$ & 0 & 0 & 4 & 2 \\
        \hline
         $\varphi^4 D^2$ & 0 & 0 & 2 & 4 \\
        \hline
         $\varphi^6$ & 0 & 0 & 0 & 6 \\
        \hline
        \hline 
         $\psi^2 X D$ & 1 (V) & 1 & 1 & 0 \\
        \hline
         $\psi^2 X \varphi$ & 1 (T) & 1 & 0 & 1 \\
        \hline
         $\psi^2 D^3$ & 1 (V) & 0 & 3 & 0 \\
        \hline
         $\psi^2 \varphi D^2$ & 1 (S, T) & 0 & 2 & 1 \\
        \hline
         $\psi^2 \varphi^2 D$ & 1 (V) & 0 & 1 & 2 \\
        \hline
         $\psi^2 \varphi^3$ & 1 (S) & 0 & 0 & 3 \\
        \hline
        \hline
        $\psi^4$ & 2 & 0 & 0 & 0 
    \end{tabular}
    \caption{Dimension-six operator classes dictated by NDA and Lorentz invariance. The expressions of the operators are deduced from a combination of gauge invariance, integration by parts, equations of motion, Bianchi and Fierz identities.}
    \label{tab:NDADim6Operators}
\end{table}

In the case of bosonic operators, Lorentz invariance requires the number of derivatives to be even without fermionic fields\footnote{Otherwise, some Lorentz indices would not be summed over. Gamma matrices associated to vector fermionic bilinears could be contracted with the remaining Lorentz indices but, without gamma matrices, unsummed indices explicitly break Lorentz invariance. }. The derivation thus starts with operators built with bosonic fields and an even number of covariant derivatives. The counting numbers used to generate dimension-six bosonic operators separate them into nine operator classes, as listed in the first nine rows of Table~\ref{tab:NDADim6Operators}.

In the following, the covariant derivatives are replaced as much as possible by SM fields either by using their commutator 
\begin{equation}
\label{eq:CovariantDerivativesCommutator}
    \left[ D_{\mu}, D_{\nu} \right] = i g X^{\alpha}_{\mu \nu} t^{\alpha},
\end{equation}
or by the equations of motion of the SM fields on which they act. The equations of motion are truncated at the $\Lambda^{0}$ order (that is the Euler-Lagrange equations derived from $\mathcal{L}_{SM}$) not to generate $\Lambda^{-4}$ effects after substitution.

If covariant derivatives do not act on SM fields or, in the case of the Higgs field, if there are not enough contracted covariant derivatives, the equations of motion are recovered using integration by parts with additional vanishing total derivatives. The Bianchi identity is also used to remove terms with covariant derivatives contracted with dual tensors $\widetilde{X}_{\mu \nu}$,
\begin{equation}
\label{eq:BianchiIdentity}
    D_\nu \widetilde{X}^{\nu \mu} = 0.
\end{equation}

The $X\varphi^4$ class actually does not contain any operator since Lorentz indices of the traceless tensor $X_{\mu \nu}$ cannot be contracted. Using Eq.(\ref{eq:CovariantDerivativesCommutator}), operators from the $X D^4$ class are equivalent to linear combinations of operators from the $X^2 D^2$ class. Lorentz indices of the antisymetric $X_{\mu \nu}$ have to be contracted with two covariant derivatives, so
\begin{equation}
\label{eq:CovDerivativesWithX}
	D_{\mu} D_{\nu} X^{\mu \nu} \sim [D_{\mu}, D_{\nu}] X^{\mu \nu}  \xrightarrow{\text{(\ref{eq:CovariantDerivativesCommutator})}} X_{\mu \nu} X^{\mu \nu},
\end{equation}
or the covariant derivatives are contracted with an antisymetric Levi-Civita tensor
\begin{equation}
\label{eq:CovDerivativesWithEpsilon}
	\epsilon_{\mu \nu \rho \sigma} D^{\mu} D^{\nu} \sim  \epsilon_{\mu \nu \rho \sigma} [D^{\mu}, D^{\nu}]  \xrightarrow{\text{(\ref{eq:CovariantDerivativesCommutator})}}   \epsilon_{\mu \nu \rho \sigma} X^{\mu \nu} .
\end{equation}
The $\varphi^2 D^4$, $X \varphi^2 D^2$ and $X^2 D^2$ classes are redundant as the equations of motion of the gauge vector bosons and the Higgs relate them to operators in classes with single-fermion-current, or the $X^3$, $X^2 \varphi^2$, $\varphi^6$ and $\varphi^4 D^2$ bosonic classes.

For the $\varphi^2 D^4$ class, covariant derivatives are either contracted with a Levi-Civita tensor which makes the operators equivalent to operators in the $X \varphi^2 D^2$ class, as demonstrated in Eq.(\ref{eq:CovDerivativesWithEpsilon}), or they are contracted with each other. Integration by parts ensures that covariant derivatives act on the same Higgs field and the Higgs equation of motion can be used to remove these operators from the basis.

Operators from the $X \varphi^2 D^2$ class cannot have contracted covariant derivatives due to the traceless $X_{\mu \nu}$ (preventing the use of the Higgs equation of motion). If both covariant derivatives act on different Higgs fields, integration by part makes them act on the same Higgs up to total derivatives.  Eq.(\ref{eq:CovDerivativesWithX}) makes operators with both covariant derivatives acting on a Higgs or $X_{\mu \nu}$ equivalent to operators of the $X^2 \varphi^2$ class. The last possibility is one covariant derivative acting on one Higgs and the other on $X_{\mu \nu}$ in which case either $X$ is dual and the operator vanished by the Bianchi identity in Eq.(\ref{eq:BianchiIdentity}), or the equation of motion associated to $X$ removes the operator from the basis.

Operators from the $X^2 D^2$ class with contracted covariant derivatives can be removed by first using the Bianchi identity
\begin{equation*}
	X_{\mu \nu} D^\rho D_\rho X^{\mu \nu} = - X_{\mu \nu} D^\rho D^\mu X^{\nu}_{~\rho} - X_{\mu \nu} D^\rho D^{\nu} X_{\rho}^{~\mu}.
\end{equation*}
After applying Eq.(\ref{eq:CovariantDerivativesCommutator}), the resulting operators are either equivalent to operators from $X^3$ or made redundant by the equation of motion of $X$. Eqs.(\ref{eq:CovDerivativesWithX}) and (\ref{eq:CovDerivativesWithEpsilon}) respectively remove operators  in which the covariant derivatives are contracted with a single $X$ or a Levi-Civita tensor. Lastly, if the covariant derivatives are contracted with different tensors, it is always possible to use the equation of motion of on $X$ to make the operators redundant.

So, the non-redundant classes of bosonic operators are
\begin{equation*}
    \{ X^3, X^2 \varphi^2, \varphi^6, \varphi^4 D^2 \}.
\end{equation*}

When a single $(\overline{\psi}\psi)$ pair is considered, it is complemented by combinations of bosonic fields and covariant derivatives. Before listing the operator classes, the left-handed fermion field $\psi$ is constructed with the left-handed fermion fields and the charge conjugate of the right-handed fermion fields, thus left-handed, such that $\psi =  (L_L, e_R^C, Q_L, u_R^C, d_R^C)$.

Three fermionic currents are possible with $\psi$: the scalar $\psi^T_i C \psi_j$, the vector $\overline{\psi}_i \gamma_\mu \psi_j$ and the tensor $\psi_i^T C \sigma_{\mu \nu} \psi_j$. The $C$ operator in between the fermions in the scalar and tensor currents allows the contraction of the left-handed spinors. The six possible operator classes allowed by the contraction of Lorentz indices are presented in rows ten to fifteen of Table~\ref{tab:NDADim6Operators}. The type of fermionic current involved in the operator class is written in brackets where S stands for the scalar current, V the vector current and T the tensor current.

Operators made of scalar and tensor currents contain an odd number of Higgs fields, so the $(\overline{\psi}\psi)$ pairs must be $SU_L(2)$ doublets. Thus, one of the fermions is left-handed while the other is right-handed. The charge conjugation of the right-handed fields becomes unnecessary in $\psi$ and the fermion pairs are Dirac products of the form $\overline{\psi}\psi$. The scalar and the tensor currents can be respectively written as $\overline{\psi_i} \psi_j$ and $\overline{\psi_i} \sigma_{\mu \nu} \psi_j$.

In the $\psi^2 D^3$ class, at least one covariant derivative is contracted with the $\gamma_\mu$ and the two remaining derivatives are contracted together. Integration by parts restricts the operators to the ones in which the single derivative acts on the fermion field $\psi$, so this class is equivalent to the $\psi^2 \varphi D^2$ by the fermion equation of motion.

The operators from the $\psi^2 \varphi D^2$ class involving a tensor current are reduced to operators from $\psi^2 X \varphi$ by Eq.(\ref{eq:CovariantDerivativesCommutator}) when both derivatives act on the same field, either a Higgs or a fermion,
\begin{equation*}
	\sigma^{\mu \nu} D_{\mu} D_{\nu} \sim \sigma^{\mu \nu} [D_{\mu}, D_{\nu}]  \xrightarrow{\text{(\ref{eq:CovariantDerivativesCommutator})}} \sigma^{\mu \nu} X_{\mu \nu} . 
\end{equation*}
Otherwise, if the derivatives act on different fields, we can integrate by parts to limit the derivation to
\begin{equation*}
\begin{split}
	(D_\mu \varphi) \overline{\psi} \sigma^{\mu \nu} D_{\nu} \psi  &  \sim  (D_\mu \varphi) \overline{\psi} (\gamma^\mu \slashed{D} - \slashed{D} \gamma^\mu) \psi , \\
		&  \sim (D_\mu \varphi) \overline{\psi} \gamma^\mu \slashed{D} \psi - (D_\mu \varphi) \overline{\psi} D^{\mu} \psi.
\end{split}
\end{equation*}
The definition of $\sigma^{\mu \nu}$ was used in the first line and the anticommutation property of gamma matrices in the second. The first resulting operator is made redundant by the equation of motion of $\psi$ and the other belongs to another operator class.

Now for operators involving a scalar current, when both derivatives acting on the Higgs, the equation of motion directly removes these operators from the basis. If both derivatives act on a single fermion, we can use 
\begin{equation}
\label{eq:GammaMatricesRelation1}
	\eta_{\mu \nu} = \gamma_\mu \gamma_\nu + i \sigma^{\mu \nu}, 
\end{equation}
such that
\begin{equation*}
	\varphi \overline{\psi} D_{\mu} D^{\mu} \psi \sim \varphi \overline{\psi} \slashed{D} \slashed{D} \psi + i \varphi \overline{\psi} \sigma^{\mu \nu} \psi .
\end{equation*}	
The first term is removed from the basis by the equation of motion and the latter belongs to operators with tensor currents which were proved to be redundant. Now, if the derivatives act on different fields, 
\begin{equation*}
\begin{split}
	(D_{\mu} \varphi) \overline{\psi} D^{\mu} \psi  & \sim  (D_\mu \varphi) \overline{\psi} (\gamma^\mu \slashed{D} + \slashed{D} \gamma^\mu) \psi , \\
	& \sim  (D_\mu \varphi) \overline{\psi} \gamma^\mu \slashed{D} \psi - \overline{\psi} \overleftarrow{\slashed{D}} \gamma^\mu \psi (D_\mu \varphi) - \overline{\psi} \gamma^\nu \gamma^\mu \psi D_{\nu} D_{\mu} \varphi .
\end{split}
\end{equation*}
The first two term can be respectively removed by the equation of motion of $\psi$ and $\overline{\psi}$, while the last term is made redundant with Eq.(\ref{eq:GammaMatricesRelation1}) and the Higgs equation of motion.

Finally, in the $\psi^2 X D$ class, the covariant derivative must be contracted with the $X_{\mu \nu}$. So, if the derivative acts on the $X_{\mu \nu}$, we can use its equation of motion to make the operators redundant, or the Bianchi identities if it is a dual tensor. If the derivative acts on the fermion, 
\begin{equation*}
\begin{split}
	X^{\mu \nu} \overline{\psi} \gamma_{\mu} D_{\nu} \psi  & \sim X^{\mu \nu} \overline{\psi} (\gamma_{\mu} \gamma_{\nu} \slashed{D} + \gamma_{\mu} \slashed{D} \gamma_{\nu} ) \psi  ,\\
	& \sim X^{\mu \nu} \overline{\psi} (\gamma_{\mu} \gamma_{\nu} \slashed{D} - \slashed{D} \gamma_{\mu} \gamma_{\nu} ) + X^{\mu \nu} \overline{\psi} \gamma_{\nu} D_{\mu} \psi \psi.  
\end{split} 
\end{equation*}
The last term is identical to the initial operator, so it is not necessary to keep it. After applying an integration by parts, we get
\begin{equation*}
	X^{\mu \nu} \overline{\psi} \gamma_{\mu} D_{\nu} \psi \sim  X^{\mu \nu} \overline{\psi} \gamma_{\mu} \gamma_{\nu} \slashed{D} \psi + \overline{\psi} \overleftarrow{\slashed{D}} \gamma_{\mu} \gamma_{\nu}  \psi X^{\mu \nu} + \overline{\psi} \gamma_{\rho} \gamma_{\mu} \gamma_{\nu} \psi D^\rho X^{\mu \nu}.  
\end{equation*}
The first two term can be respectively removed by the equation of motion of $\psi$ and $\overline{\psi}$, while the last term is made redundant by using the following gamma-matrix relation
\begin{equation}
	\label{eq:GammaMatricesRelation2}
		\gamma_\mu \gamma_\nu \gamma_\rho = \eta_{\mu \nu} \gamma_\rho + \eta_{\nu \rho} \gamma_\mu - \eta_{\mu \rho} \gamma_\nu + i \epsilon_{\mu \nu \rho \sigma} \gamma^\sigma \gamma_5 , 
\end{equation}
and the equation of motion of $X$.

As a result, the non-redundant classes of operators are
\begin{equation*}
    \{ \psi^2 \varphi^3, \psi^2 X \varphi, \psi^2 \varphi^2 D\}.
\end{equation*}

Now for the four-fermion operators, there are hundreds of possibilities. Contraction of Lorentz indices required by Lorentz invariance indicates that the fermion currents cannot be multiplied with another type. This means that scalar currents are combined with scalar currents, vector currents with vector currents, and tensor currents with tensor currents. Operators of the form $\overline{\psi} \psi \psi \psi$ or $\overline{\psi} \overline{\psi} \overline{\psi} \psi$ are never obtained with the current products and their hermitian conjugates.

Then, the content of $\psi$ is selected to generate operators with no hypercharge. If possible, the Fierz identity
\begin{equation}
    (\overline{\psi}_L \gamma_\mu \chi_L)(\overline{\chi}_L \gamma^\mu \psi_L) = (\overline{\psi}_L \gamma_\mu \psi_L)(\overline{\chi}_L \gamma^\mu \chi_L)
\end{equation}
is used to rearrange the fields in order to write currents with zero hypercharge instead of products of vector currents carrying opposite hypercharges. Therefore, four-fermion operators are written using $SU_L(2)$ doublets and singlets to form $(\overline{L}L)(\overline{L}L)$, $(\overline{L}L)(\overline{R}R)$ and $(\overline{R}R)(\overline{R}R)$ operators. Some products can be further made redundant by using the algebra generator relations
\begin{equation}
\begin{split}
    T^{A}_{ij}T^A_{kl} & = \frac{1}{2} \delta_{il} \delta_{jk} - \frac{1}{6} \delta_{ij} \delta_{kl}, \\
    \tau^I_{ij}\tau^I_{kl} & = 2 \delta_{il} \delta_{jk} - \delta_{ij} \delta_{kl}.
\end{split}
\end{equation}

However, the vector current products do not cover all combinations. The following operators are obtained from scalar and tensor currents :
\begin{multline}
\label{eq:NonCurrentFourFermion}
    \{ (\overline{L}_L \overline{e}^C_R) (d^C_R Q_L), (Q_L u^C_R) (Q_L d^C_R), (L_L e^C_R) (Q_L u^C_R), \\ 
    (Q_L Q_L) (Q_L L_L), (d^C_R u^C_R) (u^C_R e^C_R), (Q_L Q_L) (\overline{u}^C_R \overline{e}^C_R), (Q_L L_L) (\overline{u}^C_R \overline{d}^C_R) \}.
\end{multline}
The latter four operators violate the baryon number $B$\footnote{They also violate $L$ but preserve the difference $B-L$.} while the other three four-fermion operators conserve it. The number of singlets in the Lorentz, $SU_L(2)$ and $SU_C(3)$ groups allows only one operator for the three types in Eq.(\ref{eq:NonCurrentFourFermion}) made of two $\psi$ and two $\overline{\psi}$. The Fierz identity
\begin{multline}
    (\psi^T_{1L} C \sigma_{\mu \nu} \psi_{2L})(\psi^T_{3L} C \sigma^{\mu \nu} \psi_{4L}) = -4 (\psi^T_{1L} C \psi_{2L})(\psi^T_{3L} C \psi_{4L}) \\ 
    -8 (\psi^T_{1L} C \psi_{4L})(\psi^T_{3L} C \psi_{2L})
\end{multline}
transforms tensor currents into scalar currents, so the other types of Eq.(\ref{eq:NonCurrentFourFermion}) with $\psi^4$ have only one operator too.

It is important to mention that other bases and Lagrangians were constructed for dedicated studies. Notably, the Strongly-Interacting-Light-Higgs (SILH) Lagrangian is aimed at describing UV-complete theories where the Higgs field emerges from a strongly interacting gauge sector, similarly to hadrons from the QCD sector of the SM. The Hagiwara-Ishihara-Szalapski-Zeppenfeld (HISZ) basis targets NP modifications in the Higgs and weak gauge sectors.

To go from one basis to another, field redefinitions might be required. Physical observations are independent of these transformations, but operators can be shifted from one class to another by field redefinitions. They are also used to recover canonical kinetic terms when some higher-dimensional operators contribute to the propagator of a field or to some input parameter. In Ref.~\cite{Brivio:2017vri} for instance, the Higgs field is redefined since the operators $\mathcal{O}_{\varphi \square}$ and $\mathcal{O}_{\varphi D}$ add vertices to the Higgs propagator. In the unitary gauge,
\begin{equation}
    \varphi = \frac{1}{\sqrt{2}} 
    \begin{pmatrix}
        0 \\ (1 + \delta_{h}) h + v_T
    \end{pmatrix}, 
\end{equation}
where $\delta_h = (C_{\varphi \square} - C_{\varphi D}/4) v_T^2$ and $v_T=(1+3C_{\varphi} v^2 /8\lambda) v$ is the vacuum expectation value of the canonical Lagrangian.

The anomalous dimension matrix for the Wilson coefficients corresponding to operators in the Warsaw basis has been computed at one loop in Refs.~\cite{Jenkins:2013zja, Jenkins:2013wua, Alonso:2013hga} (current SMEFT RGEs involve dimension-eight operators \cite{Chala:2021pll, DasBakshi:2022mwk, Bakshi:2024wzz, Wu:2025qto, DasBakshi:2026ief}). So, the basis remains complete beyond tree-level which is a non-trivial result important for loop computations necessary for precision studies. However, Fierz identities and gamma matrices relations cannot hold simultaneously in dimensions different than 4. Thus, during the DR procedure, non-redundant operators outside of the Warsaw basis must be taken into account until the limit $\epsilon \to 0$ is applied. These operators are called \textit{evanescent}, see Ref.~\cite{Born:2026xkr} for a discussion on evanescent operators in the anomalous matrix at two loop and Ref.~\cite{Fuentes-Martin:2022vvu} in the matching.

\subsection{Counting of the $CP$-odd Operators}

The $CP$-odd operators of the Warsaw basis have been counted in Appendix A of Ref.~\cite{Alonso:2013hga}: 23 $CP$-odd operators for one fermion generation and 1149 for three fermion generations. Table~\ref{complete CPV operator basis} lists the operators in the unbroken phase for one fermion generation and sorts them in the 6 classes mentioned above. We have not represented the $CP$-odd operators for three fermion generations since they will not be necessary in the rest of this work. The left column of Table~\ref{complete CPV operator basis} is composed of the pure gauge boson $X^3$ and the Higgs-gauge boson $X^2\varphi^2$ classes. These operators belong to the $\mathcal{O}_i$ category. The four-fermion $\psi^4$ and the scalar-fermion $\psi^2 \varphi^3$ classes are presented in the central column. For one generation, only 5 operators are present in the $\psi^4$ class: one $(\Bar{L}R)(\Bar{R}L)$ operator and four $(\Bar{L}R)(\Bar{L}R)$ operators. The $\psi^2 \varphi^3$ class includes three operators that modify the interactions between the Higgs and fermions. Finally, the right column contains the dipole class $X\psi^2\varphi$ with its eight operators. The operators from the central and right columns are not hermitian thus belong to the $\hat{\mathcal{O}}_i$ category. The Wilson coefficients
\begin{equation}
    \{ C_{\Tilde{G}GG}, C_{\Tilde{W}WW}, C_{\varphi\Tilde{G}}, C_{\varphi\Tilde{W}}, C_{\varphi\Tilde{B}}, C_{\varphi\Tilde{W}B} \}
\end{equation}
are real and $CP$ is broken when they are non-vanishing, while, for one fermion generation,
\begin{multline}
    \{ C_{u\varphi}, C_{d\varphi}, C_{e\varphi}, C_{\varphi u d}, C_{ledq}, C_{lequ}^{(1)}, C_{lequ}^{(3)}, C_{quqd}^{(1)},C_{quqd}^{(8)}, \\ C_{uG}, C_{uW}, C_{uB}, C_{dG}, C_{dW}, C_{dB}, C_{eW}, C_{eB} \}
\end{multline}
are complex numbers and their imaginary parts induce new sources of $CP$ violation.

\begin{table}
\centering
\scalebox{0.81}{%
\begin{tabular}{|c|c|c|c|c|c|}
\hline 
   \multicolumn{2}{|c|}{$(X^3)$}  & \multicolumn{2}{|c|}{$(\psi^2\varphi^3)$}  &  \multicolumn{2}{|c|}{$(\psi^2\varphi^2 D)$}  \\
\hline
  $O_{\Tilde{G}GG}$ & $f^{ABC}\widetilde{G}_\mu^{A\nu} G_\nu^{B\rho} G_\rho^{C\mu}$  &  $O_{u\varphi}$ & $(\varphi^\dagger\varphi)(\overline{Q} u \Tilde{\varphi})$ & $O_{\varphi u d}$ & $i(\Tilde{\varphi}^\dagger D_\mu\varphi)(\overline{u} \gamma^\mu d)$  \\
  $O_{\Tilde{W}WW}$ & $\epsilon^{IJK}\widetilde{W}_\mu^{I\nu} W_\nu^{J\rho} W_\rho^{K\mu}$ & $O_{d\varphi}$ & $(\varphi^\dagger\varphi)(\overline{Q} d \varphi)$ & & \\
   & & $O_{e\varphi}$ & $(\varphi^\dagger\varphi)(\overline{L} e \varphi)$ & & \\
\hline \hline 
  \multicolumn{2}{|c|}{$(X^2\varphi^2)$} & \multicolumn{2}{|c|}{$(\psi^4)$} & \multicolumn{2}{|c|}{$(X\psi^2\varphi)$} \\
\hline
  $O_{\varphi \Tilde{G}}$ & $\varphi^\dagger\varphi\widetilde{G}_{\mu\nu}^A G^{A\mu\nu}$ & $O_{ledq}$ & $(\overline{L}^j e)(\overline{d} Q^j)$ & $O_{uG}$ & $(\overline{Q} \sigma^{\mu\nu}T^A u)\Tilde{\varphi}G^A_{\mu\nu}$  \\
  $O_{\varphi \Tilde{W}}$ & $\varphi^\dagger\varphi\widetilde{W}^I_{\mu\nu} W^{I\mu\nu}$ & $O_{lequ}^{(1)}$ & $(\overline{L}^j e)\epsilon_{jk}(\overline{Q}^k u)$ & $O_{uW}$ & $(\overline{Q} \sigma^{\mu\nu}u)\tau^I\Tilde{\varphi}W^I_{\mu\nu}$  \\
  $O_{\varphi \Tilde{B}}$ & $\varphi^\dagger\varphi\widetilde{B}_{\mu\nu} B^{\mu\nu}$  & $O_{lequ}^{(3)}$ & $(\overline{L}^j \sigma^{\mu\nu} e)\epsilon_{jk}(\overline{Q}^k \sigma_{\mu\nu} u)$  & $O_{uB}$ & $(\overline{Q}\sigma^{\mu\nu}u)\Tilde{\varphi}B_{\mu\nu}$ \\
  $O_{\varphi \Tilde{W}B}$ & $\varphi^\dagger\tau^I\varphi\widetilde{W}^I_{\mu\nu} B^{\mu\nu}$  & $O_{quqd}^{(1)}$ & $(\overline{Q}^j u)\epsilon_{jk}(\overline{Q}^k d)$ & $O_{dG}$ & $(\overline{Q}\sigma^{\mu\nu}T^A d)\varphi G^A_{\mu\nu}$  \\
    &  & $O_{quqd}^{(8)}$ & $(\overline{Q}^j T^A u)\epsilon_{jk}(\overline{Q}^k T^A d)$ & $O_{dW}$ & $(\overline{Q} \sigma^{\mu\nu} d)\tau^I\varphi W^I_{\mu\nu}$ \\
    &  & & & $O_{dB}$ & $(\overline{Q} \sigma^{\mu\nu} d)\varphi B_{\mu\nu}$ \\
   & & & & $O_{eW}$ & $(\overline{L} \sigma^{\mu\nu} e )\tau^I\varphi W^I_{\mu\nu}$ \\
   & & & & $O_{eB}$ & $(\overline{L} \sigma^{\mu\nu} e)\varphi B_{\mu\nu}$ \\
\hline
\end{tabular}
}
\caption{List of 23 $CP$-odd dimension-6 operators present in the Warsaw basis for one fermion generation when the $B$ and $L$ quantum numbers are conserved. The explicit chirality of the fields has been dropped for readability.}
\label{complete CPV operator basis}
\end{table}

However, one important comment is worth highlighting about $CP$-odd operators from both $\psi^4$ and $\psi^2\varphi^2 D$ classes. The $\psi^4$ class is the most sensitive to the number of generations and is by far the largest class in terms of $CP$-odd operators. As soon as more generations are taken into consideration, the number of operators increases significantly with up to 1014 $CP$-odd four-fermion operators for three generations. The reason is that the central column of Table \ref{complete CPV operator basis} would include additional four-fermion operators with different chirality configurations, \textit{i.e.} it would also contain $(\Bar{L}L)(\Bar{L}L)$, $(\Bar{R}R)(\Bar{R}R)$ and $(\Bar{L}L)(\Bar{R}R)$ operators if these operators mix fermions of different generations. For example, if the two fermion fields in one fermion pair are from different generations, the operator is no longer hermitian: the operator $(\Bar{e}\gamma_\mu e)(\Bar{e}\gamma^\mu e)$ is hermitian and therefore $CP$-even while the operator with a substitution of an electron by a muon, $(\Bar{\mu}\gamma_\mu e)(\Bar{e}\gamma^\mu e)$, is non-hermitian and can induce $CP$ violation if its Wilson coefficient has an imaginary part. The argument of the generation indices also stands in the $\psi^2\varphi^2 D$ class and provides additional $CP$-odd operators, up to 30 for three generations instead of 1 for one generation. The complete list of four-fermion and $\psi^2\varphi^2 D$ operators is given in Table 3 of Ref.~\cite{Grzadkowski:2010es}.

In the case of $CP$ violation, Section~\ref{sec:CPsymmetries} demonstrated that field redefinitions are necessary to construct $CP$ invariants in the form of the Jarlskog invariant in the SM.  For the SMEFT, Ref.\cite{Bonnefoy:2021tbt} counted 699 $CP$-odd (flavour) invariants instead of 1149. If the Jarlskog invariant was not neglected, the number of $CP$ invariants would increase to 1551 due to the interference of $CP$-even dimension-six operators with $CP$ effects from the SM as demonstrated by Ref.~\cite{Bonnefoy:2023bzx}.

\section{Simplified Models Approach}
\label{sec:BasicGeneric}

Testing all UV-complete models is extremely time consuming, computationally expensive and demanding in human resources which makes it ineffective. The SMEFT provides a good alternative where, as demonstrated in Section~\ref{sec:BasicSMEFT}, higher-order operators are built with existing degrees of freedom. However, fitting EFT parameters is difficult since there are multiple processes affected at different orders by a large number of operators. Blind directions can appear when operators contribute with opposing signs to the observable(s) of interest. Thus, global fits of EFT parameters are an important goal of particle physics but remain a non-trivial task to perform. The SMEFT relies on a large scale separation between the SM and UV-complete theories but there is still room for hidden on-shell resonances in readily accessible rarer processes.

A complementary method in between UV-complete models and EFTs exists in the form of \textit{simplified models}. Simplified models have been proposed even before the discovery of the Higgs boson \cite{LHCNewPhysicsWorkingGroup:2011mji}. At the time, physicists were exploring supersymmetry theories where a large spectrum of new particles was introduced with a large number of parameters. Simplified models allowed the reduction of the number of particles and parameters to be considered simultaneously in amplitudes and cross-sections, see Refs. \cite{Alwall:2008ag, Rentala:2011mr} for instance. Simplified models are also widely used in dark matter studies. Different dark matter candidates are added to the SM, occasionally with a mediator to bridge the SM and dark sectors, and their parameter space is constrained by the DM density and collider data. Recent analyses of such simplified models are presented in Refs. \cite{Kolay:2024wns, Arina:2023msd, PerezAdan:2023phe, Albert:2022xla, Ghosh:2022zef, Chang:2023cki, Chang:2022jgo} and the FeynRules database "Simple extensions of the SM/Simplified DM Models" contains simplified DM models for NLO. It is interesting to note that some simplified models contain higher dimension operators (in some limit) even though they are usually associated with EFTs.

\begin{figure}[t]
    \centering
    \includegraphics[width=\linewidth]{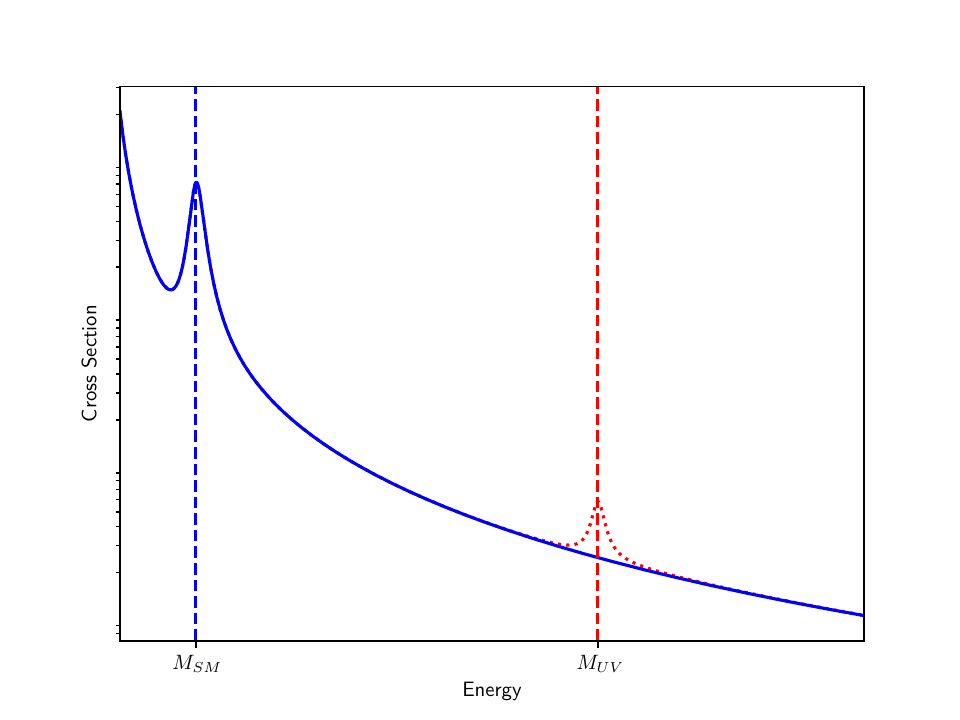}
    \caption{Representation of the cross-section with respect to the energy for the SM and a simplified model. As in Figure~\ref{fig:EFTFig}, the SM cross-section is shown in blue and the cross-section of the simplified model is shown by the red dotted line. The SM resonance is still highlighted by the vertical dashed blue line but the BSM resonance (in red) is here in the energy reach. The latter peak is small because of a small value of the coupling, and remains visible since the y axis is on a logarithmic scale.}
    \label{fig:EFTFigBis}
\end{figure}

Lately, rather than reducing the number of particles in an overcrowded UV-complete theory as in SUSY models, simplified models have found a new purpose in collider physics by extending the spectrum of the SM with new particles in a more systematic manner \cite{Baker:2024xwh}. A good example of a simplified model would be to add the singlet scalar $\phi$ introduced in the toy model presented in Section~\ref{sec:ToyModelMatching} to the SM. If $\phi$ was coupled universally to leptons then the s-channel process in a lepton collider $l \overline{l} \to \phi \to l \overline{l}$ is dominant at $\sqrt{\hat{s}} \sim m_{\phi}$. A Breit-Wigner distribution would be a sign of the presence of $\phi$. The absence of any deviation from the SM in any collider could be explained by a small coupling between the scalar and the leptons. The Breit-Wigner peak would be small in that case as displayed in Figure~\ref{fig:EFTFigBis} and be hidden in the SM background. Searches for localized deviations in differential cross-section distributions are sometimes referred to as \textit{bump searches}. They require strong statistical evaluations of the signal prediction and of the relevant backgrounds where the resonance could hide.

\begin{table}[t]
    \centering
    \begin{tabular}{|c|c|c|c|c|}
      \hline
        Index  & 1 & 2 & 3 & 4 \\
        Name   & $\mathcal{S}$ & $\mathcal{S}_1$ & $\mathcal{S}_2$ & $\varphi$ \\
        Irrep. & $(1,1)_{0}$ & $(1,1)_{1}$ & $(1,1)_{2}$ & $(1,2)_{\frac{1}{2}}$ \\
      \hline
        Index  & 5 & 6 & 7 & 8 \\
        Name   & $\Xi$ & $\Xi_1$ & $\Theta_1$ & $\Theta_3$ \\
        Irrep. & $(1,3)_{0}$ & $(1,3)_{1}$ & $(1,4)_{\frac{1}{2}}$ & $(1,4)_{\frac{3}{2}}$ \\
      \hline
        Index  & 9 & 10 & 11 & 12 \\
        Name   & $\omega_4$ & $\omega_1$ & $\omega_2$ & $\Pi_1$ \\
        Irrep. & $(3,1)_{-\frac{4}{3}}$ & $(3,1)_{-\frac{1}{3}}$ & $(3,1)_{\frac{2}{3}}$ & $(3,2)_{\frac{1}{6}}$ \\
      \hline
        Index  & 13 & 14 & 15 & 16 \\
        Name   & $\Pi_7$ & $\zeta$ & $\Omega_1$ & $\Omega_2$ \\
        Irrep. & $(3,2)_{\frac{7}{6}}$ & $(3,3)_{-\frac{1}{3}}$ & $(6,1)_{\frac{1}{3}}$ & $(6,1)_{-\frac{2}{3}}$ \\
      \hline
        Index  & 17 & 18 & 19 &  \\
        Name   & $\Omega_4$ & $\Upsilon_1$ & $\Phi$ &  \\
        Irrep. & $(6,1)_{\frac{4}{3}}$ & $(6,3)_{\frac{1}{3}}$ & $(8,2)_{\frac{1}{2}}$ &  \\
      \hline
    \end{tabular}
    \caption{Scalar part of the simplified model dictionary from Ref.~\cite{Li:2023cwy}. The respective names of the particles come from Ref.~\cite{deBlas:2017xtg}. The irreducible representations of every particle are presented in the third line following the notation $(SU_C(3), SU_L(2))_{Y}$}.
    \label{tab:GenericModelsScalars}
\end{table}

\begin{table}[t]
\centering
\begin{tabular}{|c|c|c|c|c|c|}
	\hline
	Index  & 1 & 2 & 3 & 4 & 5 \\
	Name   & $N$ & $E^c$ & $\Delta^c_1$ & $\Delta^c_3$ & $\Sigma$ \\
	Irrep. & $(1,1)_{0}$ & $(1,1)_{1}$ & $(1,2)_{\frac{1}{2}}$ & $(1,2)_{\frac{3}{2}}$ & $(1,3)_{0}$\\
	\hline
	Index  & 6 & 7 & 8 & 9 & 10 \\
	Name   & $\Sigma^c_1$ &   & $D$ & $U$ & $Q_5$ \\
	Irrep. & $(1,3)_{1}$ & $(1,4)_{\frac{1}{2}}$ & $(3,1)_{-\frac{1}{3}}$ & $(3,1)_{\frac{2}{3}}$ & $(3,2)_{-\frac{5}{6}}$ \\
	\hline
	Index  & 11 & 12 & 13 & 14 &  \\
	Name   & $Q_1$ & $Q_7$ & $T_1$ & $T_2$ &  \\
	Irrep. & $(3,2)_{\frac{1}{6}}$ & $(3,2)_{\frac{7}{6}}$ & $(3,3)_{-\frac{1}{3}}$ & $(3,3)_{\frac{2}{3}}$ & \\
	\hline
\end{tabular}
\caption{Fermion part of the simplified model dictionary from Ref.~\cite{Li:2023cwy} and the names from Ref.~\cite{deBlas:2017xtg}. The particle in the 7th position does not have a name as it was absent from Ref.~\cite{deBlas:2017xtg} and it was left empty in Ref.~\cite{Li:2023cwy}. }
\label{tab:GenericModelsFermions}
\end{table}

\begin{table}[t]
\centering
\begin{tabular}{|c|c|c|c|c|c|}
	\hline
	Index  & 1 & 2 & 3 & 4 & 5 \\
	Name   & $\mathcal{B}$ & $\mathcal{B}_1$ & $\mathcal{L}_3^\dagger$ & $\mathcal{W}$ & $\mathcal{U}_2$ \\
	Irrep. & $(1,1)_{0}$ & $(1,1)_{1}$ & $(1,2)_{\frac{3}{2}}$ & $(1,3)_{0}$ & $(3,1)_{\frac{2}{3}}$\\
	\hline
	Index  & 6 & 7 & 8 & 9 & 10 \\
	Name   & $\mathcal{U}_5$ & $\mathcal{Q}_5$ & $\mathcal{Q}_1$ & $\mathcal{X}$ & $\mathcal{Y}_1^\dagger$ \\
	Irrep. & $(3,1)_{\frac{5}{3}}$ & $(3,2)_{-\frac{5}{6}}$ & $(3,2)_{\frac{1}{6}}$ & $(3,3)_{\frac{2}{3}}$ & $(6,2)_{-\frac{1}{6}}$ \\
	\hline
	Index  & 11 & 12 & 13 & 14 &  \\
	Name   & $\mathcal{Y}_5^\dagger$ & $\mathcal{G}$ & $\mathcal{G}_1$ & $\mathcal{H}$ &  \\
	Irrep. & $(6,2)_{\frac{5}{6}}$ & $(8,1)_{0}$ & $(8,1)_{1}$ & $(8,3)_{0}$ & \\
	\hline
\end{tabular}
\caption{Vector part of the simplified model dictionary from Ref.~\cite{Li:2023cwy} and the names from Ref.~\cite{deBlas:2017xtg}.}
\label{tab:GenericModelsVectors}
\end{table}

In more realistic simplified models than the toy example presented above, they can be categorised with "increasing complexity" of the new degree of freedom. The simplified models can be organised by considering:
\begin{itemize}
    \item scalar, fermion and vector representations of the Lorentz group,
    \item singlet, doublet, triplet, $\ldots$ representations of the SM gauge subgroups,
    \item one particle, two particles, and so on.
\end{itemize}
The representations and quantum numbers help to classify particles in dictionaries, just as higher-order operators are sorted in operator classes in EFTs, as presented in Refs. \cite{deBlas:2017xtg, Li:2023cwy, Guedes:2023azv, Gargalionis:2024jaw, Guedes:2024vuf}. Tables \ref{tab:GenericModelsScalars}, \ref{tab:GenericModelsFermions} and \ref{tab:GenericModelsVectors} taken from Ref.~\cite{Li:2023cwy} respectively list the spectrum extensions in scalar, fermion and vector dictionaries\footnote{It seems important to note that an AI agent called \lstinline{FERMIACC}\cite{Agrawal:2026lvg} has recently been built to generate viable simplified models that could explain "excesses" in ATLAS and CMS observations. }

Either these new states couple to all SM particles as allowed by symmetries, or they are interacting with specific particles or groups of particles. In general, the additional operators do not have mass dimension greater than 4, but simplified models can easily be matched to higher-order operators of EFTs (dimension-six or higher). As a result, limits on parameters obtained in either approach can be translated into the other using the matching relations at tree-level or at one-loop. Constraints on models were obtained from global fits of SMEFT coefficients in Refs.~\cite{Ellis:2018gqa, Ellis:2020unq, Bagnaschi:2022whn, DasBakshi:2020pbf, Anisha:2021hgc}.

Simplified models can also be matched to EFTs just as UV theories. For instance, Ref.~\cite{Darme:2021gtt} presents four simplified models\footnote{The models are available in the FeynRules database "Simple extensions of the SM/Top-philic resonances".} of heavy scalars $S$ and vectors $V$ which can be either singlets or octets of $SU_C(3)$. Pseudoscalar singlet and octet are considered as well. Each new state has effectively one coupling to the SM in the form of either a Yukawa-type interaction with coupling $y$ for scalar states or a vector-current-type interaction with coupling $g(=g_R=g_L)$ for vector states with the top. These simplified models are matched\footnote{Analytic expressions of the Wilson coefficients are provided for each simplified models in Table 1 of Ref.~\cite{Darme:2021gtt}.} with four SMEFT operators
\begin{equation}
\begin{split}
	\mathcal{O}_{LL}^1 &= \overline{t}_L \gamma^\mu t_L ~  \overline{t}_L \gamma_\mu t_L, \\
	\mathcal{O}_{RR}^1 &= \overline{t}_R \gamma^\mu t_R ~  \overline{t}_R \gamma_\mu t_R, \\
	\mathcal{O}_{LR}^1 &= \overline{t}_L \gamma^\mu t_L ~  \overline{t}_R \gamma_\mu t_R, \\
	\mathcal{O}_{LR}^8 &= \overline{t}_L T^A \gamma^\mu t_L ~  \overline{t}_R T^A \gamma_\mu t_R, 
\end{split}
\end{equation}
and four non-SMEFT operators
\begin{equation}
\begin{split}
	\mathcal{O}_{S}^1 &= \overline{t} t ~  \overline{t} t, \\
	\mathcal{O}_{S}^8 &= \overline{t} T^A t ~  \overline{t} T^A t, \\
	\mathcal{O}_{PS}^1 &= \overline{t} t ~  \overline{t} i \gamma_5 t, \\
	\mathcal{O}_{PS}^8 &= \overline{t} T^A t ~  \overline{t} T^A i \gamma_5 t, 
\end{split}
\end{equation}
to consider whether the easier EFT treatment of these models can be used instead. $T^A$ are the $SU_C(3)$ generators defined in Section~\ref{sec:SMLag}. Figures 2 and 3 of Ref.~\cite{Darme:2021gtt} demonstrate that the EFT approach is compatible with the relevant simplified model only in the large mass limit of the BSM resonance. They also show that the limit depends on the value of the coupling as the convergence between the two models is reached at smaller masses for larger values of the coupling. In any case, at small masses and small coupling, on-shell production of BSM resonances dominates and the EFT approach is not suited to extract constraints of the parameter space.

\section{Generic Models}
\label{sec:IntroTwoAnalyses}

Generic models are BSM models separate from UV-complete theories in the sense that they are not meant to give a single description of Nature which better explains the SM shortcomings listed in Section~\ref{sec:SMIssues}. Instead, generic models aim to probe the possible deviations from the SM in order to provide a direction on which type or particle or interactions could point to the new description.
	
In this Chapter, two types of generic models have been presented to probe the theoretical landscape of UV theories more systematically. On the one hand, off-shell effects of heavy BSM resonances are covered by means of EFTs as long as the operator basis is built of the existing degrees of freedom and respects the symmetries (whether they are gauge or global symmetries). The simplified models focus on lighter BSM resonances to describe their resonant effects. 

It is always possible to recast existing analyses for different insertions of EFT operators in relevant amplitudes or by incorporating various BSM resonances one by one. However, in the next two Chapters, dedicated analyses will be presented to target specific SMEFT operators and BSM resonances. In particular, observables sensitive to contributions arising from the insertion of CP-odd SMEFT operators in the form of triple products will be compared in Chapter~\ref{chap:smeftcpv} and a new particle search fo scalar resonances in four top production will be presented in Chapter~\ref{chap:fourtops}.

%!TEX root = main.tex

%%%%%%%%%%%%%%%%%%%%%%%%%%%%%%%%%%%%%%%%%%%%%%%%%%%%
%
%      Chapter 3 :
%
%
%%%%%%%%%%%%%%%%%%%%%%%%%%%%%%%%%%%%%%%%%%%%%%%%%%%

\chapter{CP violation in the SMEFT}
\label{chap:smeftcpv}
\pagestyle{fancy}

Having defined the Warsaw basis in Section~\ref{sec:BasicSMEFT}, the basis is further reduced by considering the limit of massless light fermions. First, 10 $CP$-odd operators are relevant when the top quark is massive and 17 $CP$-odd operators when both the top and bottom quarks are massive. Then, the sign of the interference between $CP$-odd dimension-six operators with the SM is discussed to introduce $CP$ sensitive observables. This is followed by a review of $CP$ sensitive observables. Direct observables from collider searches are detailed as well as indirect observables in the form of electric dipole moments (EDMs). Other observables relying on machine learning techniques or on matrix element evaluations are included.

New observables in the form of triple products are presented in the case of $W^\pm Z$ and $W \gamma$ processes in ATLAS. The asymmetry efficiencies are provided by comparing them to the theoretical and measurable optimal asymmetries. Differential cross-sections of the triple product asymmetries and optimal asymmetries are shown for both operators in both processes. Finally, the constraints on Wilson coefficients of two relevant $CP$-odd bosonic operators are derived.

This chapter is based on Ref.~\cite{Degrande:2021zpv}. Parts of the text have been modified to fit the rest of the manuscript, especially to match the notation adopted in Chapter~\ref{chap:sm}. The review has been updated with relevant publications that appeared since it was published. The two diboson analyses by the ATLAS collaboration providing constraints on $CP$-odd operators while citing Ref.~\cite{Degrande:2021zpv} are mentioned at the end of the Chapter.

\newpage

\section{New Bases and Relevant Processes}
\label{sec:CPBases}

\subsection{Basis Reduction with $U(1)^{14}$ symmetry}
\label{subsec:basis reduction u14}

Dimensional analysis indicates that the leading energy dependence in $\mathcal{M}_{int}$ from Eq.(\ref{amplitude expansion}) can have a growth factor $E^2/\Lambda^{2}$ compared to $\mathcal{M}_{SM}$ where $E$ is the characteristic energy of the process. This energy growth is expected in amplitudes from non-renormalisable operators and begins to be more significant when $E^2/\Lambda^{2} \to 1$, that is $E \to \Lambda$. Comparatively, contributions proportional to fermion masses $m_f$ become negligible at high energy. Since the goal is to track the leading $CP$ violating contributions in high energy colliders, all the fermion masses are neglected except the top quark mass which is considered too close to the TeV range to be discarded.

To enforce this hypothesis, the flavour symmetry imposed is the 14 $U(1)$ fermion field rephasings, written as $U(1)^{14}$, at high energies\footnote{This symmetry has fewer generators than the $U(3)_l \times U(3)_e \times U(3)_d \times U(2)_q \times U(2)_u$ symmetry often used in LHC analyses to restrict the analysis to the top sector.}. The Yukawa terms in $\mathcal{L}_{SM}$ violate the new symmetry so all Yukawa couplings $Y_f$, for $f=\{Q, L, u, d, e\}$, must vanish except for the top quark, $Y_t\neq 0$,
\begin{equation}
    \mathcal{L}_{SM} \supset \mathcal{L}_{Yuk.} = Y_l \Bar{L} \varphi e + Y_d \Bar{Q} \varphi d + Y_u \Bar{Q} \widetilde{\varphi} u \xrightarrow
 {\substack{U(1)^{14}}} Y_t \Bar{Q}_3 \widetilde{\varphi} t .
\end{equation}
As a result, only the heaviest SM particles remain massive after the EWSSB: the top quark $t$, the Higgs boson $h$ and the weak vector bosons $W^{\pm}, Z$. Moreover, left-handed and right-handed light fermion fields also decouple from each other since their coupling through the Higgs boson in $\mathcal{L}_{SM}$ has been discarded. This is particularly interesting because the phase of each light fermion field can now be chosen independently.

The phases of light fermion fields in dimension-six operators of $\mathcal{L}_{SMEFT}$ are shifted such that phases of Wilson coefficients are absorbed. The $\mathcal{O}_{bG}$ operator from the dipole class and its Wilson coefficient $C_{bG}$ are taken as an example. This operator introduces a new vertex contributing to $pp \rightarrow b\Bar{b}h$. Let us define the phase $\theta_i$ of the Wilson coefficient $C_i$, $C_i \equiv e^{i\theta_i}|C_i|$. The right-handed bottom field $b_R$ is rephased with the opposite phase of $C_{bG}$, namely
\begin{equation}
\label{eq:brephasing}
    b_R \rightarrow e^{-i \theta_{bG}} b'_R,
\end{equation}
such that $\mathcal{L}_{SM}(m_f\rightarrow 0)$ is unaffected but the $\mathcal{O}_{bG}$ operator together with its Wilson coefficient becomes
\begin{equation}
\label{eq:ObGrephasing}
   e^{i\theta_{bG}} |C_{bG}| (\Bar{Q}_3 \sigma^{\mu\nu} T^A b) \Tilde{\varphi} G^A_{\mu\nu}  \rightarrow |C_{bG}| (\Bar{Q}_3 \sigma^{\mu\nu} T^A b') \Tilde{\varphi} G^A_{\mu\nu}  .
\end{equation}
Due to its real Wilson coefficient, when added to its self-conjugate in $\mathcal{L}_{SMEFT}$, the $\mathcal{O}_{bG}$ operator no longer gives a $CP$ violating contribution at $\mathcal{O}\left(\Lambda^{-2}\right)$. The rephasing can be applied to all operators one at a time to reduce the list of CP-odd operators from Table \ref{complete CPV operator basis}. It is possible to do so because amplitudes and cross-sections are restricted to $\Lambda^{-2}$ corrections, \textit{i.e.} the interference amplitude between the SM and one CP-odd operator at a time, which are linear in the Wilson coefficients. This simplification would not stand if the restriction is set at $\mathcal{O}\left(\Lambda^{-4}\right)$. As a matter of fact, the operators of the Warsaw basis are independent, so the physical effects at $\mathcal{O}\left(\Lambda^{-2}\right)$ of one operator cannot be fully reproduced by any combination of other operators of the basis. At the same time, the physical effects of an operator should be independent of the fermion rephasing.

Therefore, if all the effects at $\mathcal{O}\left(\Lambda^{-2}\right)$ of one operator vanish for one choice of fermion phase and cannot be reproduced by the effects of the other operators of the basis which have seen their phase changed by the rephasing, they should vanish for all choices of fermion phase. This means that the operator has no physical observable effects at $\mathcal{O}\left(\Lambda^{-2}\right)$, \textit{i.e.} at the interference level. This is the strength of this argument, even if all the CP-violating operators of the Warsaw basis were included in any computation, any observable would only depend on the coefficients of the operators of our reduced basis at $\mathcal{O}\left(\Lambda^{-2}\right)$ since every interference of other operators with the SM vanishes exactly over the whole phase space in the limit of massless light fermions.

As stated above, at the $\mathcal{O}\left(\Lambda^{-4}\right)$, the amplitude can simultaneously depend on two Wilson coefficients in the interference between two dimension-six operators. In this case, the extra phase cannot be absorbed and will be transferred from one operator to the other instead. For instance, still in the $pp \rightarrow b\Bar{b}h$ process, $\mathcal{O}_{bG}$ and $\mathcal{O}_{b\varphi}$ are two relevant operators and, under the transformation outlined in Eq.(\ref{eq:brephasing}), they become
\begin{eqnarray}
   && e^{i\theta_{bG}} |C_{bG}| (\Bar{Q}_3 \sigma^{\mu\nu} T^A b) \Tilde{\varphi} G^A_{\mu\nu}  \rightarrow |C_{bG}| (\Bar{Q}_3 \sigma^{\mu\nu} T^A b') \Tilde{\varphi} G^A_{\mu\nu},  \nonumber \\
   && e^{i\theta_{b\varphi}} |C_{b\varphi}| (\Bar{Q}_3  b\varphi)   \left(\varphi^\dagger \varphi\right)  \rightarrow e^{i(\theta_{b\varphi}-\theta_{bG})} |C_{b\varphi}| (\Bar{Q}_3  b'\varphi)   \left(\varphi^\dagger \varphi\right).
   \label{eq:bop}
\end{eqnarray}
A real $C_{bG}$ is still obtained as in (\ref{eq:ObGrephasing}) but this time its phase is passed to $C_{b\varphi}$, therefore transferring the $CP$ violating effects.

Sticking to the $\mathcal{O}\left(\Lambda^{-2}\right)$, the rephasing trick is used independently to every operator containing light fermion fields. The massless approximation does not affect the hermitian operators in the bosonic classes $X^3$ and $X^2\varphi^2$, and they remain present in the reduced basis: 
\begin{equation*}
    \{ \mathcal{O}_{\widetilde{G}GG}, \mathcal{O}_{\widetilde{W}WW}, \mathcal{O}_{\varphi \widetilde{G}}, \mathcal{O}_{\varphi \widetilde{W}}, \mathcal{O}_{\varphi \widetilde{B}}, \mathcal{O}_{\varphi \widetilde{W}B} \}.
\end{equation*} 
In fact, those operators are even invariant under the full $U(1)^{15}$ symmetry, associated with the 15 fermion field rephasings when all the fermions are massless including the top field. The $U(1)^{15}$ symmetry actually removes all the other CP-odd operators, just as the full $U(3)^5$ flavour symmetry of the SM \cite{Gerard:1982mm}.

However, the simplification drastically decreases the number of operators in the four other operator classes because the operators contain either light fermion fields or the right-handed bottom field. Operators with fermion fields remain only if they consist of bosons and top quark fields (the left-handed $Q_3$ and the right-handed $t$). The $\psi^4$ and $\psi^2\varphi^2 D$ classes become free of CP-odd operators, $\mathcal{O}_{t\varphi}$ is the only remaining operator in the $\psi^2\varphi^3$ class and one operator for each vector boson field associated to top quark fields remains in the $X \psi^2\varphi$ class. Therefore, the operators with fermions kept in our basis are:
\begin{equation*}
    \{ \mathcal{O}_{tG}, \mathcal{O}_{tW}, \mathcal{O}_{tB}, \mathcal{O}_{t \varphi} \}.
\end{equation*}
In the end, 10 CP-odd operators remain under $U(1)^{14}$ and they are listed in Table~\ref{New CPV operator basis u14} following the class ordering from Table~\ref{complete CPV operator basis}.

\begin{table}[t]
\centering
\scalebox{0.90}{%
\begin{tabular}{|c|c|c|c|c|c|}
\hline 
   \multicolumn{2}{|c|}{$(X^3)$}  & \multicolumn{2}{|c|}{$(\psi^2\varphi^3)$}  &  \multicolumn{2}{|c|}{$(\psi^2\varphi^2 D)$}  \\
\hline
  $O_{\Tilde{G}GG}$ & $f^{ABC}\widetilde{G}_\mu^{A\nu} G_\nu^{B\rho} G_\rho^{C\mu}$  &  $O_{t\varphi}$ & $(\varphi^\dagger\varphi)(\overline{Q}_3 t\Tilde{\varphi})$ & // & /////  \\
  $O_{\Tilde{W}WW}$ & $\epsilon^{IJK}\widetilde{W}_\mu^{I\nu} W_\nu^{J\rho} W_\rho^{K\mu}$ &  &  & & \\
\hline \hline 
  \multicolumn{2}{|c|}{$(X^2\varphi^2)$} & \multicolumn{2}{|c|}{$(\psi^4)$} & \multicolumn{2}{|c|}{$(X\psi^2\varphi)$} \\
\hline
  $O_{\varphi \Tilde{G}}$ & $\varphi^\dagger\varphi\widetilde{G}_{\mu\nu}^A G^{A\mu\nu}$ & // & ///// & $O_{tG}$ & $(\overline{Q}_3 \sigma^{\mu\nu}T^A t)\Tilde{\varphi}G^A_{\mu\nu}$  \\
  $O_{\varphi \Tilde{W}}$ & $\varphi^\dagger\varphi\widetilde{W}^I_{\mu\nu} W^{I\mu\nu}$ &  &  & $O_{tW}$ & $(\overline{Q}_3 \sigma^{\mu\nu}t)\tau^I\Tilde{\varphi}W^I_{\mu\nu}$  \\
  $O_{\varphi \Tilde{B}}$ & $\varphi^\dagger\varphi\widetilde{B}_{\mu\nu} B^{\mu\nu}$  &  &   & $O_{tB}$ & $(\overline{Q}_3 \sigma^{\mu\nu}t)\Tilde{\varphi}B_{\mu\nu}$ \\
  $O_{\varphi \Tilde{W}B}$ & $\varphi^\dagger\tau^I\varphi\widetilde{W}^I_{\mu\nu} B^{\mu\nu}$  &  &  &  &  \\
\hline
\end{tabular}
}
\caption{List of CP-odd dimension-6 operators in our reduced basis under $U(1)^{14}$. The operators of the first column are invariant under the full $U(3)^5$ flavour symmetry. }
\label{New CPV operator basis u14}
\end{table}

It is important to keep in mind that the massless fermions approximation does not affect the real part of Wilson coefficients, \textit{i.e.} the CP-even operators and the CP-even part of non-hermitian operators. Thus, a complete reduced basis at the $\Lambda^{-2}$ order would not only contain the 10 CP-odd operators from Table \ref{New CPV operator basis u14} but all CP-even operators with three generations from Table \ref{complete CPV operator basis} invariant under the chosen symmetry as well.

The leading $CP$ violating contributions come from these 10 operators which not only introduce anomalous couplings with respect to the SM but introduce new contact interactions as well, as shown in Table \ref{tab:couplings}.

\begin{table}[t]
    \centering
    \begin{tabular}{|c|c|}
        \hline
        Operator & Anomalous couplings between... \\
        \hline
        $\mathcal{O}_{\widetilde{G}GG}$ & 3 to 6 gluons \\
        %ggg/gggg/ggggg/gggggg \\
        \hline
        $\mathcal{O}_{\widetilde{W}WW}$ & 3 to 6 electroweak bosons \\
        %WWB/WWBB/WWWW/WWBBB/WWWWB \\
        \hline
        $\mathcal{O}_{\varphi \widetilde{G}}$ & 1 or 2 Higgs and 2 to 4 gluons \\
        %hgg/hggg/hgggg/hhgg/hhggg/hhgggg \\
        \hline
        $\mathcal{O}_{\varphi \widetilde{W}}$ & 1 or 2 Higgs and 2 to 3 electroweak bosons \\
        %hBB/hhBB/hWW/hhWW/hhBWW \\
        \hline
        $\mathcal{O}_{\varphi \widetilde{B}}$ & 1 or 2 Higgs and 2 $B$ bosons \\
        %hBB/hhBB \\
        \hline
        $\mathcal{O}_{\varphi \widetilde{W}B}$ & 1 to 2 $B$ bosons, 2 to 3 electroweak bosons and up to 2 Higgs \\
        %WWB/WWBh/WWBhh/BBh/BBhh  \\
        \hline
        $\mathcal{O}_{t \varphi}$ & 2 tops and 1 to 3 Higgs  \\
        \hline
        $\mathcal{O}_{t G}$ & 2 tops, 1 to 2 gluons and even 1 Higgs  \\
        \hline
        $\mathcal{O}_{t W}$ & 2 tops, 1 to 2 electroweak bosons and even 1 Higgs \\
        %ttB/ttBh/ttWW/ttWWh/tbW/tbWh/tbWBh  \\
        \hline
        $\mathcal{O}_{t B}$ &  2 tops, 1 $B$ boson and even 1 Higgs \\
        %ttB/ttBh  \\
        \hline
    \end{tabular}
    \caption{Interactions introduced by the operators in Table \ref{New CPV operator basis u14} in the unitary gauge. W stands for the charged gauge boson, B for the neutral electroweak boson, \textit{ i.e.} the Z boson and the photon, h for the Higgs, g for the gluon, t for the top quark and b for the bottom quark. The bottom quark comes from the third generation left-handed doublet $Q_3$. }
    \label{tab:couplings}
\end{table}

\subsection{Basis Reduction with $U(1)^{13}$ symmetry}
\label{subsec:basis reduction u13}

The $U(1)^{14}$ symmetry introduced in the previous section is not a subgroup of the $U(3)_l \times U(3)_e \times U(3)_d \times U(2)_q \times U(2)_u$ symmetry often used for SMEFT analyses related to the top sector contrary to $U(1)^{13}$. In this latter symmetry, both top and bottom quarks are massive. Therefore, there is less freedom to cancel the phases of CP-odd operators. For example, the phase of the operators in Eq.\eqref{eq:bop} cannot be removed anymore.

Therefore, the operators present under the $U(1)^{14}$ symmetry must remain and are supplemented by other operators containing bottom quark fields. As already mentioned, the operators from $X^3$ and $X^2\varphi^2$ classes remain unaffected by any redefinition as they are invariant under the $U(3)^5$ symmetry.
By following the same method as in Section \ref{subsec:basis reduction u14}, the non-hermitian operators of the $\psi^2\varphi^3$ and $X\psi^2\varphi$ categories including the right-handed bottom quark fields are:
\begin{equation*}
    \{ \mathcal{O}_{bG}, \mathcal{O}_{bW}, \mathcal{O}_{bB}, \mathcal{O}_{b \varphi} \}.
\end{equation*}

However, this symmetry also allows operators of the two other categories. We mentioned previously that $O_{\varphi ud}$ was the only CP-odd operator in the $\psi^2 \varphi^2 D$ class and, using the up and down right-handed fields, we now need to include :
\begin{equation*}
    \{ \mathcal{O}_{\varphi tb}  \}.
\end{equation*}
To build non-hermitian four-fermion operators with chirality configurations $(\Bar{L}L)(\Bar{L}L)$, $(\Bar{R}R)(\Bar{R}R)$ and $(\Bar{L}L)(\Bar{R}R)$, we need fields from different generations. Since we are dealing only with the third generation, no operators come from these configurations. The $(\Bar{L}R)(\Bar{R}L)$ configuration involves only leptons so we do not have CP-odd operators from this configuration either. Finally, $(\Bar{L}R)(\Bar{L}R)$ has two operators without leptons which can be CP-odd : 
\begin{equation*}
    \{ \mathcal{O}_{qtqb}^{(1)}, \mathcal{O}_{qtqb}^{(8)}  \}.
\end{equation*}

Here, 17 CP-odd operators remain under $U(1)^{13}$ and they are listed in Table~\ref{New CPV operator basis u13} following the class ordering from Table~\ref{complete CPV operator basis}.

\begin{table}[t]
\centering
\scalebox{0.8}{%
\begin{tabular}{|c|c|c|c|c|c|}
\hline 
   \multicolumn{2}{|c|}{$(X^3)$}  & \multicolumn{2}{|c|}{$(\psi^2\varphi^3)$}  &  \multicolumn{2}{|c|}{$(\psi^2\varphi^2 D)$}  \\
\hline
  $O_{\Tilde{G}GG}$ & $f^{ABC}\widetilde{G}_\mu^{A\nu} G_\nu^{B\rho} G_\rho^{C\mu}$  &  $O_{t\varphi}$ & $(\varphi^\dagger\varphi)(\overline{Q}_3 t\Tilde{\varphi})$ & $O_{\varphi tb}$ & $i( \Tilde{\varphi}^\dagger D_\mu \varphi ) ( \Bar{t} \gamma^\mu b)$  \\
  $O_{\Tilde{W}WW}$ & $\epsilon^{IJK}\widetilde{W}_\mu^{I\nu} W_\nu^{J\rho} W_\rho^{K\mu}$ & $O_{b\varphi}$ & $(\varphi^\dagger\varphi)(\overline{Q}_3 b \varphi)$ & & \\
\hline \hline 
  \multicolumn{2}{|c|}{$(X^2\varphi^2)$} & \multicolumn{2}{|c|}{$(\psi^4)$} & \multicolumn{2}{|c|}{$(X\psi^2\varphi)$} \\
\hline
  $O_{\varphi \Tilde{G}}$ & $\varphi^\dagger\varphi\widetilde{G}_{\mu\nu}^A G^{A\mu\nu}$ & $O_{qtqb}^{(1)}$ & $(\Bar{Q}^j_3 t) \epsilon_{jk} (\Bar{Q}^k_3 b)$ & $O_{tG}$ & $(\overline{Q}_3 \sigma^{\mu\nu}T^A t)\Tilde{\varphi}G^A_{\mu\nu}$  \\
  $O_{\varphi \Tilde{W}}$ & $\varphi^\dagger\varphi\widetilde{W}^I_{\mu\nu} W^{I\mu\nu}$ & $O_{qtqb}^{(8)}$ & $(\Bar{Q}^j_3 T_A t) \epsilon_{jk} (\Bar{Q}^k_3 T_A b)$ & $O_{tW}$ & $(\overline{Q}_3 \sigma^{\mu\nu}t)\tau^I\Tilde{\varphi}W^I_{\mu\nu}$  \\
  $O_{\varphi \Tilde{B}}$ & $\varphi^\dagger\varphi\widetilde{B}_{\mu\nu} B^{\mu\nu}$  &  &   & $O_{tB}$ & $(\overline{Q}_3 \sigma^{\mu\nu}t)\Tilde{\varphi}B_{\mu\nu}$ \\
  $O_{\varphi \Tilde{W}B}$ & $\varphi^\dagger\tau^I\varphi\widetilde{W}^I_{\mu\nu} B^{\mu\nu}$  &  &  & $O_{bG}$ & $(\overline{Q}_3 \sigma^{\mu\nu}T^A b)\varphi G^A_{\mu\nu}$ \\
   & & & & $O_{bW}$ & $(\overline{Q}_3 \sigma^{\mu\nu}b)\tau^I \varphi W^I_{\mu\nu}$ \\
   & & & & $O_{bB}$ & $(\overline{Q}_3 \sigma^{\mu\nu}b) \varphi B_{\mu\nu}$ \\
\hline
\end{tabular}
}
\caption{List of CP-odd dimension-6 operators in our reduced basis under the $U(1)^{13}$ symmetry. }
\label{New CPV operator basis u13}
\end{table}

\subsection{Sign of the Interference}
\label{subsec:signofinterference}

The reduced bases are effective to track leading $CP$ violating effects when considering $\mathcal{O}(\Lambda^{-2})$ interference effects from one dimension-6 operator at a time. Those interferences are always odd under $CP$ symmetry since we neglect the SM $CP$ phase. Therefore, $\mathcal{M}_{int.,i}$ does not contribute to $CP$-even observables, such as the total cross-section of C-even processes. Taking $g g \rightarrow t \Bar{t}$ as an example,
\begin{equation}
\begin{split}
    \sigma_{tot}^{g g \rightarrow t \Bar{t}} &= \int d\Pi_{LIPS} \left[ \left| \mathcal{M}_{SM}\right|^2 + 2 \mathcal{R}e \left\{ \mathcal{M}_{SM}^* \mathcal{M}_i \right\} + \mathcal{O}(\Lambda^{-4}) \right] \\
    &= \sigma_{SM}^{g g \rightarrow t \Bar{t}} + 0 + \mathcal{O}(\Lambda^{-4}),
\end{split}
\end{equation}
where $d\Pi_{LIPS}$ is the $CP$-even Lorentz-invariant phase-space. As a result, the operators in Tables~\ref{New CPV operator basis u14} and \ref{New CPV operator basis u13} have often been constrained thanks to a part of their $\mathcal{O}(\Lambda^{-4})$ contributions,  $\| \mathcal{M}_i \|^2$ in Eq.(\ref{amplitude expansion}). This has been done at LEP \cite{LEP:2003aa} and at LHC \cite{Aaboud:2019nkz,Sirunyan:2017zjc}. Those squared amplitudes are $CP$-even and do contribute to $CP$-even observables but are more suppressed in $1/\Lambda$. Analysing $CP$-odd operators with the total cross-section is expected to lead to less stringent $\mathcal{O}(\Lambda^{-4})$ constraints on their Wilson coefficients. The main drawback is that they do not test whether $CP$ is actually broken because corresponding $CP$-even operators generally give similar contributions at that order. In general, conventional $CP$-even observables are not suited to efficiently probe $CP$-violating effects since they present no or small variations from expected SM simulations by relying on $\Lambda^{-4}$-suppressed effects \cite{LEP:2003aa, Aaboud:2019nkz, Sirunyan:2017zjc}.

\begin{figure}[t]
    \centering
    \includegraphics[width=\textwidth]{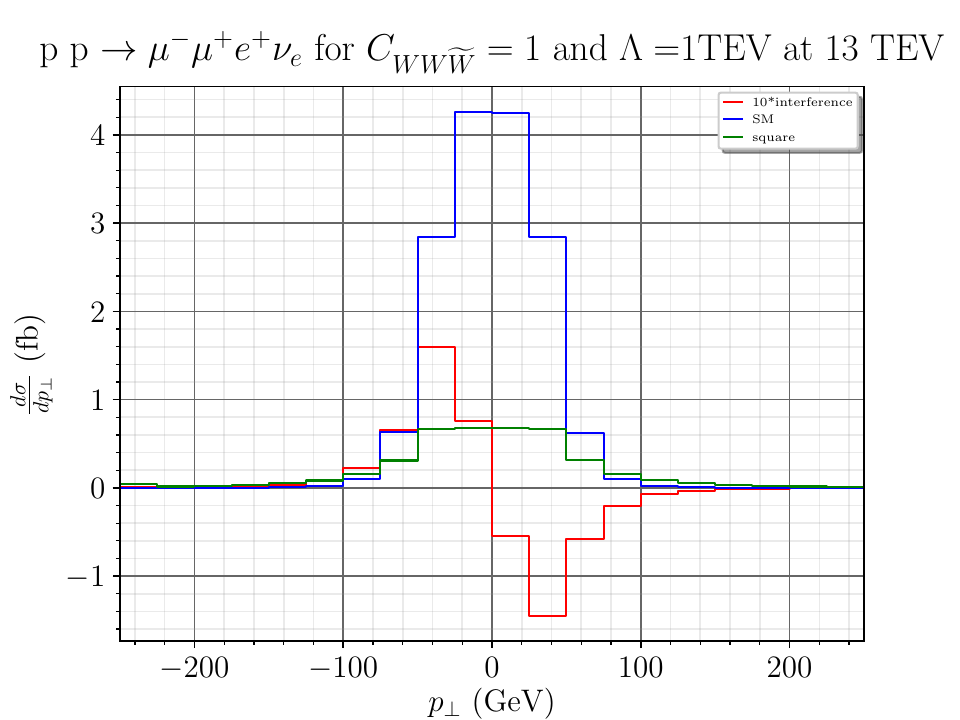}
    \caption{Differential cross-sections of $pp \rightarrow \mu^- \mu^+ e^+ \nu_e$ in ATLAS at $\sqrt{s}=13$ TeV with respect to the triple product $p_\perp$ (the observable is defined and discussed hereafter). The operator considered here is $\mathcal{O}_{\widetilde{W}WW}$ and its Wilson coefficient has been set to 1. The NP scale $\Lambda$ is set to 1 TeV. }
    \label{fig:tripmom ZW}
\end{figure}

The vanishing interference cross-section is due to flips of the sign of the interference over the phase space. On Figure~\ref{fig:tripmom ZW}, the SM and interference differential cross-sections are compared with respect to an almost $CP$-odd observable, which will be described in the review below, called the \textit{triple product} $p_{\perp}$. The relative difference in the heights of the distributions is partially due to the expected $\Lambda^{-2}$ suppression.

The interference contribution does not vanish over the whole phase space but actually modulates between positive and negative values, contrary to the SM amplitude which is only positive-definite. The symmetric profile of the pure SM contribution and the asymmetrical profile of interference are explicit. This strongly suggests using asymmetries to probe $CP$-odd operators. When integrating over the almost $CP$-odd observable, the positive and negative contributions in the differential interference cross-section almost completely cancel each other. In a C-even process, they would compensate each other exactly and the total interference would vanish.

On the contrary, when looking at the asymmetry of the $CP$-odd observable, it is the symmetric $CP$ conserving contributions that vanish. Meanwhile, asymmetric $CP$ violating effects are highlighted by keeping track of both positive and negative values of the interference, that is if the asymmetry is defined from an (almost) $CP$-odd observable. The modulation in sign of the interference displayed on Figure~\ref{fig:tripmom ZW} is around 0, so the phase space can be divided in two separate regions to define the asymmetry. This way the asymmetric contributions from the interference add up while the symmetric contributions from the SM and the pure SMEFT are suppressed. However, not all $CP$-odd observables share this particular pattern. This means that the separation of the phase space and the asymmetry must be adapted accordingly.

It is thus possible to focus on a single asymmetry for each observable/distribution for each process to keep the computation of efficiencies in reviving $CP$ violating effects and their comparisons simple. However, differential asymmetries or distributions could be looked into at a later stage to improve the sensitivities.

ATLAS has already been investigating a $CP$-odd observable in $Z+jj$ events~\cite{Aad:2020sle}. The simulations of $CP$-odd operator signals in differential distributions with respect to a $CP$-odd observable, in this case $\Delta \phi_{jj}$ presented in the review below, highlight this modulation of the interference amplitude and the phase-space cancellation in $CP$-even observables for $CP$-odd operators.

\subsection{Relevant Processes}
\label{subsec:POI}

In this section, the most interesting processes that could directly probe vertices in Table \ref{tab:couplings} are presented. As mentioned above, the $\mathcal{O}(\Lambda^{-2})$ contribution from dimension-six $CP$-odd operators to the total cross-section of C-even processes vanishes and can only be constrained using differential distributions or, in particular, asymmetries. To measure accurately asymmetries or differential distributions in general, we need a large number of events and therefore processes with quite large cross-sections. Hence, low multiplicity processes are prioritised.

In the $X^3$ class, the $\mathcal{O}_{\widetilde{G}GG}$ could be investigated by looking at multi-jet events. However, this may not be easy as even the $CP$-even version of this operator $\mathcal{O}_{GGG}$ is known for its vanishing or suppressed interference \cite{Krauss:2016ely, Hirschi:2018etq, Azatov:2016sqh}. Moreover, this operator is already well constrained by the indirect upper bound measurements of the neutron electromagnetic dipole moment (EDM) as shown in Ref.~\cite{Dekens:2013zca} (the $CP$-violating $\overline{\theta}$ angle from QCD is neglected even though the neutron EDM constrains both $\overline{\theta}$ and all relevant Wilson coefficients). The vertices with the smaller number of legs of $\mathcal{O}_{\widetilde{W}WW}$ are triple gauge boson interactions and contribute to the diboson production (VV and Vh) at LEP, the Tevatron, and the LHC (see Refs.\cite{Heister:2001qt,  doi:10.1146/annurev.nucl.48.1.33, Falkowski:2016cxu} and comparative studies in Refs.~\cite{Wang:2014uea, Grojean:2018dqj}). This operator suffers the same suppression as the $CP$-even operator $\mathcal{O}_{WWW}$ due to helicity selection rules, as shown in Ref.~\cite{Azatov:2016sqh}. Since $\mathcal{O}_{\varphi\widetilde{W}B}$ has the same vertices, it can be constrained by the same processes. Those two $CP$-odd operators affect also lower cross-section processes such as weak vector boson fusion (VBF), weak vector boson scattering (VBS) or triple gauge bosons production (VVV). Moreover, $\mathcal{O}_{\varphi\widetilde{W}}$ and $\mathcal{O}_{\varphi\widetilde{B}}$ contribute to the latter with an intermediate Higgs boson.

The operators from the $X^2\varphi^2$ class except $\mathcal{O}_{\varphi\widetilde{W}B}$ have at least one Higgs in each of their vertices. Therefore, they contribute to the production and decay of the Higgs. However, the $1 \to 2$ and $2 \to 1$ kinematics do not allow one to build P-odd observables, so it requires either adding particles in the final state (e.g. jets in Higgs production by gluon fusion) or being sensitive to the polarisation of the vector boson through their decay products or the distribution of the jets in VBF.

The only operator in the $\psi^2\varphi^3$ class, $\mathcal{O}_{t\varphi}$, is a correction to the SM top quark Yukawa. It affects the top pair production in association with a Higgs. Measurements from ATLAS and CMS \cite{ATLAS:2020ior, CMS:2020cga} offer new insights into the $CP$-nature of the Higgs, rejecting the full $CP$-odd state, and provide better constraint on $\mathcal{O}_{t\varphi}$. This operator also affects at one-loop the production of single Higgs but its interference vanishes by symmetry. The observation of $t\Bar{t}t\Bar{t}$ in ATLAS \cite{ATLAS:2020hpj} could also help to probe this operator.

The last class is the dipole operators $X\psi^2\varphi$. These operators generate couplings between various particles. This means that they appear in a large variety of processes. For example, $\mathcal{O}_{tG}$ contributes to top pair production \cite{Bernreuther:2015yna}, single top production, top pair production in association with a Higgs and Higgs production at one-loop. $\mathcal{O}_{tW}$ is well known for its effects on top decay and single top production while $\mathcal{O}_{tB}$ requires an extra photon or Z boson such as in $pp\to t\bar{t}Z/\gamma$ or $e^+e^-\to t\bar{t}$.

Nevertheless, they all contribute indirectly to EDMs. SM contributions to EDMs arise only beyond the two-loop level. Hence they are largely suppressed, whereas some of our SMEFT operators contribute already at the one loop level as demonstrated in Refs.~\cite{Panico:2018hal, Dekens:2013zca}. EDMs offer a good opportunity to constrain $CP$-odd SMEFT operators in indirect observations. As EDMs are sensitive to a single combination of operators, they probe only one direction in the parameter space and require collider observables to remove any potential blind direction.

\section{Review of CP Observables}
\label{sec:CPObsReview}

A short review is presented here on SMEFT analyses looking for $CP$ violating effects arising from the operators listed in Table \ref{New CPV operator basis u14} and focusing on which $CP$ sensitive observables are used. Other studies involving $CP$-odd operators not included in our reduced basis in Table \ref{New CPV operator basis u14} exist but are not considered. In particular, the rich literature on 4-fermion operators will not be covered here.

In the following, measurements are divided into direct, indirect and other observables. The first tracks $CP$ violating effects arising in high energy processes at colliders while the second looks into deviations in low-energy properties by high energy corrections. The latter consists of alternative approaches to probe $CP$ effects in the form of matrix-element-based observables, machine learning algorithms and global analyses.

\subsection{Direct Observables}
\label{subsec:DirectCPObs}

The different analyses and their respective observables are gathered in Table \ref{tab:directobs}. In the second column the relevant operators are listed with their respective analyses.

\begin{table}
\centering
\scalebox{0.85}{%
\begin{tabular}{|c|c|c|c|}
\hline
  Ref(s)  &  Operator(s) &  Observable(s)  &  Process(es)  \\
\hline
\cite{Panico:2017frx,Azatov:2019xxn} & $\mathcal{O}_{\widetilde{W}WW}$ &  $\sin 2\phi_Z+\sin{2\phi_W}$ & $pp\rightarrow ZW$ \\
                        &    &  $\sin{2\phi_W}$ & $pp\rightarrow W\gamma $   \\
\hline
\cite{Kumar:2008ng} & $\mathcal{O}_{\widetilde{W}WW}$ & $sign[(p_Z)^z] sign[(p_l\times p_Z)^z]$ & $pp\rightarrow ZW$   \\
                    & $\mathcal{O}_{\varphi\widetilde{W}}$,$\mathcal{O}_{\varphi\widetilde{W}B}$ &  &\\
\hline
\cite{Bernreuther:2013aga,Bernreuther:2015yna} &   $\mathcal{O}_{tG}$     &     $B_{1,2}$ and $O^{CP}_{1,2}$   & $pp \rightarrow t\Bar{t}$   \\
\hline
\cite{Bishara:2020vix} & $\mathcal{O}_{\varphi \widetilde{W}}$ & $\sin{\phi_W}$ & $pp\rightarrow Wh $  \\
\hline
\cite{Brehmer:2017lrt} & $\mathcal{O}_{\varphi\widetilde{W}}$ & $\Delta \phi_{ll}$ & $pp\rightarrow hqq'$ (VBF)     \\
                       & $\mathcal{O}_{\varphi\widetilde{B}}$ & $\Delta \phi_{ll}$ & $pp\rightarrow hZ$  \\
                       & & $\sin\Phi$ & $pp\rightarrow h \rightarrow 4l$   \\
\hline
\cite{Englert:2019xhk} & $\widetilde{O}_g \subset \mathcal{O}_{\varphi G}$ & $\Delta \phi_{ll}$ & $pp\rightarrow hqq'$ (VBF)    \\
                       & $i\Bar{t}\gamma_5 t h \subset \mathcal{O}_{t\varphi}$ & $\Delta \phi_{jj}$ & $pp\rightarrow tth$   \\
\hline
\cite{Plehn:2001nj} & $\mathcal{O}_{\varphi\widetilde{W}}$ & $\Delta \phi_{jj}$  & $pp\rightarrow hqq'$ (VBF)   \\
\hline
\cite{Beneke:2014sba}    & $\mathcal{O}_{\varphi \widetilde{W}}$  & $\sin{\Phi}$ & $pp \rightarrow h \rightarrow 4l$  \\
                          & $\mathcal{O}_{\varphi \widetilde{B}}$,$\mathcal{O}_{\varphi \widetilde{W}B}$ & $\sin{2\phi}$& \\
\hline
\cite{Banerjee:2020vtm} & $\mathcal{O}_{\varphi \widetilde{W}}$  & $\phi,\theta_1,\theta_2$ & $pp \rightarrow h \rightarrow 4l$  \\
                          & $\mathcal{O}_{\varphi \widetilde{B}}$,$\mathcal{O}_{\varphi \widetilde{W}B}$ &  & \\
\hline
\cite{Bernlochner:2018opw} & $\mathcal{O}_{\varphi \widetilde{G}}$, $\mathcal{O}_{\varphi \widetilde{W}}$  & $\Delta \phi_{ll}$ & $pp \rightarrow h \rightarrow ZZ^* / \gamma\gamma$  \\
                          & $\mathcal{O}_{\varphi \widetilde{B}}$,$\mathcal{O}_{\varphi \widetilde{W}B}$ & & \\
\hline
\cite{Banerjee:2019pks,Banerjee:2019twi} &  $\mathcal{O}_{\varphi\widetilde{B}}$  &  $\varphi,\theta,\Theta$ & $pp \rightarrow Z h$    \\
                        &  $\mathcal{O}_{\varphi\widetilde{W}}$,$\mathcal{O}_{\varphi\widetilde{W}B}$  &   & $pp \rightarrow W h$ \\
\hline
\cite{Biekotter:2020flu} &  $\mathcal{O}_{\varphi\widetilde{W}}$    &  \multirow{2}{*}{ $\frac{\Vec{p}_\gamma . (\Vec{p}_{j2} \times \Vec{p}_{bb})}{|\Vec{p}_\gamma| |\Vec{p}_{j2}| |\Vec{p}_{bb}|} $  } &   $pp \rightarrow h(\rightarrow b\Bar{b})\gamma jj$ (VBF)    \\
                    & $\mathcal{O}_{\varphi\widetilde{W}B}$ &  &  \\
\hline
\cite{DasBakshi:2020ejz}  &   $\mathcal{O}_{\varphi \widetilde{W}B}$    &    $\Delta \phi_{Zl}$      &    $p p \rightarrow WZ$ \\
                        & $\mathcal{O}_{\widetilde{W}WW}$    &    $\Delta \phi_{ll'}$      & $pp\rightarrow WW$ \\
                       &      &    $\Delta \phi_{\gamma l}$      &    $pp \rightarrow W\gamma$ \\
\hline
\cite{Christensen:2010pf}   &  $\mathcal{O}_{\varphi \widetilde{W}}$,$\mathcal{O}_{\varphi \widetilde{W}B}$     & $\phi_{ll}$   &  $p p \to Zh$  \\
\hline 
\cite{Godbole:2014cfa}  & $\mathcal{O}_{\varphi \widetilde{W}}$ & $\cos \delta^+$  & $p p \to Vh$ $(V=W,Z)$ \\
                        & $\mathcal{O}_{\varphi \widetilde{B}}$,$\mathcal{O}_{\varphi \widetilde{W}B}$  &  &  \\
\hline 
\cite{Subba:2025hxn}  & $\mathcal{O}_{\widetilde{W}WW}$  & $A_{44} (A_{16})$   &  $e^- e^+ \to WW$  \\
                        & $\mathcal{O}_{\varphi \widetilde{W}}$  & $A_{44}, A_{16} (A_7)$ &  \\
\hline 
\cite{Hankele:2006ma} & $\mathcal{O}_{\varphi \widetilde{W}}$ & $p_{T+} p_{T-} \sin{(\Delta \phi_{jj})}$ & $pp \to hjj$ (VBF) \\
                      & $\mathcal{O}_{\varphi \widetilde{B}}$,$\mathcal{O}_{\varphi \widetilde{W}B}$   &   &    \\
\hline
\cite{Klamke:2007cu} & $\mathcal{O}_{\varphi \widetilde{G}}$ & $p_{T+} p_{T-} \sin{(\Delta \phi_{jj})}$ & $pp \to hjj$ (VBF) \\
\hline
\cite{Ellis:2013yxa}  & $i\Bar{t}\gamma_5 t h \subset \mathcal{O}_{t\varphi}$ & $\Delta \phi_{l^-l^-} \sim \vec{p}_{t}.(\vec{p}_{l^-} \times \vec{p}_{l^+})$ & $pp \to t\overline{t}h$ \\
                     &  &  $\vec{p}_{l}.(\vec{p}_{j} \times \vec{p}_{h})$  &  $pp \to thj$ \\
\hline
\cite{Cao:2025fla}  &  $\mathcal{O}_{\widetilde{W}WW}$,  $\mathcal{O}_{\varphi\widetilde{W}}$  & $\hat{z}.(\vec{p}_{\nu} \times \vec{p}_{\mu})$  & $\mu^- \mu^+ \to \nu \mu^{\pm} W^{\mp}$  \\
                    & $\mathcal{O}_{\varphi \widetilde{B}}$,  $\mathcal{O}_{\varphi\widetilde{W}B}$  & $\hat{z}.(\vec{p}_{\mu^-} \times \vec{p}_{\mu^+})$  & $\mu^- \mu^+ \to \mu^- \mu^+ h$  \\
\hline
\cite{Asteriadis:2024xuk, Asteriadis:2024xts} & $\mathcal{O}_{\varphi \widetilde{W}}$  &  $\cos{\theta_{e^+ h}}$  & $e^- e^+ \to Zh$ \\
        &  $\mathcal{O}_{\varphi \widetilde{B}}$, $\mathcal{O}_{\varphi \widetilde{W}B}$ &  &  \\
\hline
\cite{Bar-Shalom:2024dav}  &  $\mathcal{O}_{t\varphi}$  &  $\epsilon(p_{e^-}, p_{e^+}, p_{t}, p_{\overline{t}})$  & $e^- e^+ \to t \overline{t} h$  \\
                            &    &  $\epsilon(p_{b}, p_{t}, p_{h}, p_{W})$   & $e^- e^+ \to t h W$  \\
                            &    &  $\epsilon(p_{b}, p_{t}, p_{h}, p_{j})$   & $e^- e^+ \to t h j$  \\
\hline
\end{tabular}
}
\caption{Summary table of direct observables to constrain operators from the reduced basis. }
\label{tab:directobs}
\end{table}

\begin{table}
\centering
\scalebox{0.85}{%
\begin{tabular}{|c|c|c|c|}
\hline
  Ref(s)  &  Operator(s) &  Observable(s)  &  Process(es)  \\
\hline
\cite{Subba:2025hkq}  & $\mathcal{O}_{\varphi \widetilde{W}}$  &  $\mathcal{A}^{l}_{2}$ &  $e^- e^+ \to Zh(\to b \overline{b})$ \\
                        &   &   $\mathcal{A}^{lj}_{86}$, $\mathcal{A}^{l}_{2}$ ($\mathcal{A}^{lj}_{61}$,$\mathcal{A}^{l}_{4}$) &  $e^- e^+ \to Zh(\to l \nu  jj)$ \\
                        &   &  $\mathcal{A}^{l}_{2}$ ($\mathcal{A}^{l}_{4}$,$\mathcal{A}^{ll}_{86}$) &  $e^- e^+ \to Zh(\to 4 l)$ \\
\hline
\cite{ElFaham:2024uop} & $\mathcal{O}_{\widetilde{W}WW}$  & $\phi^*_{e^+}, \phi^*_{\mu^-}, \theta^*_{e^+}$  &  $p p \to W^{\pm} Z$ \\
\hline
\cite{Rossia:2024rfo}  & $\mathcal{O}_{\varphi \widetilde{W}}$ & $\sin{\phi_W}$  & $p p \to Wh$ \\
\hline
\cite{Miralles:2024huv} & $\mathcal{O}_{t \varphi}, \mathcal{O}_{t W}, \mathcal{O}_{t G},$  & $\vec{p}_h.(\vec{p}_t \times \vec{p}_j), \hat{z}.(\vec{p}_t \times \vec{p}_j)|_h,$ &  $p p \to thj$ \\
                        & $\mathcal{O}_{\varphi \widetilde{W}}, \mathcal{O}_{\varphi \widetilde{G}}$  & $ \cos{\theta_e^y}, \cos{\theta_e^s}$ &   \\
                        &    &  $\vec{p}_t.(\vec{p}_{e^-}  \times \vec{p}_{e^+})|_{t\overline{t}}, \Delta \phi_{t\overline{t}}, \Delta \phi_{e^- e^+},$ &  $p p \to t \overline{t} h$ \\
                        &       &  $ \text{sign}[\vec{p}_t.(\vec{p}_{e^-}  \times \vec{p}_{e^+})](\vec{p}_{e^-} . \vec{p}_{e^-})$   &  \\
\hline
\end{tabular}
}
\caption{Continuation of Table~\ref{tab:directobs}. }
\label{tab:directobs2}
\end{table}

\paragraph{Simple differences:} Ref.~\cite{Han:2009ra} presents two $CP$ sensitive observables in the form of differences of usual observables in the transverse momenta between charged particles $\Delta p_T = p_{T}^{+} - p_{T}^{-}$ or their transverse energy $\Delta E_T = E_{T}^{+} - E_{T}^{-}$. The superscripts indicate the electric charge of the particle. The difference in transverse energies had been proposed by Ref.~\cite{PhysRevLett.69.410} for $t\overline{t}h$ production. However, Ref.~\cite{Han:2009ra} highlights that these observables require a $CP$-violating phase and a $CP$-conserving one motivating why it had not been used in their analysis. Ref.~\cite{Christensen:2010pf} used $\Delta E_T$ in $Zh$ production and demonstrated that $\Delta E_T$ probes $CP$ violation from absorptive complex phases of effective couplings equivalent to $\mathcal{O}_{\varphi \widetilde{W}}$ and $\mathcal{O}_{\varphi \widetilde{B}}$. So, these observables cannot be considered as appropriate $CP$ sensitive observables for the SMEFT operators listed in Table~\ref{New CPV operator basis u14}.

Ref.~\cite{ElFaham:2024uop} searched for $CP$ violating contributions at NLO in QCD in $W^+W^-$ with the azimuthal separation of the charged leptons from the $W$ decays $\Delta \phi_{e^+ \mu^-}$. It was proven that the $CP$ contributions to this observable were negligible, slightly increased by the fiducial cuts.

    \begin{figure}[t]
        \centering
        \includegraphics[scale=0.6]{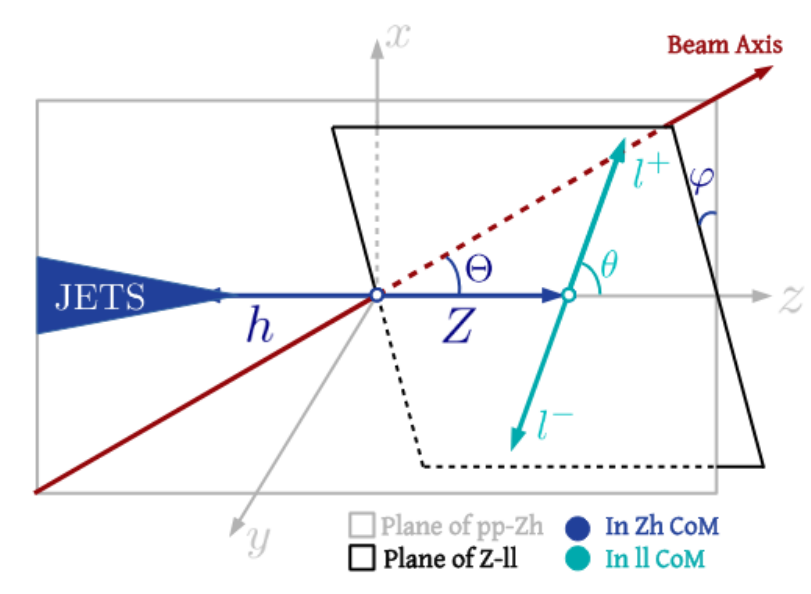}
        \caption{Description of the $Zh$ production in the centre-of-mass frame providing guidance on the different frames in which the angles are defined. }
        \label{fig:zh_angles}
    \end{figure}

\paragraph{Higgs in association with a vector boson:} In Ref.~\cite{Banerjee:2019pks}, the leading contribution of $\mathcal{O}_{\varphi\widetilde{W}B}$, $\mathcal{O}_{\varphi\widetilde{W}}$ and $\mathcal{O}_{\varphi\widetilde{B}}$ to $Zh$ production is proportional to  
\begin{equation}\label{eq:zhangle}
    \sin\varphi \sin{2\Theta}\sin{2\theta},
\end{equation}
where the angles are defined in two frames. All angles and frames are displayed in Figure \ref{fig:zh_angles}, taken from Ref.~\cite{Banerjee:2019pks} for better understanding, and are defined as follows. The right-handed basis $\{\hat{x}, \hat{y}, \hat{z} \}$ is constructed in the $Zh$ centre-of-mass frame such that the $\hat{z}$ axis is aligned with the $Z$ momentum; the $\Hat{y}$ axis is normal to the plane formed by $\hat{z}$ and the beam axis; the $\Hat{x}$ axis completes the set of normal coordinates. $\Theta$ is the angle between the $Z$ momentum and the beam axis and $\varphi$ is the angle between the $Z$ decay products plane and the $(\hat{x}, \hat{z})$ plane, both in the $Zh$ centre-of-mass frame. Finally, $\theta$ is the angle, in the $Z$ centre-of-mass frame, of the positively charged lepton momentum with the $\hat{z}$ axis. Since the $CP$-odd contribution vanishes after integration over the phase space, the constraints are obtained by convolution with the function of the angle in Eq.(\ref{eq:zhangle}). Similarly, for $Wh$ production, the same angle $\varphi_W$ is defined in Ref.~\cite{Bishara:2020vix} for the W instead of the Z boson and it was used to probe $CP$ violating effects. NLO QCD corrections to $Wh$ were analysed in Ref.\cite{Rossia:2024rfo} with the modulation of $\varphi_W$ and it was demonstrated that $CP$-violating effects in $Zh$ required more specific observables than $\varphi$ (even at NLO QCD). NLO effects in Higgs production associated with a $Z$ at an electron-positron collider were derived by Refs.~\cite{Asteriadis:2024xuk, Asteriadis:2024xts}. The $CP$ observable was the angle $\theta_{e^{+}h}$ between the incoming electron and the outgoing $h$.

In Ref.~\cite{Banerjee:2019twi}, the Method of Moments is applied to $Zh$ and $Wh$, and three angular functions probe the three operators mentioned above :
\begin{itemize}
    \item $\Tilde{f}^1_{LT} = \sin{\varphi} \sin{\theta} \sin{\Theta}$,
    \item $\Tilde{f}^2_{LT} = \sin{\varphi} \sin{\theta} \sin{\Theta} \cos{\theta} \cos{\Theta}$,
    \item $\Tilde{f}_{TT'} = \sin{2\varphi} \sin^2{\theta} \sin^2{\Theta}$,
\end{itemize}
where the angles are defined in Figure \ref{fig:zh_angles}. The indices $L$ and $T$ respectively denote the longitudinal and transverse polarisations of the weak vector bosons in the interfering amplitudes. $TT'$ stands for the interference of two transverse amplitudes where the bosons have opposite polarisations. Due to the linear dependence on the trigonometric functions, any integration over one of the angles $\{\varphi,\theta,\Theta\}$ would cancel $\Tilde{f}^2_{LT}$ whereas $\Tilde{f}_{TT'}$ is zero under the integration over $\varphi$. The corresponding coefficients $a_i$, called angular moments, of the three angular functions then provide constraints on the Wilson coefficients $C_{\varphi\widetilde{W}B}$, $C_{\varphi\widetilde{W}}$ and $C_{\varphi\widetilde{B}}$. This method was applied in $Zh$ production at a future leptonic collider in Ref.~\cite{Subba:2025hkq} where different decay channels of the Higgs boson are investigated.

\paragraph{Signed azimuthal angle difference:}
The asymmetry in the signed azimuthal angle difference $\Delta \left(\Delta \phi_{pp'} \right)$ between final state particles $p$ and $p'$ is defined as
\begin{equation}
    \Delta \left(\Delta \phi_{pp'} \right) = \sigma(\Delta \phi_{pp'} < \pi/2) - \sigma(\Delta \phi_{pp'} > \pi/2),
\end{equation}
where the final state particles are ordered by their rapidities $y_p$, $y_{p'}$ such that, if $|y_{p'}|>|y_p|$, then
\begin{equation} \label{eq: signed azim angle def}
    \Delta \phi_{pp'} = \phi_{p'} - \phi_p.
\end{equation}
By construction, $\Delta \phi_{pp'} \in [-\pi,\pi]$. This observable corresponds to the difference in azimuthal angles between the 'forward' particle and the 'backward' particles , with respect to the $\hat{z}$ axis, produced in a collider with a symmetric initial state such as the LHC.

In Ref.~\cite{Brehmer:2017lrt}, it is mentioned that this observable in Higgs production by VBF is related to a triple product but the experimental limitations in the initial partons momenta do not allow the measurement of the triple product directly with the required precision. Ref.\cite{Rahaman:2019lab} shows the effect of asymmetries in a fit of diboson data, particularly how the $CP$-even and $CP$-odd can be disentangled once the asymmetry in $\Delta \phi$ is included compared to a fit using only cross-sections.

In the context of SMEFT, $\Delta\phi_{jj'}$ was studied in Higgs production by VBF, \textit{i.e.} in the $hjj$ production channel where the signed azimuthal angle difference between the jets is probed, see Refs.~\cite{Plehn:2001nj, Hankele:2006ma, Klamke:2007cu, Brehmer:2017lrt, Englert:2019xhk}. The ATLAS collaboration produced limits on $C_{\varphi \widetilde{G}}/\Lambda^2$ and $C_{\varphi \widetilde{W}}/\Lambda^2$ based on $\Delta\phi_{jj'}$ in $h \to WW^*$. In the $h \to \tau \tau$ channel, $\Delta\phi_{jj'}$ was used with other $CP$ sensitive observables to extract constraints on $C_{\varphi \widetilde{W}}/\Lambda^2$ in Ref.\cite{ATLAS:2025bts} since it contributes in the Higgs production matrix element. One of the other observable is the sine of $\Delta\phi_{jj'}$ multiplied by the $p_T$'s of the VBF jets.

Ref.~\cite{Englert:2019xhk} further applied the observable in Higgs associated with a top pair production where the signed azimuthal angle difference is between the leptons $l^{ (\prime) }$ originating from the two top decays $\Delta\phi_{ll'}$. Ref.\cite{Miralles:2024huv} compared $\Delta\phi_{ll'}$ to the equivalent parton-level observable $\Delta\phi_{t\overline{t}}$ showing that the easily accessible observable $\Delta\phi_{ll'}$ retains comparable sensitivity to $CP$ effects compared to $\Delta\phi_{t\overline{t}}$.

Recently, this observable was applied in the leptonic decay of diboson production by Ref.~\cite{DasBakshi:2020ejz}. For $W^+W^-$, $p$ and $p'$ in $\Delta \phi_{pp'}$ are the two detectable leptons $l^+$ and $l^-$ from the decay of the two massive bosons. In case of a neutral boson, either a $Z$ boson or a photon, associated with a $W$, the momentum of the neutral boson is used with the lepton $l$ from the decay of $W$.

\paragraph{h $\to$ 4l:}
Another observable has been proven useful to the particular case of a Higgs boson decaying in two pairs of same-flavour leptons with opposite charges, $l^+ l^-$ and $l'^{+}l'^{-}$, through two decaying neutral bosons
\begin{equation}
    p~ p \rightarrow h \rightarrow l^+(p_1)~ l^-(q_1)~ l'^{+}(p_2)~ l'^{-}(q_2),
\end{equation} 
thus testing the $hZZ$ vertex. One can construct the angle $\Phi \in [-\pi, \pi]$ between the $Z$ decay planes in the Higgs frame, 
\begin{equation}
\label{eq:Phi}
    \Phi = \frac{\Vec{P} . (\hat{n}_1 \times \hat{n}_2)}{|\Vec{P} . (\hat{n}_1 \times \hat{n}_2)|} \arccos{(-\hat{n}_1 . \hat{n}_2)}.
\end{equation}
where the unit vectors $\Hat{n}_i$ are normal to each decay plane, $\Hat{n}_i = \frac{( \Vec{p}_i \times \Vec{q}_i)}{|\Vec{p}_i \times \Vec{q}_i|}$, and $\Vec{P}$ is the 3-momentum of the $Z$ boson with the largest invariant mass reconstructed from the lepton pairs. Rather than the actual value of this angle, it is the sign of $\Phi$ which is the $CP$ sensitive observable, or equivalently, the sign of $\sin{\Phi}$. Thus, an asymmetry $\Delta \Phi$ or $\Delta \sin{\Phi}$ is again helpful to extract information on $CP$ violating effects. This observable was originally designed to probe the $CP$ and tensor structure of a single-produced resonance decaying in SM particles, that is testing the properties of the Higgs \cite{Gao:2010qx, Bolognesi:2012mm}. Nevertheless, it was later used to constrain SMEFT operators in the four-lepton decay of the Higgs in Refs.~\cite{Beneke:2014sba,Brehmer:2017lrt}, despite the slightly different ways to construct the angle $\Phi$. It is worth emphasising that a triple product is the starting point to derive this observable.

\begin{figure}[t]
    \centering
    \includegraphics[scale=0.6]{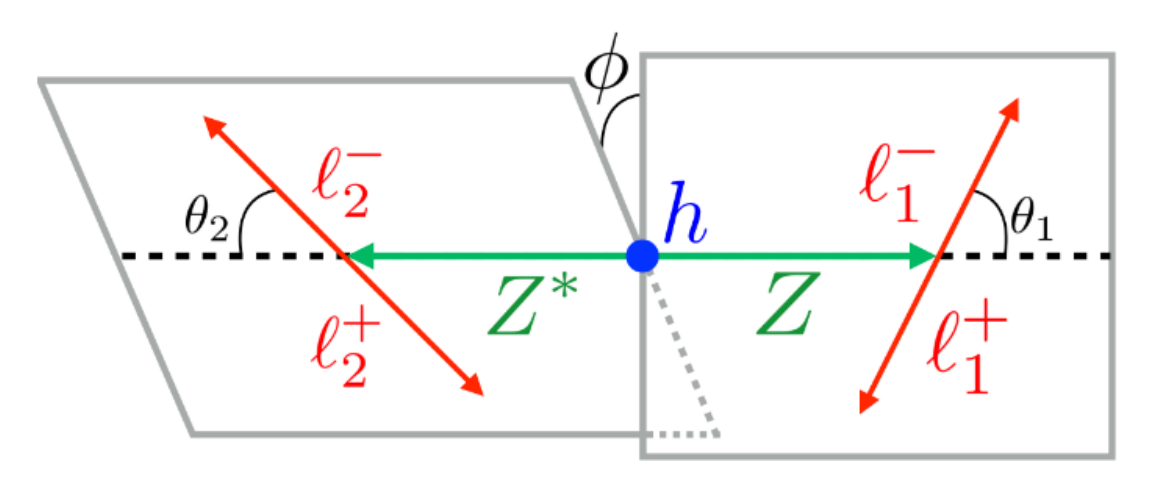}
    \caption{Description of the $h \to 4l$ production in the centre-of-mass frame providing guidance on the different frames in which the angles are defined. }
     \label{fig:hto4l_angles}
\end{figure}

Similarly to the associated Higgs production with a weak boson in Ref.\cite{Banerjee:2019twi}, Ref.\cite{Banerjee:2020vtm} studies the angular moments of this process. Here, three functions are sensitive to $\mathcal{O}_{\varphi\widetilde{W}B}$, $\mathcal{O}_{\varphi\widetilde{W}}$ and $\mathcal{O}_{\varphi\widetilde{B}}$:
\begin{itemize}
    \item $f_7 = \left( \cos^2{\theta_1} -1 \right) \left( \cos^2{\theta_2} -1 \right) \sin{2\phi}$,
    \item $f_8 = \sin{\theta_1} \sin{\theta_2}  \sin{\phi}$,
    \item $f_9 =\sin{2\theta_1} \sin{2\theta_2}  \sin{\phi}$,
\end{itemize}
where the angles and frames are displayed in Figure \ref{fig:hto4l_angles}, taken from Ref.~\cite{Banerjee:2020vtm}. The polar angles $\theta_i$ are taken between the negatively-charged lepton with the associated decaying Z boson in the Z boson rest frame and the $i$ refers to the 2 lepton pairs. The two decay planes determine the azimuthal angle $\phi$ in the $h$ rest frame (the same as the $4l$ centre-of-mass frame).  
In practice, all possible helicity configurations must be summed over after the corresponding phase shifts of the three angles and it is the charge that becomes fixed. 

\paragraph{Diboson production:}
A direct observable has been recently developed in diboson production 
\begin{equation}
    p~ p \rightarrow W^{\pm}~ Z/\gamma
\end{equation} 
in Refs.~\cite{Azatov:2019xxn,Panico:2017frx}. In particular, Ref.~\cite{Azatov:2019xxn} presents a short development to introduce the observable in $W \gamma$ production that can be translated to $WZ$ production. Considering the $2 \rightarrow 3$ process, the differential cross-section is given by 
\begin{equation}
    \frac{d\sigma}{d\Omega} = \frac{1}{2s} \frac{|\sum_{hel=0,\pm} (\mathcal{M}^{SM}_{q\Bar{q}\rightarrow \gamma+ W_{hel}} + \mathcal{M}^{dim6}_{q\Bar{q}\rightarrow \gamma+ W_{hel}}) \mathcal{M}_{W_{hel} \rightarrow l-\Bar{\nu}+}  |^2}{(p_W^2 - m_W^2)^2 + m_W^2 \Gamma_W^2}
\end{equation}
where $hel$ are the possible helicities for the intermediate W boson and 
%$\Omega=(2\pi)^4 \delta^4( \sum_i p_i - p_f) \Pi_i \frac{d^3 p_i}{2E_i (2\pi)^3}$ 
$\Omega$ is the Lorentz-invariant phase space from Eq.(\ref{eq:LorentzInvariantPhaseSpace}). $m_Z$ and $m_W$ are the masses of the Z and W bosons respectively, and $\Gamma_W$ is the width of W. In the narrow width approximation, the leading contribution of the interference term reduces to
\begin{multline}\label{eq:interfWA}
    \frac{\pi}{2s} \frac{\delta(s-m_W^2)}{\Gamma_W m_W} \mathcal{M}^{SM}_{q\Bar{q}\rightarrow \gamma+ W_{T_-}} (\mathcal{M}^{dim6}_{q\Bar{q}\rightarrow \gamma+ W_{T_+}})^* \mathcal{M}_{W_{T_-} \rightarrow l-\Bar{\nu}+} (\mathcal{M}_{W_{T_+} \rightarrow l-\Bar{\nu}+})^* \\ + h.c.
\end{multline}
The longitudinal polarisations are subleading contributions so they are not included, neither in Ref.\cite{Azatov:2019xxn} nor here. The interference would vanish due to the helicity selection rules without the last two factors in Eq.(\ref{eq:interfWA}). This is the interference resurrection. If only the $2 \rightarrow 2$ process was considered and the branching ratios were applied, the interference would vanish. The decay amplitudes prevent this from happening: it is only by looking at the complete process that the interference can be appreciated. Moreover, the decay amplitudes are such that
\begin{equation*}
    \mathcal{M}_{W_{T_-} \rightarrow l-\Bar{\nu}+} (\mathcal{M}_{W_{T_+} \rightarrow l-\Bar{\nu}+})^* \propto e^{-2i\phi_W},
\end{equation*}
with $\phi_W$ being the angle between the plane determined by the W decay products and the $W\gamma$ scattering plane. The amplitude produced by the product of the first two terms in Eq.(\ref{eq:interfWA}), 
\begin{equation*}
    P(a\rightarrow b) = \mathcal{M}^{SM}_{q\Bar{q}\rightarrow \gamma+ W_{T_-}} (\mathcal{M}^{dim6}_{q\Bar{q}\rightarrow \gamma+ W_{T_+}})^*
\end{equation*} 
has definite $CP$-properties
\begin{equation*}
    P(a \rightarrow b) = - P(b \rightarrow a),
\end{equation*}
but simultaneously the optical theorem ensures that the amplitude is equivalent to the conjugate amplitude of the reversed process. Thus,
\begin{equation*}
    P(a \rightarrow b) = -  P(a \rightarrow b)^*.
\end{equation*}
When considering $CP$-odd dimension-six operators this means that the hermitian conjugate of the interference amplitude will take a minus sign and an opposite phase with respect to the original interference amplitude, resulting in an interference cross-section depending on $\sin{\phi_W}$. This development is valid for $WZ$ production as well when applied for each decaying boson. In the end, we have
\begin{equation}\label{eq:barducciobservables}
\begin{split}
    |\mathcal{M}_{W \gamma}|^2 & ~~ \propto \sin{2\phi_W}, \\
    |\mathcal{M}_{W Z}|^2 & ~~ \propto \sin{2\phi_W} + \sin{2\phi_Z}.
\end{split}
\end{equation}
The reader is referred to Ref.~\cite{Panico:2017frx} for a rigorous development.

Practically, this observable is constructed similarly to $\Phi$ but, rather than looking at the angle between two vector boson decay planes, one looks at the angle between the decay and scattering planes for each boson. To construct the observable a reference axis is needed. Refs.~\cite{Azatov:2019xxn,Panico:2017frx} use the $\Hat{z}$-axis in the laboratory frame. The two unit vectors $\hat{n}_{scat}$ and $\hat{n}_{decay}$ are defined as
\begin{equation}
\begin{split}
     \hat{n}_{scat} &= \frac{\hat{z} \times \Vec{p}_{V}}{|\Vec{p_V}|}, \\
     \hat{n}_{decay} &=\frac{\Vec{p}_{l+} \times \Vec{p}_{l-}}{|\Vec{p}_{l+} \times \Vec{p}_{l-}|},
\end{split}
\end{equation}
where $V=\{W,Z\}$ is the considered vector boson decaying into the leptons $l_+$ and $l_-$. The index $\pm$ indicates the helicity of the lepton. The two vectors $\hat{n}_{scat}$ and $\hat{n}_{decay}$ are respectively normal to the scattering plane and the decay plane of the vector boson forming an angle $\phi_V$ between them. The angle is obtained by  
\begin{equation}
    \phi_V = \textrm{sign}[ (\hat{n}_{scat} \times \hat{n}_{decay}) . \Vec{p}_{V}] \arccos{( \hat{n}_{scat} . \hat{n}_{decay} }).
\end{equation}
In the end, the considered observable is the sine of twice the angle between the two planes, $\sin 2\phi_V$. The angle is defined to rotate counter-clockwise with respect to the boson direction of flight. Following the treatment of the amplitudes above and their final dependence in Eq.(\ref{eq:barducciobservables}), the observable is $\sin 2\phi_W$ in $W\gamma$ production or $\sin 2\phi_W + \sin 2\phi_Z$ in $WZ$ production. It is interesting to note that a triple product is again present to construct this $CP$ sensitive observable.

Ref.\cite{ElFaham:2024uop} investigated the dileptonic decay channel of $W^+ Z$ production at NLO in QCD with three angular observables. The first two are the azimuthal angle of the lepton boosted in the massive boson rest frame : $\phi^*_{e^+}$ in the $W$ boson rest frame and $\phi^*_{\mu^-}$ in the $Z$ boson rest frame. The polar angle of the positron in the $W$ rest frame $\theta_{e^+}^*$ is the third $CP$ sensitive observable considered, even though the deviations from the SM are not as large as for the two previous observables.

\paragraph{Spin correlations :}

The top quark is expected to be particularly sensitive to NP: the four $CP$-odd operators with fermion fields in the reduced basis are built with the top quark resulting in many relevant processes involving the top in Section~\ref{subsec:POI}. In particular, top pair production has a large cross section and been very accurately observed at the LHC. Therefore, it is especially interesting for SMEFT studies including $CP$ violation. The results from Ref.~\cite{Bernreuther:2015yna} are summarised below with a focus on $CP$-odd observables.

At the partonic level, the production matrix $R^I$ in top pair production can be decomposed in four terms with respect to the top and anti-top spin spaces,
\begin{equation}
\label{eq:partontoppair}
    R^I = f^I [A^I \mathbb{1} \otimes \mathbb{1} + \widetilde{B}^{I+}_i \sigma^i \otimes \mathbb{1} + \widetilde{B}^{I-}_j \mathbb{1} \otimes \sigma^j + \widetilde{C}^I_{ij} \sigma^i \otimes \sigma^j],
\end{equation}
where $\mathbb{1}$ is the $2\times2$ unit matrix and $\sigma^i$ ($\sigma^j$) are the Pauli matrices forming the top (anti-top) spin space. The index $I$ represents the initial state in top pair production which can be either gluon-gluon $(g g)$ or quark-antiquark $(q\Bar{q})$ at LO. An orthogonal basis $\{\Hat{r},\Hat{k}, \Hat{n}\}$ built from the top and parton momenta is defined as follows: $\hat{k}$ is aligned with the top quark momentum in the $\bar{t}t$ centre-of-mass frame,  $\hat{p}$ is the direction of one of the initial partons in the same frame, $\hat{r}$ and $\hat{n}$ are respectively
\begin{equation*}
    \hat{r} = \frac{1}{r} (\hat{p}-y\hat{k}), ~~ \hat{n}=\frac{1}{r}(\hat{p}\times\hat{k}),
\end{equation*}
where $ y = \hat{p}.\hat{k}$ and $r=\sqrt{1-y^2}$. Namely, $\hat n$ is orthogonal to the scattering plane while  $\hat r$ is in the scattering plane and orthogonal to $\hat k$.
This basis ensures that each energy-dependent component of $\widetilde{B}^{I\pm}_i$ and $\widetilde{C}^I_{ij}$ now respects definite properties under charge conjugation C-, parity P- and pseudo-time $T_N$-transformations. They are decomposed as follows,
\begin{eqnarray*}
    \Hat{B}_{i}^{I\pm} &=& b_{r}^{I\pm} \hat{r}_i + b_{k}^{I\pm} \hat{k}_i + b_{n}^{I\pm} \hat{n}_i , \\
    \Hat{C}_{ij}^{I} &=& c^I_{rr} \hat{r}_i \hat{r}_j + c^I_{kk} \hat{k}_i \hat{k}_j + c^I_{nn} \hat{n}_i \hat{n}_j \\
                    & & + c^I_{rk} (\hat{r}_i \hat{k}_j + \hat{k}_i \hat{r}_j) + c^I_{kn} (\hat{k}_i \hat{n}_j + \hat{n}_i \hat{k}_j) + c^I_{rn} (\hat{r}_i \hat{n}_j + \hat{n}_i \hat{r}_j) \\
                    & & + c^I_r (\hat{k}_i \hat{n}_j - \hat{n}_i \hat{k}_j) + c^I_k (\hat{n}_i \hat{r}_j - \hat{r}_i \hat{n}_j) + c^I_n (\hat{r}_i \hat{k}_j - \hat{k}_i \hat{r}_j).
\end{eqnarray*}
Table 1 in Ref.~\cite{Bernreuther:2015yna} lists the properties of the $\Hat{B}_{i}^{I\pm}$ and $\Hat{C}_{ij}^{I}$ coefficients under $CP$ and $P$. The coefficients $c_{rn}$ and $c_{kn}$ are $C$- and $P$-odd which makes them $CP$-even. However, the coefficients $c_k$, $c_r$, $b_r^+-b_r^-$ and $b_k^+-b_k^-$ are all $CP$-odd and P-odd while $c_n$ and $b_n^+-b_n^-$ are $CP$-odd but P-even. They all can constrain $CP$-odd operators. These coefficients are negligible in the SM because they arise from the CKM phase or from absorptive phase at one-loop. Since all considered $CP$-odd dimension-six operators are P-odd, they can only be constrained by the first four combinations. These coefficients defined at the partonic level are not observables per se and should be linked to actual accessible measurements, \textit{i.e.} independent of the initial partons and dependent on the top decay products: the charged leptons, the jets and the missing transverse energy in semi- or di-leptonic events. Fortunately, the charged lepton directions are highly correlated with the top spin direction.

The normalised differential cross-section expressed as a function of the angular distributions of the final state leptons, $\Omega_{\pm}=\phi_\pm \cos{\theta_\pm}$, 
\begin{equation}
\label{eq:particletoppair1}
    \frac{1}{\sigma} \frac{d\sigma}{d\Omega_{+} d\Omega_{-}} = \frac{1}{(4\pi)^2} [1 + \Vec{B}_1 .\Hat{l}_+ + \Vec{B}_2.\Hat{l}_- + \Hat{l}_+.C'.\Hat{l}_-]
\end{equation}
offers therefore an access to the top quarks spins.
In this equation, the vector $\Hat{l}_+$ ($\Hat{l}_-$) is the unit vector aligned with the lepton $l^+$ ($l^-$) direction of flight in the $t$ ($\Bar{t}$) rest frame. 
The definition of two reference axes $\Hat{a}$ and $\Hat{b}$ allows the definition of the angles
\begin{equation}
    \cos{\theta_+} = \Hat{l}_+ . \Hat{a} ~~\text{and}~ \cos{\theta_-} = \Hat{l}_- . \Hat{b}.
\end{equation}
Integrating the differential cross-section over all the other variables, we obtain
\begin{multline}
\label{eq:particletoppair2}
    \frac{1}{\sigma} \frac{d\sigma}{d\cos{\theta_+} d\cos{\theta_-}} = \frac{1}{4} [1 + B_1(\Hat{a}) \cos{\theta_+} + B_2(\Hat{b}) \cos{\theta_-} \\ + C(\Hat{a},\Hat{b}) \cos{\theta_+} \cos{\theta_-} ].
\end{multline}
The reference axes $\Hat{a}$ and $\Hat{b}$ are chosen from a non-orthogonal set of vectors $\{\Hat{p}_p, \Hat{r}_p, \Hat{n}_p \}$:  $\hat{k}$ is still aligned with the top quark momentum in the $\bar{t}t$ centre-of-mass frame, now $\hat{p}_p=(0,0,1)$ is one of the proton beam directions in the laboratory frame, then $\hat{r}_p$ and $\hat{n}_p$ are respectively
\begin{equation*}
    \hat{r}_p = \frac{1}{r_p} (\hat{p}_p-y_p\hat{k}), ~~ \hat{n}_p=\frac{1}{r_p}(\hat{p}_p\times\hat{k}), 
\end{equation*}
where $ y_p = \hat{p}_p.\hat{k}$ and $r_p=\sqrt{1-y_p^2}$. The authors define several sets of reference axes:
\begin{eqnarray*}
    n &:& \hat{a} = sign(y_p)\hat{n}_p, ~ \hat{b} = -sign(y_p)\hat{n}_p, \\
    r &:& \hat{a} = sign(y_p)\hat{r}_p, ~ \hat{b} = -sign(y_p)\hat{r}_p, \\
    k &:& \hat{a} = \hat{k}, ~ \hat{b} = -\hat{k}, \\
    r^* &:& \hat{a} = sign(\Delta|y|)sign(y_p)\hat{r}_p, ~ \hat{b} = -sign(\Delta|y|)sign(y_p)\hat{r}_p, \\
    k^* &:& \hat{a} = sign(\Delta|y|)\hat{k}, ~ \hat{b} = -sign(\Delta|y|)\hat{k},   
\end{eqnarray*}
where $\Delta |y| = |y_t|-|y_{\Bar{t}}|$, $|y|$ being the modulus of the rapidity in the laboratory frame.

The differences $C(n,r) - C(r,n)$ and $C(n,k) - C(k,n)$ (P-odd, $CP$-odd) probe $CP$ violating effects through correlations between lepton angular distributions. In Table 6 of Ref.~\cite{Bernreuther:2015yna}, these observables are actually related to the partonic level coefficients $c_k$ and $c_r$, respectively. The differences $B_1(k)-B_2(k)$ and $B_1(k^*)-B_2(k^*)$ are related to the difference in transverse polarisations $b^{I+}_{k} - b^{I-}_{k}$ defined at the partonic level. A similar situation appears for $B_1(r)-B_2(r)$  and $B_1(r^*)-B_2(r^*)$ with $b^{I+}_{r} - b^{I-}_{r}$. For these differences the SM contributions vanish at tree-level and any sizeable value would point to $CP$ violation from NP.

One particularly interesting point is that differences of correlation coefficients $C(n,r) - C(r,n)$ and $C(n,k) - C(k,n)$ are related to $CP$-odd triple product correlations 
\begin{equation}
\begin{split}
    O^{CP}_1 &=(\Hat{l}_+ \times \Hat{l}_-).\Hat{k}, \\
    O^{CP}_2 &= sign(y_p) (\Hat{l}_+ \times \Hat{l}_-).\Hat{r}_p,
\end{split}
\end{equation}
such that
\begin{equation}
\begin{split}
    C(n,r) - C(r,n) & = 9<O^{CP}_1> , \\
    C(n,k) - C(k,n) & = -9<O^{CP}_2>.
\end{split}
\end{equation}

Ref.~\cite{Subba:2025hxn} relates the spin correlation parameters to angular moments in $W^- W^+$ production at a leptonic collider. The production density matrix of the $W^{\pm}$ bosons reads as
\begin{multline}
\label{eq:SpinCorrWW}
    \rho_{W^- W^+} = \mathbb{1} \otimes \mathbb{1} + \sum_{i=1}^{8} \left( P ^{(2)}_i \mathbb{1}  \otimes J^{(2)}_i + P ^{(1)}_i J^{(1)}_i \otimes \mathbb{1} \right) \\
    + \sum_{i,j = 1}^{8} C_{ij}^{(12)}  J^{(1)}_i \otimes J^{(2)}_j .
\end{multline}
The spin correlation parameters $C^{(12)}_{ij}$ are obtained from the angular asymmetries $A_{ij}$ which are defined as  
\begin{equation}
    A_{ij} = \frac{\sigma(f_i^{(1)} f_j^{(2)} > 0) - \sigma(f_i^{(1)} f_j^{(2)} < 0)}{\sigma(f_i^{(1)} f_j^{(2)} > 0) + \sigma(f_i^{(1)} f_j^{(2)} < 0)}.
\end{equation}
The $f_i^{(1/2)}$ are angular functions of the final state visible fermions from the respective decays of the $W$ bosons. Among them, $f_2=\sin \theta \sin \phi$ is $CP$-odd, where $\theta$ is the polar angle and $\phi$ the azimuthal angle. On the other hand, $f_1=\sin \theta \cos \phi$ and $f_3=\cos \theta$ are $CP$-even. That means that the combinations of $f_1$ and $f_3$ with an odd number of occurrences of $f_2$ are $CP$-odd too. Thus, $f_4 = f_1 . f_2$ and $f_6=f_2 . f_3$ are $CP$-odd. The polarisation asymmetries $A_{i0}$ and $A_{0j}$ are obtained with the same angular functions. $\mathcal{O}_{\widetilde{W}WW}$ and $\mathcal{O}_{\varphi \widetilde{W}}$ are particularly sensitive to spin correlation asymmetries $A_{ij}$ and the polarisation asymmetries give marginal improvements. In particular, $A_{44}$ strongly constrains both $\mathcal{O}_{\widetilde{W}WW}$ and $\mathcal{O}_{\varphi \widetilde{W}}$. $A_{16}$ gets sizeable deviations from $\mathcal{O}_{\varphi \widetilde{W}}$ but does not perform as well with $\mathcal{O}_{\widetilde{W}WW}$. The polarisation asymmetry $A_{70}$ obtains subleading contributions from $\mathcal{O}_{\varphi \widetilde{W}}$ compared to the spin correlation asymmetries.

\paragraph{Triple products:}
   
Genuine triple products are older observables, first used at the Tevatron \cite{Dawson:1996ge, Donoghue1987}, where quark and anti-quark directions were statistically easier to access due to an anti-proton beam, and in hadron decays \cite{Durieux:2015zwa}.

They have been used in diboson production as well. In Ref.~\cite{Kumar:2008ng}, the triple product $\Vec{p}_q . (\Vec{k}_Z \times \Vec{p}_l)$, where $\Vec{p}_q$ is the momentum of the incoming quark, $\Vec{k}_Z$ the reconstructed momentum of the $Z$ boson and $\Vec{p}_l$ the momentum of the outgoing charged lepton, has been used to probe $CP$ violation due to the anomalous triple coupling $\widetilde{\lambda}_Z$, which is originating from $\mathcal{O}_{\widetilde{W}WW}$, in $WZ$ production. Since the incoming quark momentum is not an observable quantity, the authors replaced it with $(0,0,\Vec{k}_Z^z)$ as a proxy. They find that the replacement holds around 70\% of the time so the sensitivity to $CP$ largely remains. Their asymmetry $\Delta$ is then built by weighting the events such that
\begin{equation}
    \Delta = \int d\sigma~ \Xi^z_\pm (k_Z, p_l) .
\end{equation}
where 
\begin{equation}\label{eq: def Xi}
    \Xi^z_\pm (k_Z, p_l) = sign(k_Z^z)~ sign[(\Vec{k}_Z \times \Vec{p}_l)^z].
\end{equation}

In $W^+ W^-$ production, Ref.~\cite{Han:2009ra} also considers a genuine triple product $(\Vec{p}_{f} \times \Vec{p}_{\Bar{f}}) . \Vec{p}_q$ with $\Vec{p}_f$ $(\Vec{p}_{\Bar{f}})$ being the 3-momentum of the positively (negatively) charged lepton but faces the same issues with the quark momentum $\Vec{p}_q$. Consequently, the triple product is multiplied by a quantity that depends on the quark direction as well, which we will refer to as the direction factor. The final observable is quadratic in the beam direction and will not be affected by the uncertainty in the actual quark direction. The observable is noted here as
\begin{equation}
    O_Z = \left( (\Vec{p}_{f} \times \Vec{p}_{\Bar{f}}) . \Vec{p}_q \right) sign[(\Vec{p}_{f} - \Vec{p}_{\Bar{f}}) . \Vec{p}_q]
\end{equation}
and $\Vec{p}_q$ is substituted by the $\Hat{z}$ axis without losing any information on the sign of $O_Z$. This is a generalisation of the observable in Ref.~\cite{Dawson:1996ge}.

In $W\gamma$ production, Ref.~\cite{Dawson:2013owa} follows a similar approach and argues that the best observables to probe $CP$ violation are the asymmetries in the triple products
\begin{equation}
\begin{split}
    O_W & = (\Vec{p}_\gamma \times \Vec{p}_{beam}) . \Vec{p}_l, \\
    O_{\gamma} &= \Vec{p}_{\gamma} . \Vec{p}_{beam} O_W, \\
    O_{l} &= \Vec{p}_l . \Vec{p}_{beam} O_W.
\end{split}
\end{equation}
Just like in Ref.~\cite{Han:2009ra}, the first observable is ill-defined due initially to the quark direction and returns a zero asymmetry, as opposed to the last two which contain the direction factor and are quadratic in the beam direction. However, due to the $pp$ initial state, the $W^\pm \gamma$ processes are not exactly $CP$ eigenstates from each other and a non-zero asymmetry would not directly provide a proof of $CP$ violation. It is important to note that Ref.~\cite{Dawson:2013owa} allows for absorptive phases while it is not the case here.

In $Z\gamma$ production, the asymmetry in $O_Z$ is used to probe $CP$ violating effects in Ref.~\cite{Dawson:2013owa}. Reciprocally to the angles related to triple products above, these triple products determine different angles in the final state. As an explicit example, Ref.~\cite{Han:2009ra} practically defines an azimuthal angle $\Phi$ between the two outgoing charged leptons in the transverse plane,
\begin{equation}
    \Phi = sign[(\Vec{p}_{f} - \Vec{p}_{\Bar{f}}) . \Hat{z}] \arcsin{\left[(\Vec{p}_{f} \times \Vec{p}_{\Bar{f}}) . \Hat{z} \right]} 
\end{equation}
and provides predictions at the LHC for the asymmetry
\begin{equation}
    A_{\Phi} = \frac{ \mathcal{N}_{\Phi>0} - \mathcal{N}_{\Phi<0} }{ \mathcal{N}_{\Phi>0} + \mathcal{N}_{\Phi<0} }.
\end{equation}
An analysis of $Zh$ production by Ref.~\cite{Christensen:2010pf} also used this asymmetry where $f$ and $\overline{f}$ are the leptons from the leptonic decay of the $Z$ boson.

In $Vh$, $V=W$ or $Z$, an observable constructed from a triple product but not used as such is $\cos \delta^+$. Ref.~\cite{Godbole:2014cfa} presents $\delta^+$ as the angle between the direction of flight of the lepton from the $V$ decay and the production plane of $V$ and $h$. The angle is defined as
\begin{equation}
    \cos \delta^+ = \frac{\vec{p}_{l1}^{~(V)} . (\vec{p}_h \times \vec{p}_{V}) }{|\vec{p}_{l1}^{~(V)}| |\vec{p}_h \times \vec{p}_{V}| }.
\end{equation}
The momentum of the Higgs and the vector boson $V$ are respectively $\vec{p}_{h}$ and $\vec{p}_{V}$. $\vec{p}_{l1}^{(V)}$ is the momentum of the lepton from the vector boson in its rest frame. The lepton considered can be either the visible lepton for $Wh$ or the negatively-charged lepton for $Zh$. Ref.~\cite{Godbole:2014cfa} also introduces 
\begin{equation}
    \cos \delta^- = \frac{\vec{p}_{V} . (\vec{p}^{h}_{l_1} \times \vec{p}^{h}_{l_2}) }{|\vec{p}_{V}| |\vec{p}^{h}_{l_1} \times \vec{p}^{h}_{l_2}| },
\end{equation}
with $l_2$ being the neutrino for $Wh$ or the positively-charged lepton for $Zh$. This observable is correlated to $\cos \delta^+$, so Ref.~\cite{Godbole:2014cfa} only relies on the asymmetry of $\cos \delta^+$ to extract limits on $CP$-odd effective interactions related to $\mathcal{O}_{\varphi \widetilde{W}}$, $\mathcal{O}_{\varphi \widetilde{B}}$ and $\mathcal{O}_{\varphi \widetilde{W}B}$. A recent analysis by ATLAS further multiplied $\cos \theta^+$ with the sign of the charge of the lepton coming from the $W$ decay in $Wh$ production.

In $t\overline{t}h$ and $th$ production, Ref.~\cite{Ellis:2013yxa} describes how the asymmetries in angular observables used in their analysis arise from triple products. $(\vec{p}_j \times \vec{p}_h). \vec{p}_l$ in $th$ and the sign of $(\vec{p}_{l^-} \times \vec{p}_{l^+}). \vec{p}_t$ in $t\overline{t}h$ give respectively $\cos{\theta_{l \perp}}$ and the sign of $\Delta \phi_{ll}$. Ref.~\cite{Miralles:2024huv} considers the two triple products $(\vec{p}_t \times \vec{p}_j). \vec{p}_h$ and $\hat{z}.(\vec{p}_t \times \vec{p}_j)|_h$ in the Higgs rest frame for single top-Higgs associated production where the top decays leptonically. The latter triple product shows a different modulation in sign compared to the former triple product, the interference changes sign three times over the phase space, so it requires a specific binning to construct its asymmetry ($[-1.0, -0.1, 0.0, 0.1, 1.0]$ instead of the naive $[-1.0, 0.0, 1.0]$) as discussed in Section~\ref{subsec:signofinterference}. Polarisation observables $\theta^{y/s}_l$ are obtained with some triple product by projecting the lepton $l$ momentum on the axes $\hat{y}$ and $\hat{s}$ defined respectively as
\begin{equation}
    \hat{y} = \frac{\vec{p}_j \times \hat{z}}{|\vec{p}_j \times \hat{z}|} \text{  and  } \hat{s} = \frac{\vec{p}_j \times \vec{p}_h}{|\vec{p}_j \times \vec{p}_h|}.
\end{equation}
In the dileptonic decay channel of top-pair production with a Higgs, Ref.~\cite{Miralles:2024huv} uses the same triple product as Ref.~\cite{Ellis:2013yxa} and takes into account the challenging reconstruction of the top quarks by focusing on its sign and multiplying it with the scalar product of the lepton momenta in the $t\overline{t}$ rest frame $\text{sign}[(\vec{p}_{l^-} \times \vec{p}_{l^+}). \vec{p}_t] (\vec{p}_{l^-} . \vec{p}_{l^+})|_{t\overline{t}}$.

Triple products have also been tested in potential future leptonic colliders. Ref.~\cite{Bar-Shalom:2024dav} constructs $CP$ observables made of four particles contracted by a Levi-Civita tensor in $e^- e^+ \to t \overline{t} h$, $e^- e^+ \to t \overline{t} h W$ and $e^- e^+ \to t h j$ which are respectively
\begin{equation*}
    \epsilon(p_{e^-}, p_{e^+}, p_{t}, p_{\overline{t}}), ~~ \epsilon(p_{b}, p_{t}, p_{h}, p_{W}), ~~ \epsilon(p_{b}, p_{t}, p_{h}, p_{j}). 
\end{equation*}
These observables are equivalent to triple products if they are evaluated in the rest frame of one of the four particles. In the case of Higgs production by VBF and (single) $W$ production at muon colliders, Ref.~\cite{Cao:2025fla} also considers triple products. The two outgoing muons are used with the beam axis for the former and the reconstructed neutrino and the muon from the $W$ decay are used with the beam axis in the latter.

A linear combination of triple products has been optimised with a neural network by Ref.~\cite{Bortolato:2020zcg} to search for anomalous Yukawa interactions derived from $\mathcal{O}_{t\varphi}$ in $t\Bar{t}h$ production. A systematic construction of the $CP$-odd observables allows the definition of 22 potentially $CP$ sensitive observables, all involving a triple product of visible momenta. Good sensitivity can already be obtained in $t\Bar{t}h$ with two particular combinations, see Refs.\cite{Bortolato:2020zcg,Faroughy:2019ird},
\begin{equation}
\begin{split}
    \omega_6 &= \frac{\left[ (p_{l^+} \times p_{l^-}).(p_{b} + p_{\Bar{b}}) \right]\left[ (p_{l^+} - p_{l^-}).(p_{b} + p_{\Bar{b}}) \right]}{\left| p_{l^+} \times p_{l^-}| |p_{b} + p_{\Bar{b}} \right|\left| p_{l^+} - p_{l^-}||p_{b} + p_{\Bar{b}} \right|}, \\
    \omega_{14} &= \frac{\left[ (p_{l^+} \times p_{l^-}).(p_{b} - p_{\Bar{b}}) \right]\left[ (p_{l^+} - p_{l^-}).(p_{b} - p_{\Bar{b}}) \right]}{\left| p_{l^+} \times p_{l^-}| |p_{b} - p_{\Bar{b}} \right|\left| p_{l^+} - p_{l^-}||p_{b} - p_{\Bar{b}} \right|}.
\end{split}
\end{equation}
The reader is referred to Ref.~\cite{Bortolato:2020zcg} for the complete list and for the results using the whole set to extract the sensitivity reachable at HL-LHC, HE-LHC and FCC colliders.

For Ref.~\cite{Bernreuther:2015yna}, only the dipole operator $\mathcal{O}_{tG}$ is considered in Table \ref{tab:directobs} even though the authors also considered the two operators
\begin{equation}
\begin{split}
    \mathcal{O}_{gt} &= [\Bar{t}_R \gamma^\mu T^A D^\nu t_R] G^A_{\mu\nu},  \\
    \mathcal{O}_{gQ} &= [\Bar{Q}_L \gamma^\mu T^A D^\nu Q_L] G^A_{\mu\nu}.
\end{split}
\label{eq:aachenop}
\end{equation}
Those two operators can be re-written following the development of Eq.(6.6) in Ref.~\cite{Grzadkowski:2010es} in terms of the chromomagnetic operator\footnote{and four-fermion operators for their $CP$-even part.}. This development is independent whether it is done before or after EW symmetry breaking. In the latter case, the full (\textit{i.e.} containing the Higgs field) chromomagnetic operator is replaced by its dimension five version and is proportional to the top mass. It has been analytically checked that all the top observables introduced in Ref.~\cite{Bernreuther:2015yna} only receive contributions proportional to those of the chromomagnetic operator and not from the two operators in Eq.~\eqref{eq:aachenop}. In particular, there is no contribution to the $CP$-odd but P-even observable $b_n^+-b_n^-$ in the SMEFT. Therefore, only the chromomagnetic operator belongs to the SMEFT basis and not the two other operators in agreement with Ref.~\cite{Grzadkowski:2010es}.

An interesting point to note is, in most of the listed $CP$ sensitive observables, a triple product is either present in the construction of the observable or can be associated with the observable. As expected, triple products asymmetries are a key ingredient to track $CP$ violating effects from NP in many LHC processes.

\subsection{Indirect Observables}
\label{subsec:IndirectCPObs}

Indirect measurements are based on low-energy observations and arise from the running of Wilson coefficients from high energy scales to lower energy scales. As presented on Figure~\ref{fig:MatchingAndRunningStairway}, the SMEFT is used above the EW scale then the heavy SM particles are integrated out as we go down to electron mass. Many low-energy observables have the capacity to test the $CP$ symmetry but a lot of them investigate operators not present in the reduced basis. For instance, tree-level contributions to kaon decays involve 1st and 2nd generation fermions that have been discarded in the massless fermions hypothesis, or look at contact interactions between the fermions thus investigating 4-fermion operators absent from the reduced basis as well. In the end, among low-energy observables, EDMs have provided the strongest constraints on the operators from the reduced basis so far.

Focusing on the results of Ref.~\cite{Panico:2018hal} based on the electron EDM, it should be noted that with the $U(1)^{14}$ symmetry, no EDM can be generated. However, if this symmetry makes sense for the LHC, it is not justified for this observable. Therefore, operators outside our reduced basis are also constrained by this measurement.

Below the EW scale, the electron EDM is induced by the operator
\begin{equation}
    \mathcal{O}_{e \gamma} = \Bar{e}_L \sigma^{\mu\nu} e_R F_{\mu\nu},
\end{equation}
while above this scale, the dipole operator originates from two operators of the Warsaw basis, $\mathcal{O}_{e B}$ and $\mathcal{O}_{e W}$ defined in Table \ref{complete CPV operator basis}, at the tree-level
\begin{equation}
    C_{e\gamma} (m_W) = \frac{v}{\sqrt{2} \Lambda} \left( \sin{\theta_W} C_{eW}(m_W) - \cos{\theta_W} C_{eB}(m_W) \right) + O(\Lambda^{-3}),
\end{equation}
such that the electron EDM is given by 
    \begin{equation}
        d_e(\mu)= \frac{\sqrt{2} v}{\Lambda^2} \text{Im}[\sin{\theta_W} ~ C_{eW}(\mu) - \cos{\theta_W} ~ C_{eB}(\mu)],
    \end{equation}
where $\theta_W$ is the weak angle. Only the $CP$-odd operators proportional to the imaginary parts of the two operators coefficients contribute as expected. On the contrary, the real parts contribute to the magnetic dipole. Thanks to helicity selection rules, only a few operators mix with those operators at 1-loop. Three operators of the reduced basis are relevant for the 1-loop RGEs: 
\begin{multline}\label{eq: running op}
    \frac{d}{d \ln{\mu}} \text{Im}
    \begin{pmatrix}
        C_{eW}\\
        C_{eB}
    \end{pmatrix}
    = \\
    -\frac{y_e g}{16 \pi^2}
    \begin{pmatrix}
        0 & 2 \tan{\theta_W}(Y_l + Y_e) & \frac{3}{2}  \\
        1 & 0 & \tan{\theta_W}(Y_l + Y_e)
    \end{pmatrix}
    \begin{pmatrix}
        C_{\varphi\widetilde{W}} \\
        C_{\varphi\widetilde{B}}  \\
        C_{\varphi\widetilde{W}B}
    \end{pmatrix}.
\end{multline}
In Eq.(\ref{eq: running op}), $y_e$ is the electron Yukawa and $Y_l$ and $Y_e$ are the hypercharges of left-handed lepton $l$ and right-handed electron $e$. In addition, $\mathcal{O}_{\widetilde{W}WW}$ gives a finite contribution, one that does not evolve with the scale, which is  
\begin{equation}
    \text{Im} C_{eW} = \frac{3}{64\pi^2} y_e g^2 C_{\widetilde{W}WW}.
\end{equation}

\begin{table}[t]
    \centering
    \begin{tabular}{c|c|c}
        Operator & $\Lambda=10$TeV & $\Lambda=1$TeV \\
    \hline
       $\mathcal{O}_{\widetilde{W}WW}$  & $6.4~10^{-2}g_2^3$ & $1.7~10^{-4}$ \\
       $\mathcal{O}_{\varphi\widetilde{W}}$  & $4.7~10^{-3}g_2^2$ &  $2.0~10^{-5}$\\
       $\mathcal{O}_{\varphi\widetilde{B}}$  & $5.2~10^{-3}g_1^{2}$ & $6.7~10^{-6}$ \\
       $\mathcal{O}_{\varphi\widetilde{W}B}$  & $2.4~10^{-3}g_1g_2$ & $5.6~10^{-6}$ \\
       $\mathcal{O}_{tW}$  & $6.9~10^{-3}y_t g_2$ & $4.2~10^{-5}$ \\
       $\mathcal{O}_{tB}$  & $1.2~10^{-2}y_t g_1$ & $4.0~10^{-5}$ 
    \end{tabular}
    \caption{Constraints on the Wilson coefficients corresponding to the operators listed in the first column from Ref.\cite{Panico:2018hal}. $g_2$ and $g_1$ are respectively the $SU_L(2)$ and $U_Y(1)$ coupling constants, $y_t$ is the top Yukawa coupling constant. The second column basically reproduces the results detailed in Ref.\cite{Panico:2018hal} thus considering $\Lambda=10$ TeV. In the last column, the constraints are translated for $\Lambda=1$TeV with the numerical values of the couplings inserted. }
    \label{tab:EDMconstraints}
\end{table}

At the 2-loop level, one must take into account 1-loop mixing of the operators already present at the 1-loop level and direct 2-loop contributions to $\mathcal{O}_{eW}$ and $\mathcal{O}_{eB}$. In particular, the electroweak dipole operators of the reduced basis contribute to the electric dipole at this level. Finally, $\mathcal{O}_{\widetilde{W}WW}$ also mixes with $\mathcal{O}_{\varphi\widetilde{W}}$, $\mathcal{O}_{\varphi\widetilde{B}}$ and $\mathcal{O}_{\varphi\widetilde{W}B}$. 

Those result in the strongest constraints on the operators that we consider later in our analysis. They are displayed in Table \ref{tab:EDMconstraints}. 

Ref.~\cite{Deka:2025qjc} compares the limits on SMEFT operators obtained not only from the electron EDM but also from the muon EDM. The bounds\footnote{The results are displayed on Table 2 of Ref.~\cite{Deka:2025qjc}. The table is not reproduced here since it is too large to fit the format.  } are generally weaker with the muon EDM due to a weaker experimental result. Ref.\cite{Kley:2021yhn} considers the neutron EDM in addition to the electron and muon EDMs. The analysis considers all finite contributions at 1-loop in the calculations. The reader is encouraged to consider the tables in the latter for the constraints on the different Wilson coefficients.

Ref.~\cite{Ardu:2025rqy} argues for the inclusion of semi-leptonic $CP$ interactions between the electron and the nuclei evaluated at the electron mass scale when the SMEFT operators are run down to derive the electron EDM. Taking into account these additional contributions, the electron EDM is able to constrain more operators than previously thought. The limits on Wilson coefficients are presented in Figure 2 of the paper where it assumes $\Lambda = 10$ TeV.

\begin{table}[!ht]
\centering
\begin{tabular}{|c|c|c|c|}
\hline
  Ref  &  Op(s) &  Level & Observable   \\
\hline
\cite{Dekens:2013zca} & $\mathcal{O}_{\widetilde{G}GG}$ & 1-loop & $d_n \leq 2.9 \times 10^{-13}~e.fm$  \\
                      & $\mathcal{O}_{\widetilde{W}WW}$ & 1-loop  &          \\
                      & $\mathcal{O}_{\varphi \widetilde{G}}$ & 1-loop &     \\
                      & $\mathcal{O}_{\varphi \widetilde{W}}$ & 1-loop  &      \\
                      & $\mathcal{O}_{\varphi \widetilde{B}}$ & 1-loop  &     \\
                      & $\mathcal{O}_{\varphi \widetilde{W}B}$ &  1-loop  &     \\
\hline
\cite{Panico:2018hal} & $\mathcal{O}_{\widetilde{W}WW}$ & 2-loop  & $d_e<1.1 \times 10^{-29}~e.cm$    \\
                      &                                 & + 1-loop finite &       \\     
                      & $\mathcal{O}_{\varphi \widetilde{W}}$ & 1-loop   &                      \\
                      & $\mathcal{O}_{\varphi \widetilde{B}}$ &  1-loop &      \\
                      & $\mathcal{O}_{\varphi \widetilde{W}B}$ & 1-loop  &       \\
                      & $\mathcal{O}_{u W}$ & 2-loop   &                \\
                      & $\mathcal{O}_{u B}$ & 2-loop   &                   \\
\hline
\cite{Kley:2021yhn} & $\mathcal{O}_{\widetilde{G}GG}$ & LO in $d_n$ & $\|d_n\|<1.8 \times 10^{-26}~e.cm$  \\
                      & $\mathcal{O}_{u G}$ & 1-loop finite in $d_n$ &   $\|d_e\|<1.1 \times 10^{-29}~e.cm$  \\
                      & $\mathcal{O}_{\widetilde{}WW}$ & 2-loop  & $\|d_\mu\|<1.5 \times 10^{-19}~e.cm$ \\
                      &                                 & + 1-loop finite &  $\|d_\tau\|<1.6 \times 10^{-18}~e.cm$ \\    
                      & $\mathcal{O}_{\varphi \widetilde{G}}$ & 1-loop   &     \\
                      & $\mathcal{O}_{\varphi \widetilde{W}}$ & 1-loop   &     \\
                      & $\mathcal{O}_{\varphi \widetilde{B}}$ &  1-loop &   \\
                      & $\mathcal{O}_{\varphi \widetilde{W}B}$ & 1-loop  &     \\
\hline
\end{tabular}
\caption{List of the operators of the reduced basis with the type of their contribution to the EDM, including the references of their computations.}
\label{tab:indirectobs}
\end{table}

\subsection{Other Observables}
\label{subsec:OtherCPObs}

\paragraph{Matrix-element-based observables:}

ATLAS has produced a combined fit testing the $CP$ nature of the interaction between the Higgs and vector bosons in Ref.~\cite{ATLAS:2026urg}. The combined fit is based on several analyses from Refs.\cite{ATLAS:2025bts, ATLAS:2025hki, ATLAS:2023mqy, ATLAS:2022tan} which looked into Higgs decays. In three of these analyses, at least one optimal observable $\mathcal{OO}$ was built as
\begin{equation}
    \mathcal{OO} = 2 \frac{\text{Re}(\mathcal{M}^*_{SM} \mathcal{M}_{BSM})}{|\mathcal{M}_{SM}|^2}. 
\end{equation}
$\mathcal{M}_{SM}$ and $\mathcal{M}_{BSM}$ are respectively the matrix element amplitudes of the SM and of the BSM. One BSM matrix element was considered at a time when multiple SMEFT operators were included. So, multiple optimal observables were built if multiple operators contributed, in particular $\mathcal{O}_{\varphi \widetilde{W}}$ and $\mathcal{O}_{\varphi \widetilde{W}B}$ in Higgs production by VBF.

Optimal observables have been proposed for previous leptonic colliders in top-pair production by Ref.~\cite{Atwood:1991ka} and in $WW$ production by Ref.~\cite{Diehl:1996wm}. This approach has been brought back for potential future leptonic colliders in $Zh$ by Ref.~\cite{Bhattacharya:2025jhs} and in $W^+W^-$ by Ref.~\cite{Jahedi:2024wnw}.

%~~\\
%\underline{Machine Learning}: 
%~~\\
\paragraph{Machine Learning:}

Following the emergence of ML in particle physics, algorithms have been built to search for $CP$-violating effects. Ref.~\cite{Ren:2019xhp} presents a Message Passing Neural Network (MPNN) to track the $CP$ properties of the top quark-Higgs coupling in the semi-leptonic decay channel of the associated production of a Higgs with a top-pair. The coupling is expressed in the $\kappa$ framework but can be easily translated to the Wilson coefficient of the SMEFT Yukawa operator $\mathcal{O}_{t\varphi}$. From event graphs, the NN is able to separate the $CP$-even and $CP$-odd components of the Higgs coupling to the top quark and its sensitivity increases with the luminosity. Ref.~\cite{Esmail:2024gdc} compares the use of a Multi-Layer Perceptron (MLP) with a Graph Neural Network (GNN). Similarly to Ref.~\cite{Ren:2019xhp}, the $\kappa$ framework is used in $t\overline{t}h$ production but in the fully leptonic channel and $h \to b \overline{b}$. Three $CP$ observables are among the inputs to both NNs : $\cos \theta^*$, $\cos\theta_{lh}$ and $\cos \overline{\theta}_{lh}$. $\theta^*$ is the angle between the top quark direction of flight in the $t\overline{t}$ rest frame and the beam three-momentum such that
\begin{equation*}
    \cos \theta^* = \frac{\vec{p}_t^{~t\overline{t}} . \hat{p}_{\text{beam}} }{|\vec{p}_t^{~t\overline{t}}|~ |\hat{p}_{\text{beam}}|} .
\end{equation*}
$\theta_{lh}$ is the angle spanned by the dilepton system projected on the plane orthogonal to the Higgs boson
\begin{equation*}
    \cos\theta_{lh} = \frac{(\vec{p}_{l^+} \times \vec{p}_{h}) (\vec{p}_{l^-} \times \vec{p}_{h})}{|\vec{p}_{l^+} \times \vec{p}_{h}|~ |\vec{p}_{l^-} \times \vec{p}_{h}|},
\end{equation*}
and 
\begin{equation*}
    \cos \overline{\theta}_{lh} = \text{sign}((\vec{p}_b - \vec{p}_{\overline{b}}) . (\vec{p}_{l^-} \times \vec{p}_{l^+})) \cos \theta_{lh}.
\end{equation*}
The GNN is proven to offer better constraints than the MLP.

In Ref.~\cite{Bhardwaj:2021ujv}, two NNs are constructed to act as $CP$ observables themselves. They aim to estimate the sign of the interference amplitude. The first NN is trained on interference events and the second on interference and SM events. The derived $CP$ observable is the difference in the probabilities for the event to be a positive- or negative-interference event 
\begin{equation*}
    \text{O}_{\text{NN}} = P_+ - P_-.
\end{equation*}
The NNs are compared to the asymmetry in $\Phi$ from Eq.~(\ref{eq:Phi}) in $h \to 4l$ and $\Delta \phi_{jj}$ in Higgs production by VBF. The performance to constrain $\mathcal{O}_{\varphi \widetilde{W}}$, $\mathcal{O}_{\varphi \widetilde{B}}$ and $\mathcal{O}_{\varphi \widetilde{W}B}$ is slightly improved by the NNs compared to the $CP$ observable. The method was later used in various electroweak processes in Ref.~\cite{PhysRevD.107.016008} to probe the same operators and $\mathcal{O}_{\widetilde{W}WW}$. Ref.~\cite{Cruz:2024grk} proves that using the symmetry property of an equivariant NN to generate a $CP$ function further increases the performance. Three SMEFT operators are investigated in three processes: $\mathcal{O}_{tG}$ in $t\overline{t}$, $\mathcal{O}_{\widetilde{W}WW}$ in $WZ$ and $\mathcal{O}_{tZ}$ in $t\overline{t}\gamma$. Ref.~\cite{Barrue:2023ysk} builds a detector-level optimal observable with a NN, meaning it includes parton shower and detector effects, probing $CP$ effects from $\mathcal{O}_{\varphi \widetilde{W}}$ in $Wh$ production. Ref.~\cite{Silva:2025hzo} compares detector-level optimal observables obtained with different methods but the bounds on $\mathcal{O}_{\varphi \widetilde{W}}$ are not dramatically better than from 2D bins of $Q_l \cos{\delta^+} \times p_{T}^W$.

Another approach followed in Refs~\cite{Butter:2021rvz, Bahl:2025jtk} relies on symbolic regression. Instead of acting as the $CP$ observable, the NN derives an analytic formula from a set of constant parameters, observables and mathematical operations. Analytic expressions are optimised to be sensitive to $CP$ effects from $\mathcal{O}_{\varphi \widetilde{W}}$ in Higgs production by VBF by Refs.~\cite{Butter:2021rvz, Bahl:2025jtk} and $\mathcal{O}_{t\varphi}$ in $t\overline{t}h$ by Ref.~\cite{Bahl:2025jtk}.

%~~\\
%\underline{Global analyses}: 
%~~\\

\paragraph{Global analyses:}

Recent studies combine direct and indirect observables resulting in consistent stringent constraints on Wilson coefficients of $CP$-odd operators. However, the best constraints on $CP$-odd operators are actually obtained from EDMs \cite{Fuchs:2020uoc, Panico:2018hal, Cirigliano:2016njn, Cirigliano:2016nyn}. The LHC is currently not competitive compared to EDMs because of the accidental suppression of SM contributions in EDMs, the precision of its experimental measurements and the relative good control of the theoretical uncertainties. Nevertheless, direct $CP$-odd observables still provide highly valuable information to probe blind directions even if large cancellations are necessary to satisfy EDM constraints. Therefore, rather than being in a competition, direct and indirect observables are complementary approaches to detect new $CP$ violating effects from any $CP$-odd operator.

\section{Diboson Analysis}
\label{sec:CPDibosonAnalysis}

The case of interest here is the diboson production of a $W$ boson with a neutral boson, $Z$ or $\gamma$. Each massive gauge boson decays into leptonic channels with different flavours: $W^\pm \rightarrow e^\pm \overset{\textbf{\fontsize{2pt}{2pt}\selectfont(---)}}{\nu}_{e}$ and $Z\rightarrow \mu^-\mu^+$. There are four reasons for this choice. Firstly, those channels, even if they are not C-even processes, almost behave as such. Namely, the interference cross-section is heavily suppressed due to a cancellation between the different regions of the phase space. Therefore, the goal is to test which observables could disentangle those regions. Their efficiency is estimated by using the sign of the matrix element as proposed in Ref.\cite{Degrande:2020tno}. New observables are proposed and compared to those of previous studies on the same processes. Secondly, those processes have been measured at different centre-of-mass energies at the LHC \cite{Aad:2012twa, Aad:2016ett, Aaboud:2016yus, Khachatryan:2016tgp, Khachatryan:2016poo, Aaboud:2019gxl, Sirunyan:2019bez} and the cross-sections are relatively large. A large cross-section leads to a large expected number of events which is necessary to measure accurately $CP$-odd observables such as asymmetries. Thirdly, these final states can be easily reconstructed resulting in a quite clean signal and a relatively low background. Finally, their leptonic channels contain only one neutrino compared to the C-even process $W^+ W^-$ and the two different lepton flavours ensure that there is no confusion between the Z and W decay products\footnote{Detector effects such as misidentification are ignored.} which makes the analysis easier. The study of other leptonic decays is left for future work as well as the semi-leptonic and hadronic decays.

\subsection{$WZ$ and $W\gamma$ Processes}
\label{subsec:DibosonProcesses}

%Feynman Diagrams
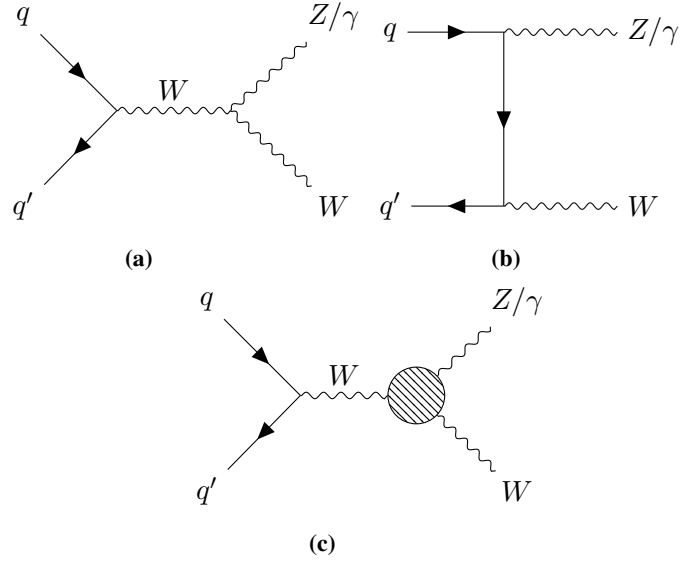
\begin{figure*}[t]
    \centering
    \begin{subfigure}[t]{0.3\textwidth}
    \begin{tikzpicture}
    \begin{feynman}
        \vertex (f1) {\(q\)};
        \vertex [below right=1.75cm of f1] (a);
        %\vertex [below left=of a] (f2) {\(q'\)};
        \vertex [below =2.5cm of f1] (f2) {\(q'\)};
        \vertex [right=of a] (c);
        \vertex [right=4.1cm of f1] (b1) {\(Z/\gamma\)};
        \vertex [right=4.1cm of f2] (b2) {\(W\)};
 
        \diagram* {
        (f1) -- [fermion] (a) ,
        (a) -- [fermion] (f2) ,
        (a) -- [boson, edge label=\(W\)] (c),
        (c) -- [boson] (b1),
        (c) -- [boson] (b2),
        };
    \end{feynman}
    \end{tikzpicture}
    \caption{
    %SM s-channel of $WZ/\gamma$ production.
    }
    \end{subfigure}
    ~~~~~~~~~~~
    \begin{subfigure}[t]{0.3\textwidth}
    \begin{tikzpicture}
    \begin{feynman}
        \vertex (f1) {\(q\)};
        \vertex [right=of f1] (a);
        \vertex [right=of a] (b1) {\(Z/\gamma\)};
        \vertex [below=2.3cm of a] (c);
        \vertex [below=2.3cm  of f1] (f2) {\(q'\)};
        \vertex [right= of c] (b2) {\(W\)};
        \diagram* {
        (f1) -- [fermion] (a) ,
        (a) -- [boson] (b1) ,
        (a) -- [fermion] (c) ,
        (c) -- [fermion] (f2) ,
        (c) -- [boson] (b2)
        };
    \end{feynman}
    \end{tikzpicture}
    \caption{
    %SM t-channel of $WZ/\gamma$ production.
    }
    \end{subfigure}
    \\
    \begin{subfigure}[t]{0.3\textwidth}
    \begin{tikzpicture}
    \begin{feynman}
        \vertex (f1) {\(q\)};
        \vertex [below right=1.75cm of f1] (a);
        \vertex [below =2.5cm of f1] (f2) {\(q'\)};
        \vertex [blob,right=1.15cm of a] (c) { };
        \vertex [right=4.1cm of f1] (b1) {\(Z/\gamma\)};
        \vertex [right=4.1cm of f2] (b2) {\(W\)};
        \diagram* {
        (f1) -- [fermion] (a) ,
        (a) -- [fermion] (f2) ,
        (a) -- [boson, edge label=\(W\)] (c),
        (c) -- [boson] (b1),
        (c) -- [boson] (b2),
        };
    \end{feynman}
    \end{tikzpicture}
    \caption{
    %NP s-channel of $WZ/\gamma$ production.
    }
    \end{subfigure}
\caption{Feynman diagrams for diboson production in the SM (the s-channel in (a) and the t-channel in (b), u-channel is not displayed) and with the new $WWZ/\gamma$ vertex from the dimension-six operators $\mathcal{O}_{\widetilde{W}WW}$ and $\mathcal{O}_{\varphi \widetilde{W}B}$ in (c). }
\label{ZW production}
\end{figure*}

Out of the list of 10 operators from Table~\ref{New CPV operator basis u14}, the relevant operators in $WZ$ and $W\gamma$ production are $\mathcal{O}_{\widetilde{W}WW}$ and $\mathcal{O}_{\varphi \widetilde{W}B}$. Those operators modify the WWZ/$\gamma$ coupling present in the s-channel diagram. As a matter of fact, no $CP$-odd dimension-six contribution arises in the light quark interaction or in the weak bosons decay thanks to the simplification in Section~\ref{subsec:basis reduction u14}. The Feynman diagrams are represented on the Figure~\ref{ZW production}.

The contribution of the interference between the SM and $\mathcal{O}_{\widetilde{W}WW}$ to the total cross-section of the 2$~\rightarrow~$2 process is suppressed by the helicity selection rules at LO in the massless limit as demonstrated in Ref.~\cite{Azatov:2016sqh}. The same method applied to $\mathcal{O}_{\varphi\widetilde{W}B}$ shows that helicity selection rules also suppress its interference\footnote{ Looking at Table II in Ref.~\cite{Azatov:2016sqh} and applying Eq.(9) with the given values results in $h\left(A^{\mathcal{O}_{\varphi\widetilde{W}B}}_4\right)=2$ while $h\left(A^{SM}_4\right)=0$.  }. However, at least one of the massive bosons has to be off-shell and, when considering the complete 2$~\rightarrow~$3 or 2$~\rightarrow~$4 processes, the interference is recovered. This is the interference resurrection carefully demonstrated in Refs.~\cite{Azatov:2019xxn,Panico:2017frx} with $\mathcal{O}_{\Tilde{W}WW}$ and reproduced in Eq.(\ref{eq:interfWA}) for W$\gamma$.

The analysis is limited at the LO and leaves NLO corrections for future work. 
The analysis is also limited to processes at the parton level. 
It will be insightful to observe the modification in the efficiencies of the observables and their asymmetries due to shower and detector effects but these important aspects are kept for future analyses
%\JT{\sout{The analysis is also limited to processes at the parton level. It will be insightful to observe the modification in the efficiencies of the observables and their asymmetries due to shower and detector effects but these important aspects are kept for future analyses.} We will start by investigating events at the partonic level to give a early assessment on the practical efficacy of asymmetries to probe CP-odd interactions.} 
This simple framework 
%\JT{also} 
allows one to test many new observables as it is faster, computationally cheaper and easier to understand the results. 
%\JT{Then, we move on to showered events to gain insight on the effects of the parton shower and hadronisation onto the asymmetries.}

The set of operators in Table~\ref{New CPV operator basis u14} has been implemented in a FeynRules model \cite{Alloul:2013bka} where, in both $X^2\varphi^2$ and $\varphi^3\psi^2$ classes definition, $v^2$ has been subtracted from $(\varphi^\dagger\varphi)$ for convenience. A UFO model was generated \cite{Degrande:2011ua} and was passed to MG5\_aMC@NLO \cite{Alwall:2014hca}. 
%\JT{Parton shower and hadronisation effects are simulated with \pyt \cite{Bierlich:2022pfr} and its default parameter card.} 
The PDF set exploited in the event generation is NNPDF2.3 \cite{Ball:2012cx} in which $\alpha_S(M_Z)=0.119$. The SM parameters are fixed at the Z pole mass :
\begin{equation*}
\begin{split}
    & m_Z=91.1876~ \text{GeV}, ~~  (\alpha_{EM})^{-1}=127.9,~~ G_F= 1.166370~10^{-5}~ \text{GeV}^{-2},\\
    & \Gamma_Z = 2.4952~ \text{GeV},~~ \Gamma_W = 2.085~ \text{GeV}. 
\end{split}
\end{equation*}
The CKM matrix $V_{CKM}$ is reconstructed using the Euler angles, respectively
\begin{equation*}
    \theta_{12} = 0.227~rad, ~~~ \theta_{23}=0.041~rad, ~~~ \theta_{13} = 0.003~rad,
\end{equation*} 
and the $CP$ angle $\delta_{13}=1.2~rad$ in the standard parametrisation of $V_{CKM}$ shown in Eq.(\ref{eq:StandardCKMmatrix}). Even if the $CP$ violating phase is included, it has been checked that it does not change the results.

In this work, the bosons decay respectively in the dileptonic channel $W^\pm Z \rightarrow \mu^+ \mu^- e^\pm \overset{\textbf{\fontsize{2pt}{2pt}\selectfont(---)}}{\nu}_{e}$ and in the leptonic decay $W^\pm \gamma \rightarrow \gamma e^\pm \overset{\textbf{\fontsize{2pt}{2pt}\selectfont(---)}}{\nu}_{e}$.
800,000 events are produced for the different final states and for each contribution, \textit{i.e.} for the SM, the interference and the squared amplitude to obtain accurate differential distributions. Cuts are taken from the ATLAS analysis at 13 TeV ~\cite{Aaboud:2016yus} for $WZ$ events and are the following:
\begin{equation*}
\begin{split}
    & p_T(\mu)> 15\text{GeV},~~~ |\eta(\mu)|<2.5,~~~  p_T(e)> 20\text{GeV},~~~  |\eta(e)|<2.5, \\
    & \Delta R(\mu^+\mu^-)>0.2,~~~  \Delta R(e\mu^-)>0.3,~~~  \Delta R(e\mu^+)>0.3,\\
    & |m_{\mu^-\mu^+}-m_Z|<10\text{GeV},~~~  m_T(e\nu_e)>30\text{GeV}. 
\end{split}
\end{equation*}

For $W\gamma$ events, the fiducial phase space corresponding to the ATLAS analysis~\cite{Aad:2012mr} is replicated by using the cuts:
\begin{equation*}
\begin{split}
    & p_T(e)>25\text{GeV},~~~  |\eta(e)|<2.47,~~~  E_T(\gamma)>15\text{GeV},~~~  |\eta(\gamma)|<2.37, \\
    & E^{miss}_T>35\text{GeV},~~~  \Delta R(e\gamma)>0.7,~~~  m_T(e\nu_e)>40\text{GeV}, \\
    & |m(e\gamma) - m_Z|>10\text{GeV}. 
\end{split}
\end{equation*} 
The cuts for the muon decay of the W boson are very similar except that the last cut is not applied.

\subsection{Formalism}

Searches aimed at studying the $O_{WW\widetilde{W}}$ operator have been optimised for the high energy tail of the distributions, as, for example, in Ref.~\cite{Azatov:2019xxn}, to take advantage of the energy growth. Here, a complementary approach is followed by looking at asymmetries over the whole fiducial space. Accurate asymmetry measurements require a high number of events which is not available in high energy tails. Therefore the large amount of events available close to threshold is used instead.

Consistency with the theoretical development provided in Section~\ref{subsec:basis reduction u14} requires that the differential cross-section with respect to a $CP$-odd observable $X$ is truncated at the $\Lambda^{-2}$ order such that
\begin{equation}
    \frac{d\sigma}{dX} = \frac{d\sigma(SM)}{dX} +  \frac{c_i}{\Lambda^2}\frac{d\sigma \big( \obothi \big)}{dX}.
\end{equation}
The square of the amplitude containing the dimension-six operator is used as a mean to check the perturbative behaviour of the analysis and to understand the phase-space suppression of the interference cross-section.

The asymmetry in observable $X$ is defined as
\begin{equation}\label{eq:genericdiff}
    \Delta X \equiv  \sigma(X > 0) - \sigma(X < 0) = \Delta X(SM)+  \frac{c_i}{\Lambda^2} \Delta X (\obothi),
\end{equation}
with $\sigma(X>0)=\int_{0}^{b_+} \frac{d\sigma}{dX}~dX$ and $\sigma(X<0)=\int^{0}_{b_-} \frac{d\sigma}{dX}~dX$. $b_\pm$ are the upper and lower bounds of integration of the variable $X$. One could normalise it by taking the ratio of the asymmetry with the total cross-section, 
\begin{equation}\label{eq:genericasym}
    A_{X} \equiv \frac{ \sigma(X>0) - \sigma(X<0) }{  \sigma(X>0) + \sigma(X<0) } = \frac{\Delta X}{\sigma(SM)+ \frac{c_i}{\Lambda^2} \sigma(\obothi)}.
\end{equation}
The denominator in the normalised asymmetry is the full cross-section which corresponds to the sum of the SM cross-section $\sigma(SM)$ and the contribution of a $CP$-odd operator $\sigma(\obothi)$. The NP contribution introduces another parameter dependence. However, the denominator is already quite constrained by the measurements of the total cross-section which are in agreement with the SM prediction. Although, the results will mainly be presented in terms of the asymmetry to avoid those extra dependencies, the SM cross-section value will be used to provide a sensitivity estimate later on.

The method defined in Ref.\cite{Degrande:2020tno} is followed to quantify the interference suppression over the phase spaces and the efficiencies of the asymmetries.
Namely, the integral of the absolute value of the interference is computed with
\begin{equation}
    \sigma^{|int|} \equiv \int d\Phi \left| \frac{d\sigma_{int}}{d\Phi }\right|,
\end{equation}
to understand how much the suppression of the interference is due to the sign flips over the phase space. Practically, it is computed from the sum of the absolute value of the weights. The best asymmetry experimentally measurable is given by the measurable absolute value cross-section,
\begin{equation}
    \sigma^{|meas|} \equiv \int d\Phi_{meas} \left|\sum_{\{um\}}\frac{d\sigma_{int}}{d\Phi } \right|
\end{equation}
where the sum runs over the set of unmeasurable quantities $\{um\}$:
\begin{itemize}
    \item the side where the quark and antiquark are coming from;
    \item the polarisations of the different final leptons, photons and initial quarks;
    \item the flavours of the quarks and their PDFs;
    \item the longitudinal momentum of the neutrino.
\end{itemize} 
In this last case, the sum is replaced by the integral over the longitudinal component of the neutrino momentum. This quantity, $\sigma^{|meas|}$, is obtained by integrating and summing the interference matrix element over the unmeasurable quantities for each event generated according to this same interference. Appendix~\ref{sec:appendixAsymmetryDeterioration} shows how the asymmetry decreases when the unobservables effects are accumulated. Although this is the best measurable asymmetry, it is computationally expensive to compute, especially on real events where transfer functions have to be taken into account. Therefore, additional asymmetries built from simple observables are compared to it. The comparison will quantify their efficiency in order to find simple and efficient observables to constrain the $CP$ violating operators.

\subsection{Triple Products}
\label{subsec:DibosonTripleProducts}

The asymmetries are built from triple products in order to improve the sensitivity to $CP$-odd operators. The triple products require the inclusion of the initial quark momenta as well as the momenta of the photon and one of the decay product of the W boson to be non-vanishing in the case of W$\gamma$ production. Moreover, including the initial quark and the neutral gauge boson also leads to larger asymmetries in WZ production. Therefore, the triple product is defined as 
\begin{equation}\label{eq:tripmomdef}
    p_\perp (p_l, \Hat{n}_{ref}) \equiv \Hat{n}_{ref} . (\Vec{p}_l \times \Vec{p}_V)
\end{equation}
where $\Vec{p}_l$ is the 3-momentum of one visible lepton, i.e. an electron or a muon, $\Vec{p}_V$ is the reconstructed 3-momentum of the Z boson in $WZ$ production or of the photon in $W\gamma$ production and $\Hat{n}_{ref}$ will be a reference axis approximating the quark momenta. In $WZ$ production, the asymmetry $\Delta {p_\perp}(p_{\mu^+}, \Hat{n}_{ref})$ will not be presented in the following because, by definition, it is always opposite to $\Delta {p_\perp}(p_{\mu^-}, \Hat{n}_{ref})$.

The direction of the quark is not experimentally available. However, due to the PDF, quarks are on average more energetic than anti-quarks. To take advantage of this, multiple approximations of the quark momentum $\Hat{n}_{ref}$ are explored: the laboratory axis $\Hat{z}=(0,0,1)$, the longitudinal momentum of neutral boson $\vec n_{Z/\gamma}=(0,0,p^z_{Z/\gamma})$, of the electron $\vec{n}_e=(0,0,p_{e}^z)$ and of the sum of visible particles $\vec{n}_{\sum}=(0,0,p_{\sum}^{z})$ where $p_{\sum}^z$ is the third spatial component of the sum in the momentum of the visible particles. All vectors are taken in the laboratory frame.

Note that some observables mentioned in Section~\ref{subsec:DirectCPObs} are equivalent to some of triple product configurations defined here. The triple product defined for $WZ$ production in Ref.~\cite{Kumar:2008ng}, listed in the review in Eq.\eqref{eq: def Xi}, corresponds to 
\begin{equation} \label{eq:Xi}
     \Xi = p_\perp (p_l, \vec n_z).
\end{equation}
In $W\gamma$ production, the observables in Ref.~\cite{Dawson:2013owa} have the same sign as some triple products producing equivalent asymmetries:
\begin{equation*}
\begin{split}
    \mathcal{O}_W & \sim p_\perp (p_e, \Hat{z}), \\
    \mathcal{O}_\gamma & \sim p_\perp(p_e,p_\gamma), \\
    \mathcal{O}_l & \sim p_\perp(p_e,p_e).
\end{split}
\end{equation*}
As a result, the results from previous analyses cited above can be reproduced by simply exploring the different estimations of the quark momentum in $p_\perp$ in the triple product and its asymmetry.

\subsection{Angular Observables} 
\label{subsec:DibosonAngularObservables}

Looking in Table \ref{tab:directobs}, two other observables, that are not triple products, have already been investigated in diboson production: the observable from Ref.~\cite{Azatov:2019xxn} and the signed azimuthal angle difference.

For the first, the construction of the observable follows the one described in Section~\ref{subsec:DirectCPObs}.  Namely, the $W$ momentum is needed to reconstruct the scattering plane normal $\Hat{n}_{scat}$ and requires the neutrino momentum $p_\nu$. The reconstruction method in Ref.~\cite{Azatov:2019xxn} is replicated here to recover $p_\nu$. Assuming a massless neutrino, its transverse momentum is identified with the missing transverse momentum of the event. The energy of the neutrino momentum is then calculated such that the $W$ boson is on-shell. If two real solutions are possible one of them is randomly chosen or, if no real solution exists, $p_z^{\nu}$ is taken as the value which minimises the electron-neutrino invariant mass. The angles $\phi_W$ and, if applicable, $\phi_Z$ are then derived as in Section~\ref{subsec:DirectCPObs}. From now on, the observable is noted as $\sin{\phi}_{WZ/\gamma}$ with a different definition in each process to ease the notation:
\begin{equation}\label{eq:barducci observable definition}
\begin{split}
    \sin{\phi}_{WZ} & = \sin{2\phi_W}+ \sin{2\phi_Z}, \\
    \sin{\phi}_{W\gamma} & = \sin{2\phi_W}.
\end{split}
\end{equation}
For the second, the two particles of interest are the electron momentum and the neutral boson momentum.

Therefore, Eq.~(\ref{eq:genericasym}) dictates that the asymmetries $\Delta \sin{\phi}_{WZ/\gamma}$ and $\Delta \left( \Delta \phi_{eZ/\gamma} \right)$ are respectively defined as
\begin{equation}\label{eq:sinphiVVasym}
    \Delta \sin{\phi}_{WZ/\gamma} = \sigma (\sin{\phi}_{WZ/\gamma}>0) - \sigma (\sin{\phi}_{WZ/\gamma}<0) ,
\end{equation}
where the upper and lower bounds are $\pm$ 1 in this asymmetry, and 
\begin{equation}
    \Delta \left( \Delta \phi_{eZ/\gamma} \right) = \sigma \left( \Delta \phi_{eZ/\gamma}>0 \right) - \sigma \left( \Delta \phi_{eZ/\gamma}<0 \right) ,
\end{equation}
with the upper and lower bounds respectively $\pm \pi$.

\section{Results and Projections}
\label{sec:CPDibosonResults}

\subsection{$WZ$ Results }
\label{subsec:WZResults}

\begin{table}[t]
    \centering
    \begin{tabular}{|c|c|c|}
    \hline
          Process & $W^+Z \rightarrow \mu^-\mu^+ e^+ \nu_e$ & $W^-Z \rightarrow \mu^-\mu^+ e^- \Tilde{\nu_e}$  \\
    \hline
          $\sigma(SM)$ & 15.74(2) fb & 9.88(1) fb  \\
    \hline
          $\delta_{PDF}$ & 3.45\% & 3.78\% \\
    \hline
    \hline
          $\sigma(\mathcal{O}_{\widetilde{W}WW})$ & 0.047(4)  fb & -0.033(3) fb \\
    \hline
          Schwarz Bound & 16.13 fb & 8.85 fb \\
    \hline
          $\sigma^{|int|}(\mathcal{O}_{\widetilde{W}WW})$ & 3.302(4) fb &  2.028(3) fb \\
    \hline
          $\sigma^{|meas|}(\mathcal{O}_{\widetilde{W}WW})$ & 1.084(4) fb & 0.634(3) fb \\
    \hline
          $\sigma_{\Lambda^{-4}}(\mathcal{O}_{\widetilde{W}WW})$ & 4.133(5) fb &  1.982(3) fb \\
    \hline
    \hline
          $\sigma(\mathcal{O}_{\varphi\widetilde{W}B})$ & 0.0086(7) fb & -0.0066(4) fb  \\
    \hline
          Schwarz Bound & 1.21 fb & 0.76 fb \\
    \hline
          $\sigma^{|int|}(\mathcal{O}_{\varphi\widetilde{W}B})$ & 0.5467(7) fb & 0.3533(4) fb \\
    \hline
          $\sigma^{|meas|}(\mathcal{O}_{\varphi\widetilde{W}B})$ & 0.1807(7) fb & 0.1100(4) fb \\
    \hline
          $\sigma_{\Lambda^{-4}}(\mathcal{O}_{\varphi\widetilde{W}B})$ & 0.0231(3) fb & 0.0145(2) fb \\
    \hline
    \end{tabular}
    \caption{cross-sections in 2 dileptonic decay channels of $WZ$ production for the ATLAS fiducial phase space at $\sqrt{s}=13$TeV for the SM, the interference with one dimension-six operator, $\sigma(\mathcal{O}_i)$ and for the square of the $\mathcal{O}\left(\Lambda^{-2}\right)$ amplitudes, $\sigma_{\Lambda^{-4}}(\mathcal{O}_i)$. Errors are from the numerical integration and written in the brackets. For the interferences, we also display the absolute value cross-sections $\sigma^{|int|}$ and the measurable absolute value cross-sections $\sigma^{|meas|}$.  $\delta_{PDF}$ represents the uncertainty associated with PDFs taken as the envelope of the replicas for the SM. The Wilson coefficients $C_{\widetilde{W}WW}$ and $C_{\varphi\widetilde{W}B}$ are set to 1 and the NP scale $\Lambda$ is 1 TeV but the results can be re-scaled for any other value. 
    }
    \label{tab:xsecWZATLAS}
\end{table}

The interference suppression is first checked. The cross-sections for the SM $\sigma(SM)$, for the square of the $\mathcal{O}\left(\Lambda^{-2}\right)$ amplitudes, $\sigma_{\Lambda^{-4}}(\obothi)$, and for the interference $\sigma(\obothi)$ with the two dimension-six operators in $WZ$ production are presented in Table~\ref{tab:xsecWZATLAS}. The interference cross-sections show a suppression by about two orders of magnitude of $\sigma(\obothi)$ compared to the Schwarz bound\footnote{The Schwarz bound is obtained with twice the geometric mean of $\sigma(SM)$ and $\sigma_{\Lambda^{-4}}(\obothi)$.}. The absolute value interference cross-section, $\sigma^{|int|}$, and the largest measurable asymmetry, $\sigma^{|meas|}$, are displayed in the same table. They assess that the origin of this suppression is mainly due to the cancellation over the phase space as expected. Additionally, their ratio implies that at most about a third of the interference can be recovered from measured distributions. The large phase space cancellation is also responsible for the poor numerical precision of $\sigma(\obothi)$. All cross-sections, in particular $\sigma(SM)$ and $\sigma_{\Lambda^{-4}}(\obothi)$, are given here at LO and using the same setting as for the interference in order to understand the interference suppression and to check the validity of the scale expansion. $WZ$ production cross-section in the SM is known at NNLO~\cite{Grazzini:2016swo}, and is about a factor 2 bigger than the LO prediction at 13TeV.  The large value of $\sigma^{|meas|}$ for $C_i/\Lambda^2\sim1\,\text{TeV}^{-2}$ compared to $\sigma(SM)$ and the fact that it is quite close to $\sigma^{|int|}$ show that differential distributions could improve significantly the sensitivity to the interference for both operators compared to the total cross-section and that a large part of the phase space cancellation is experimentally accessible.

Before heading to the triple products with the four possible replacements of the quark momentum, it is worth looking into the asymmetries of the triple products displayed in Appendix~\ref{sec:appendixTripleProducts}. The goal is to show the various configurations, measurable and unmeasurable, that have been explored. The first line is the ideal triple product with the quark momentum acting as a reference point. It is compared to triple product with the electron momentum substituted with the muon momentum. The asymmetry of the latter is at least one order below the asymmetry of the former thus disqualifying the muon as the final state fermion to build the best measurable triple product. Then, the best measurable configuration for the triple product is shown: the one with the longitudinal component of the sum of visible particles as a proxy for the quark momentum, the Z momentum and the electron momentum.

After that, using the leptons with W turns out quite close to the best measurable configuration ($\sim$85\% of the best configuration). However, it is impossible to measure the W momentum and any approximation reduces further the asymmetry. Several combinations of the final three leptons are investigated. The longitudinal component of one lepton must be used as the triple product using the three complete momenta almost cancels. By taking the sum of the electron and either the muon or anti-muon an efficiency of 50\% of the best configuration can be approached. The difference does not improve the asymmetries either.

Henceforth the rest of the triple products are only built with the electron and the Z momenta and the three aforementioned approximations for the quark momenta: the longitudinal component of the electron, of the Z boson and of the sum over all the final visible lepton. The largest asymmetries are obtained with the latest for both operators and for both channels as shown in Table~\ref{tab:WZasymmetryobservablesATLAS}.

\begin{table}[ht]
    \centering
    \begin{tabular}{|c|c|c|c|}
    \hline 
        Process & \multicolumn{3}{c|}{$W^+ Z \rightarrow \mu^- \mu^+ e^+ \nu_e $} \\
    \hline
        Operators & $SM$ & $\mathcal{O}_{\widetilde{W}WW}$ & $\mathcal{O}_{\varphi\widetilde{W}B}$  \\
    \hline
       $\Delta p_\perp (p_e,p_q)$ & -0.04(2) & -1.612(4)  & -0.3888(7)  \\
    \hline
       $\Delta p_\perp (p_e,p_{\sum}^z)$  & -0.02(2) & -0.628(4)  & -0.1207(7)   \\
    \hline
       $\Delta p_\perp (p_e,p_e^z)$ & 0.0(2)  & -0.535(4) &  -0.1173(7) \\
    \hline
       $\Delta p_\perp (p_e,p_Z^z)$ & -0.01(2)  & -0.527(4) &  -0.0874(7) \\
    \hline
       $\Delta \sin{\phi_{WZ}}$ & -0.03(2)  & -0.321(4) &  0.0031(7)  \\
    \hline
       $\Delta \left( \Delta \phi_{eZ} \right)$ & 0.07(2)  & 0.196(4) &  0.0688(7)  \\
    \hline
      SM stat err $30~\text{fb}^{-1}$ & \multicolumn{3}{c|}{0.7}  \\
    \hline
      SM stat err $100~\text{fb}^{-1}$ & \multicolumn{3}{c|}{0.4}  \\
    \hline
      SM stat err $3000~\text{fb}^{-1}$ & \multicolumn{3}{c|}{0.07}  \\
    \hline
    \hline 
        Process & \multicolumn{3}{c|}{$W^- Z \rightarrow \mu^- \mu^+ e^- \Tilde{\nu}_e $} \\
    \hline
        Operators & $SM$ & $\mathcal{O}_{\widetilde{W}WW}$ & $\mathcal{O}_{\varphi\widetilde{W}B}$  \\
    \hline
       $\Delta p_\perp (p_e,p_q)$ & -0.08(1) & 1.006(3) & 0.2522(4)  \\
    \hline
       $\Delta p_\perp (p_e,p_{\sum}^z)$  & -0.03(1) & 0.331(3) & 0.0810(4) \\
    \hline
       $\Delta p_\perp (p_e,p_e^z)$ & -0.01(1) & 0.295(3) & 0.0514(4) \\
    \hline
       $\Delta p_\perp (p_e,p_Z^z)$ & 0.00(1) & 0.295(3) & 0.0627(4)  \\
    \hline
       $\Delta \sin{\phi_{WZ}}$ & -0.02(1) & -0.190(3) & 0.0013(4)   \\
    \hline
       $\Delta \left( \Delta \phi_{eZ} \right)$ & -0.05(1) & 0.022(3) &  0.0109(4) \\
    \hline
      SM stat err $30~\text{fb}^{-1}$ & \multicolumn{3}{c|}{0.6}  \\
    \hline
      SM stat err $100~\text{fb}^{-1}$ & \multicolumn{3}{c|}{0.3}   \\
    \hline
      SM stat err $3000~\text{fb}^{-1}$ & \multicolumn{3}{c|}{0.06} \\
    \hline
    \end{tabular}
    \caption{Asymmetries in fb in the $WZ \rightarrow \mu^-\mu^+ e^+ \nu_e$ and $WZ \rightarrow \mu^-\mu^+ e^- \widetilde{\nu}_e$ channels by using different reference axes for the triple product and $\sin{\phi_{WZ}}$, in the ATLAS fiducial phase space at $\sqrt{s}=13$TeV at the LHC. The statistical errors are displayed using the LO SM cross-sections and several integrated luminosities. }
    \label{tab:WZasymmetryobservablesATLAS}
\end{table}

As mentioned in Subsection \ref{subsec:DibosonAngularObservables}, the generic triple product asymmetries are also compared to the asymmetry of $\sin{\phi_{WZ}}$ from Ref.~\cite{Azatov:2019xxn} and of the signed azimuthal angle difference which has also been studied in $WZ$ and $W\gamma$ production. The values of the asymmetries are presented in Table~\ref{tab:WZasymmetryobservablesATLAS} as well.

The asymmetry in $p_\perp (p_e, p_{\sum})$ for $\mathcal{O}_{\widetilde{W}WW}$ is a factor 2 larger than the one obtained from $\sin\phi_{WZ}$. $\sin\phi_{WZ}$ is almost insensitive to the $\mathcal{O}_{\varphi\widetilde{W}B}$ contribution. Therefore, the two operators can be distinguished with this channel by measuring the asymmetries in $p_\perp (p_e, p_{\sum})$ and $\sin\phi_{WZ}$. On the contrary, $\Delta \phi_{eZ}$ is less sensitive to $\OWWW$ than $\OBW$ in $W^+Z$ and therefore probes other combinations of the operators.  In particular, $\Delta \phi_{eZ}$ is almost blind to $\OWWW$ in $W^-Z$. In the end, the triple product asymmetries with the longitudinal component of sum of the visible as reference axis display better sensitivities.

By comparing with $\sigma^{|meas|}$, the efficiency of the best triple product is about 50\% and 70\% for the operators $\mathcal{O}_{\widetilde{W}WW}$ and $\mathcal{O}_{\varphi\widetilde{W}B}$ respectively. Those observables are not purely $CP$-violating since the processes are not C-even. Taking the sum of the two final states is not enough as also the initial state is not C-even at the LHC. Therefore, the SM contribution to those asymmetries does not have to vanish even if effects from the CKM phase are negligible. However, the asymmetries are well below the percent level and consistent with zero with the numerical precision. This would have to be checked in higher order computations. Similarly, the asymmetries from the square of the $\mathcal{O}(\Lambda^{-2})$ amplitudes are consistent with zero, so they are not shown in Table \ref{tab:WZasymmetryobservablesATLAS}. Overall $\mathcal{O}_{\widetilde{W}WW}$ produces larger $CP$ violating effects than $\mathcal{O}_{\varphi\widetilde{W}B}$. However, the normalisation of the operators is arbitrary as long as no UV complete model is introduced.

\begin{figure}[p]
	\centering
	\includegraphics[width=0.87\textwidth,trim={0 0 20pt 20pt} ,clip]{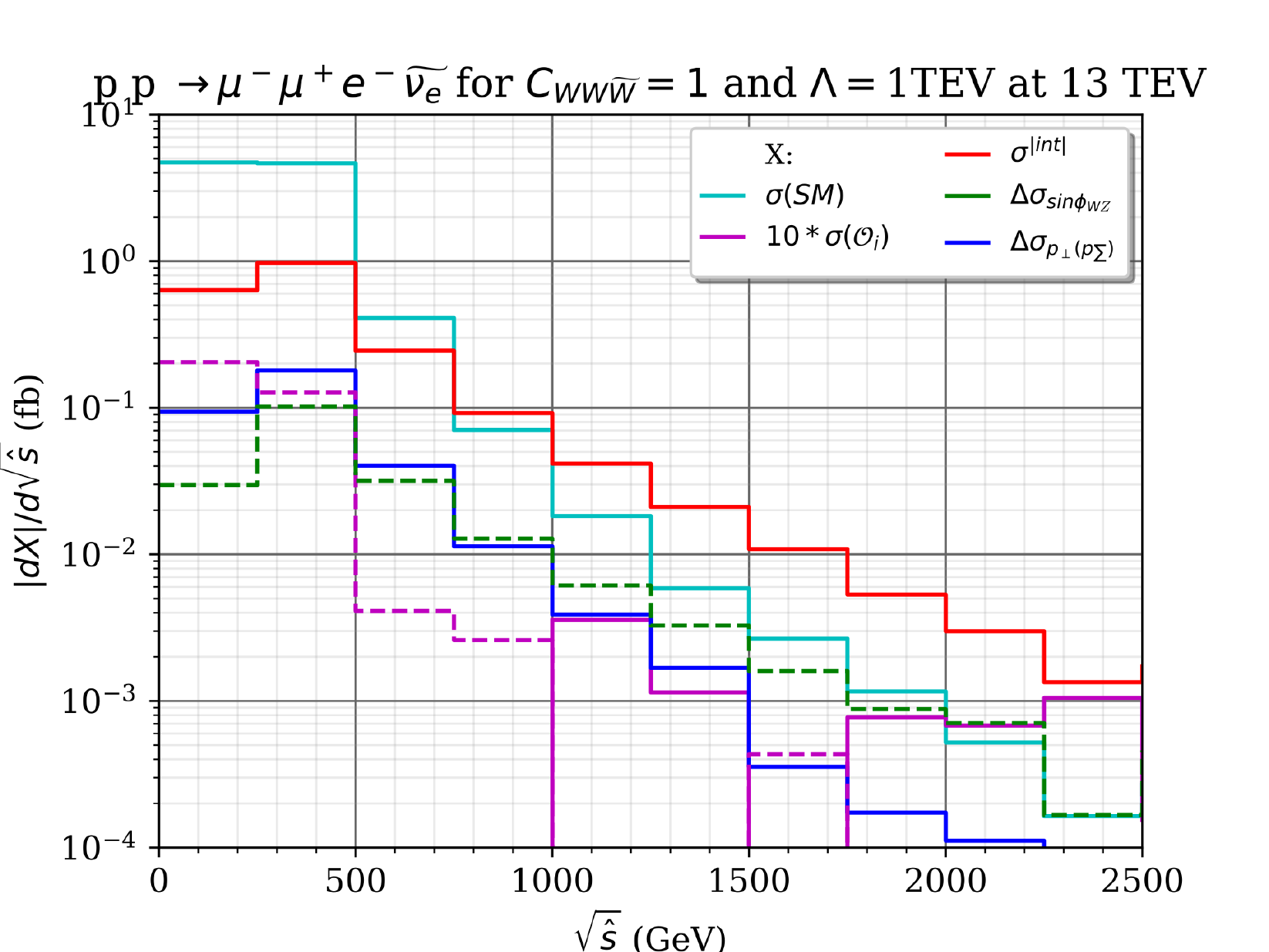}
	\hspace{3cm}
	\includegraphics[width=0.87\textwidth,trim={0 0 20pt 20pt} ,clip]{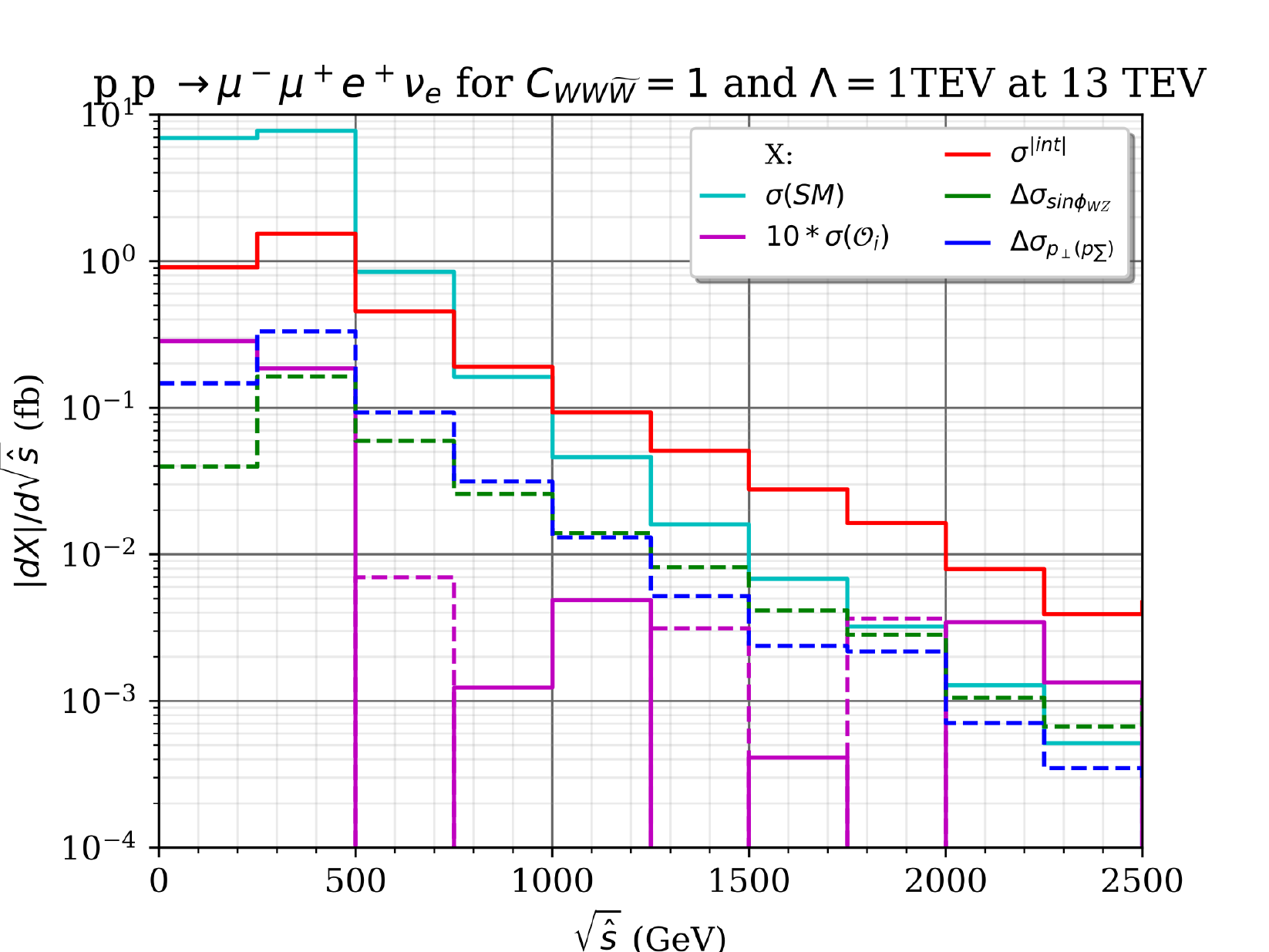}
	\caption{Differential cross-sections of SM and interferences with respect to $\sqrt{\Hat{s}}$ in $WZ$ production in the ATLAS fiducial phase space are displayed in light blue and purple respectively while we represent the asymmetries following the true matrix elements in red. The differential triple product asymmetries are drawn in blue and the differential asymmetries for $\sin\varphi_{WZ}$ in green. The dashed lines correspond to negative values. }
	\label{fig:ECMZWOWWW}
\end{figure}

\begin{figure}[p]
	\centering
	\includegraphics[width=0.87\textwidth,trim={0 0 20pt 20pt} ,clip]{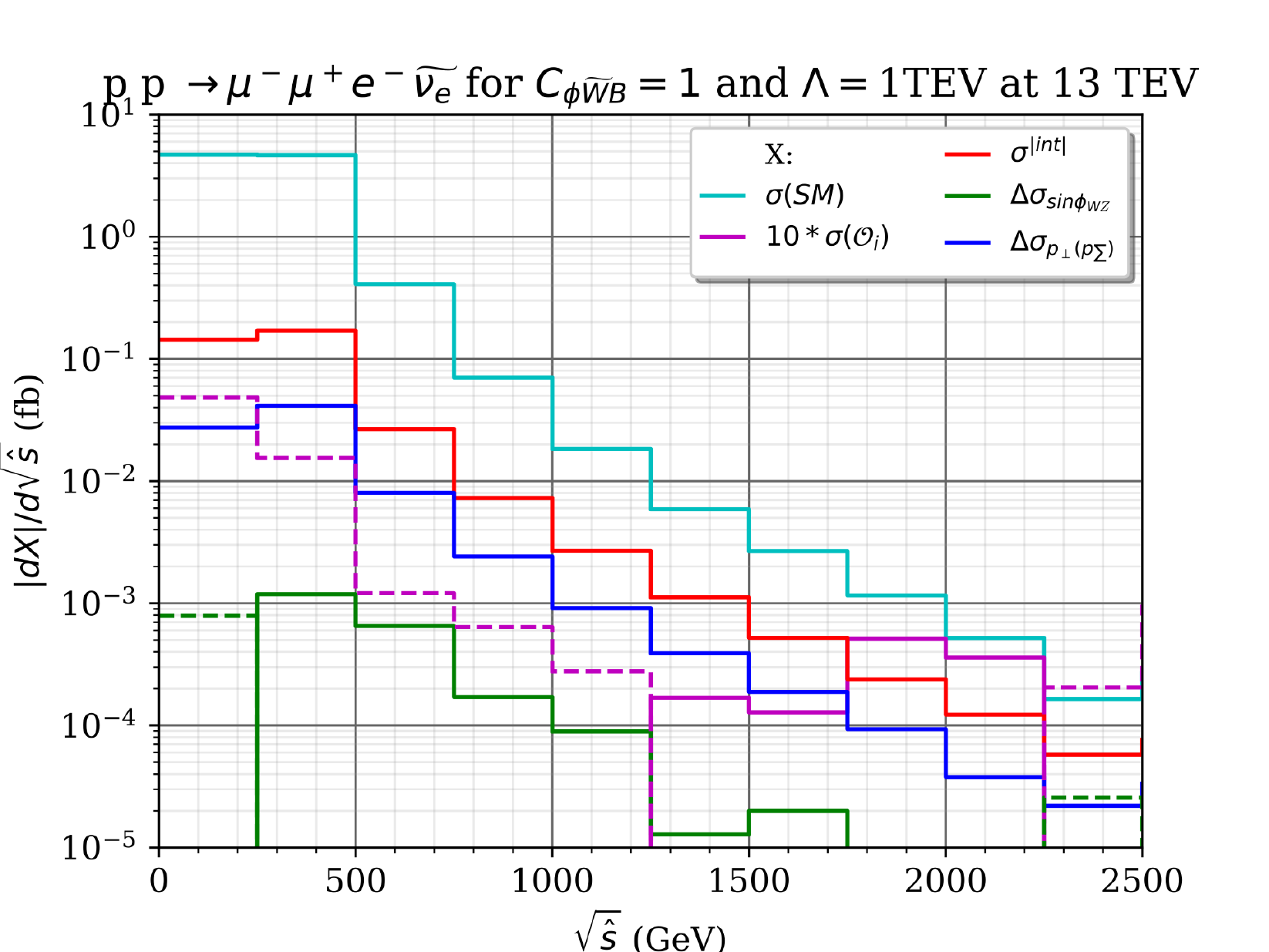}
	\includegraphics[width=0.87\textwidth,trim={0 0 20pt 20pt} ,clip]{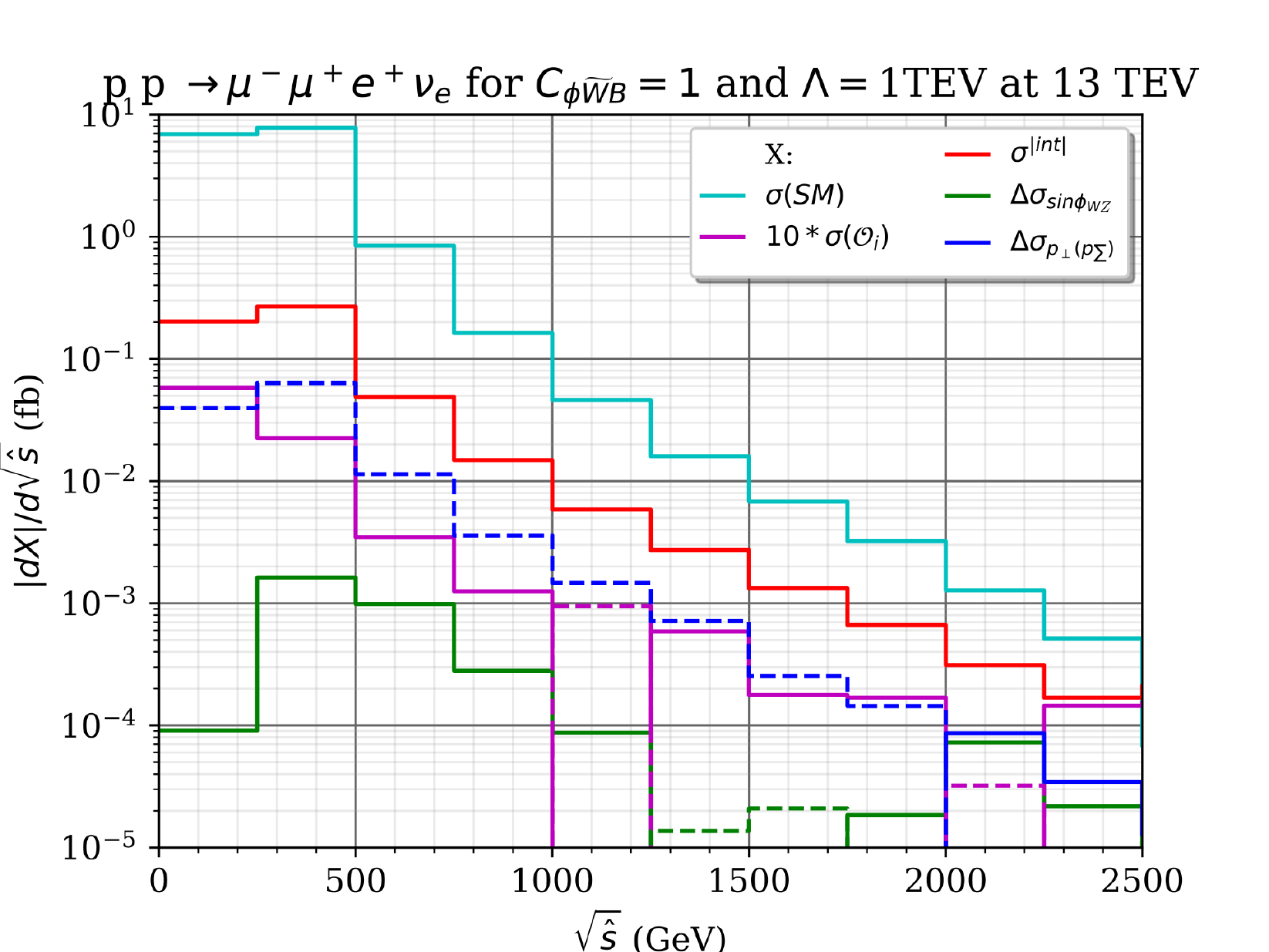}
	\caption{Same as Figure~\ref{fig:ECMZWOWWW} but for $\OBW$. These asymmetries are too small and dominated by numerical errors. }
	\label{fig:ECMZWOphiWB}
\end{figure}

Finally, the energy dependence of the asymmetries and the interference in general is checked. The differential distributions as a function of the centre-of-mass energy $\sqrt{\hat{s}}$ are displayed in Figures~\ref{fig:ECMZWOWWW} and \ref{fig:ECMZWOphiWB}. Positive values are shown with straight lines and negative ones with dashed lines. The interference cross-section has been multiplied by 10 to appear alongside the different asymmetry distributions.

The interference cross-sections change sign after the first few bins showing they are quickly dominated by numerical fluctuations. The $\sigma^{|int|}$ distributions have harder slopes than the SM as expected from higher dimensional operators contributions especially for $\OWWW$ due to the higher power of momenta in its vertices\footnote{Based on Ref.~\cite{Azatov:2017kzw} the amplitude with $\OWWW$ scales with the centre-of-mass energy as $\hat{s}/\Lambda^2 (\propto E^2)$ and the amplitude with $\OBW$ scales as $m_W \sqrt{\hat{s}}/\Lambda^2 (\propto E^1)$, while the SM amplitude scales as $m_W^2/\hat{s} (\propto E^{-2})$.}. The asymmetry of the selected triple product is more effective close to threshold while the one based on $\sin{\phi_{WZ}}$ wins at high energy for $\OWWW$ as expected. This further demonstrates the complementarity of the two observables for $\OWWW$. At high energy, $\sigma^{|int|}$ for $\OWWW$ decreases more slowly than the SM and suggests that a higher sensitivity could be reached in high energy tails. On the contrary, $\sigma^{|int|}$ has the same $1/\hat{s}$ behaviour for $\OBW$ as the SM far from threshold such that its sensitivity is not improved at high energy. This further justifies the approach based on the overall phase space for this operator rather than performing a high energy approximation similar to the one in Ref.~\cite{Azatov:2019xxn}.

The dimension-six square differential cross-sections $d\sigma_{\Lambda^{-4}}(\mathcal{O}_i)/d\sqrt{\hat{s}}$ are not displayed not to overcrowd the plots. In the context of the study of CP-odd effects, the dimension-six square differential distribution is not decisive as it could be replicated by the corresponding CP-even operators and fail to distinguish them. However, we have checked that, near threshold, $\sigma_{\Lambda^{-4}}(\OWWW)$ sits almost two orders of magnitude below $\sigma(SM)$ and one orders of magnitude below $\sigma^{|int|}$ as expected by the $1/\Lambda$ suppression of the SMEFT expansion. Then, $\sigma_{\Lambda^{-4}}$ decreases slower than $\sigma^{|int|}$ and much slower than $\sigma(SM)$ as expected of the energy scale dependence of the respective amplitudes. This is further confirmed by $\sigma_{\Lambda^{-4}}(\OBW)$ which does not scale as $\sigma_{\Lambda^{-4}}(\OWWW)$ with energy and remains consistently below $\sigma^{|int|}$.

\subsection{$W\gamma$ Results}
\label{subsec:WGammaResults}

\begin{table}[ht]
    \centering
    \begin{tabular}{|c|c|c|}
    \hline
          Process & $W^+ \gamma \rightarrow \gamma e^+ \nu_e$ & $W^- \gamma \rightarrow \gamma e^- \Tilde{\nu_e}$  \\
    \hline
          $\sigma(SM)$ & 715.1(8) fb & 589.1(7) fb  \\
    \hline
          $\delta_{PDF}$ & 2.99\% & 3.43\% \\
    \hline
    \hline
          $\sigma(\mathcal{O}_{\widetilde{W}WW})$ & -2.07(4) fb & 1.61(6) fb \\
    \hline
          Schwarz Bound & 337.3 fb & 209.0 fb \\
    \hline
          $\sigma^{|int|}(\mathcal{O}_{\widetilde{W}WW})$ & 33.83(4) fb & 24.76(6) fb \\
    \hline
          $\sigma^{|meas|}(\mathcal{O}_{\widetilde{W}WW})$ & 6.07(4) fb & 6.57(6) fb \\
    \hline
          $\sigma_{\Lambda^{-4}}\left(\OWWW\right)$ & 39.78(5) fb & 18.54(6) fb \\
    \hline
    \hline
          $\sigma(\mathcal{O}_{\varphi\widetilde{W}B})$ & 2.75(4) fb & -2.09(3) fb  \\
    \hline
          Schwarz Bound & 96.3 fb & 82.4 fb \\
    \hline
          $\sigma^{|int|}(\mathcal{O}_{\varphi\widetilde{W}B})$ & 34.00(4) fb & 26.37(3)  fb \\
    \hline
          $ \sigma^{|meas|}(\mathcal{O}_{\varphi\widetilde{W}B})$ & 9.43(4) fb & 9.53(3) fb \\
    \hline
          $\sigma_{\Lambda^{-4}}\left(\OBW\right)$ & 3.239(4) fb & 2.878(3) fb \\
    \hline
    \end{tabular}
    \caption{Cross-sections in the dileptonic decay channel of $W\gamma$ production for the ATLAS fiducial phase space at $\sqrt{s}=13$TeV for the SM, the interference with each dimension-six operator, $\sigma(\mathcal{O}_i)$, and for the square of the $\mathcal{O}\left(\Lambda^{-2}\right)$ amplitudes , $\sigma_{\Lambda^{-4}}(\mathcal{O}_i)$. Errors are from the numerical integration. For the interferences, we also display the absolute value cross-sections $\sigma^{|int|}$ and the measurable absolute value cross-section $\sigma^{|meas|}$. The Wilson coefficients $C_{\widetilde{W}WW}$ and $C_{\varphi\widetilde{W}B}$ are set to 1 and the NP scale $\Lambda$ is 1 TeV but the results can be re-scaled for any other value.}
    \label{tab:xsecWAATLAS}
\end{table}

Similarly to $WZ$, the results for $W\gamma$ production obtained for the cross-sections and the asymmetries are displayed in Table~\ref{tab:xsecWAATLAS}. The $\OWWW$ interference contribution is much smaller compared to the SM even considering the absolute value cross-section. This suppression can be understood by the lower energies probed by this process and therefore a harder cut could improve the new physics contribution relatively to the SM. As a result, the contributions from both operators are comparable in this process and suggest already that the combination of the two processes can be used to distinguish them. The ratios between the interference cross-section and $\sigma^{|int|}$ in $W\gamma$ are not as large as in $WZ$ but still signal the presence of strong phase space suppression. Additionally, the ratios between $\sigma^{|meas|}$ and $\sigma^{|int|}$ are smaller. This may be partially due to quark and anti-quark PDFs and the unmeasured photon helicity. If $W$ decays into a neutrino and a muon instead of an electron, the last cut in $W\gamma$ from Section \ref{subsec:DibosonProcesses} would no longer apply but results are not expected to vary much.

\begin{table}[t]
    \centering
    \begin{tabular}{|c|c|c|c|}
    \hline 
        Process & \multicolumn{3}{c|}{$W^+\gamma \rightarrow \gamma e^+ \nu_e $} \\
    \hline
        Operators & $SM$ & $\mathcal{O}_{\widetilde{W}WW}$ & $\mathcal{O}_{\varphi\widetilde{W}B}$  \\
    \hline
       $\Delta p_\perp (p_e,p_q)$ & 7.7(8) & -13.81(4) & 22.23(4)  \\
    \hline
       $\Delta p_\perp (p_e,p_{\sum}^z)$ & 0.8(8) & -4.60(4) & 5.59(4)   \\
    \hline
       $\Delta p_\perp (p_e,p_\gamma^z)$ & 0.5(8) & -5.62(4) & 7.59(4)   \\
    \hline
       $\Delta p_\perp (p_e,p_e^z)$ & 0.6(8) & 1.11(4) & 0.42(4) \\
    \hline
       $\Delta \sin{\phi_{W\gamma}}$ & -0.1(8) & -0.31(4) & -0.79(4)  \\
    \hline
       $\Delta \left( \Delta \phi_{eZ} \right)$ & -4.5(8) & -5.85(4) & 7.16(4) \\
    \hline
      SM stat err $30~\text{fb}^{-1}$ & \multicolumn{3}{c|}{4}  \\
    \hline
      SM stat err $100~\text{fb}^{-1}$ & \multicolumn{3}{c|}{2}  \\
    \hline
      SM stat err $3000~\text{fb}^{-1}$ & \multicolumn{3}{c|}{0.4}  \\
    \hline
    \hline 
        Process & \multicolumn{3}{c|}{$W^-\gamma \rightarrow \gamma e^- \Tilde{\nu}_e $} \\
    \hline
        Operators & $\Delta\sigma(SM)$ & $\Delta\sigma(\mathcal{O}_{\widetilde{W}WW})$ & $\Delta\sigma(\mathcal{O}_{\varphi\widetilde{W}B})$ \\
    \hline
       $\Delta p_\perp (p_e,p_q)$ & 5.3(7) & 10.65(3) & -17.27(3) \\
    \hline
       $\Delta p_\perp (p_e,p_{\sum}^z)$ & 1.2(7) & 2.34(3) & -4.15(3) \\
    \hline
       $\Delta p_\perp (p_e,p_\gamma^z)$ & 0.1(7) & -1.68(3) & 1.48(3) \\
    \hline
       $\Delta p_\perp (p_e,p_e^z)$ & 0.9(7) & 5.09(3) & -7.07(3) \\
    \hline
       $\Delta \sin{\phi_{W\gamma}}$ & -0.4(7) & -1.87(3) & 1.22(3) \\
    \hline
       $\Delta \left( \Delta \phi_{eZ} \right)$ & 1.2(7) & -6.17(3) & 8.46(3)   \\
    \hline
      SM stat err $30~\text{fb}^{-1}$ & \multicolumn{3}{c|}{4}  \\
    \hline
      SM stat err $100~\text{fb}^{-1}$ & \multicolumn{3}{c|}{2}  \\
    \hline
      SM stat err $3000~\text{fb}^{-1}$ & \multicolumn{3}{c|}{0.4} \\
    \hline
    \end{tabular}
    \caption{Asymmetries in fb in the $W \gamma \rightarrow \gamma e^+ \nu_e$ and $W\gamma \rightarrow \gamma e^- \widetilde{\nu}_e$ channels by using different reference axes for the four observables, the triple product, the two triple products with the beam direction correction factor and the Barducci observable, in the ATLAS fiducial phase space at $\sqrt{s}=13$TeV at the LHC. The statistical errors are displayed using the SM cross-sections and several integrated luminosities.     }
    \label{tab:WAasymmetryobservables}
\end{table}

In $W\gamma$ production the triple product can only be constructed with the electron momentum and the photon momentum. Therefore, all that is left is to look for the best proxy for the quark momentum as the same argument about the non-measurability of the quark momentum stands. As in the $WZ$ production, the longitudinal component of the electron, of the photon and of the sum over all the final visible particles are considered and shown in Table~\ref{tab:WAasymmetryobservables}.

The triple product asymmetries with the quark momentum are first checked: this particular triple product produces asymmetries around 40\% of $\sigma^{|int|}$ for $\mathcal{O}_{\widetilde{W}WW}$ and 65\% for $\mathcal{O}_{\varphi\widetilde{W}B}$. The asymmetry in the triple product with the quark momentum is a bit less effective in $W\gamma$ production than in $WZ$ production but remains quite good. One drawback is that the SM asymmetry is more important. The largest asymmetry for $W^+\gamma$ production is obtained with the photon longitudinal momentum while the electron longitudinal momentum gives the largest asymmetry for $W^-\gamma$. Those asymmetries have quite large efficiencies as their ratios with $\sigma^{|meas|}$ are respectively around 80\% and even above 90\% for $\OWWW$ in $W^+\gamma$. Those large efficiencies are in contrast with the poor sensitivity obtained in this channel for the observable proposed in Ref~\cite{Azatov:2019xxn}, especially for $\OWWW$. Unlike in $WZ$, $\sin{\phi_{W\gamma}}$ appears ineffective in tracking $CP$-violating operators in $W\gamma$ while $\Delta \phi_{eZ}$ provides results almost similar or even better than the best triple product in $W^+\gamma$ depending on the operator and the sign of the W boson. By taking the best observable for each channel and each operator, the efficiencies are all very close to 90\% except for $\OBW$ in $W^+\gamma$ which is at 80\%. The asymmetries derived from the squared amplitudes are consistent with zero and always at least more than one order of magnitude below those of the interferences.

\begin{figure}[p]
\centering
\includegraphics[width=0.87\textwidth,trim={0 0 30pt 20pt} ,clip]{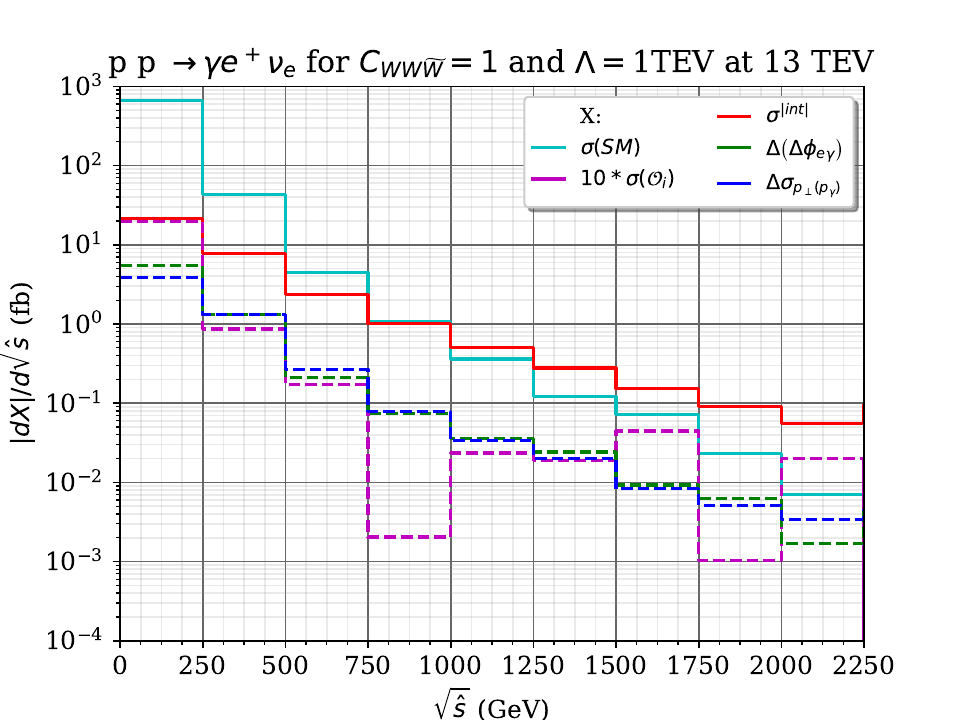}
\includegraphics[width=0.87\textwidth,trim={0 0 30pt 20pt} ,clip]{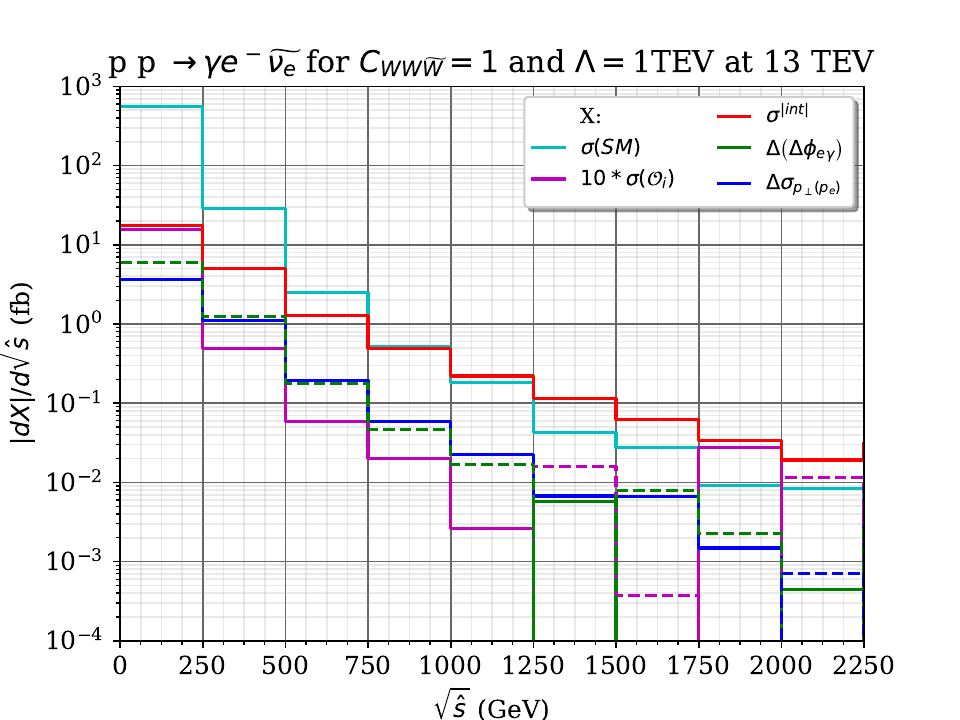}
\caption{Differential cross-sections of SM and interferences with respect to $E_{c.o.m.}$ in $W\gamma$ production in the ATLAS fiducial phase space are displayed in light blue and purple respectively while we represent the asymmetries following the true matrix elements in red. The differential triple product asymmetries are drawn in blue and the differential asymmetries of $\sin\varphi_{W\gamma}$ in green. The bins in dashed lines correspond to negative values. }
\label{fig:ECMAWOWWW}
\end{figure}

\begin{figure}[p]
\centering
\includegraphics[width=0.87\textwidth,trim={0 0 30pt 20pt} ,clip]{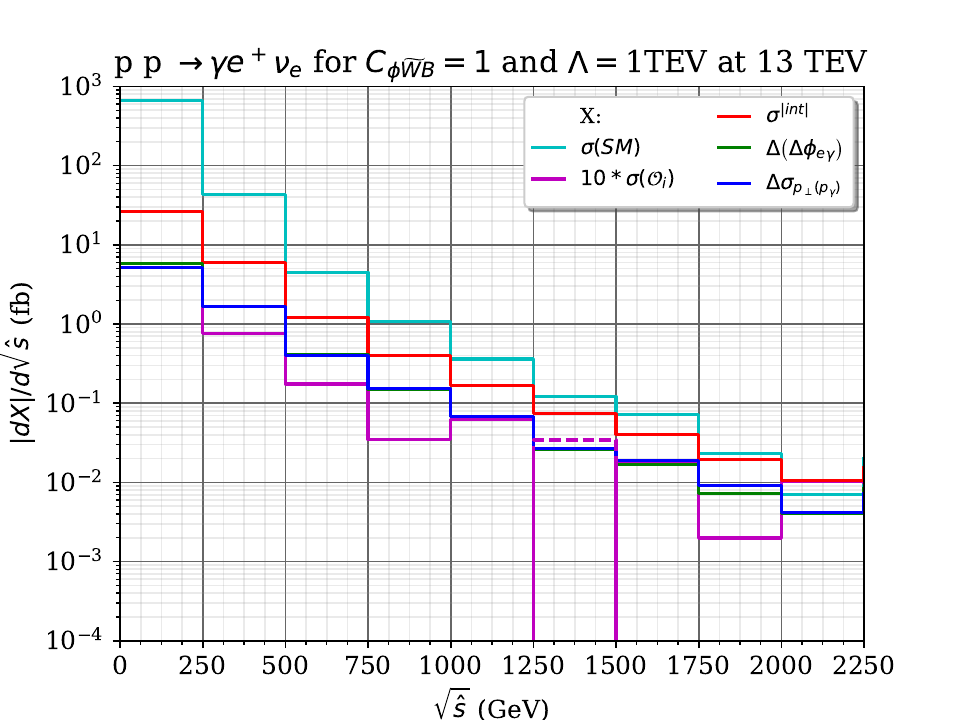}
\includegraphics[width=0.87\textwidth,trim={0 0 30pt 20pt} ,clip]{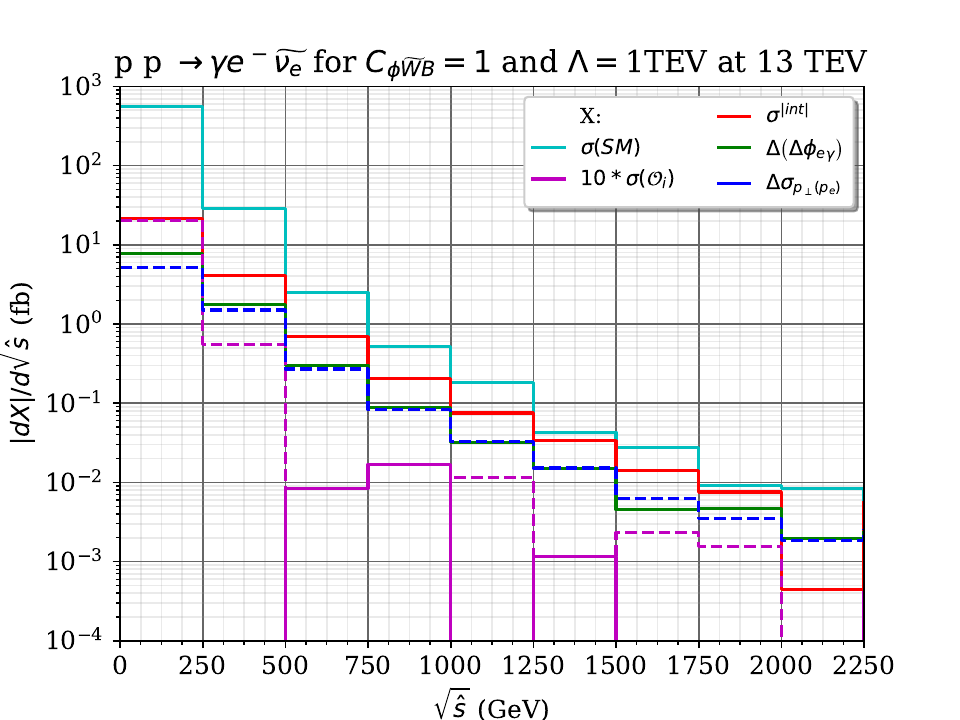}
\caption{Same as Figure~\ref{fig:ECMAWOWWW} but for $\mathcal{O}_{\varphi\widetilde{W}B}$.}
\label{fig:ECMAWOphiWB}
\end{figure}

Finally, the distributions of the asymmetries as a function of $\sqrt{\Hat{s}}$ are displayed on Figures~\ref{fig:ECMAWOWWW} and \ref{fig:ECMAWOphiWB}. The interference cross-sections are again multiplied by 10, the distributions have the same colours as in Figure~\ref{fig:ECMZWOWWW} and show a similar behaviour as for $WZ$ production: interference cross-sections begin to fluctuate in high-energy bins and even change sign, the $\sigma^{|int|}$ asymmetries have a harder slope than the SM and the $p_\perp(p_e, p_{\sum})$ asymmetry follows quite closely $\sigma^{|int|}$. The distributions of $\Delta \phi_{e\gamma}$ follow closely the triple product ones over the whole energy range irrespective of the channel or the operator.

\subsection{Sensitivity}
\label{subsec:DibosonSensitivity}

As the goal has always been to motivate observables that could be used in future  analyses to draw the most stringent constraints on operators, the method described in Ref.\cite{Kumar:2008ng} is followed to estimate the sensitivities to the two operators. For those estimates, the SM process is assumed to be the only background and the LO  cross-sections as the total measured cross-section. The SM is thought to be exactly symmetric, which does not have to be the case even without $CP$ violation as our observables are not pure $CP$-odd. The sensitivities are obtained by fixing the signal over background ratio $(S/\sqrt{B} = 2) \sim 2 \sigma$.

Firstly, the result for $p_\perp (p_e, p_Z)$, which is the variable they use in Ref.\cite{Kumar:2008ng}, is compared to the result therein. The result presented, $\Delta \sigma = \Tilde{\lambda}\times (3\times 10^3~ \text{fb})$, actually considers a coupling constant $\Tilde{\lambda}_Z$ associated to an anomalous interaction rather than a Wilson coefficient from an operator of an EFT. Using $\Tilde{\lambda}_Z = 6  C_{\widetilde{W}WW} M_W^2 / g \Lambda^2$ from Ref.~\cite{Bohm:1987ck}, it is possible to translate the result such that
\begin{equation*}
    \Delta \sigma = C_{\widetilde{W}WW} \times (1.8 \times 10^2)~\text{fb}.
\end{equation*} 
Then, we apply the decay fractions of the $W$ and $Z$ bosons into a single leptonic channel as provided by the PDG \cite{ParticleDataGroup:2024cfk}, which are $10.7\%$ and $3.37\%$ respectively, to match the final states between our analysis and Ref.\cite{Kumar:2008ng}. This results in $\Delta \sigma = C_{\widetilde{W}WW} \times 0.644~\text{fb}$. 
By adding our values of $p_\perp(p_e, p_Z)$ in both $W^\pm Z$, we get $\Delta \sigma = C_{\widetilde{W}WW} \times 0.798~\text{fb}$. There is a small difference which can be understood mainly from the different cuts.

\begin{figure}[ht]
    \centering
    \includegraphics[scale=0.75]{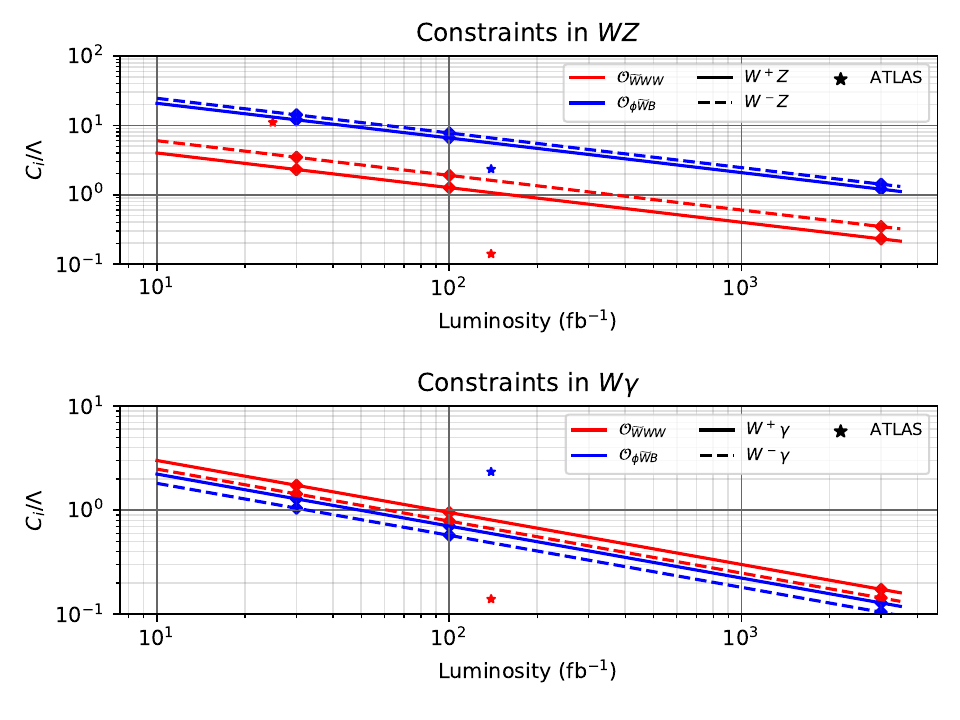}
    \caption{Sensitivity at 95\%CL as a function of the integrated luminosity for each process and each operator using the best observable in each case and the SM at LO as the only background and assuming it to be symmetric. }
    \label{fig:wilson coeff constr}
\end{figure}

Secondly, this method is applied to $p_\perp(p_e, p_{\sum})$ for the $WZ$ process, with $p_\perp(p_e, p_\gamma)$ for $W^+\gamma$, and with $\Delta \phi_{e\gamma}$ for $W^-\gamma$. Different luminosities are considered. The limits are given in Figure~\ref{fig:wilson coeff constr}. For comparison, other recent constraints are displayed in Table \ref{tab:const comp}. As usual, the LHC constraints are still several orders of magnitude less stringent than the constraints from the electron EDM such that the observation of a deviation would imply a cancellation by a few orders of magnitude for the SMEFT contribution to the EDM. The estimated constraints of the analysis are similar in both processes for $\mathcal{O}_{\widetilde{W}WW}$ and lie between the semi-hadronic $WZ$ and VBF measurements by ATLAS. On the contrary, $W\gamma$ gives much better constraints on $\mathcal{O}_{\varphi\widetilde{W}B}$ than $WZ$ which makes them competitive with VBF. As a result, they could be used to confirm or disprove the deviation seen in this process.

While the results do not seem much better than previous analyses, they can be improved in several ways. First, only one leptonic channel has been considered so adding the other leptonic channels will increase the statistics. For $WZ$, the observable is independent of the charge of the Z decay product. Therefore it suggests that it could be used also for the hadronic decay of the Z boson. Secondly, all the four processes can be combined. Thirdly, the NLO correction increases the SM cross-sections by roughly a factor 2. If this is true also for the SMEFT contributions to the $CP$-sensitive observables, this could further enhance the sensitivity. Moreover, for the large contribution with an extra radiation, the jet could be used to better approximate the quark direction. Finally, the cross-sections are sufficiently large either to use differential asymmetries or to cut the phase space in order to improve the sensitivities.

\begin{table}[ht]
    \centering
    \begin{tabular}{c|c|c|c}
       Operators & $\sigma(pp\rightarrow l\nu jj)$ \cite{ATLAS:2017luz} & $\Delta \phi_{jj}$ \cite{Aad:2020sle} & EDM \cite{Panico:2018hal} \\
        \hline
        $\mathcal{O}_{\widetilde{W}WW}$ & [-14, 14] (expected) & [-0.12, 0.12] (expected) & $\leq 1.74~10^{-4}$ \\
                  & [-11, 11] (measured) &  [-0.11, 0.14] (measured) &  \\
        $\mathcal{O}_{\varphi\widetilde{W}B}$ & // & [-1.06, 1.06] (expected) & $\leq 5.57~10^{-6}$ \\
                  & // &  [-0.23, 2.34] (measured) &  \\
    \end{tabular}
    \caption{Collection of the constraints on the two dimension-six operators with $\Lambda$ = 1TeV at 95\% CL. }
    \label{tab:const comp}
\end{table}

Since the publication of Ref.~\cite{Degrande:2021zpv}, ATLAS has released two analyses in diboson production using $CP$ sensitive observables in Refs.~\cite{ATLAS:2025edf, ATLAS:2026xuq}. The analysis of $W^\pm Z$ production from Ref.~\cite{ATLAS:2025edf} relies on the combination of three boosted decision trees (BDTs) trained to separate SMEFT contributions from the SM ones and to track the sign of the SMEFT contributions. One of the variables used in the BDTs is the triple products advocated for in the results $p_\perp (p_{l^W}, p_{\sum}^z)$. All BDTs also depend on the triple products $p_\perp(W, l^+, l^W)$ and $p_\perp(W, l^-, l^W)$. The BDT separating the SMEFT and SM contributions used $p_\perp(l^{-,z}, l^W, l^+)$ and $p_\perp(l^{+,z}, l^-, l^W)$. The observed limits on the linear contributions are $[-0.186, 0.139]$ TeV$^{-2}$ for $C_{\widetilde{W}WW}/\Lambda^2$ and $[-1.625, 1.332]$ TeV$^{-2}$ for $C_{\varphi\widetilde{W}B}/\Lambda^2$. The analysis of $W\gamma$ in Ref.~\cite{ATLAS:2026xuq} is different as NN discriminants are constructed for each operator, $O_{NN}^{\mathcal{O}_{\widetilde{W}WW}}$ and $O_{NN}^{\mathcal{O}_{\varphi\widetilde{W}B}}$. The NNs are trained to recognise positive or negative interference of the SMEFT operators with the SM, such that the constructed observable is the difference between the two probabilities $O_{NN} = P_+ - P_-$. It is argued that the discriminants improve the limits by a factor of 2 compared to the best performing observable $\Delta \phi_{l\gamma}$. The observed limits on the linear contributions are $[-0.74, 0.23]$ TeV$^{-2}$ for $C_{\widetilde{W}WW}/\Lambda^2$ and $[-0.59, 0.13]$ TeV$^{-2}$ for $C_{\varphi\widetilde{W}B}/\Lambda^2$.

\chapter{BSM scalar resonances in Four-Top Production}
\label{chap:fourtops}
\pagestyle{fancy}

\hfill

%\begin{minipage}{10cm}

%{\small\it 
%``No matter how many instances of white swans we may have observed, this does not justify the conclusion that all swans are white.''}

%\hfill {\small Karl Popper, \textit{The Logic of Scientific Discovery}}
%\end{minipage}

\vspace{0.5cm}

Having discussed off-shell effects generating CP violating contributions, this Chapter shifts to on-shell BSM resonances in four-top production. Two simplified models are investigated in the four-top process that has been recently observed at the LHC. The new degrees of freedom are the scalar singlet $S_1$ and octet representations $S_8$ of $SU_C(3)$, and their interactions are restricted exclusively to the top quark.

The resonance search in four-top production of the two top-philic simplified models is motivated in Section~\ref{sec:FourTopMotivation}. The improvement of jet tagging, the observation of four-top production and relevant UV-complete theories are discussed. The two models are presented in Section~\ref{sec:FourTopTheory} with their possible UV origins and the different benchmarks. The renormalisation procedure of the models and the subtleties of the loop diagrams selection for NLO cross-sections are thoroughly described as well. The generation of the simulated events is outlined in Section~\ref{subsec:FourTopEvtGen} which ends with the K-factors calculated for four benchmark points. The simulation of the detector response and the reconstruction of the boosted four-tops system are presented in Section~\ref{sec:FourTopReco}. The backgrounds considered in this work are listed in Section~\ref{sec:bkd} with the description of their simulation. Sections~\ref{sec:OctetAnalysis} and \ref{sec:SingletAnalysis} present the signal regions, the pairing strategies of (leptonic or hadronic) top quark candidates and the improvements of constraints on the parameter space of the octet and singlet resonances respectively.

This chapter is based on Ref.~\cite{Darme:2025leu}. As in Chapter~\ref{chap:smeftcpv}, parts of the text have been modified to fit the rest of the manuscript when necessary. My contributions to the collaborative effort was first the creation of the UFO models suited for NLO calculations. Then, I generated the shower-level events of BSM processes at LO and NLO. The pole cancellation in the amplitudes of the latter was the most challenging part of my contribution. Thirdly, I provided the background event simulations except for the $t\overline{t}jj$ with the MLM matching and merging.

\newpage

\section{Motivations}
\label{sec:FourTopMotivation}

The ability to reconstruct the nature of new coloured particles from the detailed observation of jet substructure has become a cornerstone of many analyses at the LHC. In recent years, the deployment of machine-learning-based techniques that exploit the full set of jet constituents and their properties has led to significant advances in jet tagging performance, these developments representing one of the most exciting innovations of the latest experimental runs at the LHC~\cite{Kasieczka:2017nvn, Larkoski:2017jix, Guest:2018yhq}. By dramatically reducing mistagging rates by factors compared to traditional methods~\cite{Gerbush:2007fe, ATLAS:2015ddu, CMS:2017wyc, ATLAS:2018wis, ATLAS:2022qby, Cagnotta:2022hbi, ATLAS:2024rua}, these algorithms open the door to novel search strategies, especially in regimes where large Standard Model (SM) backgrounds have previously limited sensitivity. 
In light of these new approaches, final states with high object multiplicities, such as those originating from the production of multiple top quarks possibly induced by physics beyond the SM, are particularly promising targets. It has hence been found that in such scenarios, the ability to resolve and identify individual substructures within jets is crucial to improving signal efficiency and enabling robust discrimination from background processes~\cite{Englert:2016aei, Aguilar-Saavedra:2019ptp, Belyaev:2021zgq, Araz:2023axv, Chowdhury:2023jof, Darme:2024epi, Sahu:2024fzi}.

Four-top final states represent one of the highest-multiplicity SM processes accessible at the LHC, with each top quark decaying to a $W$ boson and a $b$ quark. Owing to their complexity and rarity, four-top events have the potential to offer a powerful probe of both SM and BSM physics. Dedicated analyses from both ATLAS and CMS~\cite{CMS:2017ocm, ATLAS:2018alq, ATLAS:2018kxv, CMS:2019jsc, CMS:2019rvj, ATLAS:2020hpj, ATLAS:2021kqb, ATLAS:2023ajo, ATLAS:2023taw, CMS:2023ftu,CMS:2023zdh, ATLAS:2024jja} have hence developed increasingly sophisticated strategies, typically based on reconstructed objects such as $b$-tagged jets and isolated leptons.  These studies, leveraging data from the LHC Run~2 dataset, have provided valuable insights, notably enabling complementary constraints on the SM top Yukawa coupling through comparisons with $t\bar{t}H$ production. Until recently~\cite{ATLAS:2024jja}, most four-top searches have concentrated on the kinematic regime expected from SM predictions~\cite{Bevilacqua:2012em, Alwall:2014hca, Maltoni:2015ena, Frederix:2017wme, Jezo:2021smh, vanBeekveld:2022hty, Dimitrakopoulos:2024qib, Dimitrakopoulos:2024yjm, vanBeekveld:2025ghw}, characterised by relatively soft and isolated top decay products. However, this regime poses challenges for full event reconstruction due to the limited Lorentz boost of the individual tops, yielding a high combinatorial background associated with the large amount of well-isolated decay products of the four-top system. As a result, much of the available phase space, particularly that involving energetic and boosted top quarks as predicted in many BSM scenarios, remains under-explored.

In this context, a wide variety of new physics models predict final states with multiple top quarks, with four-top production emerging as a particularly compelling signature. In particular, such top-rich final states arise naturally when new heavy particles strongly coupled to the top quark are pair-produced via QCD interactions, and subsequently decay to top-antitop pairs~\cite{Battaglia:2010xq, Dev:2014yca, Greiner:2014qna, Alvarez:2016nrz, Kim:2016plm, Fox:2018ldq, Alvarez:2019uxp, Blasi:2023hvb}. Composite Higgs models constitute a prominent class of such setups, offering a solution to the hierarchy problem yielding composite resonances such as vector-like top partners or coloured pseudo-Nambu-Goldstone bosons with enhanced couplings to the top quark~\cite{Lillie:2007hd, Pomarol:2008bh, Zhou:2012dz, Cacciapaglia:2015eqa, Belyaev:2016ftv, Liu:2019bua, Cacciapaglia:2020vyf, Cornell:2020usb, Cacciapaglia:2024wdn}. Other well-motivated frameworks giving rise to four-top final states include minimal flavour violation models~\cite{Gerbush:2007fe, Hayreter:2017wra}, extended supersymmetric models~\cite{Salam:1974xa, Fayet:1974pd, Fayet:1975yi, AlvarezGaume:1996mv, Fox:2002bu, Plehn:2008ae, Choi:2008ub, GoncalvesNetto:2012nt, Fuks:2012im, Calvet:2012rk, Benakli:2014cia, Beck:2015cga, Kotlarski:2016zhv, Darme:2018dvz, Carpenter:2020hyz, Carpenter:2020evo} and constructions with an extended Higgs sector~\cite{Branco:2011iw, Arcadi:2019lka, Cheng:2017tbn, Cheng:2018mkc, Coloretti:2023yyq, Anisha:2023xmh}. Additionally, recent phenomenological studies have also exploited four top-quark production in an effective field theoretical framework~\cite{Agram:2013koa, Khatibi:2014via, Durieux:2014xla, Guo:2016kea, Shen:2018mlj, DHondt:2018cww, Hartland:2019bjb, Degrande:2020evl, Liu:2020bem, Banelli:2020iau, Darme:2021gtt}.

In this work, we focus on the direct production of heavy top-philic states decaying on-shell to top-antitop pairs and leading to a distinctive four-top signature. To capture the essential phenomenology while retaining model independence, we employ the same simplified model framework as the one developed in earlier studies~\cite{Fuks:2012im, Calvet:2012rk, Beck:2015cga, Darme:2018dvz, Darme:2021gtt, Darme:2024epi}. These works have demonstrated that four-top final states provide a robust and complementary probe of top-philic sectors, both in the resonant regime and in an effective construction when the mediators are too heavy to be produced. As highlighted in Ref.~\cite{Darme:2024epi}, reconstructing and tagging all four boosted top quarks in the final state dramatically suppresses the Standard Model background and opens the door to novel and more sensitive search strategies, notably in the zero-lepton, one-lepton and same-sign di-lepton channels. In particular, we have found that using modern top-tagging techniques enables a potential improvement of up to an order of magnitude over existing limits, especially in the regime of high-mass resonances. This enhanced sensitivity results from two dominant effects: a significantly increased signal efficiency in the boosted-top regime compared to traditional $b$-jet-based selections targeting a resolved four-top signal, and a strong reduction of the irreducible QCD background allowing the exploration of the more challenging fully hadronic final states.
This strategy parallels recent approaches developed in the context of four-bottom final states for di-Higgs production~\cite{ATLAS:2022hwc, CMS:2024ymd}, where the full reconstruction of boosted $b$-jets systems has yielded substantial gains in sensitivity. It also connects naturally with recent theory-driven efforts to apply machine-learning techniques to improve four-top searches with a jet substructure analysis~\cite{Choudhury:2024mox, Kvita:2024ooa, Flacke:2025xwl}.

\section{Theoretical and numerical framework}
\label{sec:FourTopTheory}

In this section, we present the theoretical and numerical framework used throughout our analysis. We focus on two classes of simplified models that extend the SM with top-philic scalar particles: a colour-singlet scalar $\sing$ state of mass $M_{\sing}$ and a colour-octet scalar $\oct$ state of mass $M_{\oct}$. These simplified extensions provide a minimal and general framework for capturing the key features of a broad class of UV completions such as those mentioned in Section~\ref{sec:FourTopMotivation}, and the corresponding interactions can be described by a compact Lagrangian formulation which we introduce in Section~\ref{subsec:FourTopLags} along with a selection of representative benchmark points. We then exploit the fact that the simplified model approach offers a flexible path for phenomenological interpretation and numerical simulation in Section~\ref{subsec:FourTopEvtGen}. To this aim, we implement these models in the UFO format~\cite{Degrande:2011ua, Darme:2023jdn} using the \fr~\cite{Christensen:2009jx, Alloul:2013bka}, {\sc MoGRe}~\cite{Frixione:2019fxg} and \nloct~\cite{Degrande:2014vpa} packages. This setup then allows us to perform signal simulations at NLO in QCD, the corresponding cross sections and $K$-factors computed using \MG~\cite{Alwall:2014hca} being discussed in detail.

\subsection{Simplified models for top-philic new physics}
\label{subsec:FourTopLags}

The new physics contributions to the Lagrangian of the two simplified models considered here and featuring top-philic scalar resonances include three terms: gauge-invariant kinetic and mass terms for the new resonance, as well as a Yukawa-like coupling to a top-antitop pair. This yields the singlet and octet Lagrangians $\mathcal{L}_{\sing}$ and $\mathcal{L}_{\oct}$ given by
\begin{equation}
\label{eq:lags}
\begin{split}
    \scr{L}_{\sing} =&\ \scr{L}_{\mathrm{SM}} + \frac{1}{2} \partial_\mu \sing \partial^\mu \sing- \frac{1}{2} M_{\sing}^2 \sing^2 + y_{\sing} \sing  \, \bar{t} t \ ,\\
  \scr{L}_{\oct} =&\  \scr{L}_{\mathrm{SM}} + \frac{1}{2} D_\mu \oct^A D^\mu \oct^A - \frac{1}{2} M_{\oct}^2 \oct^A \oct^A + y_{\oct}  \oct^A  \, \bar{t}\, T^A\, t  \ .
\end{split}
\end{equation}
where the additional scalar fields are assumed to be real and the superscript $A$ indicates (summed) colour-adjoint indices. Here, $\scr{L}_{\mathrm{SM}}$ denotes the SM Lagrangian, $y_{\sing}$ and $y_{\oct}$ are the new Yukawa couplings, while $M_{\sing}$ and $M_{\oct}$ represent the masses of the singlet and octet states respectively. The widths of the scalar octet $\Gamma_{\oct}$ and singlet $\Gamma_{\sing}$ are respectively given by
\begin{equation*}
\begin{split}
	& \Gamma_{\oct} = \frac{y_{\oct}^2 (M_{\oct}^2 - 4 m_t^2)^{3/2} }{16 \pi m_{\oct}^2} , \\
	& \Gamma_{\sing} = \frac{3 y_{\sing}^2 (M_{\sing}^2 - 4 m_t^2)^{3/2} }{8 \pi m_{\sing}^2} .
\end{split}
\end{equation*}
where $m_t$ is the top quark mass. For identical mass and coupling values, $\Gamma_{\sing}$ is 6 times larger than $\Gamma_{\oct}$. As a result, width effects play a more significant role in processes involving the singlet than the octet.

Each of these simplified models can be naturally connected to UV-complete new physics constructions. In particular, colour-octet states, regardless of their spin or CP quantum numbers, frequently arise in UV models addressing the hierarchy problem of the SM. For example, minimal supersymmetric scenarios feature colour-octet fermions known as gluinos which have been extensively searched for at the LHC and that are now constrained to be heavier than approximately 2~TeV, depending on the specific model. In extended supersymmetric frameworks, gluinos belong to supermultiplets that also include scalar or pseudo-scalar octet fields (generally organised into complex scalar fields) commonly referred to as sgluons. These scalar resonances have attracted considerable attention as they can alleviate the tension that too heavy gluinos pose on the Higgs sector~\cite{Salam:1974xa, Fayet:1974pd, Fayet:1975yi, AlvarezGaume:1996mv, Fox:2002bu, Plehn:2008ae, Choi:2008ub, GoncalvesNetto:2012nt, Fuks:2012im, Calvet:2012rk, Benakli:2014cia, Beck:2015cga, Kotlarski:2016zhv, Darme:2018dvz, Carpenter:2020hyz, Carpenter:2020evo}. If the pseudo-scalar octet is the lightest coloured BSM particle, then it is stable at tree level and couples to the SM quarks at one loop with a strength proportional to the quark mass, thus becoming top-philic. Composite models offer another motivation for top-philic scalar resonances. Analogous to QCD, such models predict meson-like composite states arising from a new strongly-coupled gauge sector, and scenarios typically exhibit a rich spectrum of bound states including pseudo-scalar colour-charged mesons that emerge as pseudo Nambu-Goldstone bosons below the new confinement scale (much like the pions in QCD)~\cite{Lillie:2007hd, Pomarol:2008bh, Zhou:2012dz, Belyaev:2016ftv, Cacciapaglia:2015eqa, Liu:2019bua, Cacciapaglia:2020vyf, Cornell:2020usb, Cacciapaglia:2024wdn}. While a wide variety of couplings are possible, sizeable interactions with the top quark can often arise~\cite{Belyaev:2016ftv}. In addition, colour-singlet scalars also commonly appear in such extensions, as well as in a broad class of other new physics models.

More generally, the BSM top-quark couplings introduced in Eq.~\eqref{eq:lags} are expected to be generated, in a UV-complete model, only after EWSSB. Since the Higgs vacuum expectation value is the sole source of EWSSB in the SM, it is reasonable to expect that any scalar singlet not involved in EWSSB would inherit a coupling structure to quarks proportional to their SM Yukawa couplings, thereby favouring a top-philic scenario. This mechanism is in fact common in dark sector constructions~\cite{Arcadi:2019lka}. Similarly, in constructions with an extended Higgs sector such as the Two-Higgs-Doublet Model, stringent flavour constraints often enforce alignment in the Yukawa sector~\cite{Branco:2011iw}, again leading to enhanced couplings to the top quark.

Finally, we note that as the scalar or pseudo-scalar nature of the top-philic resonance does not affect the analysis proposed in this study, we focus on scalar top-philic states in the remainder of this work. Our results would nevertheless remain valid if the scalar interactions in Eq.~\eqref{eq:lags} were replaced with pseudo-scalar ones. Likewise, although vector particles would lead to different projected bounds on their Lagrangian parameters, the analysis strategy presented here would remain applicable. We leave a dedicated study of alternative spin and parity assignments, as well as of the colour-sextet case which shares many features with the octet scenario at the analysis level, for future work.

From a bottom-up perspective, the main phenomenological difference between the two simplified models in Eq.~\eqref{eq:lags} lies in the possibility of producing pairs of colour-octet states via QCD interactions, as illustrated by the representative Feynman diagrams on the right of Figure~\ref{fig:FeynDiagram}. Apart from this, in both cases the new physics particle is expected to predominantly decay into top quarks, leading to a signal comprising four top quarks. The projections that will be made below can however be straightforwardly rescaled to account for a reduced branching ratio, as may occur in next-to-minimal or UV-complete models featuring multiple decay modes. The parameter space of both simplified models is thus two-dimensional and defined by the scalar mass $M_X$ and the Yukawa-like coupling $y_X$ with $X = \oct$ or $\sing$. For the colour-octet case, the dominant production mechanism (associated or pair production, respectively illustrated in the left and right of Figure~\ref{fig:FeynDiagram}) depends sensitively on the values of these parameters. As the mass $M_{\oct}$ increases, on-shell production becomes suppressed and the relative importance of pair production diminishes compared to the associated production. A similar effect arises when increasing the Yukawa coupling $y_{\oct}$, as associated production scales with $y_{\oct}^2$ whereas pair production remains dominated by QCD.

To explore these features, we define four benchmark points (BPs) that probe different regions of the parameter space and highlight contrasting topologies in four-top production. For the colour-octet model, we consider:
\begin{itemize}%[topsep=2pt,itemsep=0pt,parsep=3pt,partopsep=2pt]
    \item \textbf{BP1:} $M_{\oct} = 2$~TeV, $y_{\oct} = 1$ (yielding $\Gamma_{\oct} = 38$~GeV);
    \item \textbf{BP2:} $M_{\oct} = 3$~TeV, $y_{\oct} = 3$ (yielding $\Gamma_{\oct} = 527$~GeV).
\end{itemize}
BP1 corresponds to an intermediate case where neither production mechanism dominates, allowing us to study a mixed production regime. In contrast, BP2 implies an increased associated production contribution while pushing the particle width into a regime ($\Gamma_{\oct} / M_{\oct} \sim 0.18$) where finite-width effects and off-shell kinematics become non-negligible~\cite{Denner:1999gp, Denner:2005fg}. For the colour-singlet scenario in which only the Yukawa-induced associated production is present, we define two benchmarks focused on lower and intermediate masses with Yukawa couplings of $\mathcal{O}(1)$, reflecting the limited reach of the LHC for colour-neutral resonances:
\begin{itemize}%[topsep=2pt,itemsep=0pt,parsep=3pt,partopsep=2pt]
    \item \textbf{BP3:} $M_{\sing} = 1.5$~TeV, $y_{\sing} = 1$ (with $\Gamma_{\sing} = 165$~GeV);
    \item \textbf{BP4:} $M_{\sing} = 2$~TeV, $y_{\sing} = 1.5$ (with $\Gamma_{\sing} = 513$~GeV).
\end{itemize}
The adopted mass range $M_X \sim 1-3$~TeV is particularly relevant for LHC searches, as most coloured BSM particles are now excluded below $\sim 2$~TeV in simplified scenarios. Colour-octet scalars are a notable exception with current constraints reaching only up to about $1.3$~TeV~\cite{Darme:2018dvz, ATLAS:2024jja}. As we will show, this limit can be significantly improved by efficiently tagging the boosted top quarks arising from heavy scalar decays.

\subsection{Event generation tool chain}
\label{subsec:FourTopEvtGen}

\begin{figure}[p]
	\centering
	\includegraphics[width=0.35\textwidth]{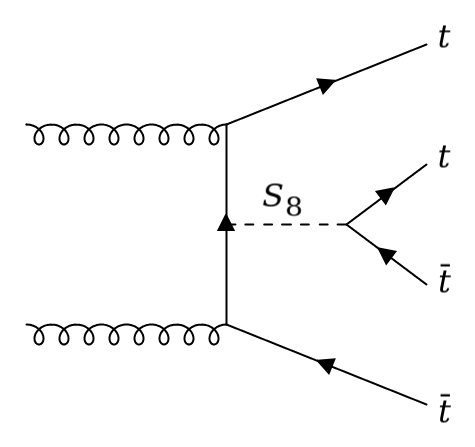}
	\includegraphics[width=0.35\textwidth]{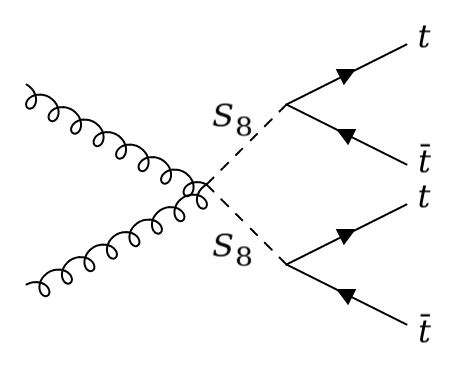}
	\includegraphics[width=0.35\textwidth]{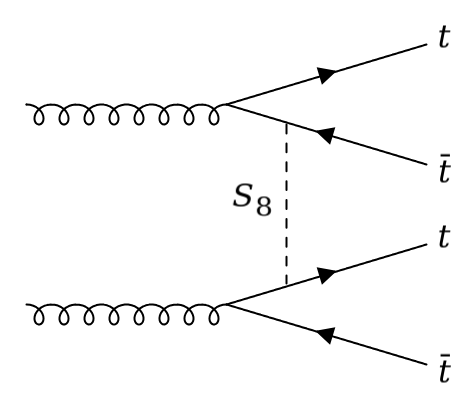} \includegraphics[width=0.35\textwidth]{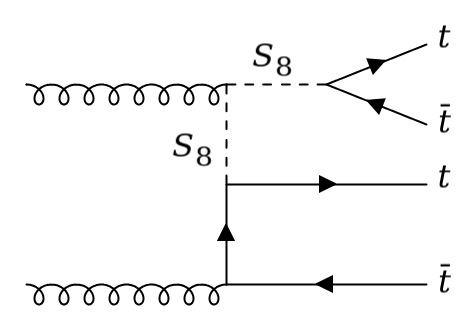}
	\includegraphics[width=0.35\textwidth]{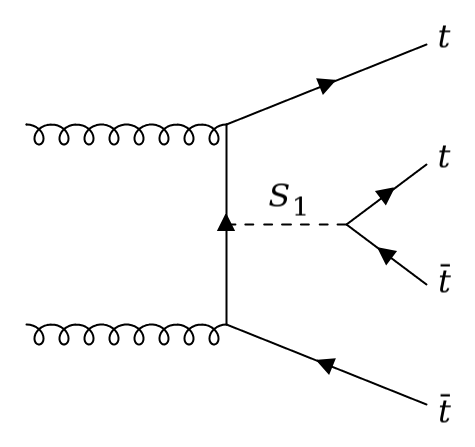}
	\includegraphics[width=0.35\textwidth]{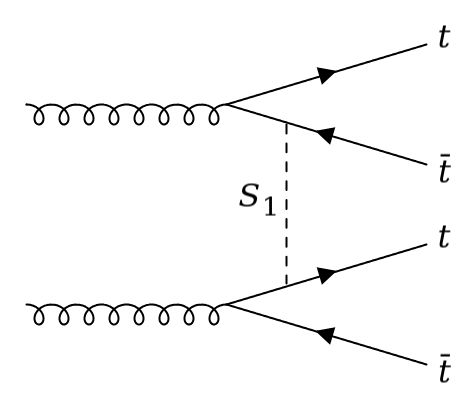}
	\caption{Representative Feynman diagrams contributing to four-top production in a simplified model with an additional scalar octet field $\oct$ or scalar singlet field $\sing$. When the scalar particles are on-shell, the associated production of the scalar octet with a top-antitop pair (top-left) is proportional to the square of the Yukawa coupling from the second Lagrangian of Eq.~\eqref{eq:lags}, while the QCD-driven pair production of two scalar octets (top-right) depends only on the octet mass $M_{\oct}$ after assuming that $\mathrm{Br}(\oct \to t\bar{t}) = 1$. Non-resonant diagrams of the scalar octet are presented in the middle row with the t-channel process (middle-left) which generate a dependence on $y_{\oct}^4$ in four-top production and another diagram contributing to the associated production that involves both a resonant octet and a non-resonant one (middle-right). An on-shell singlet state can only be produced in associated production (bottom-left) with a dependence on $y_{\sing}^2$ and off-shell contributions come from the t-channel process (bottom-right) depending on $y_{\sing}^4$. See Appendix~\ref{sec:appendixFourTopSignal} for a visual representation of the coupling dependence of the cross-section.
	}
	\label{fig:FeynDiagram}
\end{figure}

As illustrated by the representative Feynman diagrams in Figure~\ref{fig:FeynDiagram}, the set of BSM four-top production processes considered in this work is
\begin{equation}
\label{eq:processes}
pp \to t\bar{t}t\bar{t} \supset
\begin{cases}
    & pp \to t\bar{t} X \to t\bar{t}t\bar{t} \quad \text{with } X = \sing \text{ or } \oct, \\
    & pp \to \oct \oct \to t\bar{t}t\bar{t}.
\end{cases}
\end{equation}
In the following calculations, we neglect all electroweak-induced amplitudes involving, for instance, Higgs or $W/Z$ boson exchange as they are expected to be subdominant relative to the new physics or QCD contributions\footnote{See Appendix~\ref{sec:appendixFourTopSignal} contains a discussion on the definition of the signal by demonstrating the negligible contributions of electroweak interactions, the importance of the interference with the SM over the parameter space and the different contributions of the diagrams presented on Figure~\ref{fig:FeynDiagram}.}. Moreover, due to the high multiplicity of coloured final state particles, NLO QCD corrections are anticipated to be significant and are therefore included in the modelling of the BSM signal for all channels. This section details the simulation pipeline used to generate the corresponding BSM four-top signal event samples at NLO accuracy in QCD.

%%%%%%%%%%%%%%%%%%%%%%%%%%%%%%%%
%%%%%%%%%%%%%%%%%%%%%%%%%%%%%%%%

As a first step, we implement the simplified models defined by the Lagrangians of Eq.~\eqref{eq:lags} in the UFO format~\cite{Degrande:2011ua, Darme:2023jdn}. The model implementation is carried out with \fr~\cite{Christensen:2009jx, Alloul:2013bka}, starting from a reduced version of the SM Lagrangian containing only terms involving QCD-charged fields, the top Yukawa interaction, and additionally assuming a unit CKM matrix and five active quark flavours. The Lagrangian is next extended with the top-philic scalar interactions of Eq.~\eqref{eq:lags}, and we generate two UFO models, one for each scalar representation ($\sing$ and $\oct$)\footnote{For consistency with the modified version of \MG~\cite{Alwall:2014hca} described in Appendix~\ref{sec:appendixFourTopNLO}, the new particles are assigned PDG code 9000001, although this choice can be adjusted if needed provided that the required changes in \MG\ are modified accordingly.}.

\subsection{Full renormalisation of the simplified models}
\label{subsec:FourTopReno}

We use the {\sc MoGRe} package (introduced in Appendix~B of Ref.~\cite{Frixione:2019fxg}) to renormalise the Lagrangian of the simplified models considered in Eq.~\eqref{eq:lags}. {\sc MoGRe} automatically introduces the renormalisation constants associated with all fields and external parameters. It then derives those related to the internal parameters of the model, following the conventions of \fr\ by truncating their dependence on other renormalised quantities at one-loop order. The code also generates counterterms for all interactions present in the Lagrangian using the full set of introduced renormalisation constants. During this process, the user retains full control over which physical quantities are renormalised and may define custom renormalisation conditions to implement a specific scheme. This last feature was however not used in this work.

For our models, the quarks, the gluon and the BSM resonance are renormalised in the on-shell scheme. Among these, only the top quark and the BSM resonance are massive and thus require both wave-function and mass renormalisation constants instead of only a wave-function renormalisation constant. Among external parameters, we renormalise the two BSM couplings to top quarks $y_{\sing}$ and $y_{\oct}$ in the $\overline{\mathrm{MS}}$ scheme. In the case of the strong coupling constant $\alpha_s$, the five massless quark contributions are renormalised in the $\overline{MS}$ scheme too but we subtract, at zero momentum transfer, all massive-particle contributions to $\alpha_s$. The counterterms and their analytic expressions are listed in Appendices~\ref{sec:appendixOctetCTs} and \ref{sec:appendixSingletCTs} for the octet and the singlet simplified models respectively.

By construction, electroweak bosons and the SM Higgs are unaffected by this renormalisation procedure as their interactions are absent from the SM sector of the Lagrangian. This choice enables a fully consistent NLO QCD computation of BSM four-top production, and is justified by previous findings~\cite{Darme:2021gtt} showing that interferences of new physics and electroweak amplitudes become phenomenologically relevant only for very heavy resonances and/or in the non-perturbative regime of the model, where an effective field theory description would be more appropriate than a resonance-based one. We emphasise that the generation of a suitable UFO library requires the \lstinline{dev-bsm} version of \fr,\footnote{See \url{https://github.com/FeynRules/FeynRules/tree/feynrules-dev-bsm}.} as the \lstinline{current} version does not support our renormalisation procedure.

To generate the one-loop counterterms and $R_2$ rational terms required for NLO calculations in four dimensions, we make use of \nloct~\cite{Degrande:2014vpa} which relies on dimensional regularisation and {\sc FeynArts}~\cite{Hahn:2000kx} . Since the BSM scalars always decay into top pairs, we assume throughout that the scalar mass satisfies $M_X > 2 m_t$ with $X=\sing$ or $\oct$. Moreover, in order to consistently describe unstable particles with potentially large widths in our simulation chain, we adopt the complex mass scheme~\cite{Denner:1999gp,Denner:2005fg} and produce UFO models accordingly, ensuring that gauge invariance is preserved in all undertaken calculations.

For interested readers, both the full QCD+BSM renormalisation procedure detailed in this section and the more conventional QCD-only renormalisation described in the following Section~\ref{subsec:FourTopRenoQCD} have been implemented in a {\sc Mathematica} notebook publicly available on Zenodo~\cite{darme_2025_15783920}.

\subsection{Event Generation by \MG}
\label{subsec:EventGen}

Hard-scattering four-top events are then generated at NLO in QCD using \MG, with virtual corrections handled via the \textsc{MadLoop} routines~\cite{Hirschi:2011pa}. However, the latter discard by default all loop diagrams involving the scalar singlet as it is not coloured. This is problematic since we renormalised both the QCD and BSM sectors of the model so that the set of generated loop diagrams should include contributions involving the $\sing$ resonance. We consequently modify the default \MG\ behaviour following the procedure of Refs.~\cite{Borschensky:2020hot, Borschensky:2021hbo}, also detailed in the following Section~\ref{subsec:FourTopRenoQCD} and Appendix~\ref{sec:appendixCodeModifs} for the models considered in this study. This leads to a consistent cancellation of both IR and UV poles across the virtual, counterterms and real-emission contributions. In contrast, for the scalar octet case no modification to \MG\ is required as the particle is colour-charged and automatically included in the generated loop diagrams. For readers interested in using standard UFO models built with the default QCD renormalisation provided by \nloct\ and \fr, we outline in the following Section~\ref{subsec:FourTopRenoQCD} why this approach, which we do not use in this work, generally fails for BSM processes involving off-shell particles coupling to quarks. We also describe the necessary modifications to achieve consistent IR and UV divergence cancellation in this case. We have explicitly verified that both approaches (the full QCD+BSM renormalisation and the minimal QCD-only one) lead to identical NLO cross sections across our benchmark points. This reflects the fact that the additional diagrams included in the extended QCD+BSM renormalisation approach correspond to subleading corrections.

In addition, at large scalar octet masses and couplings, the numerical reduction of loop integrals becomes increasingly delicate. In this regime, we observe occasional failures of the pole cancellation checks at specific phase-space points. We traced these instabilities to the default usage of the {\sc Collier} library~\cite{Denner:2016kdg} in {\sc MadLoop}. To overcome this, we switch to alternative reduction tools such as {\sc Ninja}~\cite{Peraro:2014cba} and {\sc CutTools}~\cite{Ossola:2007ax}.

In all simulations relevant for our study, we use the NNPDF2.3NLO set~\cite{Ball:2012cx, Ball:2013hta} of parton distribution functions (PDF) and estimate theoretical uncertainties via variations in the renormalisation and factorisation scales and PDF replicas. The central scale is fixed to half the total hadronic activity in the event $H_{T}/2$, and we apply a minimum transverse momentum cut of $p_{T}>10$~GeV on all parton-level jets. Once parton-level events are generated, we model the inclusive decays of the top quarks using \mspin~\cite{Artoisenet:2012st} and \MW~\cite{Alwall:2014bza}, and the resulting decayed events are then matched to parton showers and hadronisation as implemented in \pyt \cite{Bierlich:2022pfr} using the default parameters.\footnote{\pyt~ is also used for the handling of inclusive tau-lepton decays.} While this final step has a very limited impact on the BSM signal due to the high transverse momentum of the tops produced in the BSM particle decays, it turns out to be essential for accurately modelling the dominant $t\bar{t}+$jets background that requires the extra modifications discussed in Section~\ref{sec:bkd}. In total, we generate fully showered and hadronised NLO signal event samples comprising 300,000 events for each benchmark point considered. In addition, we prepare a sparse grid of 18 intermediate points with 10k events each to enable interpolation and limit setting following the procedures described in Sections~\ref{sec:OctetAnalysis} and \ref{sec:SingletAnalysis}. All model files and Monte Carlo configuration cards are publicly available on Zenodo ~\cite{darme_2025_15783920}.

\subsection{QCD-only renormalisation of the simplified models}
\label{subsec:FourTopRenoQCD}

If one wishes to use UFO models generated with the standard QCD-only renormalisation procedure implemented in \fr\ and \nloct, the main challenges arise during the process generation step and the construction of the relevant one-loop diagrams in \MG. Ensuring a consistent automated NLO calculation that properly handles both UV and IR divergences indeed requires specific care.

\begin{figure}[t]
	\centering
	\subfloat[]{\includegraphics[width=0.38\linewidth]{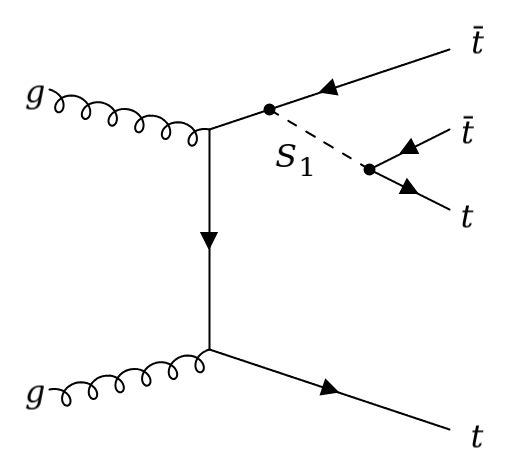}}%\hfill
	\subfloat[]{\includegraphics[width=0.38\linewidth]{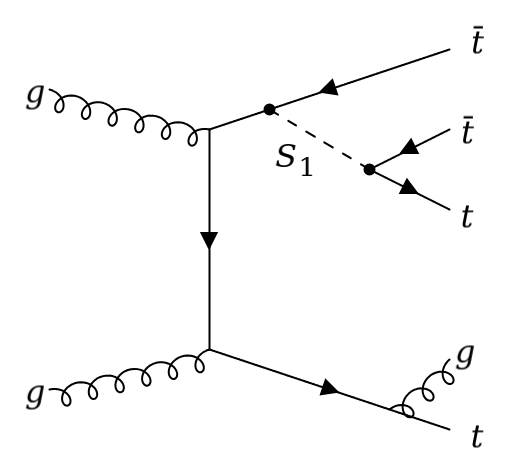}}\hfill
	\\[\smallskipamount]
	\subfloat[]{\includegraphics[width=0.38\linewidth]{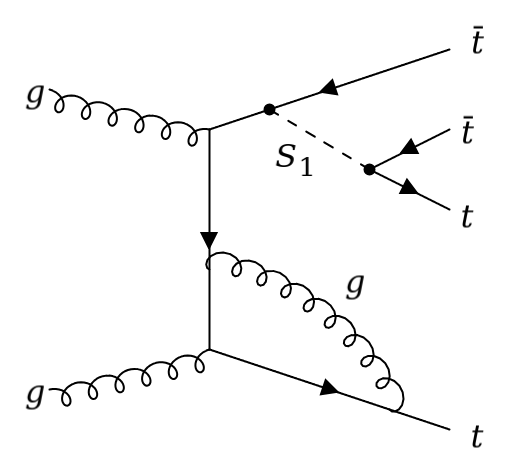}}
	\subfloat[]{\includegraphics[width=0.38\linewidth]{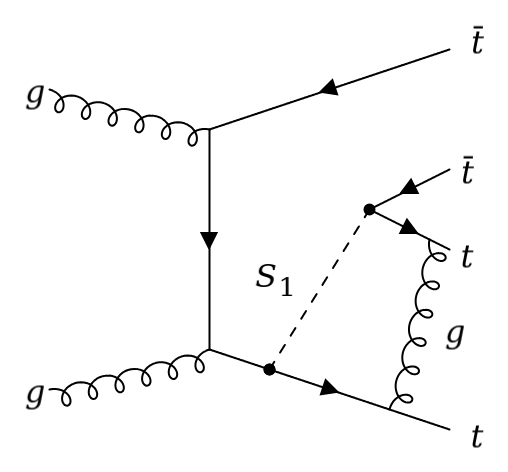}}\hfill
	%\subfloat[]{\includegraphics[width=0.3\linewidth]{figures/feynman_singlet_tttt_pentagon.png}}\hfill
	%\subfloat[]{\includegraphics[width=0.3\linewidth]{figures/feynman_singlet_tttt_hexagon.png}}
	\caption{Examples of Feynman diagrams contributing to the associated production process $pp\to t\bar{t}S_1$ at tree level (a) and with soft gluon radiation from a top quark (b). Relevant one-loop diagrams arise by inserting a gluon between two top quark lines. If the scalar singlet is absent from the loop, the resulting diagram is the usual QCD triangle involving only the $t\bar{t}g$ vertex (c) When a scalar propagator is instead attached to the same top quark lines as the gluon, a box diagram is obtained (d). 
	}    
	\label{fig:singletloops}
\end{figure}

In the scalar singlet model, the key subtlety concerns the treatment of IR divergences. While we could compute the process $pp \to t\bar{t}S_1$ at NLO without issue, our study focuses on the full $pp \to t\bar{t}t\bar{t}$ process where the scalar $S_1$ may appear through intermediate off-shell exchanges. In this case, the emission of a soft gluon from a top quark line in the tree-level amplitude introduces an IR divergence (see Figure~\ref{fig:singletloops}(a,b)). This divergence is cancelled by one-loop diagrams of two types. The first consists of QCD triangle diagrams with gluon exchange between top quarks (Figure~\ref{fig:singletloops}(c)), which are correctly generated along with the associated counterterms by \MG\ and \nloct. The second involves box, pentagon and hexagon diagrams with internal scalar singlet propagators (he box diagram is displayed on Figure~\ref{fig:singletloops}(d) and the other two are constructed similarly by moving the emission point of the singlet along the top quark line) which are not generated by default by \MG\ since the singlet is uncharged under QCD, and which have been consistently ignored by \nloct\ (especially as they do not yield any UV divergence). However, their inclusion is necessary to cancel IR divergences. These diagrams must therefore be manually included.

\begin{figure}[t]
	\centering
	\subfloat[]{\includegraphics[width=0.48\linewidth]{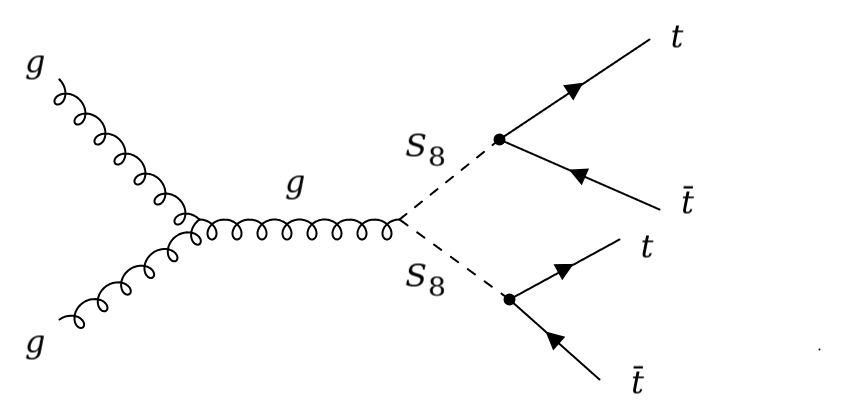}}\hfill
	\subfloat[]{\includegraphics[width=0.48\linewidth]{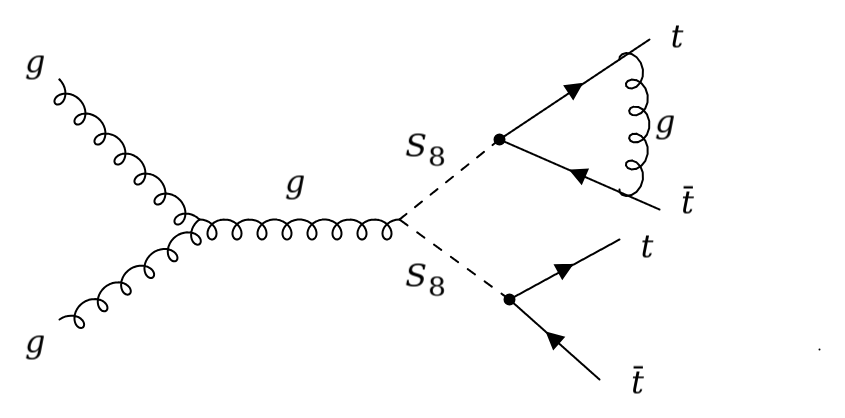}}\\[\smallskipamount]
	\subfloat[]{\includegraphics[width=0.48\linewidth]{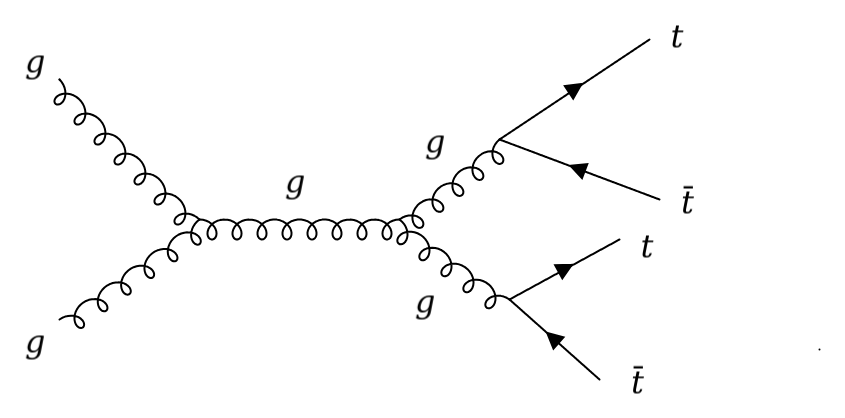}}\hfill
	\subfloat[]{\includegraphics[width=0.48\linewidth]{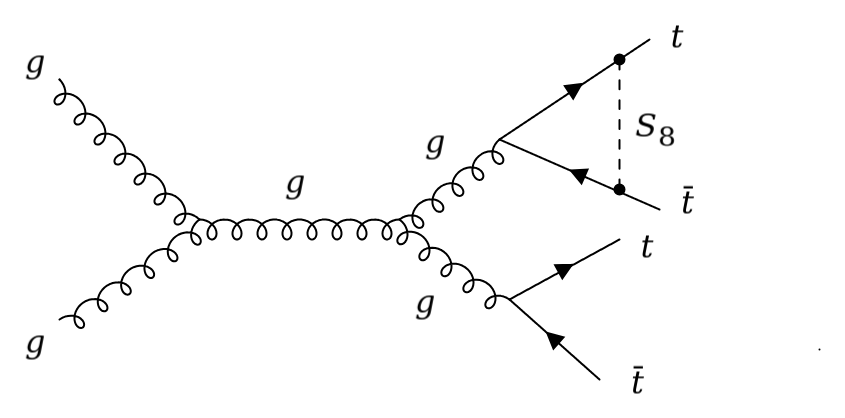}}
	\caption{Examples of Feynman diagrams for scalar octet pair production via gluon fusion at tree level (a), and with a QCD virtual gluon exchange between two top quarks (b). We additionally show a diagram for the pure SM QCD production of four top quarks (c), and with a BSM virtual exchange of a scalar octet between two top quarks (d).} 
	\label{fig:trianglettG}
\end{figure}

In contrast, the scalar octet model presents a challenge in the treatment of UV divergences. For instance, let us focus on a triangle diagram such as the one shown in Figure~\ref{fig:trianglettG}(b), which features a gluon exchange between two top quark lines and consists in a QCD correction to the tree-level diagram in Figure~\ref{fig:trianglettG}(a). At the same order, \MG\ also generates one-loop diagrams like the one in Figure~\ref{fig:trianglettG}(d), which involves a scalar octet exchange between top quarks and that corresponds to a BSM correction to the pure QCD topology of Figure~\ref{fig:trianglettG}(c). Such a diagram contains BSM vertices that have not been renormalised by \nloct, which indeed only derived the QCD counterterms. This mismatch results in uncancelled UV poles and imposes that such triangle diagrams should be removed.

We stress that the inclusion or removal of loop diagrams involving BSM resonances does not guarantee a consistent calculation. Our goal here is solely to highlight how to resolve pole cancellation issues that arise when attempting a QCD-only renormalisation of a model featuring BSM scalars. The responsibility of selecting a consistent set of diagrams lies with the user. As mentioned above, our main results do not rely on the QCD-only renormalisation procedure detailed in this section, but instead on the full renormalisation of both QCD and BSM sectors described in Section~\ref{subsec:FourTopReno}.

%%%%%%%%%%%%%%%%%%%%%%%%%%%%%%%%

\subsection{Signal NLO cross sections and K-factors}
\label{sec:Kfact}

Running large event samples at NLO accuracy demands significant computational resources. We therefore restrict full NLO computations to the benchmark points introduced in Section~\ref{subsec:FourTopLags} and validate that the corresponding NLO signal distributions remain sufficiently close to their LO counterparts. This enables us to rely on LO simulations to derive constraints while correcting the total rate using $K$-factors. The relevant kinematic distributions, the shape of the invariant mass $M_{t\overline{t},1}$ discussed in Sections~\ref{sec:OctetAnalysis} and \ref{sec:SingletAnalysis}, confirm that this approximation is justified: LO shapes provide a reliable estimate of the signal features while the NLO effects after the selection are largely captured by a global normalisation factor. In this approach, the approximate number of selected BSM events at NLO accuracy ($N_{\rm K-NLO}$) is given by
\begin{align}\label{eq:multscheme}
  N_{\rm K-NLO} \equiv \mathcal{L} \, \varepsilon_{\rm LO} \, \left( \frac{ \sigma_{\rm NLO} }{\sigma_{\rm LO} } \right) \, \sigma_{\rm LO} \equiv \mathcal{L} \, \varepsilon_{\rm LO} \, K \, \sigma_{\rm LO} \ ,
\end{align}
where $\mathcal{L}$ denotes the integrated luminosity, $\varepsilon_{\rm LO}$ is the selection efficiency evaluated using LO simulations and $K$ is defined as the ratio of the NLO and LO cross sections $\sigma_{\rm NLO}$ and $\sigma_{\rm LO}$ when they are computed using renormalisation and factorisation scales set to the mass of the BSM resonance. The strength of this method lies in its computational efficiency: both $\varepsilon_{\rm LO}$ and $K$ can be determined across a sparse grid in the mass/coupling parameter space at a fraction of the cost required for full NLO event generation, and then fitted linearly. The LO cross section $\sigma_{\rm LO}$ is finally evaluated on a finer grid since it can be obtained with reduced computational resources. Moreover, this approach is motivated by the observation that both the efficiencies and the $K$-factors exhibit slow variation in terms of the model's parameters compared to the total production cross section.

\begin{figure}
    \centering
    \includegraphics[width=\linewidth]{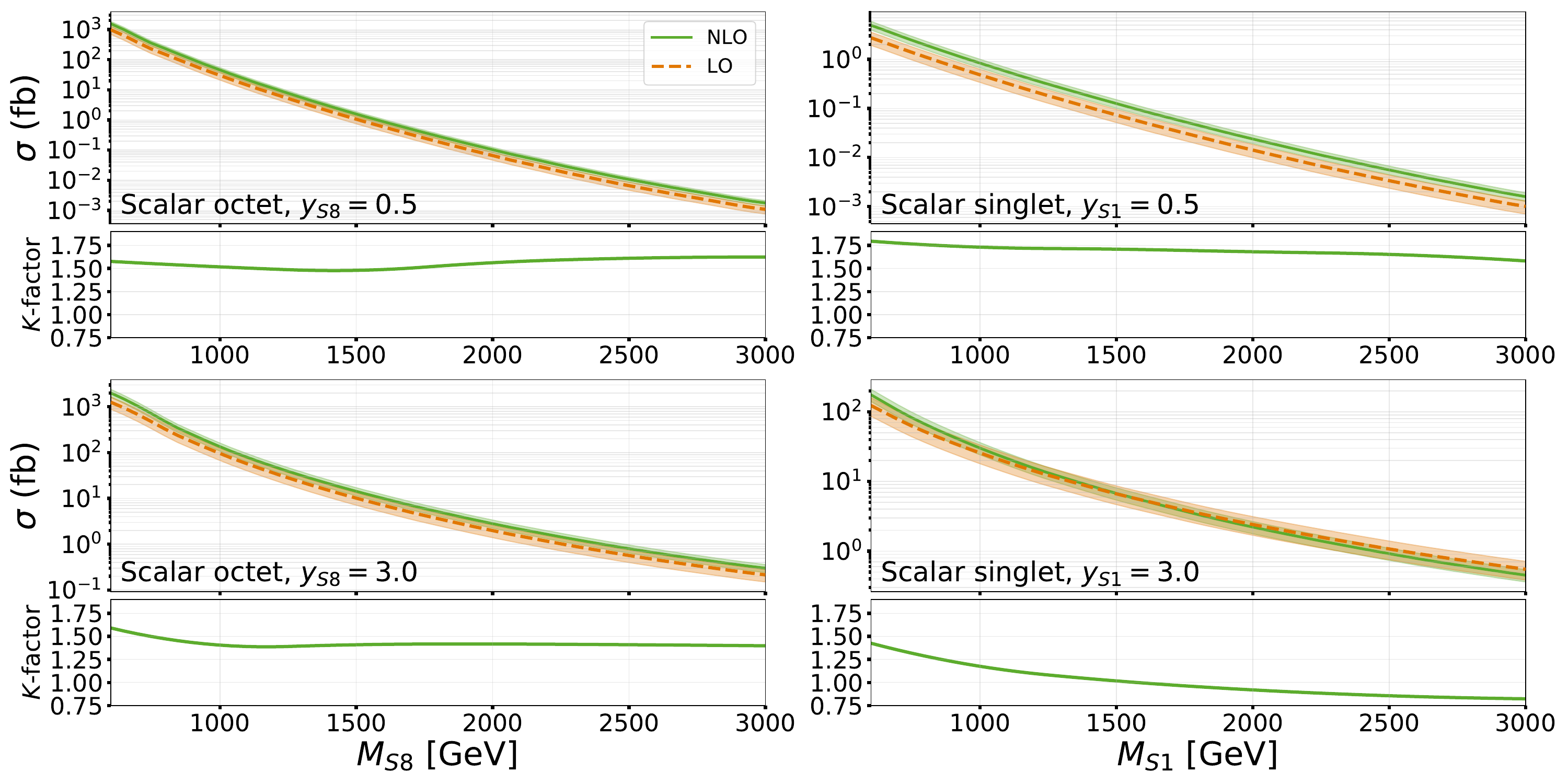}
    \caption{Comparison of LO and NLO pure BSM cross sections, that is associated to the relevant Feynman diagrams from Figure~\ref{fig:FeynDiagram}, as functions of the BSM resonance mass. Results are shown for the scalar octet ({\it left}) and singlet ({\it right}) cases, assuming couplings to top quarks of 0.5 ({\it top}) and 3 ({\it bottom}) . The lower panels of the figures display the associated $K$-factors defined as the ratio of the NLO cross section to the LO prediction computed with NLO PDFs.}
    \label{fig:CS}
\end{figure}

Figure~\ref{fig:CS} shows the four-top production cross sections induced by the inclusion in the field content of the theory of a scalar octet (left) and a scalar singlet (right), for two specific values of the corresponding Yukawa coupling to the top quark that we choose equal to 0.5 (top row of the figure; weak coupling) and 3 (bottom row of the figure; strong coupling). The shaded bands indicate theoretical uncertainties estimated using the standard seven-point scale variation procedure: the renormalisation and factorisation scales are independently varied by factors of two around the central scale, excluding the extreme combinations where one scale is multiplied by 0.5 and the other by 2. This conventional definition of scale uncertainties is further adopted throughout the rest of this work. In agreement with the results of~\cite{Darme:2021gtt}, we find that QCD-induced scalar octet pair production, thus independent of the top-quark coupling to the new resonance as illustrated with the representative diagram shown in Figure~\ref{fig:FeynDiagram}, dominates the total cross section up to resonance masses of about 2~TeV. Beyond this point, phase-space suppression becomes significant and the less-suppressed contribution from associated production overtakes, before eventually dominating at higher masses. For the scalar singlet where only associated production contributes, the total cross section therefore falls more slowly with increasing mass compared to the octet case. Finally, across the entire region of the parameter space experimentally accessible at the LHC, the $K$-factors, that we show in the lower panels of the different figures, typically lie in the range $1.4-1.8$. These moderately large values show that NLO corrections can play a critical role in improving the reliability of the total rate prediction. Altogether, after accounting for the modest effect on the shapes of the signal differential distributions (see Sections~\ref{sec:octetana} and \ref{sec:singletana}), our hybrid approach using LO shapes with NLO-corrected normalisation is found to offer an optimal balance between accuracy and computational cost for the parameter scans required in our study.

\section{Characterisation of the boosted four-top system} 
\label{sec:FourTopReco}

\subsection{Simulation of the detector response and object definitions}
\label{sec:objects}

We simulate the response of a typical LHC detector using the \sfs\ framework~\cite{Araz:2020lnp, Araz:2021akd} implemented within \ma~\cite{Conte:2012fm, Conte:2014zja, Conte:2018vmg}. Since the reconstruction of top quarks plays a central role in this study, we recalibrated the default ATLAS detector parametrisation to improve agreement with reference experimental studies~\cite{ATLAS:2018rvc, ATLAS:2019qmc, ATLAS:2020lks}, focusing in particular on the invariant mass reconstruction of top-quark candidates. Electrons and muons are reconstructed following the medium working-point performance described in Refs.~\cite{ATLAS:2019qmc, ATLAS:2020auj}, respectively. Additionally, two jet collections are defined, both using a reconstruction based on the anti-$k_T$ algorithm~\cite{Cacciari:2008gp} as implemented in \fj~\cite{Cacciari:2011ma}, but with two different radius parameters $R=0.4$ (AK4 jets) and $R=1.0$ (AK10 jets). For AK4 jets, $b$-tagging is applied probabilistically using the $p_T$- and $\eta$-dependent efficiencies of Ref.~\cite{ATLAS:2016gsw}. To avoid double-counting between the AK4 jet and lepton collections, we implement a series of overlap removal procedures following the prescription of Ref.~\cite{ATLAS:2020hpj}. AK4 jets are first removed if they are within $\Delta R < 0.2$ of a lepton, although in the case of a muon the jet must also have three or fewer associated tracks. Next, electrons and muons are respectively removed if they lie within $\Delta R < 0.4$ or $0.04 + 10\,\mathrm{GeV}/p_T$ of any of the remaining AK4 jets, while finally electrons within $\Delta R < 0.1$ of a muon are also discarded. These isolation criteria produce isolated electrons and muons.

For boosted top-quark identification, we apply different tagging strategies depending on the context. For the validation of our reconstruction procedure, we use the jet-mass and N-subjettiness classifier from Ref.~\cite{ATLAS:2015ddu}, while for our main four-top analysis, we rely on more advanced constituent-based top-tagging algorithms applicable to AK10 jets~\cite{ATLAS:2022qby}. In this last case, we specifically adopt the performance of the \lstinline{HlDNN} and \lstinline{ParticleNet} classifiers for jets satisfying $p_T > 350$~GeV, $|\eta| < 2.0$ and invariant mass $M_j > 40$~GeV. AK10 jets matched to a partonic top within $\Delta R < 0.75$ are top-tagged with an efficiency of 80\%, whereas jets not consistent with a top quark are mis-tagged at an average rate of either 10\% (conservative, \lstinline{HlDNN}-like) or 5\% (optimistic, \lstinline{ParticleNet}-like), the exact value depending on the jet $p_T$.

In our validation procedure, we have obtained excellent agreement with the expectations of an ATLAS search for resonant $t\bar{t}$ production in the semi-leptonic channel~\cite{ATLAS:2018rvc}, recovering the reconstructed resonance mass distribution, the signal selection efficiencies and the exclusion limits within 20\% for both the resolved and boosted signal regions of the ATLAS analysis. We have also recovered the fact that for $t\bar{t}$ resonance masses lying between 1 and 2~TeV, the production rate at the LHC is high enough that systematic uncertainties dominate. Subsequently, potential improvements in the higher-luminosity LHC runs are not foreseen. Moreover, as observed in Refs.~\cite{Carena:2016npr, Djouadi:2019cbm} and more recently in a CMS public note~\cite{CMS:2025rnx}, interference effects between the SM and BSM amplitudes for gluon-initiated $t\bar{t}$ production become significant above 1~TeV. These effects can create a dip rather than a peak in the top-antitop invariant mass distribution, complicating the statistical interpretation of the search. These two limitations provide an additional motivation for studying four-top final states as a probe of top-philic new physics in the high-mass regime. In particular, in this channel, the impact of interference is indeed negligible within the parameter space of interest, allowing a more robust interpretation of potential deviations from the SM prediction.

We now turn to detailing in the following subsection the analysis selection criteria imposed on the reconstructed objects introduced here.

\subsection{Reconstruction of a boosted four-top system} \label{sec:topreco}
\begin{figure}
    \centering
    \includegraphics[width=\linewidth]{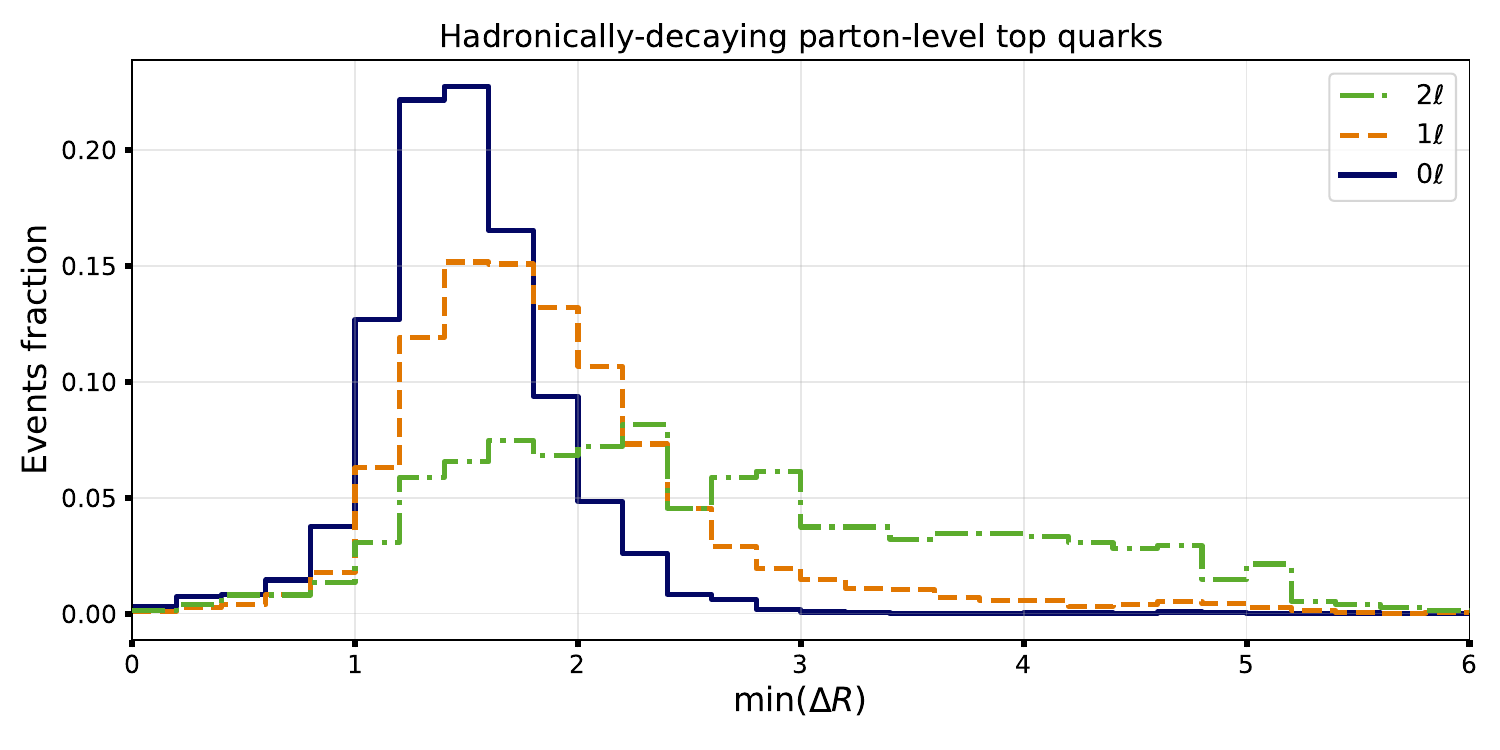}
    \caption{Minimum $\Delta R$ separation between (parton-level) top quarks, computed for events that pass at least one signal region selection after reconstruction. Events are categorised according to the number of leptons produced in the top quark decays.}
    \label{fig:minDR_tHadr}
\end{figure}

Since our main interest is the exploration of BSM resonances with masses above the TeV scale, the resulting top quarks are typically boosted enough to be accurately reconstructed and tagged using standard top-tagging algorithms such as those described in Section~\ref{sec:objects}. We have verified that for BSM resonances in the TeV range, thus viable relatively to current exclusion limits, our signal samples feature four top quarks that tend to decay in isolation from each other. As a result, the reconstruction procedure validated on BSM-induced top-antitop events remains reliable even in the case of a higher-multiplicity $t\bar{t}t\bar{t}$ final state. In particular, since AK10 jets clustered by \fj\ serve as proxies for hadronically-decaying top quarks, we have checked that the typical angular distance $\Delta R$ between them remains larger than 1, allowing thus for the independent reconstruction of each top jet. As shown in Figure~\ref{fig:minDR_tHadr}, fewer than 5\% of the selected events contain two hadronic top quarks separated by $\Delta R < 1$ at parton level.

Based on this observation, we implement a set of selection criteria starting from the classification of events according to the number of leptons and AK10 jets that they contain. Leptons must satisfy $p_T > 20$~GeV and $|\eta| < 2.5$ to be considered as potential decay products of leptonically-decaying top quarks. In addition, all reconstructed AK10 jets are ordered by decreasing $p_T$ while giving priority to those that are top-tagged. However, any AK10 jet overlapping with an isolated lepton within $\Delta R < 1$ is discarded. $b$-jets included in the reconstruction are taken to be AK4 jets with $p_T > 25$~GeV and $|\eta| < 2.4$ that pass our $b$-tagging requirements. We refer to these $b$-jets as `isolated' when they do not overlap with any of the selected AK10 jets within $\Delta R < 1$ so that they will be allowed to serve as ingredients for the reconstruction of leptonically-decaying top quarks. Non-$b$-tagged AK4 jets are retained only if they have $p_T > 40$~GeV and $|\eta| < 2.4$. The missing transverse momentum $p_T^\text{miss}$ is calculated as the negative vector sum of the transverse momenta of all leptons and jets (both $b$-tagged and non-$b$-tagged) that pass the above cuts, thereby minimising contamination from neutrinos produced in the parton shower~\cite{Darme:2024epi}. At this stage, light jets are thus only used in the $p_T^\text{miss}$ calculation.

The subsequent reconstruction procedure depends on the number of isolated leptons present. If no such lepton is found, the event is considered fully hadronic. For the analysis targeting the colour-octet case, at least four AK10 jets are required without imposing top-tagging requirements. For the colour singlet case, we instead require at least two AK10 jets along with a minimum of two isolated $b$-jets in order to suppress background.

When exactly one isolated lepton is present, the event is interpreted as containing a single leptonically-decaying top quark and three hadronically-decaying ones. In the colour octet case, at least three AK10 jets are required to represent the three hadronic tops, along with at least one isolated $b$-tagged AK4 jet. In the singlet case, we instead again require at least two AK10 jets and at least two isolated $b$-jets. In both scenarios, the isolated $b$-jet closest in $\Delta R$ to the lepton is selected for reconstructing the leptonic top. In addition, we require $p_T^\text{miss} > 25$~GeV in consistency with the presence of a neutrino. The longitudinal component of this neutrino is then estimated by assuming that it is produced together with the lepton in the decay of an on-shell $W$ boson. Up to two real solutions may be obtained from the kinematic fit resulting from this assumption, and we select the one that yields a reconstructed top quark mass closest to the expected value when the $W$ boson and $b$-jet momenta are combined. If no real solution is available, we retain the real part of the complex solutions to proceed.

In events with two isolated leptons, we require them to have the same electric charge to suppress the dominant $t\bar{t}+$jets background. We further demand two AK10 jets to account for the two hadronically-decaying tops and two isolated $b$-jets to reconstruct the leptonically-decaying ones. If more than two such $b$-jets are available, the ones closest in $\Delta R$ to the leptons are used. Furthermore, we require $p_T^\text{miss} > 50$~GeV and estimate the transverse momenta of the two neutrinos under the assumption that the event originates from the decay of heavy parent particles, which tends to favour configurations featuring a low transverse mass when combining the leptons and the neutrinos~\cite{Lester:1999tx}. Letting $(\ell_1, \nu_1)$ and $(\ell_2, \nu_2)$ denote the lepton-neutrino pairs, we define the transverse mass of each $W$ boson as
\begin{equation}
    M_T^{(i)} = \sqrt{2 |\vec{p}_T^{\,\ell_i}| |\vec{p}_T^{\,\nu_i}| \, (1-\cos\Delta \phi_{\ell\nu})},
\end{equation}
where $i = \{1, 2\}$ and $\Delta \phi_{\ell\nu}$ is the azimuthal angle between the lepton and the neutrino. The unknown neutrino transverse components are estimated by using the so-called stransverse mass $M_{T2}$~\cite{Lester:1999tx, Cheng:2008hk} that is defined through the minimisation condition
\begin{equation}
    M_{T2} = \min_{\vec{p}_T^{\,\nu_1} + \vec{p}_T^{\,\nu_2} = \vec{p}_T^\text{miss}} \max\Big(M_T^{(1)}, M_T^{(2)}\Big).
\end{equation}
The longitudinal components of the two neutrinos are then inferred from kinematic fits assuming an on-shell $W$ boson decay. Among all solutions obtained for each leptonic top quark, we select the one that leads to a reconstructed top mass closest to the physical value, using the $b$-jet closest in $\Delta R$ to the associated lepton. As before, if no real root exists, we use the real part of the complex solution.

\begin{table}[t]
  \setlength\tabcolsep{10pt}\renewcommand{\arraystretch}{1.2}
  \resizebox{0.98\textwidth}{!}{
  \begin{tabular}{cccccc} 
  \hline
    \multicolumn{6}{l}{\textbf{\rule{0pt}{1.25em}Basic kinematic requirements}} \\[0.75em]
      & Electrons & Muons & AK4 Jets& AK10 Jets& $b$-jets \\
      $p_T$ (GeV) & $>20$ &  $>20$ &  $>20$ &$>350$ &  $>25$ \\
      $|\eta|$ & $<2.47$ &  $<2.5$ &  $<2.5$ & $<2.0$ & $<2.4$ \\[0.75em]
   \hline
    \multicolumn{6}{l}{\textbf{\rule{0pt}{1.25em}Object definitions}} \\[0.75em]
    Non $b$-tagged AK4 jets & \multicolumn{5}{l}{$p_T > 40 $ GeV, $|\eta| \le 2.4$} \\
    Isolated leptons & \multicolumn{5}{l}{$p_T \ge 20$ GeV, $|\eta| \le 2.5$ + isolation criteria}\\
    Isolated AK4 jets & \multicolumn{5}{l}{ $\Delta R > 1$ from any AK10 jet} \\
    Missing energy & \multicolumn{5}{l}{$p_T^{\textrm{miss}} > 25$ GeV (one isolated lepton) or 50 GeV (two isolated leptons)}\\[0.75em]
  \hline
    \multicolumn{6}{l}{\textbf{\rule{0pt}{1.25em}Boosted four-top selection for colour-octet resonances}}\\[0.75em]
    No isolated lepton & \multicolumn{5}{l}{$\ge4$ AK10 jets} \\
    One isolated lepton & \multicolumn{5}{l}{$\ge3$ AK10 jets, $\ge1$  isolated $b$-tagged AK4 jet} \\
    Two isolated leptons & \multicolumn{5}{l}{$\ge2$ AK10 jets, $\ge2$ isolated $b$-tagged AK4 jets} \\[0.75em]
    
    \multicolumn{6}{l}{\textbf{\rule{0pt}{1.25em}Boosted four-top selection for colour-singlet resonances}}\\[0.75em]
    $\leq$ Two leptons & \multicolumn{5}{l}{$\ge2$ AK10 jets, $\ge2$ isolated $b$-tagged AK4 jets} \\[0.75em]
  \hline
  \end{tabular}
  }
  \caption{Summary of the preselection cuts. See main text for details.  }
  \label{tab:preselection}
\end{table}

At the end of this procedure that we summarise in Table~\ref{tab:preselection}, each event is associated with a set of reconstructed top quark candidates classified according to their decay modes (leptonic or hadronic) and top-tagging status. These are thus ready to be used in further analysis steps that we will describe in Section~\ref{sec:bkd}.

\section{Background description}
\label{sec:bkd}

Our signal selection strategy is primarily based on the reconstruction and the identification of boosted top quarks. This contrasts with existing experimental four-top searches, including the most recent analyses by ATLAS~\cite{ATLAS:2024jja} and CMS~\cite{CMS:2023ftu}, which do not rely on this feature. As a result, the relevant background composition differs significantly and requires a dedicated reassessment as compared to \cite{CMS:2023ftu, ATLAS:2024jja}.

\begin{table}
	\centering%\renewcommand{\arraystretch}{0.75}\setlength\tabcolsep{14pt}
	\begin{tabular}{c|ccc|ccc}
		Process & $\sigma$(LO) & Scale & PDF & $\sigma$(NLO) & Scale & PDF \\ 
		\hline
		$t \bar{t} j j $ & 354 &  $^{+ 62\%}_{- 35\%}$ & $\pm~ 5.8\%$ &
		352 & $^{+ 3.7\%}_{- 13\%}$ & $\pm~ 2.6\%$ \\   
		$t \bar{t} W $ & 0.376 & $^{+23\%}_{-17\%}$ & $\pm~3.9\%$ & 
		0.565 & $^{+ 8.3\%}_{- 8.3\%}$ & $\pm~ 1.8\%$ \\
		$t \bar{t} W j $ & 0.329 & $^{+39\%}_{-26\%}$ & $\pm~2.1\%$ & 
		0.452  & $^{+ 8.1\%}_{- 12\%}$ & $\pm~ 1.2\%$ \\
		$t \bar{t} Z $ & 0.563 &  $^{+ 31\%}_{- 22\%}$ & $\pm~4.8\%$ &
		0.756 & $^{+ 9.2\%}_{- 11\%}$ & $\pm~ 2.1\%$ \\
		$t \bar{t} Z j $ & 0.639 & $^{+ 47\%}_{- 30\%}$ & $\pm~ 6.5\%$ & 
		0.672 & $^{+ 2.6\%}_{- 9\%}$ & $\pm~ 2.5\%$ \\
		% $t \bar{t} Z j $ & 0.639 & $^{+ 47\%}_{- 30\%}$ & $\pm~ 6.5\%$ & 
		%     0.672 & $^{+ 2.6\%}_{- 9\%}$ & $\pm~ 2.5\%$ \\
		$t \bar{t} t \bar{t}$ & 0.00612 & $^{+ 65\%}_{- 37\%}$ & $\pm~ 13\%$ & 
		0.00920 & $^{+ 28\%}_{- 24\%}$ & $\pm~ 6.0\%$ \\
		
		$t \bar{t} t + t\bar{t}\bar{t}$ & 0.00155  & $^{+ 22\%}_{-17\%}$ & $\pm~ 13\%$ &
		0.00201 &  $^{+20 \%}_{-19 \%}$ &  $\pm~7.5 \%$ \\
	\end{tabular}
	\caption{LO and NLO cross sections (in pb) for the dominant SM background contributions relevant to our analysis, computed at LO and NLO in QCD for a centre-of-mass energy of 13~TeV. The predictions are obtained using the NNPDF2.3NLO parton density set~\cite{Ball:2012cx}, and include theory uncertainties from scale and PDF variations. The renormalisation and factorisation scales are centrally set to half the total hadronic transverse energy in the event ($H_T/2$), and a minimal transverse momentum of $p_T > 20$~GeV is required for each parton-level jet.\label{tab:Xsec}}
\end{table}

The dominant background in our study arises from the production of a top-antitop pair in association with additional jets and/or electroweak bosons (that are decayed inclusively in our simulation chain). In particular, the process $pp \to t \bar{t} jj$ where the extra jets can mimic boosted top quarks constitutes its main contribution. Subleading background components include the $pp \to t \bar{t} V$ and $pp \to t \bar{t} V j$ processes with $V = W, Z$ that we treat independently due to our analysis requirements. The latter indeed favour contributions where additional jets are highly energetic and could be mistagged as top quarks. As such jets are better modelled at the matrix-element level and not by parton showering, this allows us to consider two non-overlapping background samples for the $pp \to t \bar{t} V$ and $pp \to t \bar{t} Vj$ processes. SM four-top ($pp \to t \bar{t} t \bar{t}$) and three-top ($pp \to t \bar{t} t$, $t \bar{t} \bar{t}$) production have cross sections at least two orders of magnitude smaller. Despite their reduced impact, we include these processes for completeness. Conversely, we have verified that multijet and $t \bar{t} VV$ backgrounds become negligible after applying our selection and top-tagging procedure. Cross sections for the most relevant background processes, both at LO and NLO in QCD, are collected in Table~\ref{tab:Xsec} for a centre-of-mass energy $\sqrt{s} = 13$ TeV.

Background simulations are achieved with the toolchain introduced earlier but with a differing configuration for the colour-octet and colour-singlet analyses. For the octet case, we require at least three final-state parton-level objects (\ie\ prior to decay) with $p_T > 300$ GeV to enhance the chance of reconstructing four boosted top proxies, tagged or not. In the singlet case where only two top candidates are needed, we instead impose this condition on just two parton-level objects. All background samples are generated at LO due to computational constraints, as the strong background rejection stemming from our analysis selections makes the generation of sufficient statistics at NLO particularly costly and not necessary. However, we have explicitly verified that the NLO cross sections remain compatible with their LO counterparts (as listed in Table~\ref{tab:Xsec}), at least when the hard $p_T$ cuts are replaced by a minimal cut of 20~GeV.

For the dominant $t \bar{t}\ + $ jets background, we generate $t \bar{t}$ events in association with one, two or three jets and use the MLM~matching and merging procedure~\cite{Mangano:2006rw, Alwall:2008qv} to combine the resulting event samples. This is especially important given our reliance on large-radius AK10 jets and the necessity for such jets to have a high transverse momentum to pass the selection. Typically, we indeed require at least three parton-level objects to have $p_T > 300$ GeV, which leads to a partonic centre-of-mass energy above 1~TeV. Our analyses however also target the associated production of a new physics resonance with a top-antitop pair as well as channels involving leptonic top decays, where soft jets may be present. Without merging, the large scale separation between the high partonic centre-of-mass energy and the low $p_T$ threshold of 20~GeV for the subleading jets could thus lead to large logarithms and an apparent breakdown of perturbativity. In addition, such resulting high-energy events favour hard initial-state radiation, which increases the possibility that a light jet or radiation product mimics a boosted top and gets mistagged. To avoid double counting and ensure proper treatment of QCD emissions, we thus tune the simulation accordingly. In particular, we combine matrix elements with up to two and three additional jets in the singlet and octet cases respectively, while fixing the \lstinline{xqcut} parameter of \MG\ to 80~GeV and the \lstinline{Qcut} parameter of \pyt\ to 120~GeV to smoothly regulate the separation of the matrix-element and shower regimes while preserving the high-$p_T$ behaviour of the leading jets. We have checked explicitly that the final (fiducial) cross section is stable under moderate variations of these parameters. All corresponding cards are available in our Zenodo repository~\cite{darme_2025_15783920}.

Next, we generate a $t \bar{t} b \bar{b}$ background sample as associated events could pass the colour-singlet selection where we require two top-tagged jets plus two additional isolated $b$-jets. We found that this process contributes only modestly, at most 10\% (20\%) of the $t \bar{t}\ +$ jets background in the SR1 (SR2) signal region. Consequently, matching and merging are not applied, as for any other subleading background contributions, since this would yield a marginal impact while entailing a high computational cost. Lastly, we note that the $t \bar{t} W$ and $t \bar{t} W j$ background contributions are dominant for neither the SR1 nor the SR2 regions. However, we include the $t \bar{t} W j$ contribution in our analysis as the extra final-state jet may lead to a small number of events passing the SSL selection. In this case, we do not apply any matching and merging procedure again, so the overall background yield in the SSL analysis is likely conservatively overestimated.

\section{Pair production analysis}
\label{sec:OctetAnalysis}

As outlined in the previous sections, we develop two distinct analyses targeting the production of four top quarks via intermediate colour-octet and colour-singlet resonances. In the colour-octet scenario, pair production of resonances can dominate the cross section in certain regions of the parameter space, requiring the reconstruction of four objects that serve as proxies for the four top quarks produced in the resonance decays. In the colour-singlet case, only a single resonance is produced, allowing the selection to solely focus on two boosted top quark candidates, while the remaining two top quarks are expected to arise from QCD interactions and thus to feature different properties. Finally, in both scenarios, we also implement a same-sign dilepton selection strategy, which provides a complementary handle on the signal albeit with reduced efficiency due to the low branching ratio of this final state.

\subsection{Signal region definition and pairing strategy in the colour-octet model} 
\label{sec:octetana}

In the search strategy designed for the colour-octet simplified model, we aim to reconstruct two on-shell BSM particles of equal mass by pairing four identified top quark candidates. When an event contains four such objects, it is assigned to one or more overlapping signal regions based on the number of top-tagged AK10 jets.
\begin{itemize}%[topsep=2pt,itemsep=0pt,parsep=3pt,partopsep=2pt]
	\item The \textbf{SR1} region includes fully hadronic events with at least three top-tagged AK10 jets, as well as single-lepton events featuring at least two top-tagged AK10 jets. This region thus gathers events with at least three reconstructed and tagged (hadronic or leptonic) top quarks, without any top-tagging requirements on the fourth AK10 jet.
	\item The \textbf{SR2} region is defined by stricter conditions so that four top quarks are reconstructed and tagged. Fully hadronic events must thus contain at least four top-tagged AK10 jets, while one-leptonic events must feature at least three top-tagged AK10 jets.
	\item The \textbf{SSL} region encompasses events with two leptons of the same electric charge, benefiting hence from reduced background contamination so that no additional top-tagging of any AK10 jet is required.
\end{itemize}
A summary of these requirements is provided in Table~\ref{tab:octet_requirements}, which also highlights that SR1 and SR2 are not mutually exclusive: any events satisfying SR2 conditions automatically populate the SR1 region too.

\begin{table}%\renewcommand{\arraystretch}{1.3}\setlength\tabcolsep{14pt}
	\centering
	\begin{tabular}{c|cccc}
		\multicolumn{5}{c}{Colour-octet analysis} \\ \hline
		Signal Region & \# $\ell$& \# $b_\ell$ & \# AK10 & \# top-tag. \\ \hline
		\multirow{2}{*}{SR1} & 0 &  - & $\ge4$ & $\ge3$ \\
		& 1 &  $\ge1$ & $\ge3$ & $\ge2$ \\ \hline
		\multirow{2}{*}{SR2} & 0 & - & $\ge4$ & $\ge4$ \\
		& 1 & $\ge1$ & $\ge3$ & $\ge3$ \\ \hline
		SSL & 2 (same-sign) &  $\ge2$ & - & - \\
	\end{tabular}
	\caption{Summary of the selection criteria defining each signal region in the colour-octet analysis. The table lists the required number of isolated leptons $\ell$, the number of isolated AK4 $b$-jets associated with leptonically-decaying top quarks $b_\ell$, as well as the number of AK10 jets and top-tagged AK10 jets. We recall that AK10 jets are imposed not to overlap with any isolated lepton within $\Delta R = 1$, while isolated $b$-jets are similarly defined as not overlapping with any AK10 jet within the same angular distance.}
	\label{tab:octet_requirements}
\end{table}

Each selected event thus contains four reconstructed objects. Some are explicitly top-like, like for instance a leptonically-decaying top or a top-tagged AK10 jet, while others may be less clearly identified, like a non-tagged AK10 jet. These four objects are enforced to be paired into two groups to estimate the mass of the resonances that might have produced them. While experimental searches often use machine-learning techniques to optimise this pairing (like a boosted decision tree in a recent ATLAS study~\cite{ATLAS:2022hwc}), our goal here is to provide a simple and transparent illustration of the sensitivity of four-top final states to BSM top-philic resonances. Therefore, we adopt a minimalistic invariant-mass matching approach, similar in spirit to the distance metric employed by CMS in~\cite{CMS:2024ymd}. This may seem surprising as in principle, both the pair and associated production mechanisms compete for the colour-octet model, with relative rates depending on the underlying model parameters. However, while these two topologies can be distinguished at parton level, the reconstruction process tends to smear their kinematic features, making this distinction practically ineffective at reconstructed level (see Appendix~\ref{sec:appendixFourTopObs} for a further discussion on this point). For this reason, we choose to consistently pair the four top candidates by minimising the absolute difference between the two invariant masses of the pair, an approach equivalent to assuming pure pair production. This strategy, that is analogous to the one implemented in standard di-Higgs searches where each Higgs boson decays into a $b\bar{b}$ pair, will be justified \textit{a posteriori} by the competitive bounds that we will derive on the total BSM-induced four-top production cross section. 

\begin{figure}
	\centering
	\includegraphics[width=\linewidth]{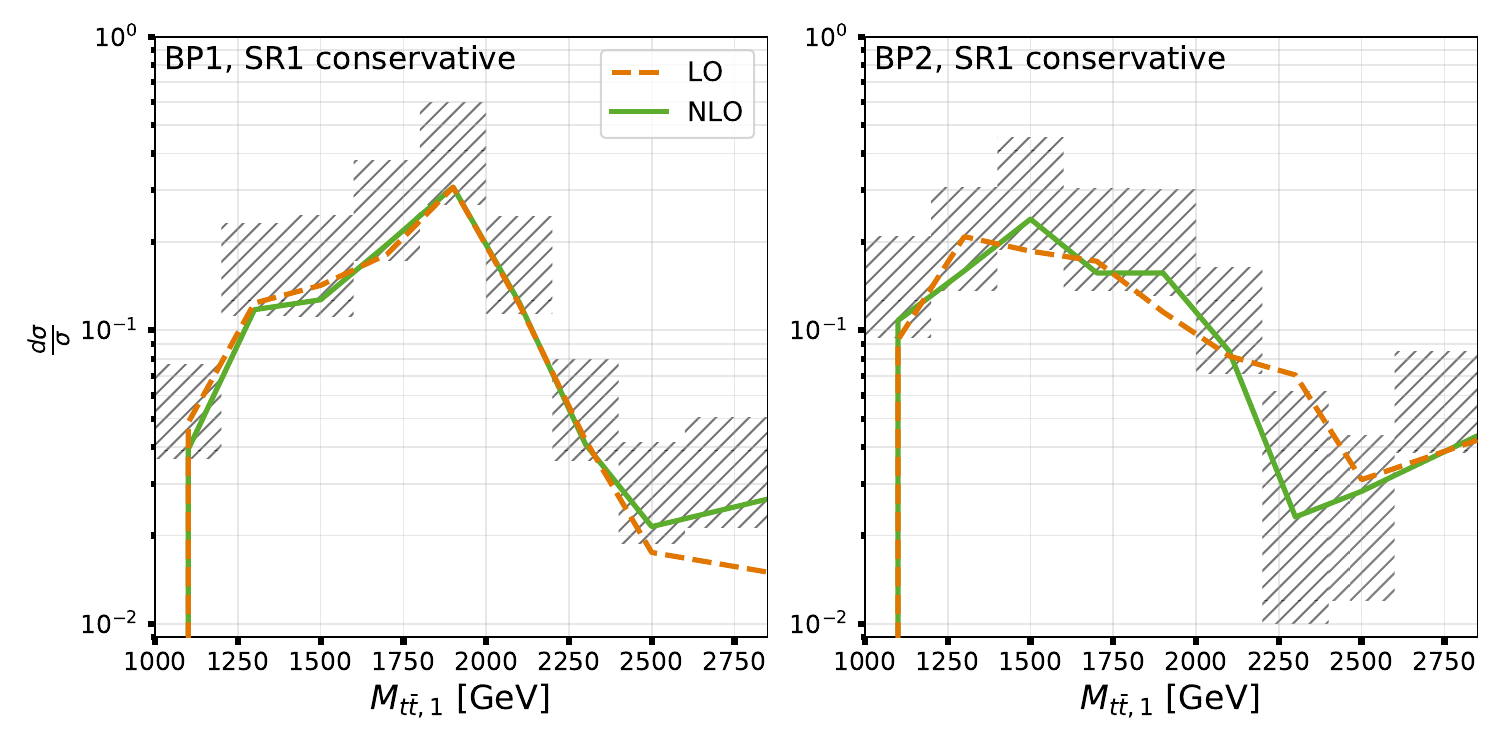}
	\caption{LO (dashed orange) and NLO (solid green) distribution of the largest reconstructed resonance mass for the colour-octet benchmark points BP1 (left) and BP2 (right), for the SR1 signal region with conservative top-tagging performance. The predictions are normalised to 1, the hatched bands represent the NLO scale variation and statistical uncertainties and the last bin includes the overflow. }
	\label{fig:loVsNlo_octet}
\end{figure}

In addition, no information about the top quark electric charge is used during the pairing since this information is not experimentally accessible. The sole exception is in the SSL region, where the presence of two same-sign leptons ensures they cannot originate from the same resonance, yielding thus a natural constraint to impose during pairing. Due to momentum smearing, the two reconstructed resonance masses typically differ. We denote them by $M_{tt,1}$ and $M_{tt,2}$, ordered such that $M_{tt,1} > M_{tt,2}$. We then use $M_{tt,2}$ for further background suppression by requiring $M_{tt,2} > 1$~TeV. Only events passing this cut contribute to the $M_{tt,1}$ distribution that we will further exploit, after binning it in 200~GeV intervals ranging from 1 to 2.6~TeV with the final bin including any overflow.

Since our reconstruction and selection efficiencies are derived from LO simulations (see Section~\ref{sec:Kfact}), it is crucial to verify that LO and NLO predictions remain consistent. Figure~\ref{fig:loVsNlo_octet} shows the $M_{tt,1}$ distributions at LO and NLO for the two colour-octet benchmark scenarios BP1 and BP2 defined in Section~\ref{sec:FourTopTheory}. The two shapes are compatible within the NLO scale uncertainties shown as hatched bands. While minor differences can be seen in the tails of the distributions, they are not significant and solely reflect the limited statistics of our NLO Monte Carlo samples.

\begin{table}[t]
	\centering\renewcommand{\arraystretch}{1.5}\setlength{\tabcolsep}{6pt}
	\resizebox{\textwidth}{!}{
		\begin{tabular}{c|ccccc}
			\multicolumn{6}{c}{Colour-octet selection [fb]} \\
			& $t \bar t +$jets & $t \bar t W$ & $t \bar t Wj$ & $t \bar t Z$ & $t \bar t Zj$  \\ \hline
			SR1 & 5.8$\cdot 10^{-2}{}^{\ +22\%}_{\ -17\%}$ & 7.1$\cdot 10^{-5}{}^{\ +16\%}_{\ -13\%}$ & 1.2$\cdot 10^{-3}{}^{\ +26\%}_{\ -19\%}$ & 2.8$\cdot 10^{-4}{}^{\ +24\%}_{\ -18\%}$ & 1.1$\cdot 10^{-3}{}^{\ +27\%}_{\ -20\%}$  \\
			SR2 & 1.6$\cdot 10^{-3}{}^{\ +2\%}_{\ -17\%}$ & 0 & 1.0$\cdot 10^{-4}{}^{\ +25\%}_{\ -19\%}$ & 3.5$\cdot 10^{-5}{}^{\ +28\%}_{\ -20\%}$ & 1.5$\cdot 10^{-4}{}^{\ +27\%}_{\ -20\%}$  \\
			SSL & 0 & 1.1$\cdot 10^{-5}{}^{\ +17\%}_{\ -13\%}$ & 2.1$\cdot 10^{-4}{}^{\ +25\%}_{\ -19\%}$ & 0 & 0  \\ [.2cm] 
			 & $t \bar t t \bar t$ & $t \bar t t + t \bar t \bar t$ & & &  \\ \hline
			SR1 & 1.1$\cdot 10^{-3}{}^{\ +36\%}_{\ -25\%}$ & 7.7$\cdot 10^{-5}{}^{\ +18\%}_{\ -14\%}$ & & &  \\ 
			SR2 & 2.8$\cdot 10^{-4}{}^{\ +35\%}_{\ -25\%}$ & 7.2$\cdot 10^{-6}{}^{\ +18\%}_{\ -15\%}$ & & &  \\ 
			SSL & 1.7$\cdot 10^{-4}{}^{\ +36\%}_{\ -25\%}$ & 2.1$\cdot 10^{-5}{}^{\ +18\%}_{\ -14\%}$ & & &  \\ 
		\end{tabular}
	}
	\caption{Cross section values after the colour-octet analysis cuts for the main background contributions together with the associated scale uncertainties. These predictions assume a conservative top-tagging performance. The null entries do not mean that contributions from the corresponding background are expected to vanish, rather no events passed the cuts. }
	\label{tab:xsects_after_cuts_octet}
\end{table}

To conclude this section, we present in Table~\ref{tab:xsects_after_cuts_octet} the fiducial cross sections of our background samples after the full reconstruction and selection procedure in the different signal regions with a conservative top-tagging performance. While the $t\bar t +$ jets contribution remains dominant, it is suppressed by more than seven orders of magnitude in the octet analysis.

\subsection{Coloured-octet resonances}

\begin{figure}
	\centering
	\includegraphics[width=\linewidth]{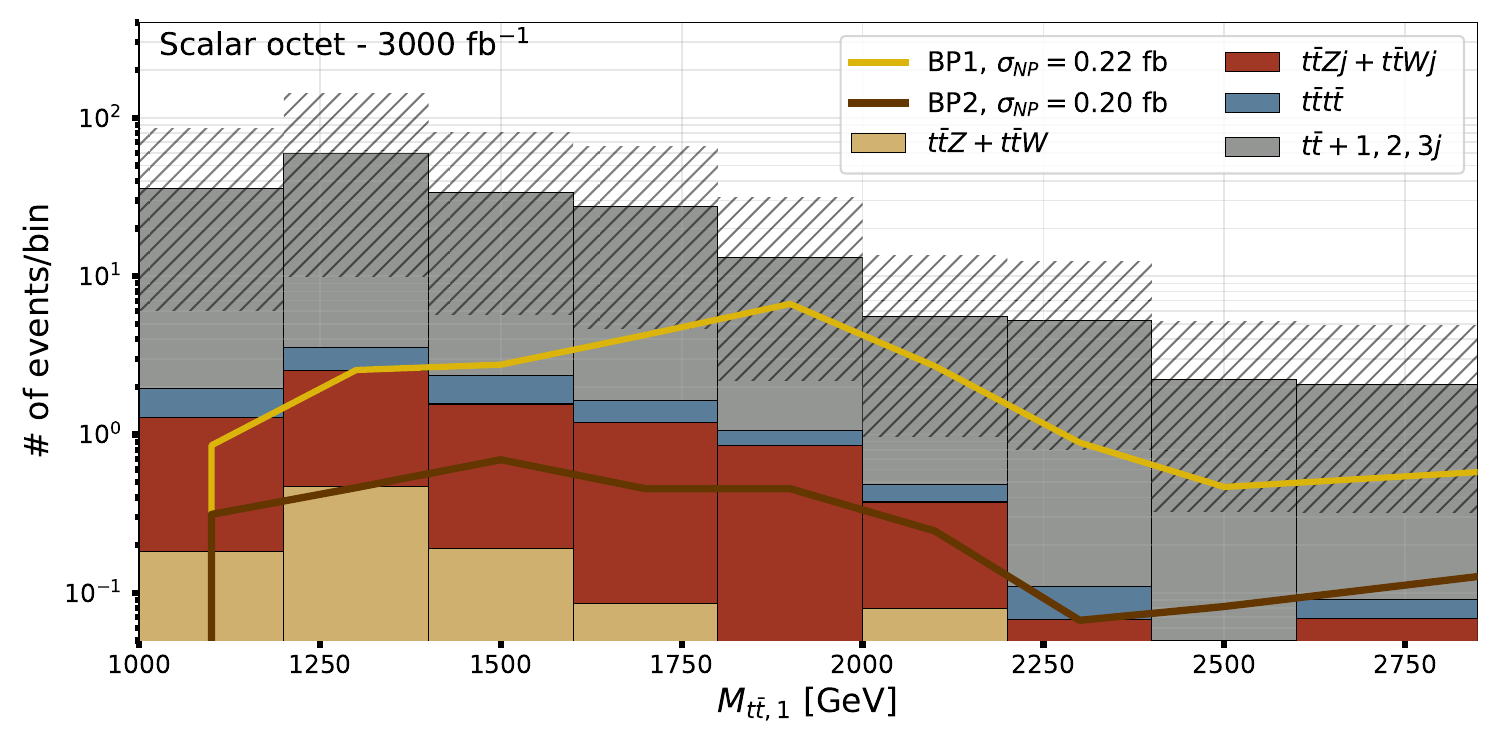}
	\caption{Signal and selected background distributions of the largest reconstructed resonance mass, for the SR1 region with conservative top-tagging and after applying our colour-octet analysis. Backgrounds are generated at LO while the signal is simulated at NLO, and the results are normalised to the HL-LHC luminosity. The hatched bands represent the cumulative scale variation uncertainties on the background and $\sigma_{\text{NP}}$ denotes the benchmark signal  total cross section before the selection cuts. The last bin includes the overflow. \label{fig:octet_with_bkd}}
\end{figure}

For the colour-octet case, Figure~\ref{fig:octet_with_bkd} shows the projected distributions of the largest reconstructed resonance mass $M_{t\bar{t},1}$ at the HL-LHC, in the SR1 signal region and assuming conservative top-tagging performance. The histogram includes signal predictions (simulated at NLO) for the two colour-octet benchmark scenarios BP1 and BP2, as well as the dominant background contributions (generated at LO) along with the cumulative scale variation uncertainties. The latter have been obtained by adding the different contributions linearly across the different background components, though the dominant contribution arises from $t\bar{t}+\text{jets}$ events. As discussed in Section~\ref{sec:bkd}, $t\bar{t}+\text{jets}$ production remains the leading background contribution despite being suppressed by over four orders of magnitude after our selection, owing to their initially large production cross section. This underlines the crucial role of improved top mistagging rejection as enabled by constituent-based top-tagging algorithms in order to enhance the sensitivity. 

Subdominant backgrounds include contributions from SM four-top, $t\bar{t}Z+\text{jets}$ and $t\bar{t}W+\text{jets}$ production. The SM four-top component is irreducible but has a small cross section, while the latter two processes have comparatively larger rates but are efficiently suppressed by our tagging and kinematic requirements. These backgrounds are primarily relevant for $M_{t\bar{t},1}$ values near 1~TeV, but their impact decreases at higher mass, particularly around 2~TeV where the HL-LHC bounds are expected to be found. The background distribution indeed peaks slightly above 1~TeV due to the hard $p_T$ requirements on the reconstructed AK10 jets which disfavour softer events. Consequently, TeV-scale signals are somewhat harder to distinguish from the background, though this is mitigated by the signal cross section being significantly larger than the background in this region (with rates at the fb level).

\begin{table}%\renewcommand{\arraystretch}{1.25}\setlength{\tabcolsep}{10pt}
	\centering
	\begin{tabular}{c|cccc|cccc}
		Top-tag. & \multicolumn{2}{c}{Optimistic} & \multicolumn{2}{c|}{Conservative} & \multicolumn{2}{c}{Optimistic} & \multicolumn{2}{c}{Conservative} \\ 
		$\mathcal{L}$ [fb$^{-1}$] & 500 & 3000 & 500 & 3000 & 500 & 3000 & 500 & 3000 \\[.2cm] 
		& \multicolumn{4}{c|}{ \textbf{LO} } & \multicolumn{4}{c}{ \textbf{LO} } \\ \hline 
		SR1 & 0.55 & 0.21 & 0.65 & 0.25 & 10.29 & 4.21 & 11.98 & 4.59 \\
		SR2 & 0.52 & 0.13 & 0.59 & 0.17 & 5.93 & 1.88 & 6.56 & 2.03 \\
		SSL & 1.65 & 0.37 & 1.64 & 0.37 & 9.11 & 2.20 & 9.09 & 2.14 \\[.2cm] 
		& \multicolumn{4}{c|}{ \textbf{NLO}  } & \multicolumn{4}{c}{ \textbf{ NLO}  } \\ \hline
		SR1 & 0.59 & 0.21 & 0.72 & 0.27 & 8.28 & 3.01 & 8.66 & 3.49 \\
		SR2 & 0.59 & 0.15 & 0.67 & 0.19 & 4.16 & 1.32 & 4.59 & 1.52 \\
		SSL & 2.31 & 0.52 & 2.28 & 0.52 & 6.73 & 1.52 & 6.71 & 1.51 \\
	\end{tabular}
	%}
\caption{Upper limits on the new physics cross section (in fb) for the colour-octet benchmark scenarios (BP1 with $y_{\oct}=1.0$ and $M_{\oct}=2$ TeV in the left column and BP2 with $y_{\oct}=3.0$ and $M_{\oct}=3$ TeV in the right column), derived from our analysis strategy across the three signal regions. Results are presented for both optimistic and conservative top-tagging assumptions and for integrated luminosities of 500 and 3000~fb$^{-1}$.  }
\label{tab:CSlimOct}
\end{table}

An important feature revealed by our analysis is that despite the considerable smearing associated with top-quark reconstruction, the signal retains a visible bump structure near the mass of the BSM resonance. This allows for a shape-based analysis directly on the $M_{t\bar{t},1}$ spectrum, without requiring a precise normalisation of the SM background. Theoretical uncertainties, particularly for the $t\bar{t}Z+\text{jets}$ and $t\bar{t}W+\text{jets}$ contributions, are therefore less critical to the overall sensitivity. Projected 95\% confidence level (C.L.) bounds on the signal are hence obtained by fitting the $M_{t\bar{t},1}$ distribution using the \pyhf\ framework~\cite{Heinrich:2021gyp}, with a correlated scale factor applied to account for theoretical uncertainties on the dominant $t\bar{t}+\text{jets}$ background. The resulting LO and NLO projected cross section limits for the two colour-octet benchmark points are listed in Table~\ref{tab:CSlimOct}, at integrated luminosities of $500~\mathrm{fb}^{-1}$ and $3000~\mathrm{fb}^{-1}$ and for the different signal regions defined in Section~\ref{sec:octetana}. Several qualitative trends emerge from these results. First, the search performs best when on-shell BSM resonances can be pair-produced (\ie\ for masses around or below 2~TeV). Second, the full reconstruction of the four-top final state yields stronger limits compared to the other explored strategies. In contrast the SSL channel, while benefiting from low background, suffers from lower signal yields and hence reduced sensitivity. Finally, the performance of the top tagger has a sizeable impact, with more optimistic mistagging assumptions improving the exclusion limits by up to 20\%.

\begin{figure}
\centering
\includegraphics[width=0.45\columnwidth]{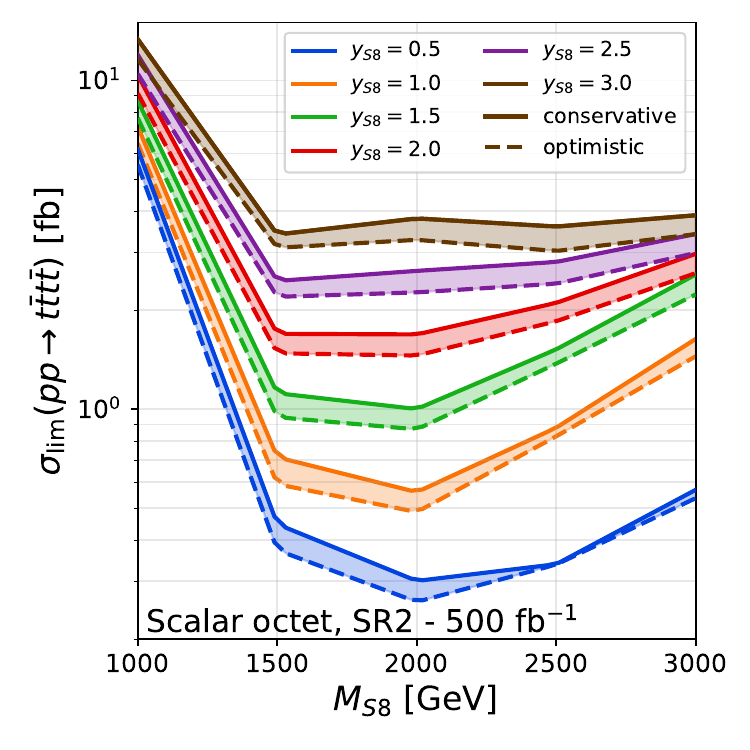}\hfill
\raisebox{0.4cm}{\includegraphics[width=0.54\linewidth]{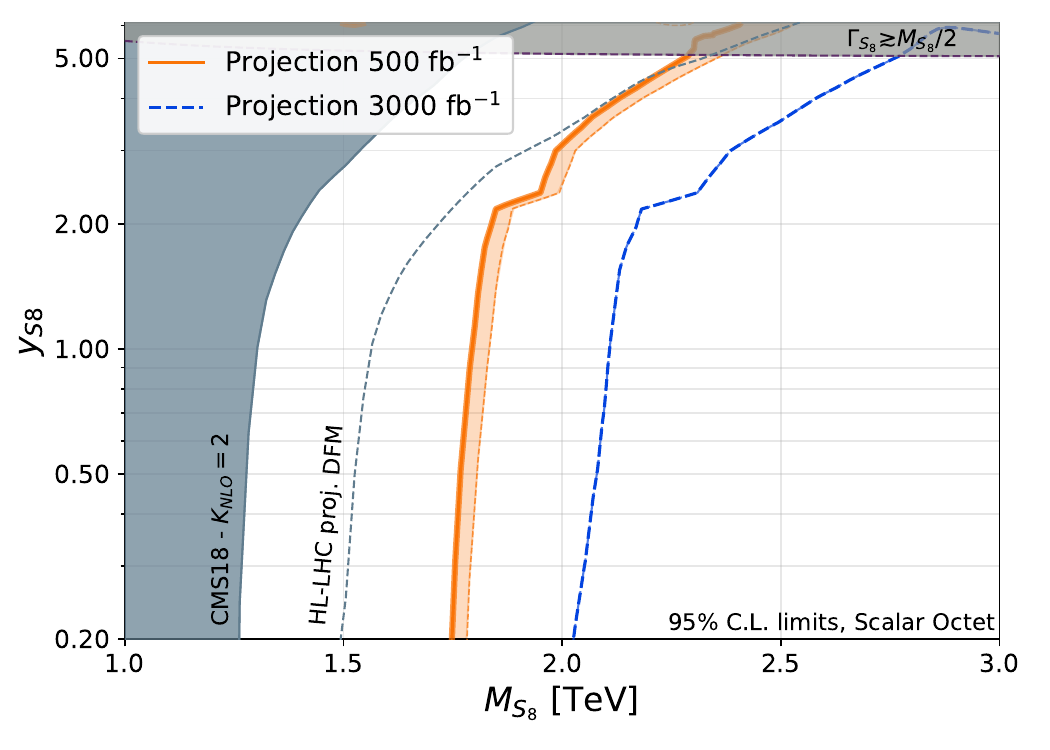}}
\caption{\textit{Left panel} -- Expected cross section limits as a function of the colour-octet resonance mass, for several values of the coupling to top quarks and assuming a branching ratio of 1. Results are shown for the SR2 signal region at 500~fb${}^{-1}$, with the solid lines corresponding to the conservative top-tagging assumption and the dashed lines to the optimistic one. \textit{Right panel} -- Projected 95\% C.L. exclusion regions in the colour-octet mass-coupling plane for integrated luminosities of 500~fb$^{-1}$ (orange) and 3000~fb$^{-1}$ (dashed blue). Solid and dashed orange lines correspond to conservative and optimistic top-tagging assumptions, respectively, and we compare these projections to the currently excluded region (dark shading) obtained from a recast of the CMS-TOP-18-003 analysis~\cite{Darme:2021gtt, Fuks:2021zbm, DVN/OFAE1G_2020} and its naive extrapolation to 3000~fb$^{-1}$~\cite{Araz:2019otb} (dashed grey) using an approximate and overestimated $K$-factor of 2 and LO simulations for the signal. The light grey area at large couplings indicates the region where the resonance width becomes large, making our approach unreliable.}
\label{fig:limits_scalarOctet}
\end{figure}

In the left panel of Figure~\ref{fig:limits_scalarOctet}, we present the expected cross section limits in the SR2 signal region as a function of the resonance mass for different values of the coupling to top quarks. As expected, the optimal sensitivity is achieved when pair production of the colour octet dominates, leading to a final state topology typically featuring four highly boosted top quarks. For larger values of the top-quark coupling, both the resonance width and the contribution from the associated production mode ($pp\to t\bar{t} \oct$) increase, which in turn degrades the efficiency of the search. The right panel of the same figure shows the projected 95\% C.L. exclusions in the mass-coupling plane representing the parameter space of the scalar colour-octet simplified model. Here, the NLO signal cross section is obtained by a linear interpolation as described in Section~\ref{sec:Kfact}.

Our NLO projections are in good agreement with previous LO estimates from Ref.~\cite{Darme:2024epi}, although the latter neglected correlations between the invariant masses of the two scalar octets. In this work, we adopt a more conservative approach by considering only the largest reconstructed invariant mass per event, which partially compensates the increase in signal strength from the inclusion of NLO corrections. As a result, our analysis excludes colour-octet scalars with masses up to approximately 2~TeV, even for moderately small Yukawa couplings. As is typical for limits on BSM resonances produced via QCD-driven pair production, the projected exclusions extend down to arbitrarily small values of $y_{\oct}$ provided that the branching ratio to top quarks remains dominant and that the resonance decays promptly. Projections for more realistic scenarios where the resonance also decays into other SM particles can be readily obtained by rescaling the predicted cross section in Figure~\ref{fig:CS} and comparing it to the cross section limits in the left panel of Figure~\ref{fig:limits_scalarOctet}. Finally, we emphasise that the search strategy employed here is not fully optimised, particularly regarding the pairing method used to reconstruct the resonance mass from four-top final states. Given the substantial advances made by experimental collaborations in analogous contexts (such as double Higgs production with a $b\bar{b}b\bar{b}$ final state), we anticipate that future HL-LHC analyses could significantly improve upon our simplified approach, potentially extending the sensitivity well beyond the 2~TeV mass range.

\section{Single resonance analysis}
\label{sec:SingletAnalysis}
\subsection{Signal region definition and resonance reconstruction in the colour-singlet model}
\label{sec:singletana}

\begin{table}[t] %\renewcommand{\arraystretch}{1.3}\setlength\tabcolsep{14pt}
	\centering
	\begin{tabular}{c|cccc}
		\multicolumn{5}{c}{Colour-singlet analysis} \\ \hline
		Signal Region & \# $\ell$ &  \# $b_\ell$ & \# AK10 & \# top-tag. \\ \hline
		\multirow{2}{*}{SR1} & 0 &  $\ge2$ & $\ge2$ & $\ge1$ \\
		& 1 & $\ge2$ & $\ge2$ & $\ge1$ \\ \hline
		\multirow{2}{*}{SR2} & 0 & $\ge2$ & $\ge2$ & $\ge2$ \\
		& 1 &  $\ge2$ & $\ge2$ & $\ge2$ \\ \hline
		SSL & 2 (same-sign) &  $\ge2$ & - & - \\
	\end{tabular}
	\caption{Summary of the selection criteria defining each signal region in the colour-singlet analysis. The table lists the required number of isolated leptons $\ell$, the number of isolated AK4 $b$-jets associated with leptonically-decaying top quarks $b_\ell$, as well as the number of AK10 jets and top-tagged AK10 jets. We recall that AK10 jets are imposed not to overlap with any isolated lepton within $\Delta R = 1$, while isolated $b$-jets are similarly defined as not overlapping with any AK10 jet within the same angular distance. \label{tab:singlet_requirements}}
\end{table}

The search strategy targeting the colour-singlet model builds on the fact that only one pair of top quarks is expected to originate from the decay of a heavy resonance. As a consequence, only two top quark candidates are required to reconstruct the BSM resonance mass, and we correspondingly ask for at least two AK10 jets in the event preselection regardless of the lepton multiplicity. If more than two AK10 jets are found, we focus on the two with the highest transverse momentum, giving precedence to those that are top-tagged, and these two AK10 jets then serve as proxies for hadronically-decaying top quarks. 

As detailed in Section~\ref{sec:topreco}, to further suppress the background we additionally require each event to contain at least two isolated $b$-tagged jets (defined as jets not overlapping with any AK10 jets within a distance $\Delta R = 1$). These $b$-jets may then be used to reconstruct leptonically-decaying top quarks, depending on the lepton content of the event. More precisely, in fully hadronic events (zero leptons), the two selected AK10 jets are directly used for the reconstruction of the BSM resonance. In the single-lepton case, the isolated $b$-jet closest in $\Delta R$ to the lepton is combined with it to reconstruct a leptonically-decaying top quark following the strategy outline above and relying on a kinematic fit of the event. Finally, in the dilepton case, each lepton is paired with its nearest $b$-tagged jet to reconstruct two such objects, the four-momenta of the two neutrinos being reconstructed from the $M_{T2}$-based strategy discussed previously.

Signal regions are next defined based on the number of top-tagged AK10 jets observed.
\begin{itemize}%[topsep=2pt,itemsep=0pt,parsep=3pt,partopsep=2pt]
	\item The \textbf{SR1} region includes all fully-hadronic and single-lepton events with at least one top-tagged AK10 jet. These events thus contain at least one reconstructed and tagged object accompanied by either an additional non-tagged AK10 jet or a leptonically-decaying top quark.
	\item The \textbf{SR2} region contains events with at least two top-tagged AK10 jets, independently of the number of leptons or other non-tagged AK10 jets.
	\item The \textbf{SSL} region includes all same-sign dilepton events, regardless of the number of tagged AK10 jets.
\end{itemize}
A summary of the requirements for each signal region is given in Table~\ref{tab:singlet_requirements}.

Once the preselection and signal region assignment are completed, we proceed to reconstruct the resonance mass. We pair the two top quark candidates with the highest transverse momentum (with precedence being given to the top-tagged AK10 jets), assuming that they originate from the BSM resonance decay in an associated production topology. Their invariant mass is finally computed and stored in a histogram. For the SR1 and SR2 regions, we use bins of 200~GeV from 400 to 4000~GeV with the last bin containing the overflow, while for the SSL region the range is limited to 1600~GeV. Since four top-like objects are not always available in these events, we do not attempt to compute a second invariant mass or apply any additional background rejection based on extra top candidates.

\begin{figure}
	\centering
	\includegraphics[width=\linewidth]{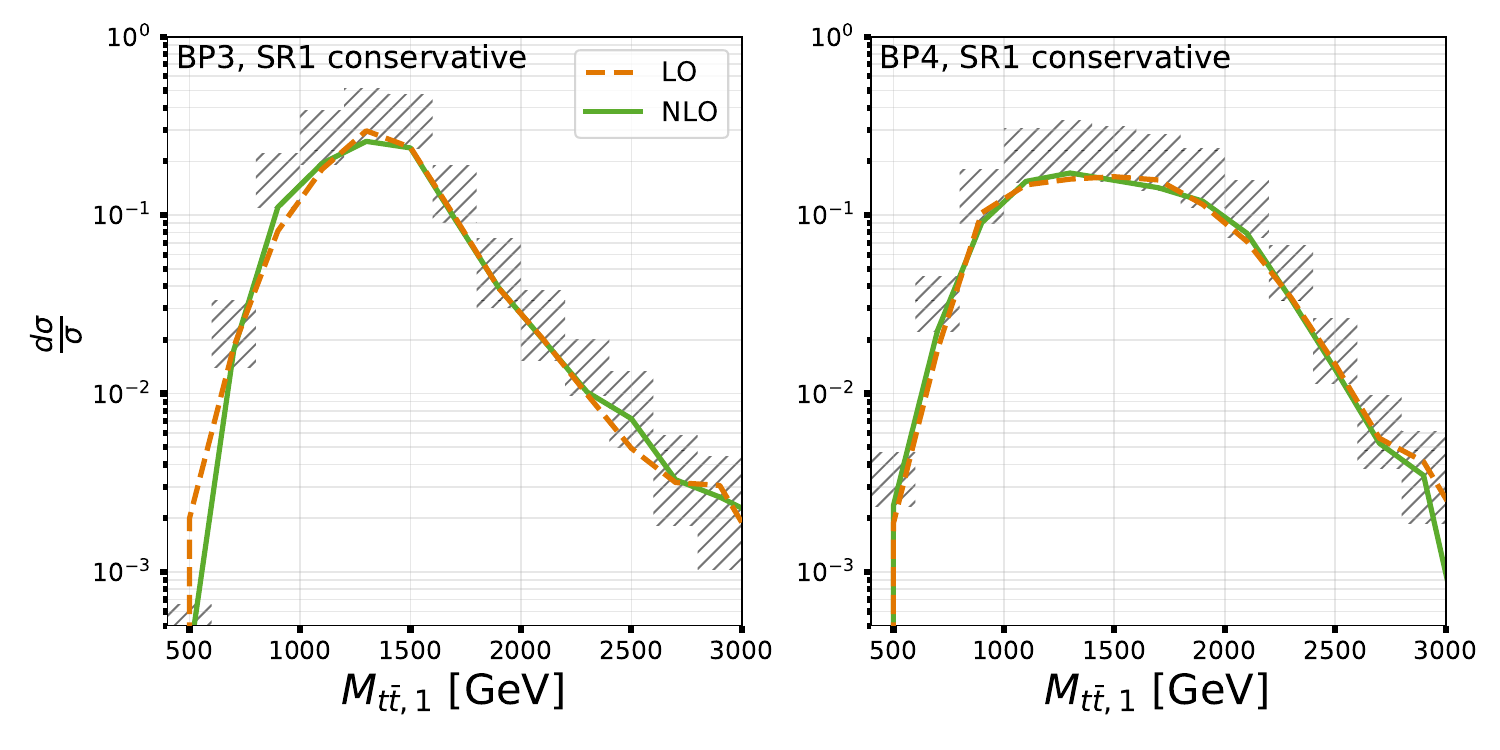}
	\caption{LO (dashed orange) and NLO (solid green) distributions of the reconstructed resonance mass for the colour-singlet benchmark points BP3 (left) and BP4 (right), for the SR1 signal region with conservative top-tagging performance. The predicted differential cross sections $\mathrm{d}\sigma$ are normalised to 1, the hatched bands represent the NLO scale variation and statistical uncertainties and the last bin includes the overflow. \label{fig:loVsNlo_singlet}}
\end{figure}

As discussed in Section~\ref{sec:Kfact}, reconstruction and selection efficiencies will be obtained from LO simulations. It is therefore important to validate that LO and NLO differential distributions are in reasonable agreement. To this aim, the reconstructed resonance invariant mass distributions at LO and NLO for the benchmark points BP3 and BP4 are presented in Figure~\ref{fig:loVsNlo_singlet}. The shapes of the LO and NLO distributions  are consistent within the scale variation uncertainties of the NLO prediction, showing that the LO simulation strategy introduced in Section~\ref{subsec:FourTopLags} can be safely used. As for the octet case, the deviations observed in the high-mass tails are not significant, and are attributable to limited statistics in the NLO Monte Carlo samples.

\begin{table}[t]
	\centering\renewcommand{\arraystretch}{1.5}\setlength{\tabcolsep}{6pt}
	\resizebox{1.0\textwidth}{!}{
		\begin{tabular}{c|cccccc}
			\multicolumn{7}{c}{Colour-singlet selection [fb]} \\
			& $t \bar t +$jets & $t \bar t W$ & $t \bar t Wj$ & $t \bar t Z$ & $t \bar t Zj$ & $t \bar t t \bar t$ \\ \hline
			SR1 & 35${}^{\ +20\%}_{\ -15\%}$ & 8.7$\cdot 10^{-2}{}^{\ +18\%}_{\ -14\%}$ & 3.2$\cdot 10^{-1}{}^{\ +27\%}_{\ -20\%}$ & 3.5$\cdot 10^{-1}{}^{\ +19\%}_{\ -15\%}$ & 9$\cdot 10^{-1}{}^{\ +27\%}_{\ -20\%}$ & 1.2$\cdot 10^{-1}{}^{\ +37\%}_{\ -25\%}$ \\
			SR2 & 5.6$^{\ +19\%}_{\ -15\%}$ & 1.8$\cdot 10^{-2}{}^{\ +18\%}_{\ -14\%}$ & 4.3$\cdot 10^{-2}{}^{\ +27\%}_{\ -20\%}$ & 9.8$\cdot 10^{-2}{}^{\ +19\%}_{\ -15\%}$ & 2.0$\cdot 10^{-1}{}^{\ +27\%}_{\ -20\%}$ & 6$\cdot 10^{-2}{}^{\ +37\%}_{\ -25\%}$ \\
			SSL & 0 & 0 & 1.4$\cdot 10^{-3}{}^{\ +27\%}_{\ -20\%}$ & 2.6$\cdot 10^{-4}{}^{\ +24\%}_{\ -18\%}$ & 8$\cdot 10^{-4}{}^{\ +27\%}_{\ -20\%}$ & 1.1$\cdot 10^{-3}{}^{\ -37\%}_{\ -25\%}$ \\
		\end{tabular}
	}
	\caption{Same as in table \ref{tab:xsects_after_cuts_octet} but for the colour-singlet analysis.}
	\label{tab:xsects_after_cuts_singlet}
\end{table}

Table~\ref{tab:xsects_after_cuts_singlet} presents the fiducial cross sections of our background samples after the full reconstruction and selection procedure in the different signal regions with a conservative top-tagging performance. The $t\bar t +$ jets contribution is again dominant, it is suppressed by about four orders of magnitude in the singlet case.

\subsection{Colour-singlet resonance}

\begin{figure}
	\centering
	\includegraphics[width=\linewidth]{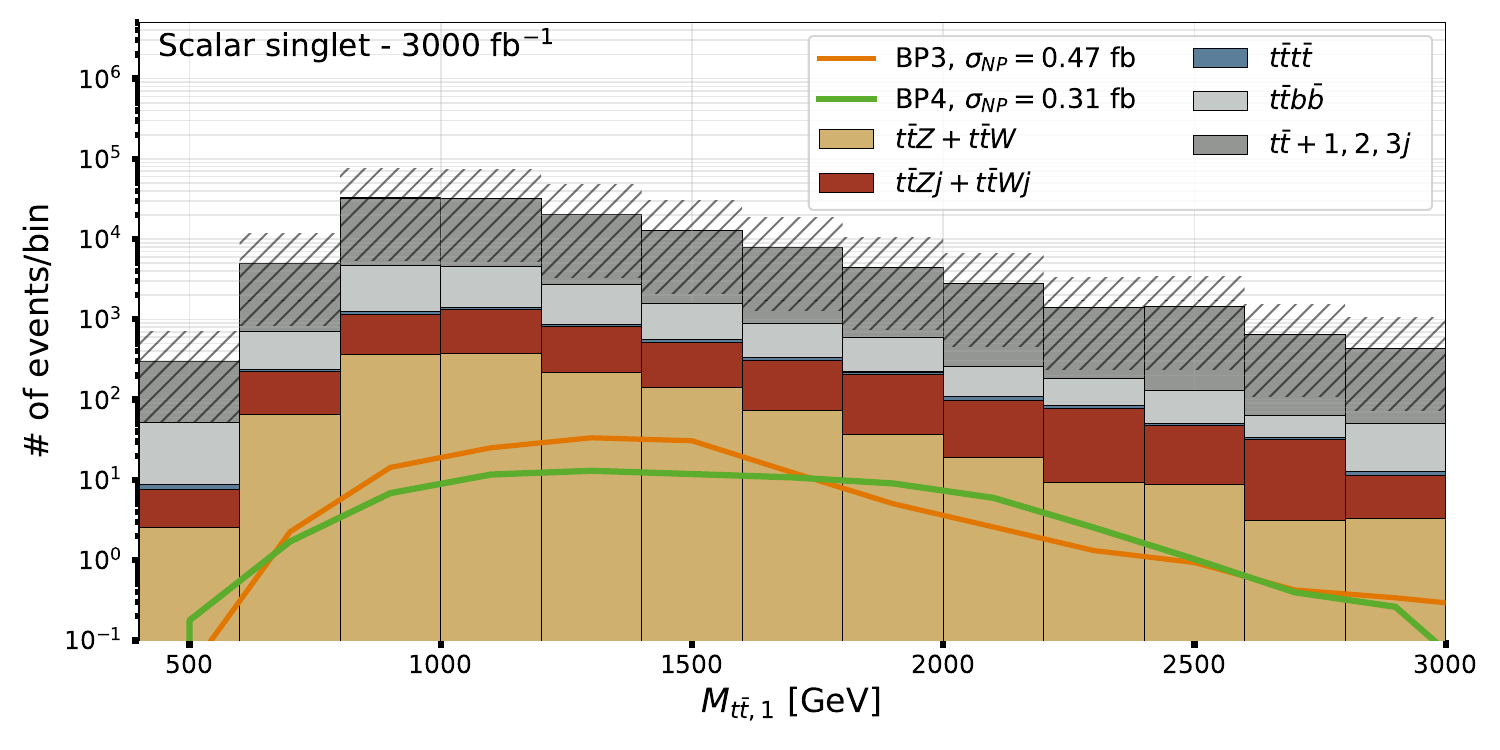}\\ \vspace{.2cm}
	\includegraphics[width=\linewidth]{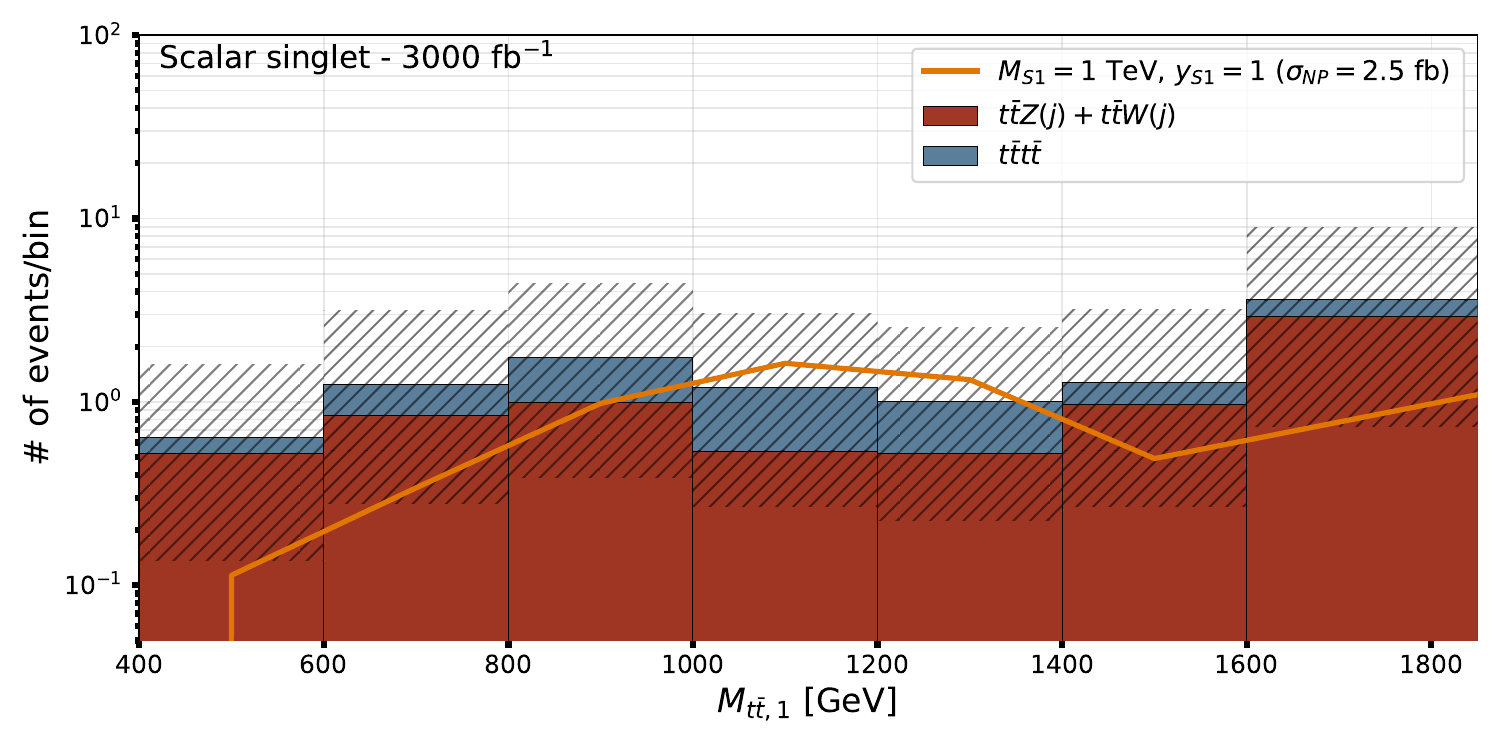}
	\caption{Signal and selected background distributions of the largest reconstructed resonance mass, for the SR1 (top) and SSL (bottom) regions with conservative top-tagging and after applying our colour-singlet analysis. Backgrounds are generated at LO while the signal is simulated at NLO, and the results are normalised to the HL-LHC luminosity. The hatched bands represent the cumulative scale variation uncertainties on the background and $\sigma_{\text{NP}}$ denotes the benchmark signal cross section before the selection cuts. The last bin includes the overflow.
		\label{fig:singlet_with_bkd}}
\end{figure}

For the colour-singlet analysis, predictions for the reconstructed invariant mass distribution of the signal at the HL-LHC are shown in Figure~\ref{fig:singlet_with_bkd} for the SR1 (top row) and SSL (bottom row) signal regions, using conservative top-tagging. The main backgrounds are also displayed, along with their cumulative scale variation uncertainties. In the SR1 region, the dominant background stemming from $t\bar{t}+\text{jets}$ production remains, like in the colour-octet case. However, the overall background level is significantly higher since the looser top-tagging requirements motivated by the typically lower $p_T$ of top quarks in the signal allow more background events to pass the selection. As in the octet analysis, our strategy nevertheless suppresses low-$p_T$ background events, which results in a peak around 1~TeV in the $M_{t\bar{t},1}$ invariant mass distribution. We emphasise that due to the specific selection cuts used in the singlet analysis, we additionally include the $t\bar{t}b\bar{b}$ background although it contributes at only the $\sim 10\%$ level compared to the dominant $t\bar{t}+\text{jets}$ background. As expected, the signal resonance is more strongly smeared in this analysis compared to the octet case, reflecting the difficulty of selecting the correct top quarks originating from the BSM decay. This suggests that more advanced reconstruction strategies may be necessary to identify a scalar top-philic singlet resonance efficiently. As a first step in this direction, we present in Appendix~\ref{sec:appendixFourTopObs} a series of differential distributions for relevant kinematic variables.

\begin{table}%\renewcommand{\arraystretch}{1.25}\setlength{\tabcolsep}{10pt}
	\centering
	\begin{tabular}{c|cccc|cccc}
		Top-tag. & \multicolumn{2}{c}{Optimistic} & \multicolumn{2}{c|}{Conservative} & \multicolumn{2}{c}{Optimistic} & \multicolumn{2}{c}{Conservative} \\
		$\mathcal{L}$ [fb$^{-1}$] & 500 & 3000 & 500 & 3000 & 500 & 3000 & 500 & 3000 \\[.2cm] 
		& \multicolumn{4}{c|}{\textbf{LO}  } & \multicolumn{4}{c}{\textbf{LO} } \\ \hline
		SR1 & 9.53 & 3.89 & 9.61 & 3.92 & 6.83 & 2.77 & 6.97 & 2.83 \\
		SR2 & 5.45 & 2.20 & 5.89 & 2.39 & 4.10 & 1.67 & 4.33 & 1.76 \\[.2cm] 
		& \multicolumn{4}{c|}{\textbf{NLO} } & \multicolumn{4}{c}{\textbf{NLO}} \\ \hline
		SR1 & 10.43 & 4.26 & 10.56 & 4.31 & 8.78 & 3.57 & 8.83 & 3.60 \\
		SR2 & 6.60 & 2.57 & 8.17 & 2.91 & 4.62 & 1.87 & 5.30 & 2.16 \\
	\end{tabular}
	\caption{Same as in Table~\ref{tab:CSlimOct} but for the colour-singlet benchmark scenarios  (BP3 with $y_{\sing}=1.0$ and $M_{\sing}=1.5$ TeV in the left column and BP4 with $y_{\sing}=1.5$ and $M_{\sing}=2$ TeV in the right column).} \label{tab:CSlimSing}
\end{table}

The projected 95\% C.L. cross section limits at $500~\text{fb}^{-1}$ and $3000~\text{fb}^{-1}$ are reported in Table~\ref{tab:CSlimSing} for the two BP3 and BP4 signal scenarios introduced in Section~\ref{subsec:FourTopLags}. Like for the octet analysis, the best sensitivities for resonance masses above 1.5~TeV are obtained in SR2 where more top quarks are fully reconstructed compared to SR1. Furthermore, we do not report the obtained limits from the SSL region in the table due to the large associated Monte Carlo statistical uncertainties: generating sufficiently large samples proved computationally intensive given the low signal acceptance. However, it is not necessary in light of the slightly stronger limits obtained for the SR2 signal region. We also verify that the differences between LO and NLO signal shapes are negligible in terms of their impact on the projected limits. 

\begin{figure}
	\centering
	\includegraphics[width=0.45\columnwidth]{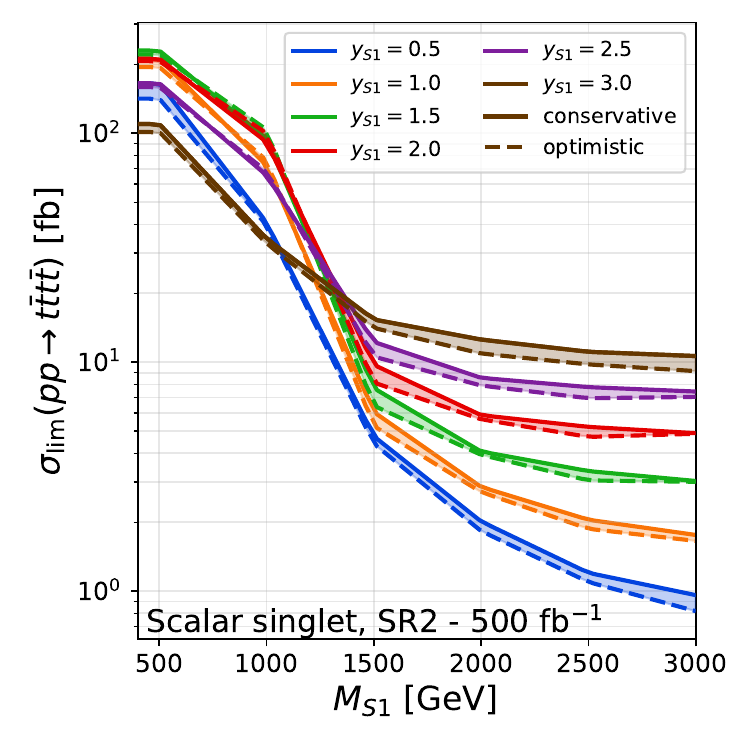}\hfill
	\raisebox{0.8cm}{\includegraphics[width=0.54\linewidth]{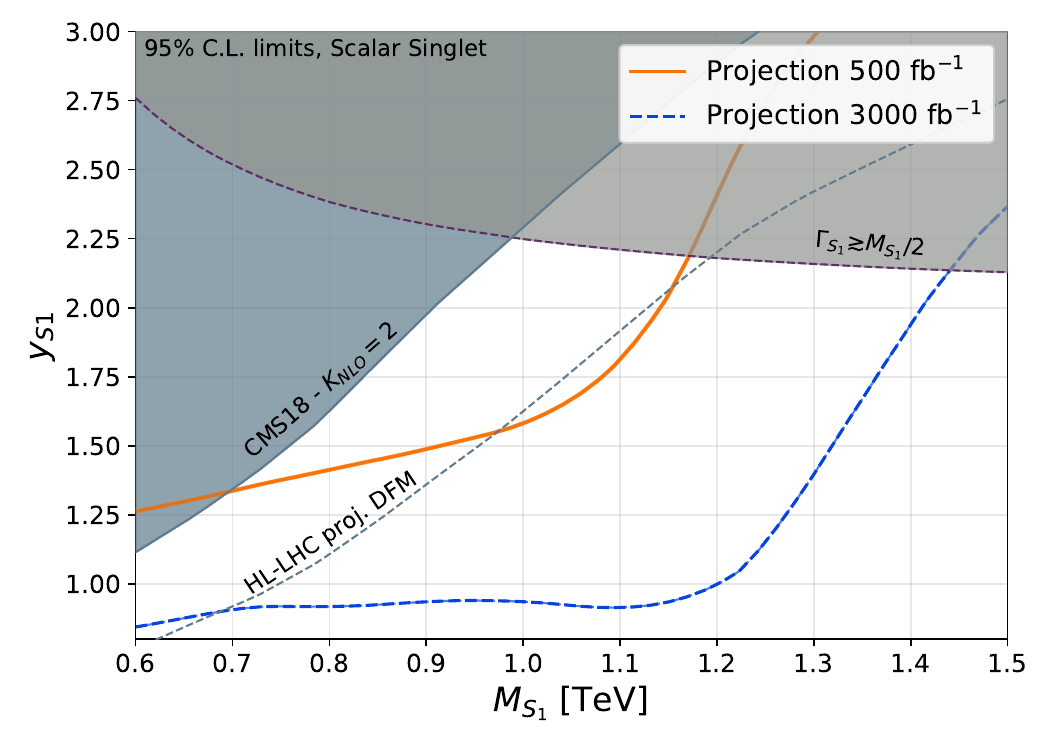}}
	\caption{\textit{Left panel} -- Expected cross section limits as a function of the colour-singlet resonance mass, for several values of the coupling to top quarks and assuming a branching ratio of 1. Results are shown for the SR2 signal region at 500~fb${}^{-1}$, with the solid lines corresponding to the conservative top-tagging assumption and the dashed lines to the optimistic one. \textit{Right panel} -- Projected 95\% C.L. exclusion regions in the colour-singlet mass-coupling plane for integrated luminosities of 500~fb$^{-1}$ (orange) and 3000~fb$^{-1}$ (dashed blue) and conservative top-tagging assumptions, using the SSL signal region. We compare these projections to the currently excluded region (dark shading) obtained from a recast of the CMS-TOP-18-003 analysis~\cite{Darme:2021gtt, Fuks:2021zbm, DVN/OFAE1G_2020} and its naive extrapolation to 3000~fb$^{-1}$~\cite{Araz:2019otb} (dashed grey), using a $K$-factor of 2 and LO simulations for the signal. The light grey area at large couplings indicates the region where the resonance width becomes large, making our approach unreliable.\label{fig:limits_scalarSinglet}}
\end{figure}

The projected cross section limits as a function of the singlet mass and coupling strength are shown in the left panel of Figure~\ref{fig:limits_scalarSinglet}. At large masses, the signal topology features increasingly boosted top quarks, which enhances the search efficiency and significantly improves the exclusion reach. For instance, the projected limit is stronger by nearly two orders of magnitude when the singlet mass increases from 1~TeV to 2~TeV. Conversely, larger values of the top-quark coupling lead to a broader resonance and enhance the relative contributions of off-shell production channels, which both degrade the sensitivity. Below the 1.5~TeV threshold, the analysis becomes markedly less effective. This is partly due to the reduced boost of the top quarks which makes their reconstruction more difficult, and partly because of the background driven by the $t\bar{t}+\text{jets}$ contribution which yields a peak in the $M_{t\bar{t},1}$ distribution around 1~TeV, thus mimicking the expected signal shape. In this region, we therefore choose to revert to the SSL search strategy which benefits from nearly background-free conditions. The lower panel of Figure~\ref{fig:singlet_with_bkd} shows the predicted invariant mass distribution for the signal in this channel, that we thus use for limit setting.

The projected 95\% C.L. exclusions for the scalar colour-singlet simplified model are displayed in the right panel of Figure~\ref{fig:limits_scalarSinglet}. As described in Section~\ref{sec:Kfact}, the limits are obtained by interpolating the LO cross section logarithmically and multiplying it by a global $K$-factor interpolated linearly from an NLO-to-LO ratio grid. As mentioned above, we use the SSL analysis as it provides the best sensitivity in the relevant region of the parameter space, where perturbative values of $y_{\sing}$ only allow resonance masses up to about 1.5~TeV to be probed at the HL-LHC. Notably, even with a luminosity of $500~\text{fb}^{-1}$, our analysis improves upon existing searches for resonance masses near 1~TeV. At higher luminosities, the reach increases further, potentially allowing BSM particles with top-quark couplings comparable to the SM Higgs to be excluded up to $\sim 1.2$~TeV.

\begin{table}%\renewcommand{\arraystretch}{1.35}\setlength{\tabcolsep}{10pt}
	\centering
	\begin{tabular}{c|cccc|cccc}
		& \multicolumn{4}{c|}{Singlet analysis} & \multicolumn{4}{c}{Octet analysis} \\
		Top-tag. & \multicolumn{2}{c}{Optimistic} & \multicolumn{2}{c|}{Conservative} & \multicolumn{2}{c}{Optimistic} & \multicolumn{2}{c}{Conservative} \\ 
		$\mathcal{L}$ [fb$^{-1}$] & 500 & 3000 & 500 & 3000 & 500 & 3000 & 500 & 3000 \\ \hline
		SR1 & 53.69 & 22.17 & 52.13 & 21.31 & 51.65 & 23.76 & 63.79 & 29.26 \\
		SR2 & 76.17 & 28.69 & 72.27 & 29.33 & 34.38 & 10.37 & 34.70 & 12.08 \\
		SSL &  9.8 & 3.3 & 10.1 & 3.2 & 15.6 &  3.7 & 15.8 & 3.7 \\
	\end{tabular}
	\caption{Upper limits on the colour-singlet production cross section (in fb) for a scenario with $M_{\sing}=1$ TeV and $\Gamma_{\sing}/M_{\sing}= 0.1 $, derived using both analysis strategies discussed in this work. Results are presented for optimistic and conservative top-tagging performances at integrated luminosities of 500 and 3000 fb$^{-1}$. \label{tab:CSlimSingCompare}}
\end{table}

The SR1 and SR2 conservative singlet analyses discussed earlier assume that top-tagging is applied only to the most boosted AK10 jets. In principle, if full top-tagging could be achieved, a significant improvement of the analysis strategies would be expected,  even for resonance masses of about 1~TeV. To illustrate this point, we perform a comparison using a more optimistic reconstruction strategy similar to the one used in the octet case, and apply it to the signal originating from a scalar singlet scenario with $M_{\sing} = 1$~TeV and $y_{\sing} = 1$ (or equivalently $\Gamma_{\sing}/M_{\sing}=0.1$). The corresponding results are shown in Table~\ref{tab:CSlimSingCompare}. In this scenario, the SR2 limit improves by a factor of more than two, while the SR1 and SSL ones show no significant change. Despite this, the SSL region remains the most promising strategy across all benchmarks, further justifying its use in our final limit projections. We remind that the SSL results in Table~\ref{tab:CSlimSingCompare} carry a statistical uncertainty of approximately 20\% owing to the limited number of Monte Carlo events passing all selection cuts.

% Conclusion
%!TEX root = main.tex

%%%%%%%%%%%%%%%%%%%%%%%%%%%%%%%%%%%%%%%%%%%%%%%%%%%%
%
%    Conclusion
%
%
%%%%%%%%%%%%%%%%%%%%%%%%%%%%%%%%%%%%%%%%%%%%%%%%%%%

\chapter{Discussion and Conclusion}
%\addcontentsline{toc}{chapter}{Conclusion}
\label{chap:conclusion}

\hfill
\begin{minipage}{8cm}
{\it 
``You miss 100\% of the shots you don't take.''} 

\hfill saying popularised by Wayne Gretzky. 
\end{minipage}

\vspace{0.5cm}

Particle physics is experiencing one of its most challenging periods. No BSM model candidate is currently preferred to solve the issues of the SM presented at the end of Chapter~\ref{chap:sm}. Sakharov's condition on the presence of $CP$ violation is already met in the SM but the tension between the SM estimation and the measured baryon asymmetry is large. With the description of the Higgs in the SM as it is, a first-order phase transition is not realised. New particle resonances could point to a model where these issues are solved. However, it is possible that the energy reach of the LHC is not sufficiently large to observe new resonances or that more events need to be collected to statistically discard resonances at current energies.

Nevertheless, there are still opportunities to search for BSM effects. Generic models proposed in Chapter~\ref{chap:bsm} aim at testing the two possibilities in complementary approaches without committing to a specific UV model. On the one hand, off-shell contributions are described by EFTs either in the top-down or bottom-up approach depending on whether the UV-complete model is specified or not. In the UV model-agnostic bottom-up approach, the leading BSM contributions of dimension-six operators are provided by the operators from the Warsaw basis. The constraints on the Wilson coefficients can be translated into bounds on UV model parameters by the matching and running procedure, as it did for the 4-Fermi theory. On the other hand, simplified models reproduce on-shell effects of BSM resonances. These resonances are organised in dictionaries with increasing "complexity" in their quantum numbers.

Additional sources of $CP$ violation arising from off-shell effects are considered in the form of $CP$-odd dimension-six SMEFT operators in Chapter~\ref{chap:smeftcpv}. The 6 dimension-six $CP$-odd operators from the $X^3$ and $X^2 \varphi^2$ classes are invariant under flavour transformations. However, the 1143 dimension-six $CP$-odd operators containing at least a pair of fermions can be reduced. With 14 $U(1)$ flavour symmetries, they are reduced to 10 operators when all light fermions are deemed massless (top excluded), and to 17 operators with 13 $U(1)$ flavour symmetries for massive top and bottom quarks. The leading effects from these operators in the $1/\Lambda$ expansion are in the modulation of their interference with the SM amplitudes. $CP$-even observables like the cross-section are only sensitive to the square contributions at $1/\Lambda^{4}$ so they receive only subleading contributions. That is why a review on the $CP$ sensitive observables is presented. We note that triple products are usually present or equivalent to direct observables.

As seen in Chapter~\ref{chap:smeftcpv}, we give a comparative study of $CP$ observables in diboson production to find the best one. In $W^{\pm}Z$, the efficiency of the best observable, $p_{\perp}(p_e, p_{\sum})$, is 50\% for $\mathcal{O}_{\widetilde{W}WW}$ and 70\% for $\mathcal{O}_{\varphi \widetilde{W} B}$. In $W^{\pm} \gamma$, there is not one best observable we found for both channels despite $p_{\perp}(p_e, p_{\sum})$ offering a middle ground option. For $W^+ \gamma$, $p_{\perp}(p_e, p_{\gamma})$ gives the largest asymmetries for $\mathcal{O}_{\varphi \widetilde{W} B}$ while $\Delta \phi_{e \gamma}$ gives the largest asymmetries in all the other cases. For $W^{\pm} \gamma$, the efficiencies are all above 80\%. Additionally, the asymmetries based on $\sin{\phi_{WZ}}$ show good results for $\mathcal{O}_{\widetilde{W}WW}$ in $W^{\pm} Z$, especially in high-energy bins where they become more efficient than our triple product. However, this observable is almost blind to contributions from $\mathcal{O}_{\varphi \widetilde{W} B}$. Therefore, we could disentangle the two operators in this process by measuring both $\sin{\phi_{WZ}}$ and $p_{\perp}(p_e, p_{\sum})$. Since the direction of the initial quarks is very influential to build effective asymmetries, leptonic colliders, even if their centre-of-mass energy is lower, could have a good sensitivity as the direction of the initial leptons is known.

Our results can be improved in several ways. First, we have considered only one leptonic channel and adding the other leptonic channels will increase the statistics. For $W^{\pm} Z$, our observable is independent of the charge of the Z decay product and therefore suggests that it could be used also for its hadronic decay. Secondly, all the four processes can be combined in a more inclusive fit. Thirdly, NLO corrections increase the SM cross-sections by roughly a factor 2. If this is true also for the SMEFT contributions to our observables, this could further enhance the sensitivity. Moreover, for the large contribution with an extra radiation, the jet could be used to better approximate the quark direction. Finally, the cross-sections are sufficiently large either to use differential asymmetries or to cut the phase space in order to improve the sensitivities.

In Chapter~\ref{chap:fourtops}, two simplified models are presented where the SM is extended with heavy top-philic scalars, the singlet $S_1$ and the octet $S_8$. Their coupling to the top quark allows us to generate a four-top final state via associated production and $S_8$-pair production. The NLO cross-sections of the complete $pp \to t\overline{t} t \overline{t}$ processes are computed for the first time with two custom UFO models. The calculations necessitated selecting NP loop corrections involving $S_8$ to the pure QCD tree-level diagram to cancel the UV divergences and NP loop corrections with $S_1$ to remove IR divergences due to soft gluon emission off the top quark. NLO corrections significantly increase the cross-section, up to 70\% for some parameter values. Thus, their inclusion is necessary in bump searches. The $t\overline{t}+$jets process is the main background and receives large contributions from soft gluon emission and initial state radiation. Thus, it is generated with MLM matching and merging procedure of matrix elements.

Then, we investigate the fully hadronic, the leptonic and the same-sign dilepton decay channels. The four-top system is reconstructed from the particle momenta at the detector level by using a validated ATLAS-like simulation. In particular, when the top quark decays hadronically, current top-tagging algorithms now allow the selection of large AK10 jets with better efficiency. For its leptonic decay, we assume that the $W$ boson decays on-shell and the neutrino momentum is reconstructed. The goal is to estimate the invariant mass of top pairs. The four reconstructed objects are paired depending on the targeted topology. For octet-pair production, we select the pair minimising their invariant mass difference and impose a hard cut on the one with smaller invariant mass to suppress the background. For the associated production of a singlet, the top pair with the largest $p_T$ is chosen.

The invariant mass distribution of the selected top pairing shows a localized increase from the signal so a search for the two resonances over the backgrounds is applicable. However, the smearing of the peak is large, worse in the singlet analysis compared to the octet one, due to the reconstruction procedure. Masses of the octet below 2 TeV are excluded for moderate values of the coupling with the top. For large $y_{S_8}$ values, sensitivity decreases rapidly. Singlet masses are excluded below 1.5 TeV when $y_{S_1}$ takes moderate values. However, when $y_{S_1}<0.8$, the analysis becomes insensitive to the scalar resonance.

We have demonstrated one example of each complementary use of generic models: asymmetries of triple products in diboson production searching for new $CP$ parameters of an EFT and a new particle search for coloured resonances in four top production. Comparisons of these new approaches to their respective existing analyses have demonstrated improvements in the limits on the model parameters. In the future, efforts to find specific triple products for each process affected by CP-odd SMEFT operators and their inclusion in data analyses should be pursued as was the case for diboson production in ATLAS. These should be complemented with ML techniques which have proven to increase sensitivity in multiple processes. However, keeping the interpretability of the observables is important, even at the cost of sensitivity. ML approaches should complement the work of particle physics, not substitute them. Similarly, limit dictionaries on the many BSM resonances from the simplified model dictionaries can be created, not only from the matching relations to SMEFT operators, but also from dedicated studies, with ML potentially improving the limits. Lastly, in a broader application to particle physics, serendipity should not be underestimated. If particle physicists are able to continue testing, failing and ultimately finding new theoretical tools and creative analyses, we will not be short of opportunities to uncover the (last?) pieces beyond the current SM still playing hide-and-seek in the theoretical landscape.

\begin{comment}

We further present in this article the first complete next-to-leading order (NLO) projections for the signal, offering a more realistic estimate of the experimental sensitivity to BSM-induced four-top production. The signal predictions are obtained using custom UFO~\cite{Degrande:2011ua, Darme:2023jdn} models for simplified top-philic scenarios, developed with the {\sc MoGRe} framework~\cite{Frixione:2019fxg} and recent extensions of \fr~\cite{Christensen:2009jx, Alloul:2013bka, Degrande:2014vpa, Frixione:2019fxg}. To our knowledge, this constitutes the first theoretical NLO estimate of new physics processes leading to four-top final states including both resonant and non-resonant contributions. We find that the NLO cross sections can exceed the leading-order (LO) predictions by up to 75\% in certain benchmark scenarios, significantly enhancing the projected reach of future LHC analyses. This highlights the importance of consistent NLO modelling when forecasting the sensitivity of multi-top searches. For the background, we place particular emphasis on the precise simulation of the $t\bar{t} + \text{jets}$ QCD background, which remains the dominant contribution. To accurately model it, we implement a full matching and merging scheme with multi-leg LO matrix elements, ensuring a reliable and consistent description of both the hard scattering and the parton shower processes. This is found essential for faithfully accounting for jets that may fake boosted top quarks. Consequently, appropriate background rejection procedure could be put in place.

\end{comment}

\printindex

\begin{appendices}
\addtocontents{toc}{\protect\setcounter{tocdepth}{-1}}
%!TEX root = main.tex

%%%%%%%%%%%%%%%%%%%%%%%%%%%%%%%%%%%%%%%%%%%%%%%%%%%%
%
%      Chapter 1 :
%
%
%%%%%%%%%%%%%%%%%%%%%%%%%%%%%%%%%%%%%%%%%%%%%%%%%%%

%\chapter{Appendices}
%\label{chap:appendices}
%\pagestyle{fancy}

%\hfill
%\begin{minipage}{10cm}

%{\small\it 
%``No matter how many instances of white swans we may have observed, this does not justify the conclusion that all swans are white.''}

%\hfill {\small Karl Popper, \textit{The Logic of Scientific Discovery}}
%\end{minipage}

%\vspace{0.5cm}

%\section{From a Single Classical Harmonic Oscillator to Quantum Field Theory}
%\label{sec:appendixQFTintro}
%\input{Appendices/QFTintro.tex}

%\section{Distinguishing Bosons and Fermions}
%\label{sec:appendixBosonsFermions}
%\input{Appendices/bosonsfermions.tex}

%\chapter{General Expression of the Higgs Field}
%\label{sec:appendixRGaugeHiggs}
%\pagestyle{fancy}
%\input{Appendices/HiggsR.tex}

%\chapter{\texorpdfstring{$CP$}~ transformations of Fields}
%\label{sec:appendixCPTransformations}
%\pagestyle{fancy}
%\input{Appendices/CPtransformations.tex}

%\chapter{Reduction Formulas for the Warsaw Basis}
%\label{sec:appendixReductionFormulas}
%\pagestyle{fancy}
%\input{Appendices/ReductionFormulas.tex}

\chapter{Asymmetries in WZ}
\label{sec:appendixTripleProducts}
\pagestyle{fancy}

\section{Asymmetry Deterioration}
\label{sec:appendixAsymmetryDeterioration}

\begin{figure}[h!]
\centering
\includegraphics[width=\textwidth]{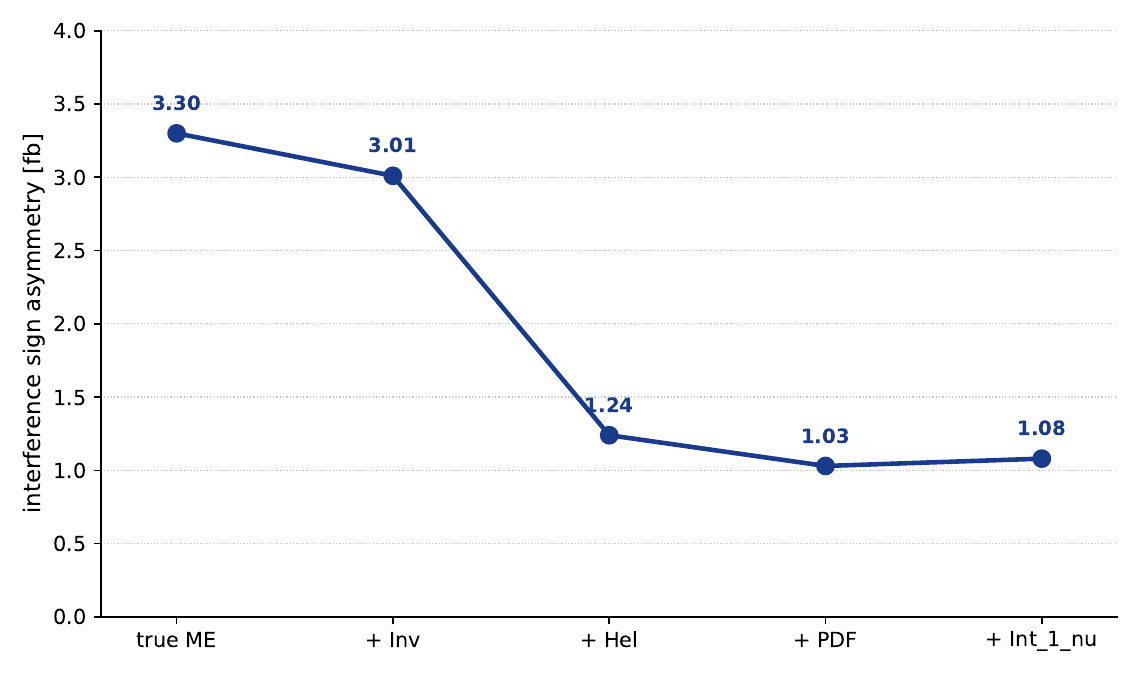}
\caption{ Asymmetry computed from the accumulation of unobservable effects in the interference between the SM amplitude of $pp \to W^+ Z$ and the single insertion of $\OWWW$. The first value corresponds to the theoretical asymmetry $\sigma^{|int|}$ in Table~\ref{tab:xsecWZATLAS} then the inversion of the initial partons, the sum over helicities, the PDFs of all possible initial partons and the integral over possible neutrino momenta are consecutively taken into account to reach the best measurable asymmetry $\sigma^{meas}$ .}
\end{figure}

\newpage

\section{Triple Products in $W^- Z$}
\label{sec:appendixTripleProduct}

\begin{table}[h]
    \centering
    \begin{tabular}{c|c|c}
        Triple products configurations & $\mathcal{O}_{\widetilde{W}WW}$ & $\mathcal{O}_{\varphi\widetilde{W}B}$ \\
    \hline
        $\left(\vec{p}_q, \vec{p}_Z, \vec{p}_e \right)$ & -1.612(4)  & -0.3888(7) \\
        $\left(\vec{p}_q, \vec{p}_Z, \vec{p}_{\mu^-} \right)$ & -0.184(4)  & -0.0271(7) \\
    \hline
        $\left([0,0,p_{\sum}^z], \vec{p}_Z, \vec{p}_e \right)$ & -0.628(4)  & -0.1207(7) \\
    \hline
        $\left(\vec{p}_W, \vec{p}_{\mu^-}, \vec{p}_e \right)$ & 0.535(4)  & 0.0965(7) \\
        $\left(\vec{p}_W, \vec{p}_{\mu^+}, \vec{p}_e \right)$ & 0.511(4)  & 0.1009(7) \\
        $\left([0,0,p_W^z], \vec{p}_e, \vec{p}_{\mu^-} \right)$ & -0.227(4)  & -0.0594(7)  \\
        $\left(\vec{p}_W, \vec{p}_{\mu^-}, \vec{p}_{\mu^+} \right)$ & -0.080(4)  & -0.0110(7) \\
        $\left([0,0,p_{\sum}^z], \vec{p}_W, \vec{p}_Z \right)$ & -0.045(4)  & -0.0086(7) \\
        $\left([0,0,p_e^z], \vec{p}_{\mu^-}, \vec{p}_W \right)$ & 0.028(4)  & 0.0061(7) \\
    \hline
        $\left(\vec{p}_e, \vec{p}_{\mu^-}, \vec{p}_{\mu^+} \right)$ & -0.025(4)  & -0.004(7) \\
        $\left([0,0,p_e^z], \vec{p}_{\mu^-}, \vec{p}_{\mu^+} \right)$ & -0.029(4)  & -0.0061(7) \\
        $\left([0,0,p_{\mu^-}^z], \vec{p}_e, \vec{p}_{\mu^+} \right)$ & -0.213(4)  & -0.0244(7) \\
        $\left([0,0,p_{\mu^+}^z], \vec{p}_{\mu^-}, \vec{p}_e \right)$ & 0.252(4)  & 0.0327(7) \\
        $\left([0,0,p_{\sum}^z], \vec{p}_e+\vec{p}_{\mu^-}, \vec{p}_{\mu^+} \right)$ & -0.362(4) & -0.0582(7) \\
        $\left([0,0,p_{\sum}^z], \vec{p}_e+\vec{p}_{\mu^+}, \vec{p}_{\mu^-} \right)$ & -0.300(4)  & -0.0481(7) \\
        $\left([0,0,p_{\sum}^z], \vec{p}_e-\vec{p}_{\mu^-}, \vec{p}_{\mu^+} \right)$ & -0.047(4)  & -0.0097(7) \\
        $\left([0,0,p_{\sum}^z], \vec{p}_e-\vec{p}_{\mu^+}, \vec{p}_{\mu^-} \right)$ & -0.160(4)  & -0.0279(7) 
    \end{tabular}
    \caption{Table of the asymmetries measured with respect to different configurations of the triple product observable in $W^+Z$ production. The first column displays the momenta chosen to build the observable and the last two columns show the asymmetries in fb obtained for each operator. The numerical errors are put in brackets.}
    \label{tab:comparison triple products}
\end{table}

\begin{comment}
\section{Asymmetry and cross-section distributions}

\begin{figure}%[p]
	\centering
	\includegraphics[width=\textwidth]{figures/asym_WmZ_OWWW_comE.pdf}
	\hspace{3cm}
	\includegraphics[width=\textwidth]{figures/asym_WpZ_OWWW_comE.pdf}
	\caption{ \JT{Same as Figure~\ref{fig:ECMZWOWWW} but with the dimension-six square differential cross-section included in orange. } }
	\label{fig:ECMZWOWWWwithSquare}
\end{figure}

\begin{figure}%[p]
	\centering
	\includegraphics[width=\textwidth]{figures/asym_WmZ_OphiWB_comE.pdf}
	\hspace{3cm}
	\includegraphics[width=\textwidth]{figures/asym_WpZ_OphiWB_comE.pdf}
	\caption{ \JT{Same as Figure~\ref{fig:ECMZWOphiWB} but with the dimension-six square differential cross-section included in orange. } }
	\label{fig:ECMZWOphiWBithSquare}
\end{figure}

\end{comment}

\chapter{Four Top Signal}
\label{sec:appendixFourTopSignal}
\pagestyle{fancy}

\section{NP cross-section with QCD interaction only}

We first compare the two LO BSM cross-sections with the octet and singlet simplified models presented in Section~\ref{subsec:FourTopLags} to SM four-top production at LHC. Both LO four-top cross-sections are generated with the same syntax in \MG :
\begin{lstlisting}[language=Python]
generate p p > t t~ t t~ QED<=0 QCD<=2 NP<=2 
\end{lstlisting}
Throughout this Appendix, two sets of 30 different data points, one for each BSM resonance, are computed with masses between 1 TeV to 3 TeV (separated by 500 GeV steps) and coupling values that range from 0.5 to 3 (separated by 0.5 steps). The event simulation parameters are kept identical to the ones in Section~\ref{subsec:EventGen}.

\begin{figure}[p]
	\centering
	\includegraphics[width=\textwidth]{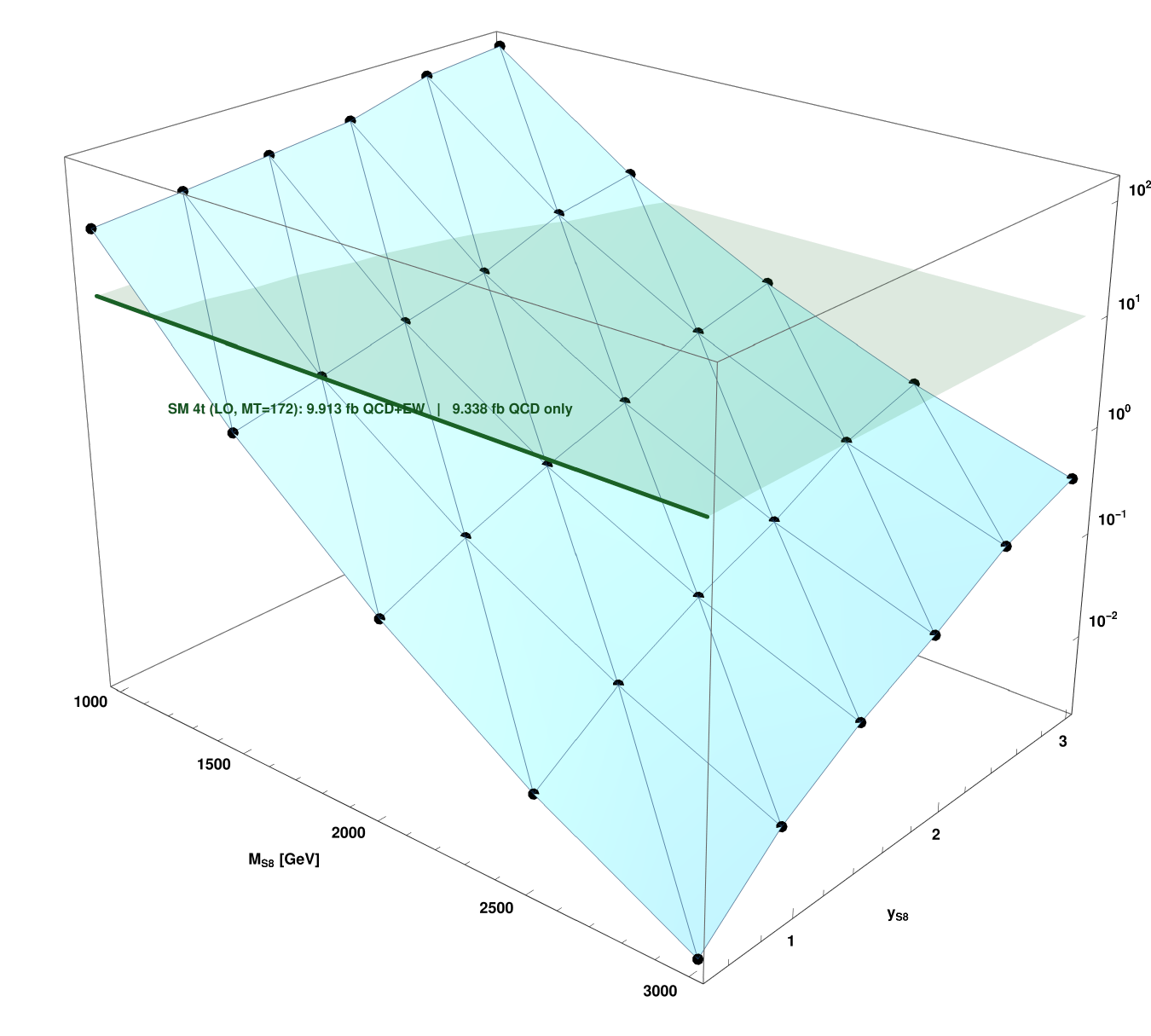}
	\includegraphics[width=\textwidth]{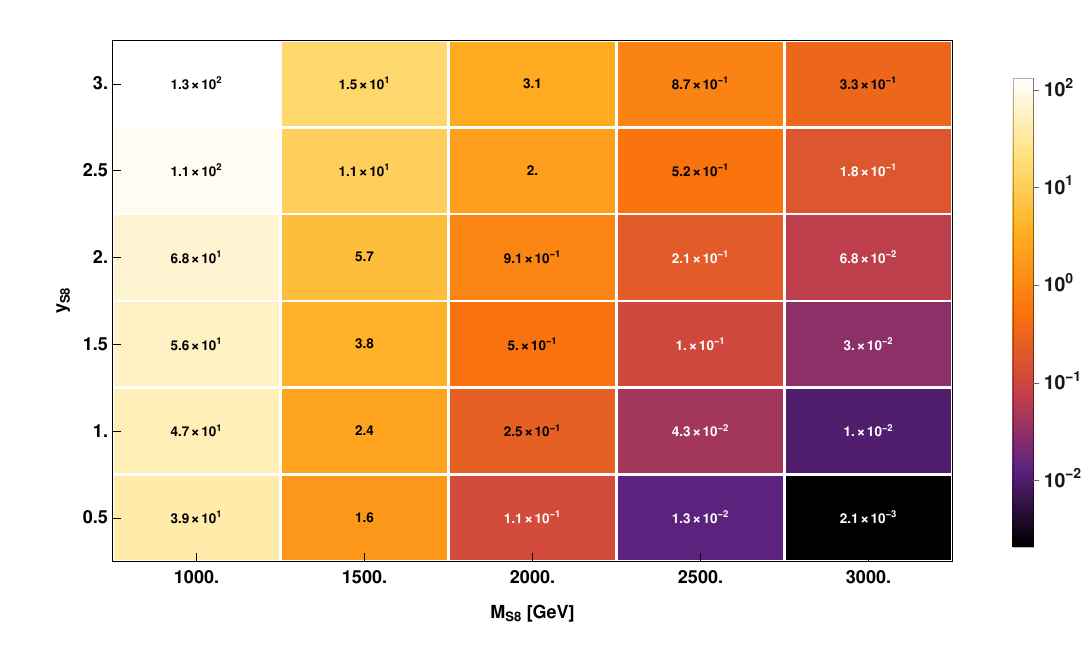}
	\caption{ Three-dimensional representation of the octet cross-section without electroweak interactions in light blue and of the SM cross-sections as a green plane since the SM cross-sections with and without electroweak processes cannot be distinguished on this plot (top). The two-dimensional representation of the BSM cross-section is also provided (bottom).  }
	\label{fig:OctetNP}
\end{figure}

\begin{figure}[p]
	\centering
	\includegraphics[width=\textwidth]{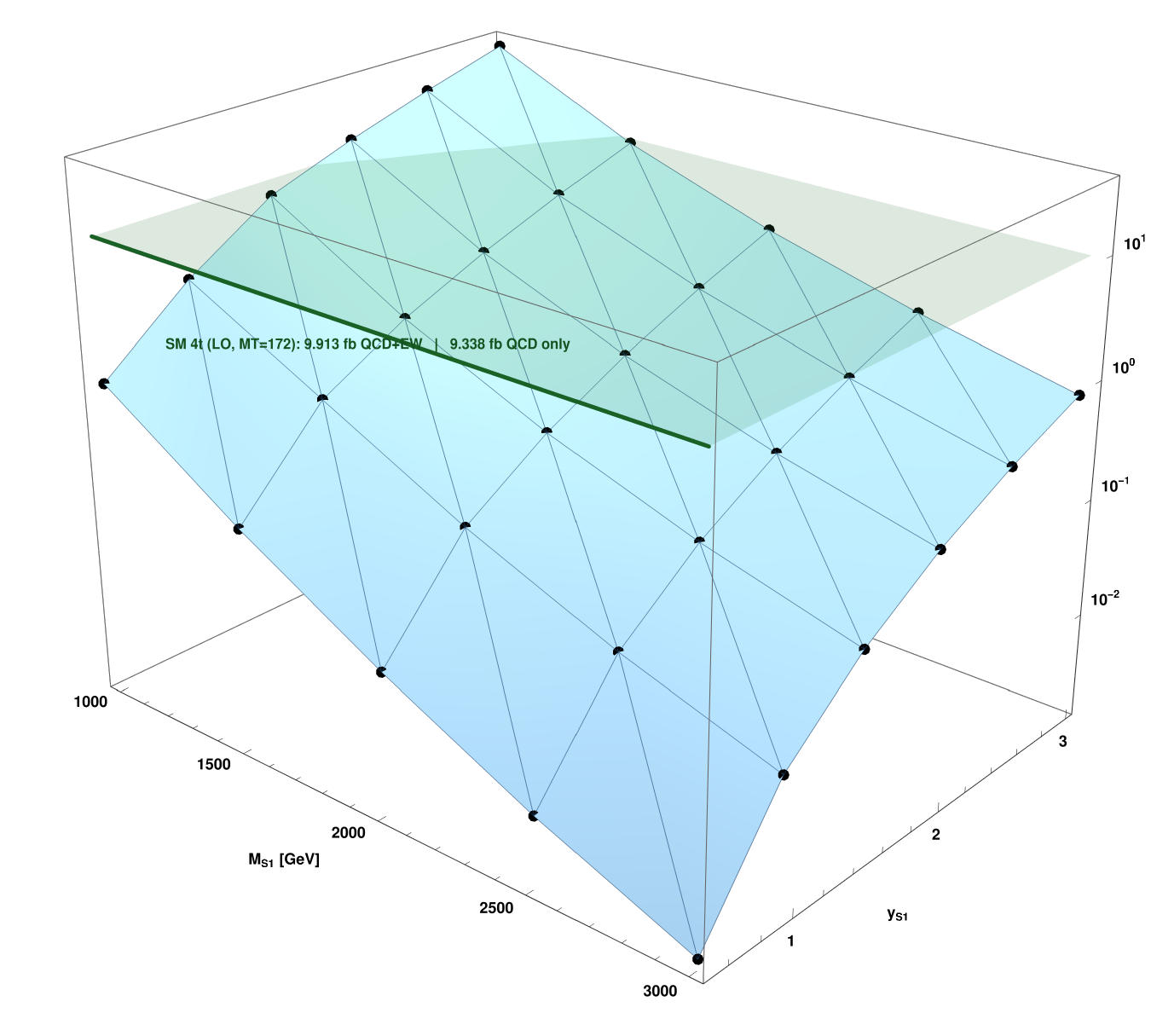}
	\includegraphics[width=\textwidth]{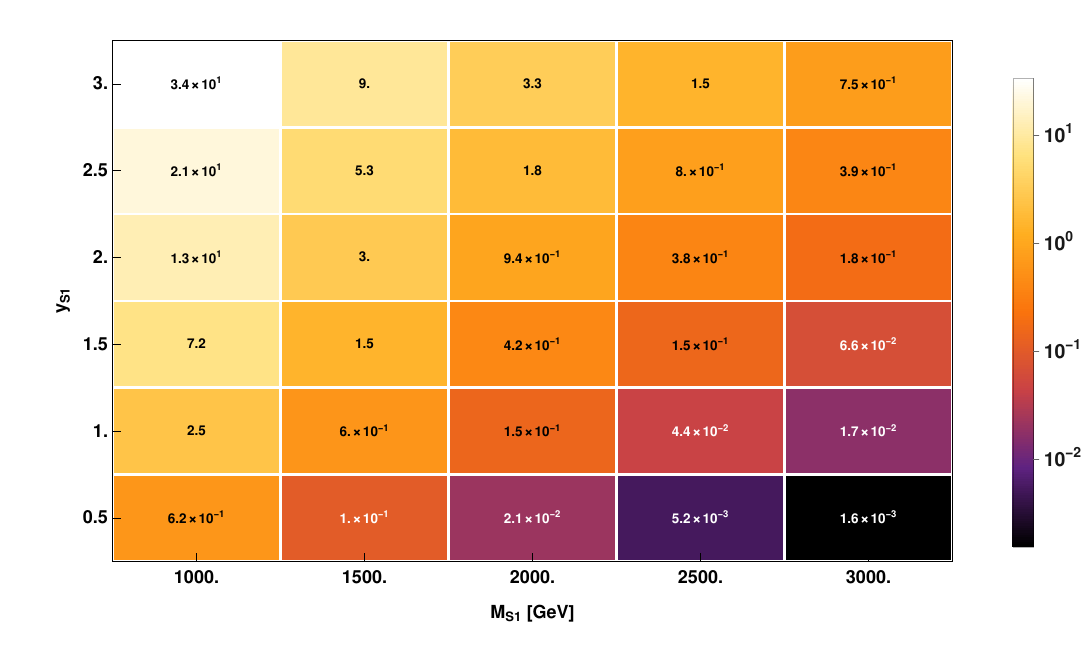}
	\caption{ Same as Figure~\ref{fig:OctetNP} for the singlet.} 
	\label{fig:SingletNP}
\end{figure}

Figures~\ref{fig:OctetNP} and \ref{fig:SingletNP} show the 3-dimensional and 2-dimensional representations of the BSM cross-section of the octet and the singlet respectively. The SM cross-section is displayed on the 3D plots as a flat plane for comparison.

%\JT{We see that the octet cross-section is always larger than the SM cross-section for masses smaller than 1.5 TeV with a mild dependence on the coupling with the top quark. Then, the cross-section rapidly decreases with larger masses, being overcome by the SM cross-section beyond 1.5 TeV, and the dependence on the strength of the coupling gets stronger. The decreases is less steep with larger coupling reaching $0.33$ fb from $130$ fb for $y_{\oct}=3.0$ but only $2.1~ 10 ^{-3}$ fb from $39$ fb for $y_{\oct}=0.5$. }

%\JT{Regarding the singlet, the BSM cross-section is larger than the SM one in a narrower region of the parameter space compared to octet, only for $y_{\sing} \gtrsim 2$ and $M_{\sing} < 1.5$ TeV, and the dependence on the coupling appears across the parameter space. The decrease of the cross-section with respect to mass is milder than in the octet especially at small value of the coupling. The cross-section loses two orders of magnitude at $y_{\sing}=3.0$ compared to three for the octet and three order of magnitude at $y_{\sing}=0.5$ instead of five for the octet. This mainly explained by the much larger values of the octet cross-section at small masses compared to the singlet. }

\section{Inclusion of QED interaction in signal}

In the SM, electroweak interactions can be safely neglected as they increase the pure-QCD cross-section by only 6\%. We that the inclusion of electroweak interactions also give negligible contributions in both octet and singlet cross-sections. The LO four-top cross-sections with the octet and singlet simplified models that includes the electroweak contributions are generated with the following syntax in \MG :
\begin{lstlisting}[language=Python]
generate p p > t t~ t t~ QED<=2 QCD<=2 NP<=2 
\end{lstlisting}
The additional diagrams are the associated production of the BSM resonance with top pair via $Z$ or $\gamma$ s-channel. Since we consider the five-flavour scheme, Higgs interactions are not considered and should be even smaller than electroweak interactions due to the small mass of initial partons.

\begin{figure}[p]
	\centering
	\includegraphics[width=\textwidth]{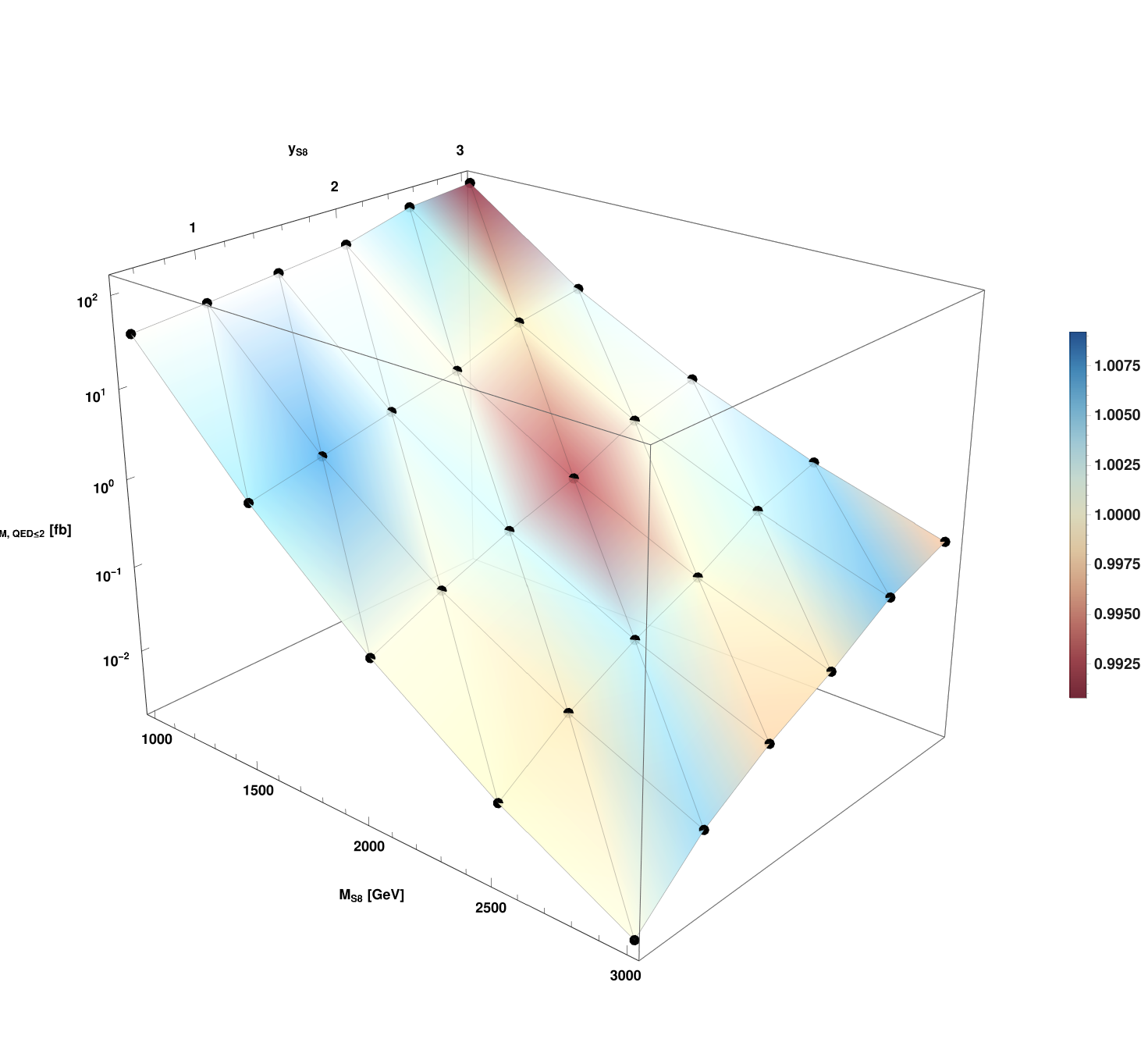}
	\includegraphics[width=\textwidth]{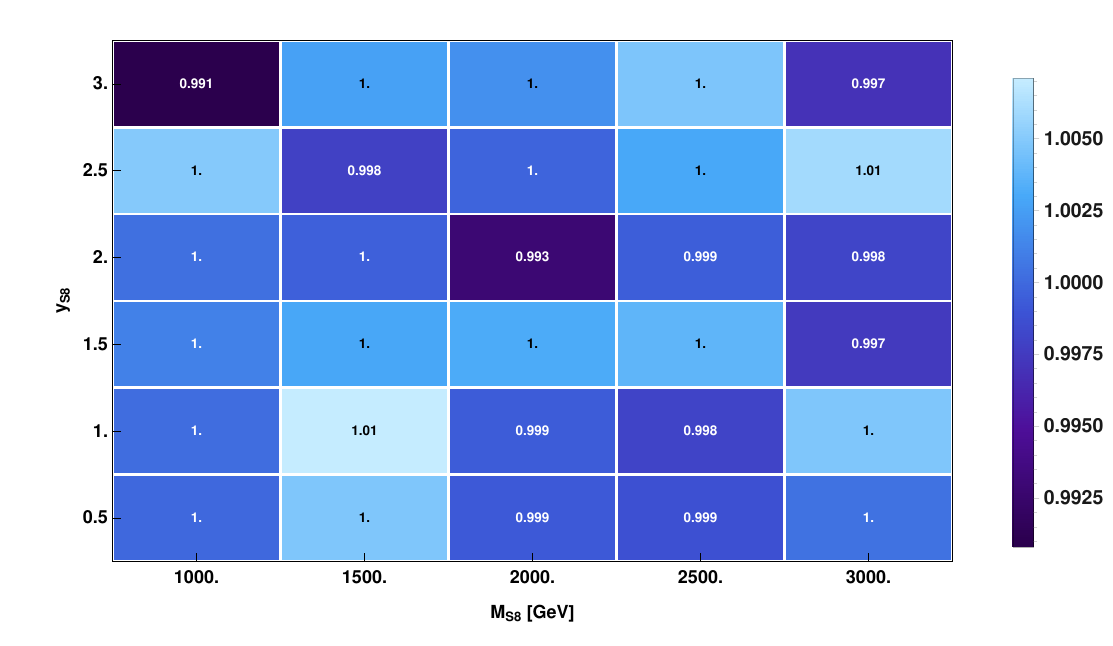}
	\caption{ Three-dimensional representation of the octet cross-section with electroweak interactions (top). The colour legend represents its ratio with the cross-section without electroweak interactions and it is shown in the two-dimensional plot (bottom) for better understanding. } 
	\label{fig:OctetQEDInclusionNP}
\end{figure}

\begin{figure}[p]
	\centering
	\includegraphics[width=\textwidth]{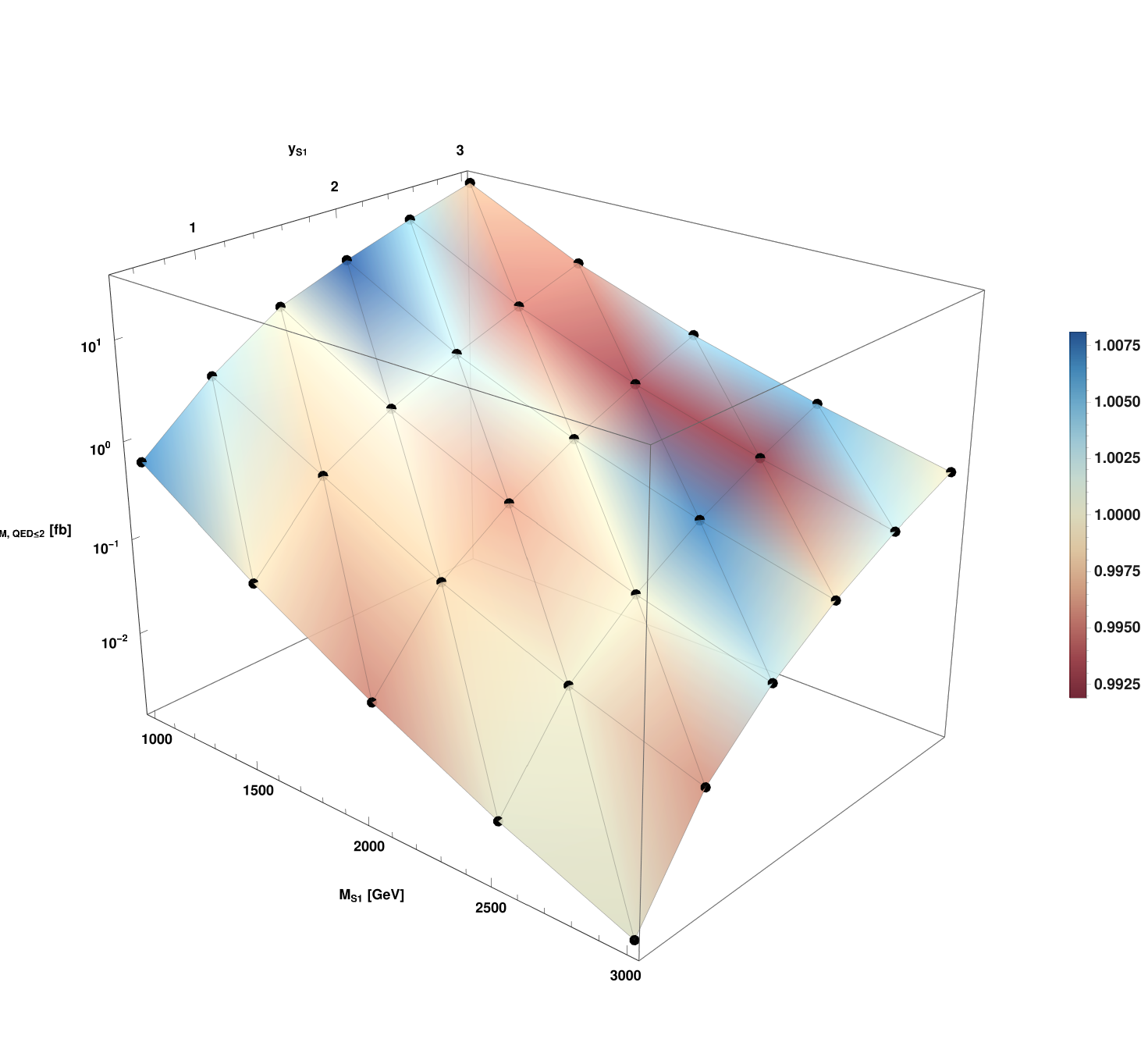}
	\includegraphics[width=\textwidth]{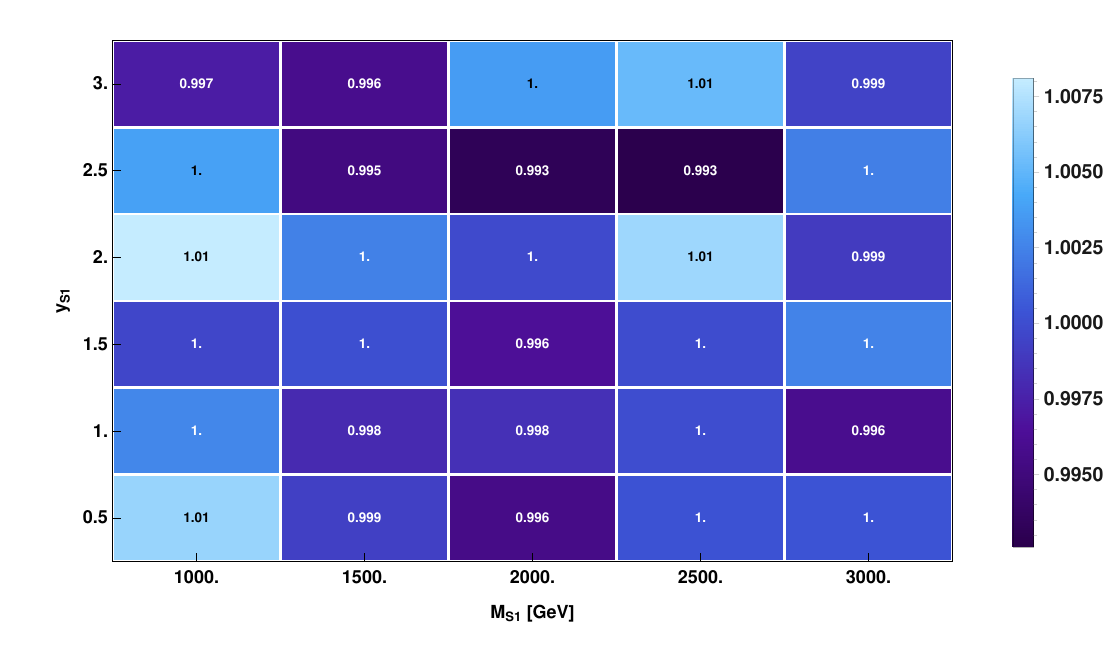}
	\caption{ Same as Figure~\ref{fig:OctetQEDInclusionNP} for the singlet.} 
	\label{fig:SingletQEDInclusionNP}
\end{figure}

Figures~\ref{fig:OctetQEDInclusionNP} and \ref{fig:SingletQEDInclusionNP} show the 3-dimensional representation of the BSM cross-section of the octet and the singlet including electroweak interactions respectively and the 2-dimensional representation of the ratio with the BSM cross-section without the electroweak interactions. The shape of the cross-section plane on the 3D plot is identical to the ones in Figures~\ref{fig:OctetNP} and \ref{fig:SingletNP}. This further confirmed by the 2D plots. The ratios are all close to 1 with a bit more variation for the singlet even though they remain negligible.

%\newpage

\section{QCD-BSM Interference}

\begin{figure}[p]
	\centering
	\includegraphics[width=\textwidth]{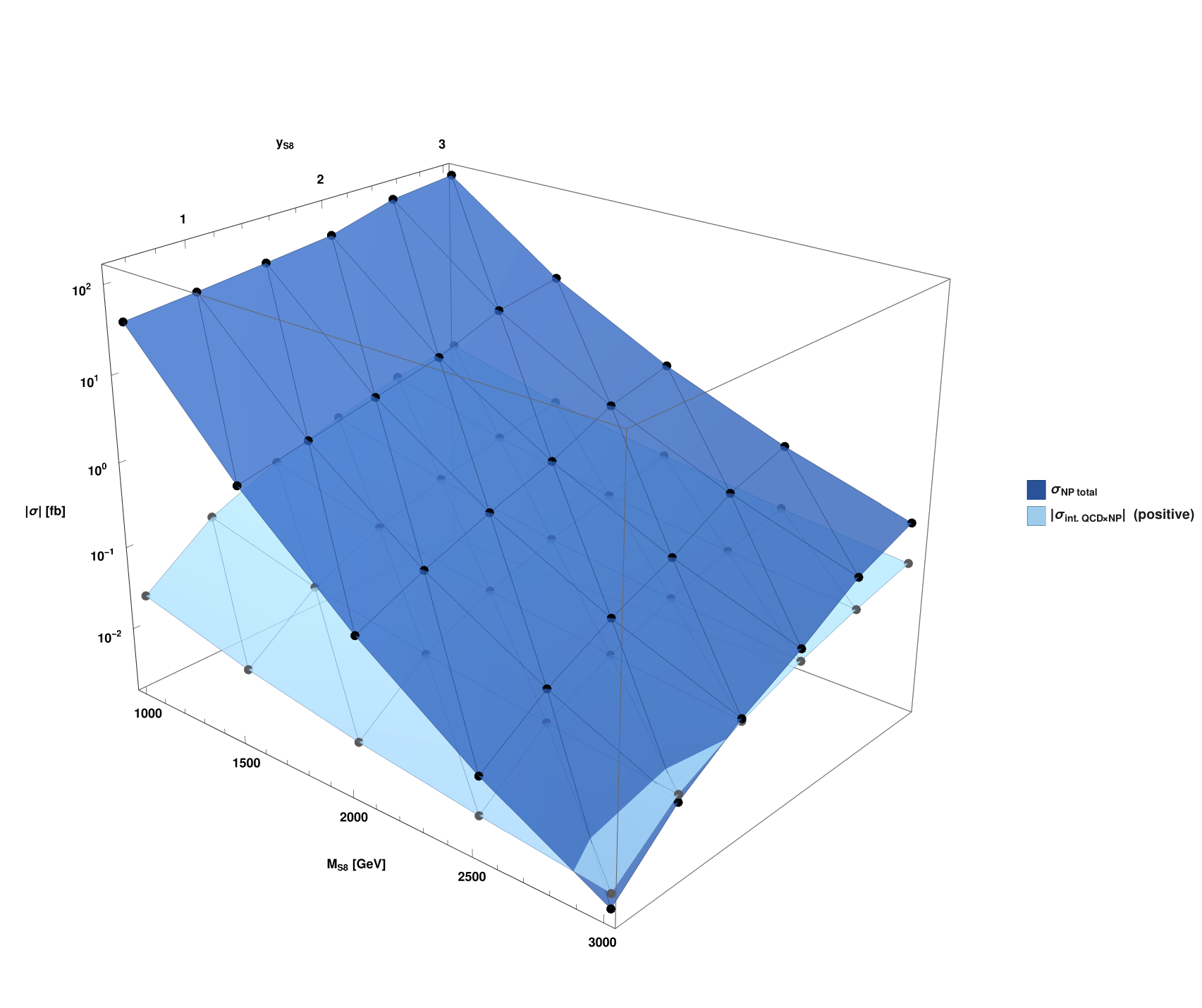}
	\includegraphics[width=\textwidth]{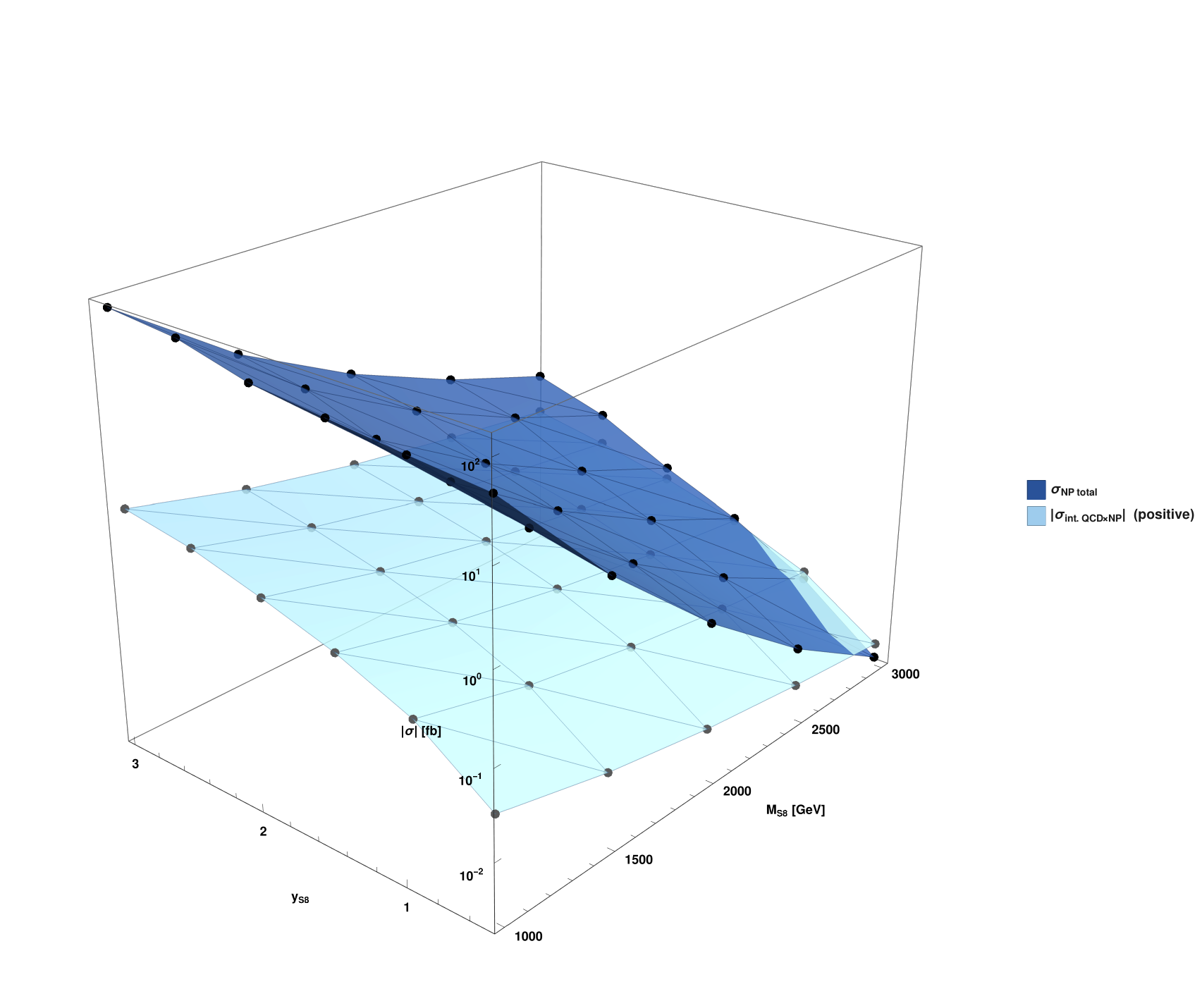}
	\caption{ Three-dimensional representation of the NP octet cross-section (dark blue) and the QCD-NP interference (light blue). }
	\label{fig:OctetInterferenceSM3D}
\end{figure}

The interference between QCD and BSM processes of $\oct$ and $\sing$ are respectively displayed on Figures~\ref{fig:OctetInterferenceSM3D} and \ref{fig:SingletInterferenceSM3D}. The \MG command used in this case was 
\begin{lstlisting}[language=Python]
generate p p > t t~ t t~ QED<=0 QCD^2==4 NP^2==2 
\end{lstlisting}
	
%For the octet case, the interference is positive over the whole parameter space and it is negligible at small values of mass and coupling compared to the BSM cross-section. When $M_{\oct}$ becomes larger, the relative contribution of the interference increases until $M_{\oct}$ reaches 3 TeV where it becomes of the same order as the BSM cross section. At $y_{\oct}=0.5$ and $1.0$ with $M_{\oct}=3$ TeV, the interference overcomes the BSM cross-section. With increasing coupling, the ratio of the interference over the BSM cross-section gets larger at small masses and smaller at large masses. 

%For the singlet case, the interference is negative and overcomes the BSM cross-section from $M_{\sing}=2$ TeV and $y_{\sing}=0.5$ to $M_{\sing}=3$ TeV and $y_{\sing}=2$. Thus, the interference is subleading at small masses and couplings but becomes non-negligible more rapidly than for the octet.  

%\JT{In the large mass regions of parameter spaces, where the interference overcomes the BSM cross-section, an EFT treatment of the signal should be preferred as was suggested in Ref.\cite{Darme:2021gtt}.}
In the large mass regions of parameter spaces, where the interference overcomes the BSM cross-section, the definition of the signal should be updated and include the interference. However, neither the octet or the singlet analyses were able to probe these regions in Figures~\ref{fig:limits_scalarOctet} and \ref{fig:limits_scalarSinglet}. 
An EFT treatment of the signal could also be used instead of the simplified models as was suggested in Ref.\cite{Darme:2021gtt}.

\begin{figure}[p]
	\centering
	\includegraphics[width=\textwidth]{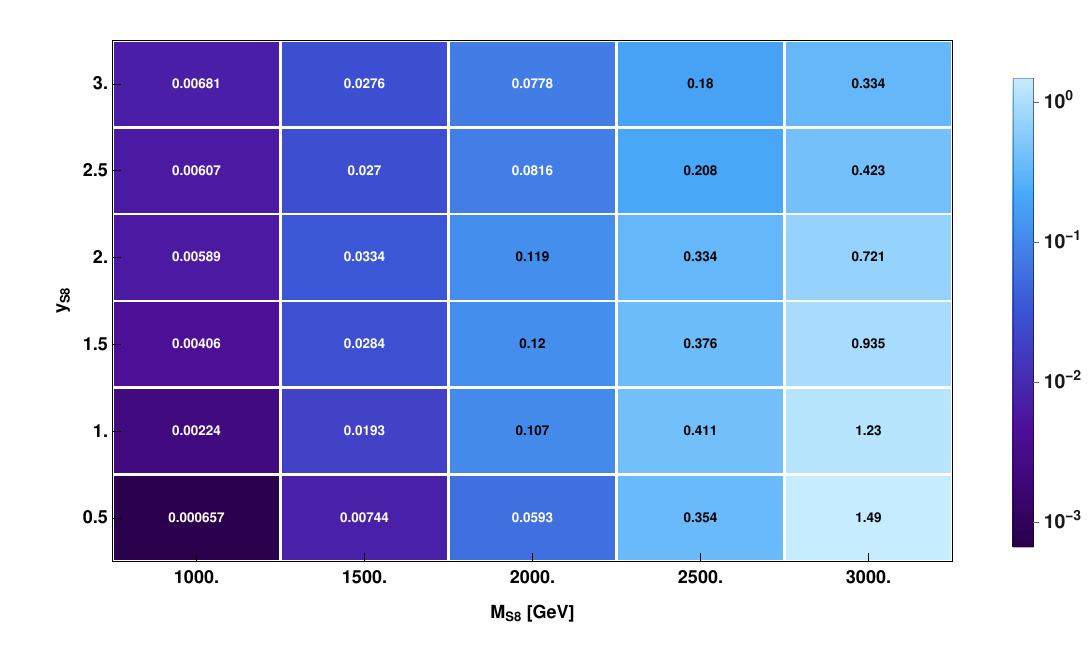}
	\includegraphics[width=\textwidth]{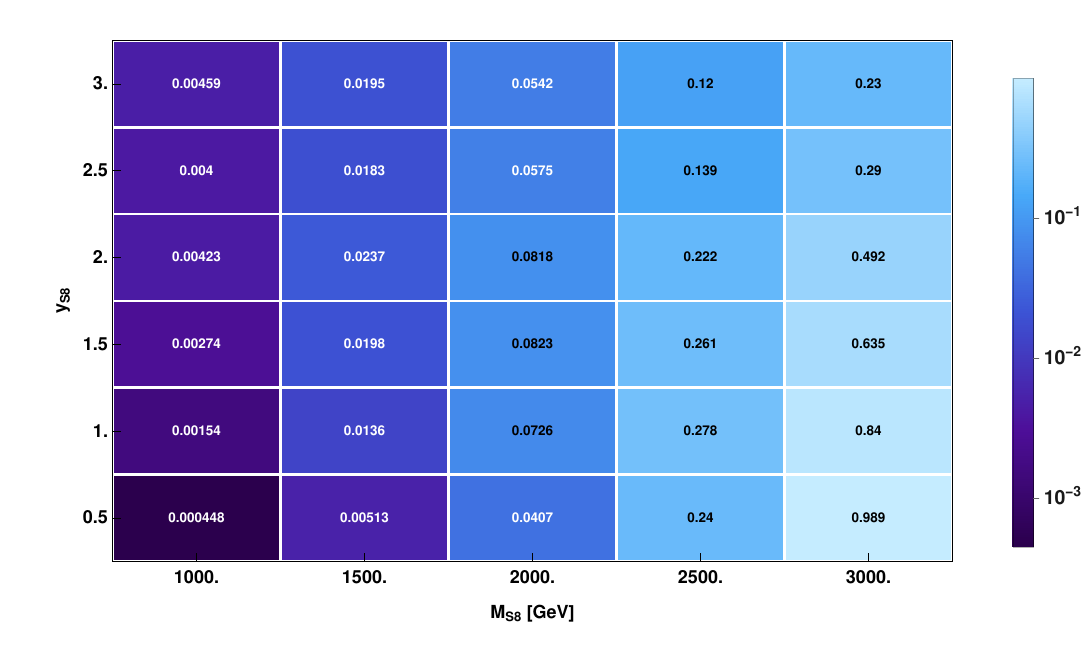}
	\caption{ Two-dimensional representation of the octet QCD-NP interference (top) and its ratio with respect to the NP cross-section.} 
	\label{fig:OctetInterferenceSM2D}
\end{figure}

\begin{figure}[p]
	\centering
	\includegraphics[width=\textwidth]{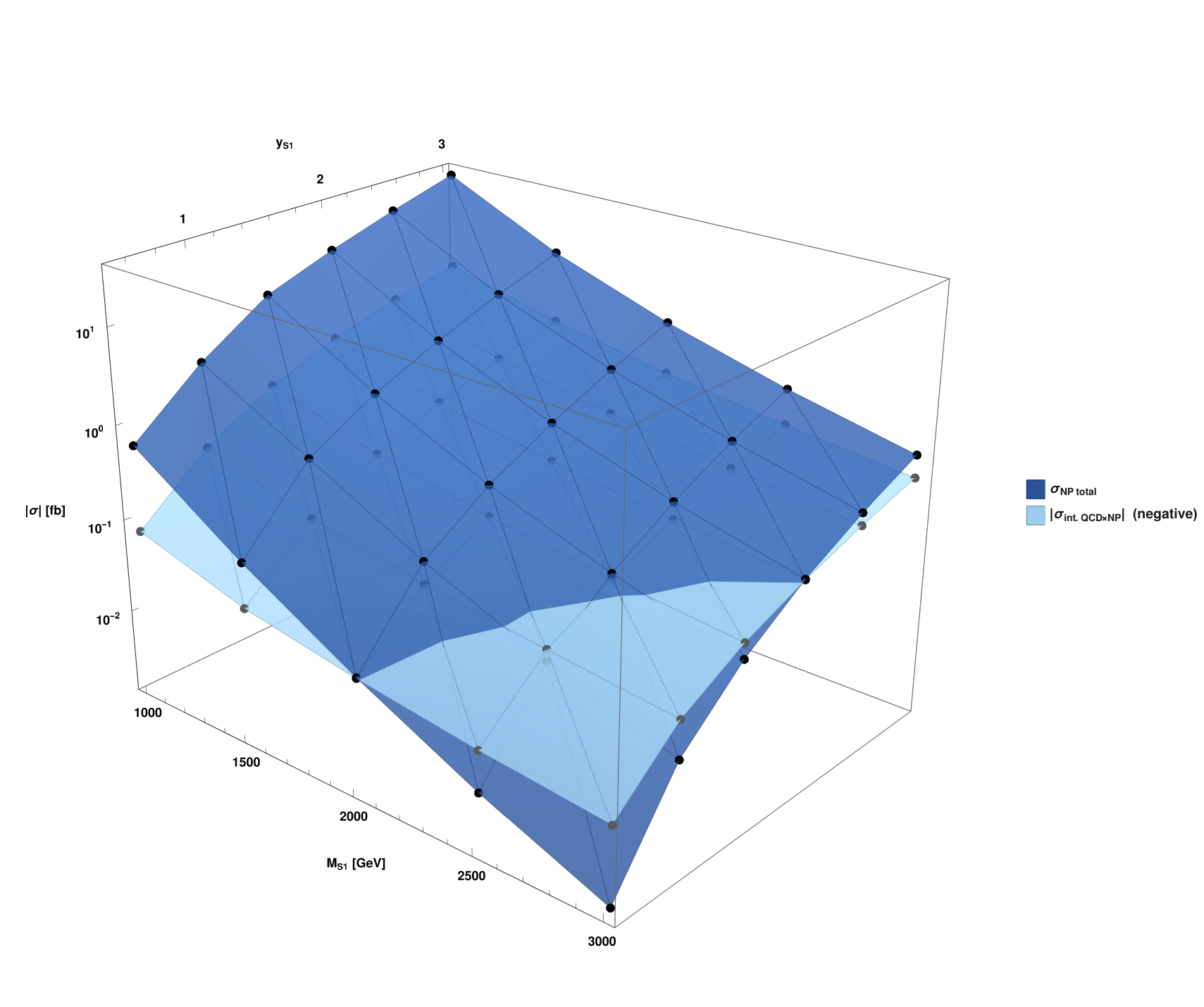}
	\includegraphics[width=\textwidth]{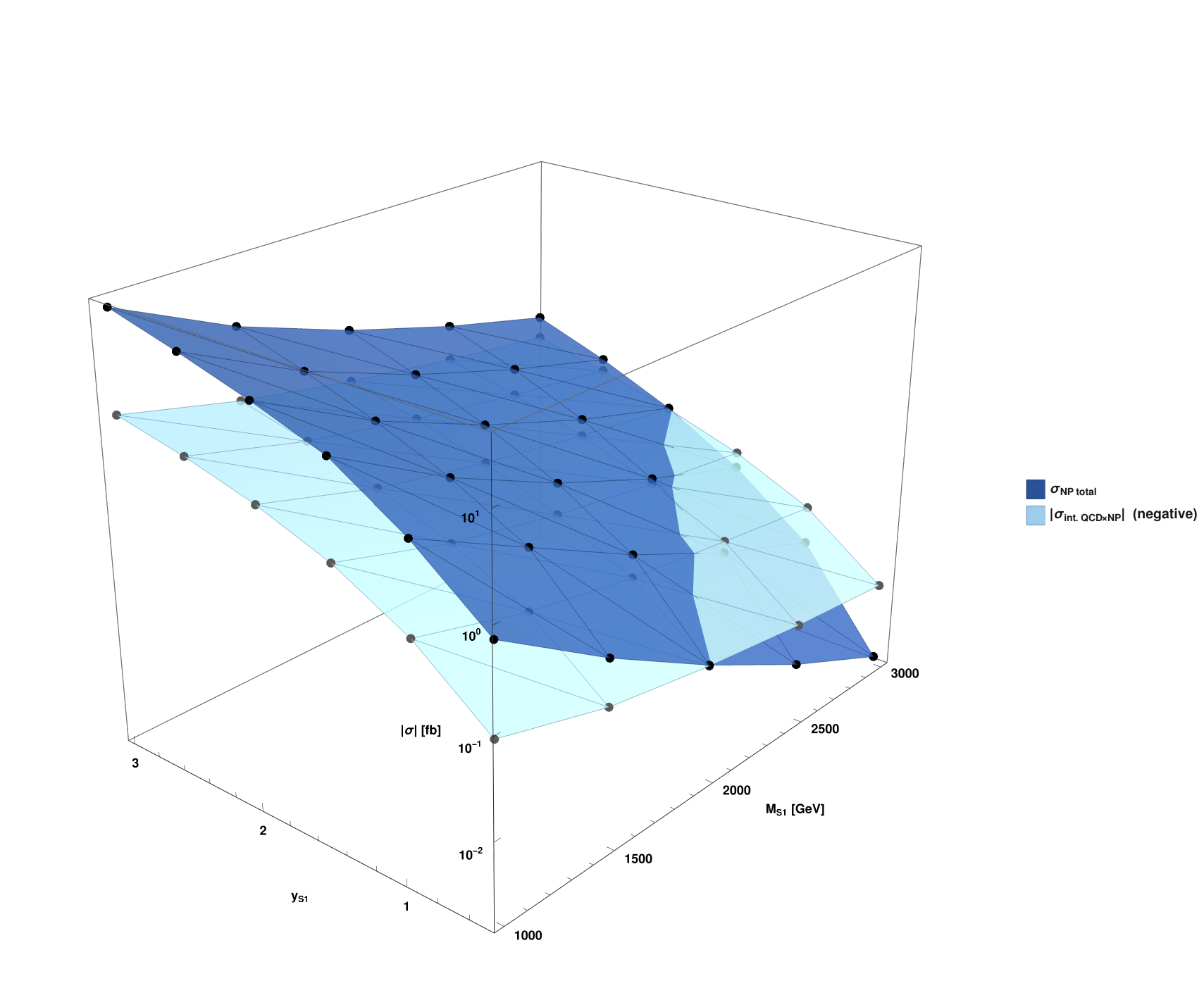}
	\caption{ Same as Figure~\ref{fig:OctetInterferenceSM3D} but for the singlet.} 
	\label{fig:SingletInterferenceSM3D}
\end{figure}

\begin{figure}[p]
\centering
\includegraphics[width=\textwidth]{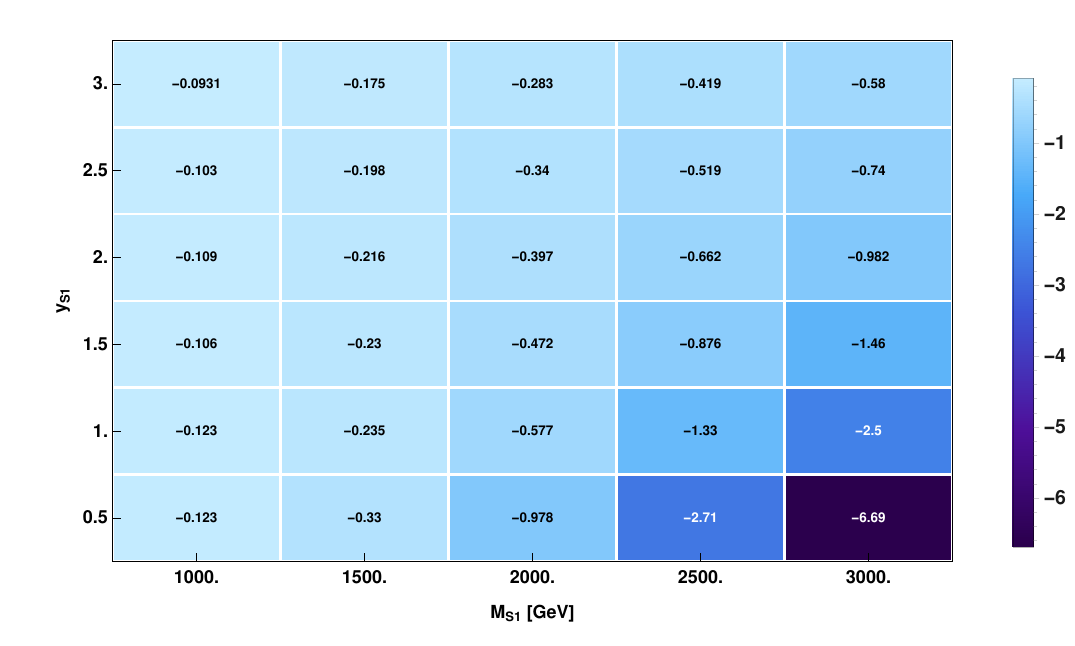}
\includegraphics[width=\textwidth]{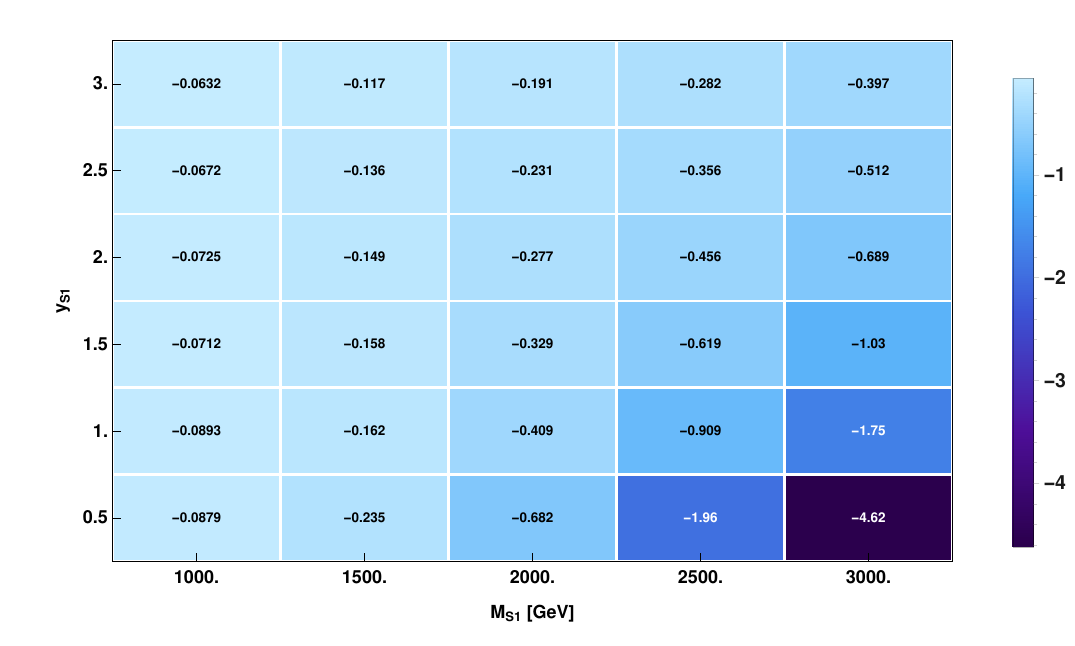}
\caption{ Two dimensional representation of the octet QCD-NP interference (top) and its ratio with respect to the NP cross-section (bottom).} 
\label{fig:SingletInterferenceSM2D}
\end{figure}

%\newpage

\section{Double resonant contributions}

\begin{figure}[p]
	\centering
	\includegraphics[width=0.85\textwidth]{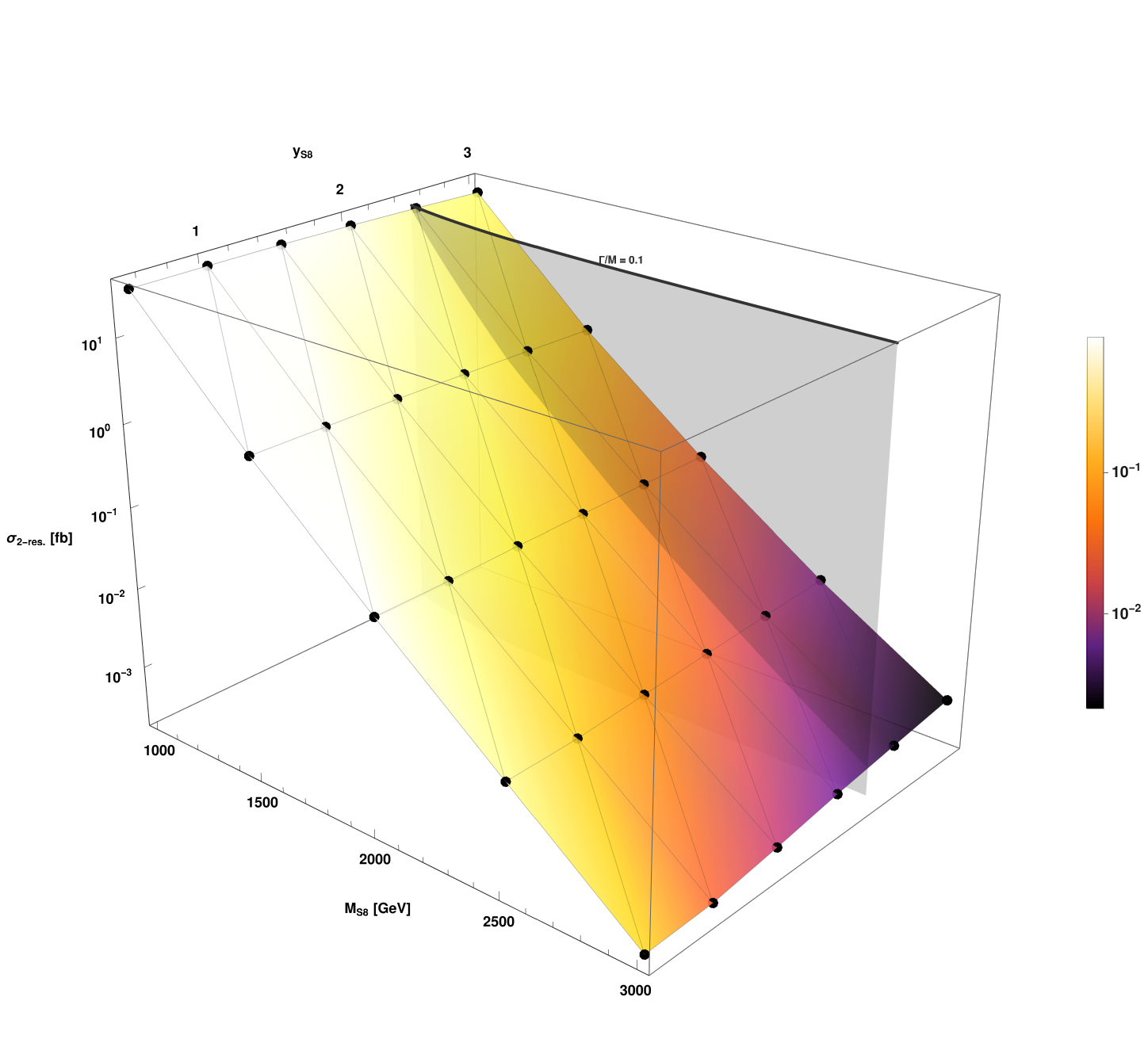}
	\includegraphics[width=0.85\textwidth]{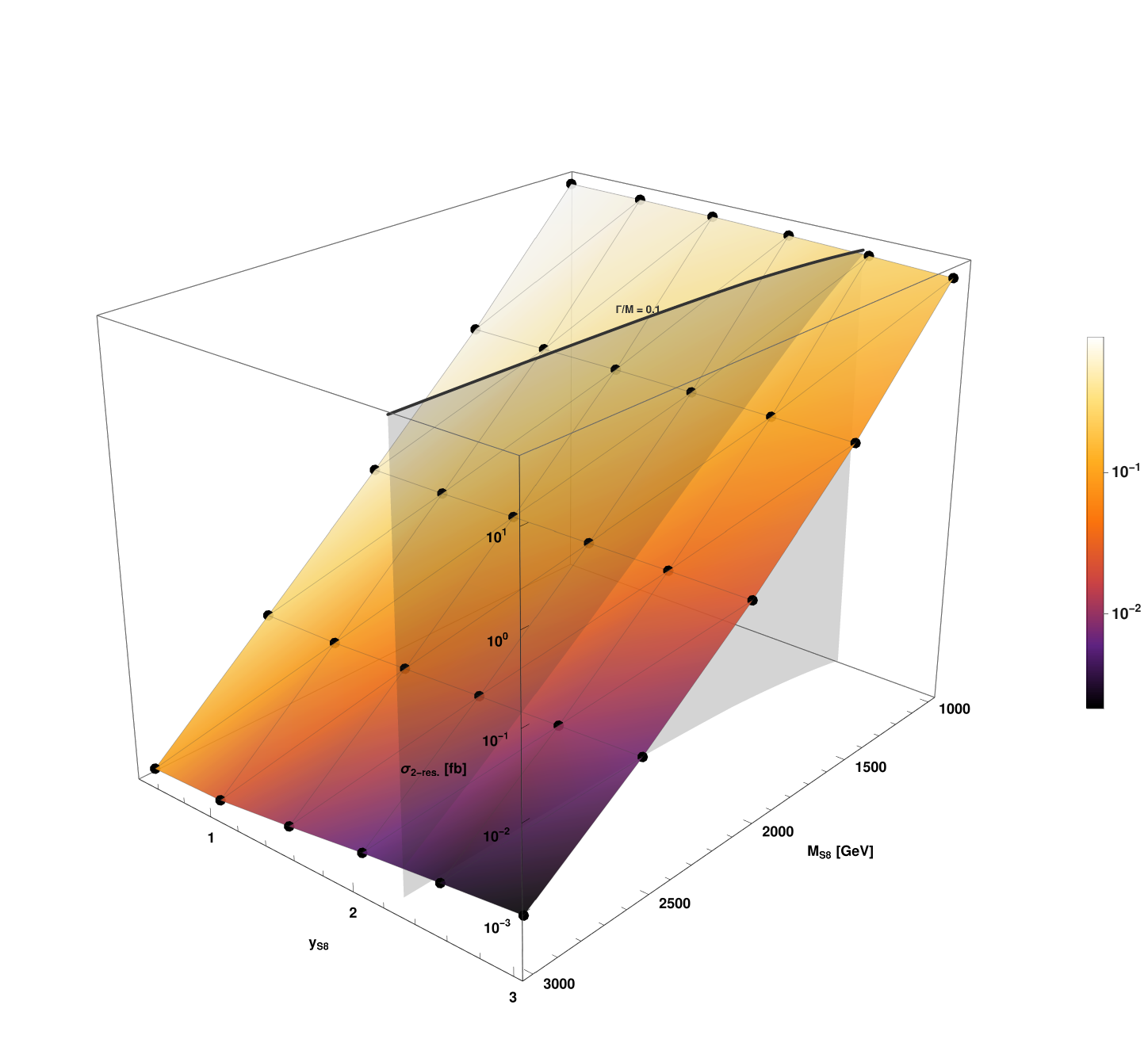}
	\caption{ Three-dimensional representation of the double resonant cross-section with a colour legend showing its ratio with the BSM cross-section. A front view of the plot is presented on the top figure and side view on the figure at the bottom. } 
	\label{fig:Octet2Resonant3D}
\end{figure}

\begin{figure}[p]
	\centering
	\includegraphics[width=\textwidth]{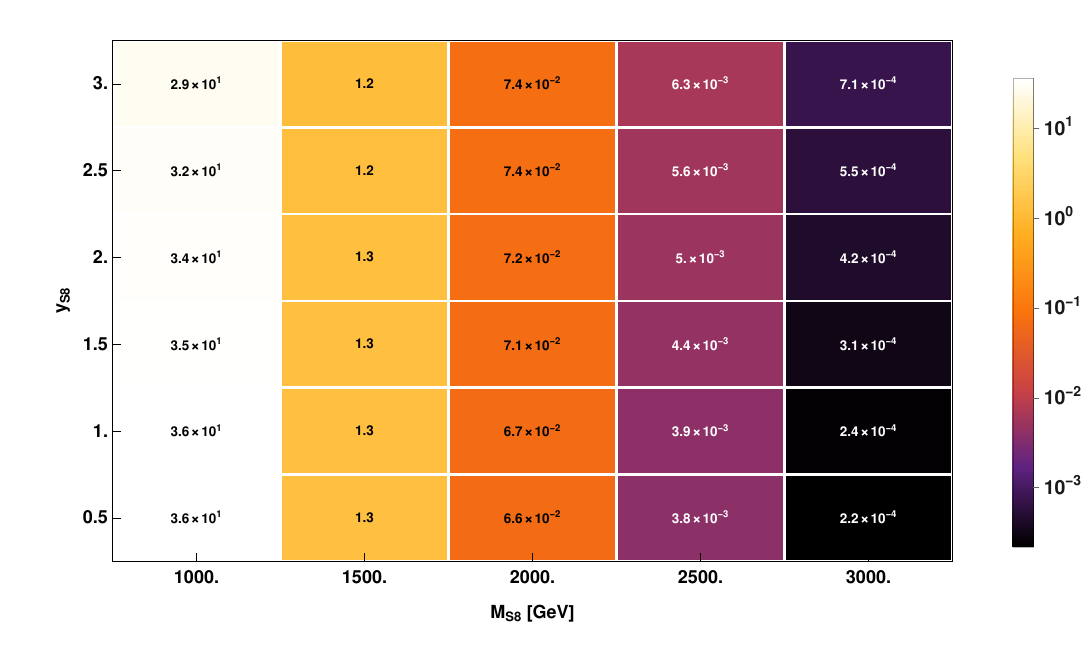}
	\includegraphics[width=\textwidth]{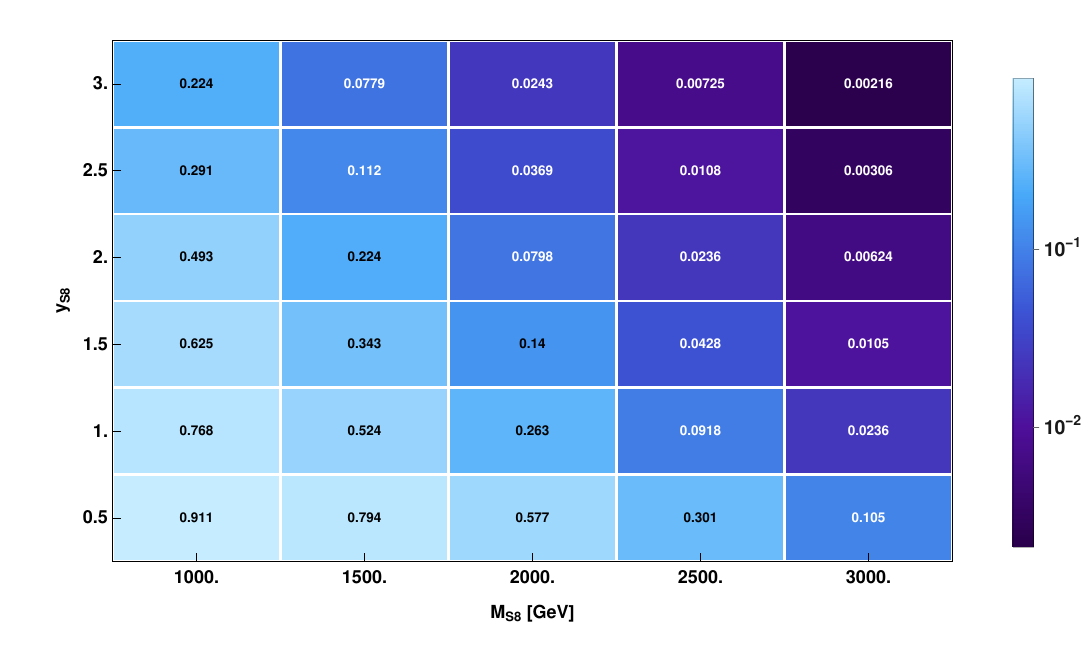}
	\caption{ Two-dimensional representation of the double resonant cross-section (top) and its ratio to the total BSM cross-section (bottom).} 
	\label{fig:Octet2Resonant2D}
\end{figure}

The double resonant production depicted on top-right diagram of Figure~\ref{fig:FeynDiagram} is only present for the octet due to its colour charge. This LO process is generated by \MG~ with the syntax :
\begin{lstlisting}[language=Python]
generate p p > sig8 sig8 QED<=2 QCD<=0 NP<=0 , sig8 > t t~ 
\end{lstlisting}

Figure~\ref{fig:Octet2Resonant3D} shows the 3-dimensional representation of the cross-section for the double resonant process defined above from a front and side views for clarity. An horizontal black plane shows the width over mass ratio set to 1. Figure~\ref{fig:Octet2Resonant2D} displays the double resonant cross-section and its ratio with the BSM cross-section.

%\JT{The double resonant cross-section is independent of the coupling at all masses. This is explained by the branching ratio of the octet to tops set to 1 which makes the double resonant production proportional to $\alpha_S$ only. Once the width starts to increase we can see contributions from the coupling to appear (beyond the width over mass plane).}

%\JT{The contribution of the double resonant production in the total cross-section is maximal at small values of $M_{\oct}$ and $y_{\oct}$. Then, its relative contribution rapidly decreases for increasing mass and coupling where other production channels start to dominate.  }

%\JT{The double resonant production explains the large values of the BSM cross-section compared to the SM cross-section for $M_{\oct}<1.5$ TeV irrespective of the coupling value on Figure~\ref{fig:OctetNP}. }

These contributions drives the limits that are set on the octet parameter space explaining the almost vertical bound on Figure~\ref{fig:limits_scalarOctet} between $y_{\oct}=0.2$ and $y_{\oct}=2$.

%\newpage

\section{Single resonant contributions}

Single resonant production is present for both the octet and the singlet. Due to the different Feynman diagrams that can contribute for each BSM resonance different \MG  syntax must be used for each process. The singlet is the most straightforward:
\begin{lstlisting}[language=Python]
generate p p > t t~ sig1 QED<02 QCD<=2 NP<=1 , sig1 > t t~ 
\end{lstlisting}
This way only the bottom-left Feynman diagram of Figure~\ref{fig:FeynDiagram} is selected and the t-channel is not generated. For the octet case, we must be careful not to let the double resonant diagram contaminate the single resonant production. The single resonant process with the octet is generated by \MG with the syntax :
\begin{lstlisting}[language=Python]
generate p p > t t~ sig8 $\textdollar \textdollar$ sig8 QED<=0 QCD<=2 NP<=1 , sig8 > t t~ 
\end{lstlisting}
This syntax allows one not to select the t-channel and remove all double resonant diagram keeping the top-left and middle-right diagrams of Figure~\ref{fig:FeynDiagram}.

\begin{figure}[p]
	\centering
	\includegraphics[width=0.85\textwidth]{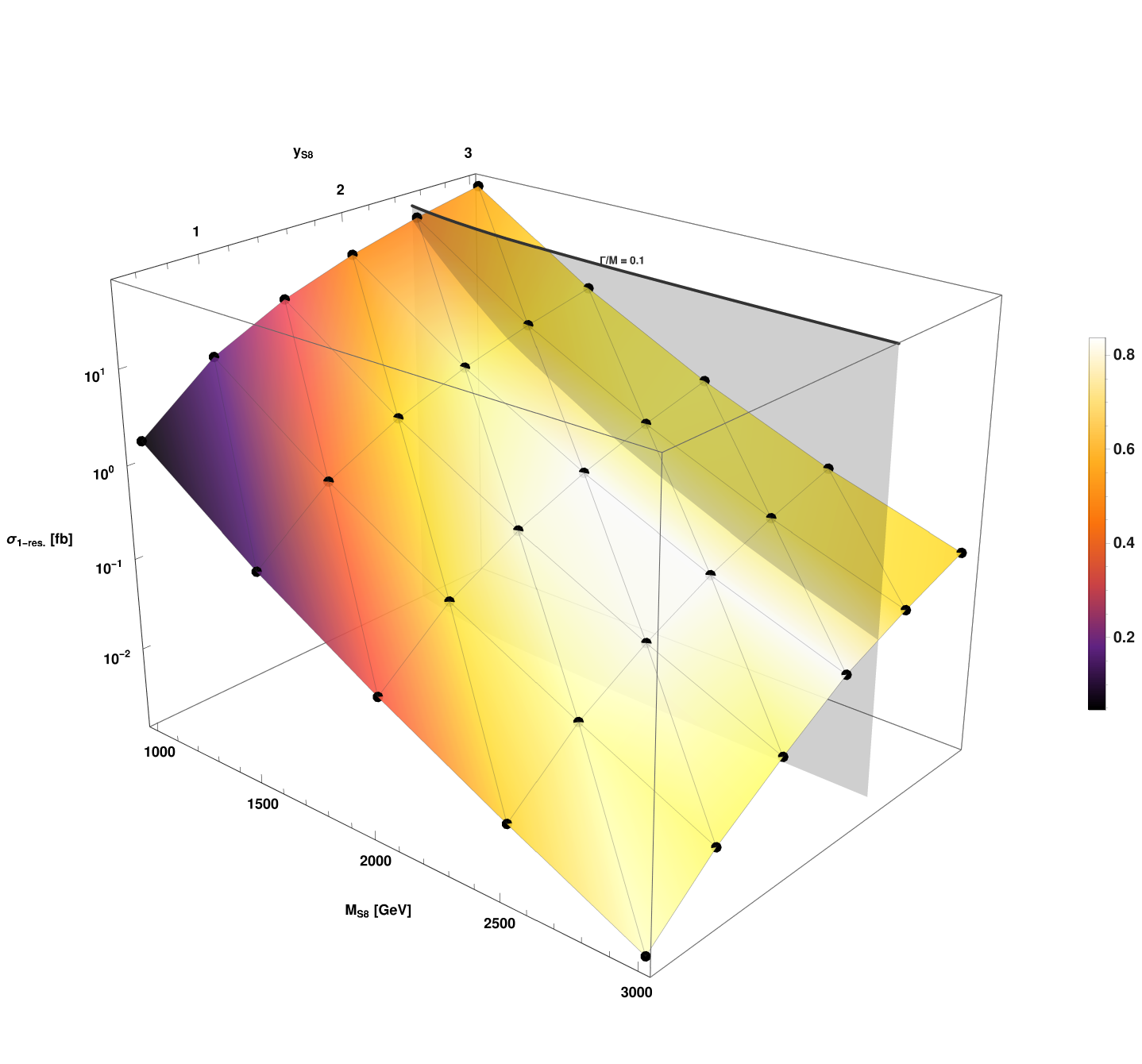}
	\includegraphics[width=0.85\textwidth]{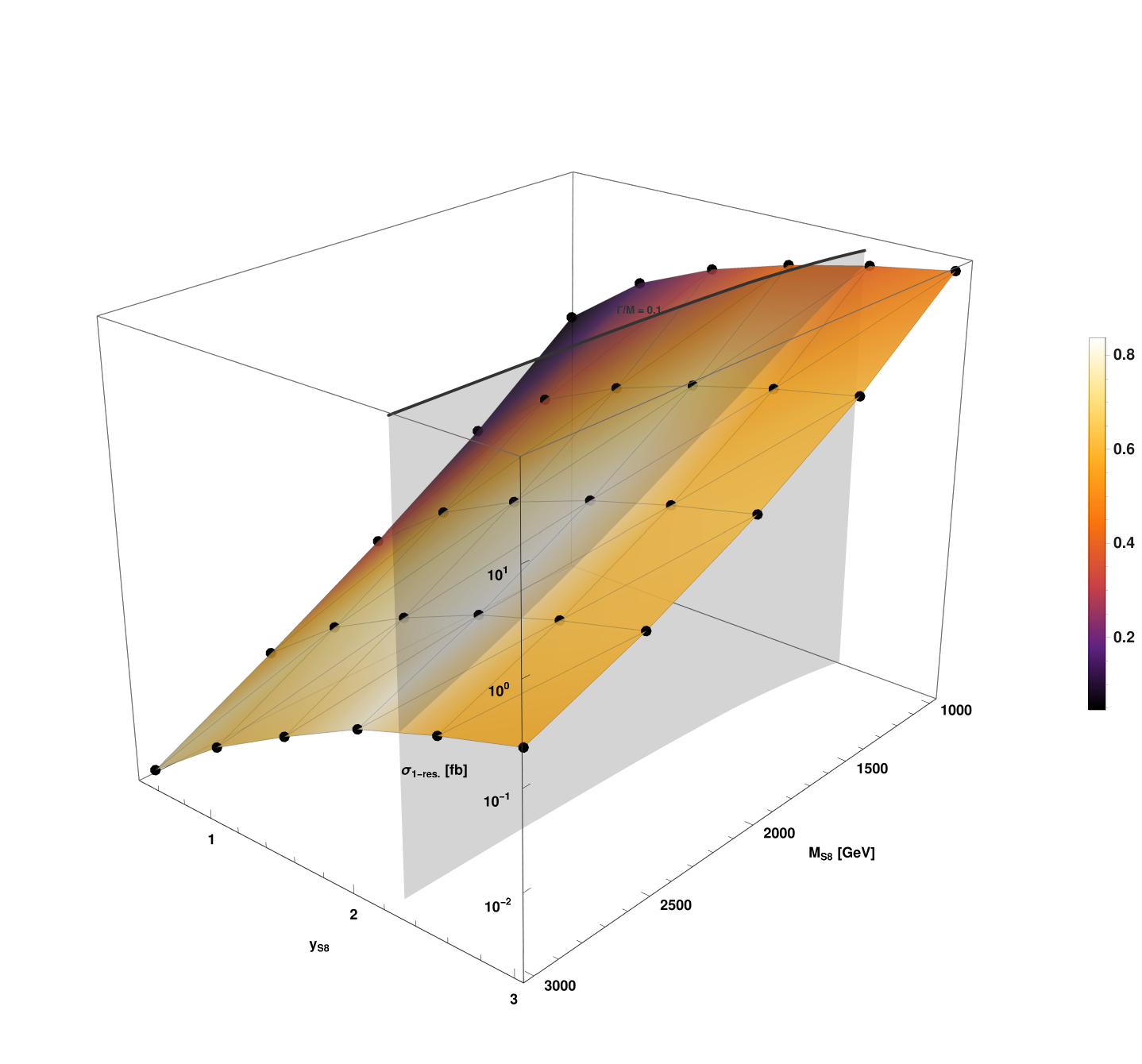}
	\caption{ Three-dimensional representation of the octet single resonant cross-section with a colour legend showing its ratio with the BSM cross-section. A front view of the plot is presented on the top figure and side view on the figure at the bottom. } 
	\label{fig:Octet1Resonant3D}
\end{figure}

\begin{figure}[p]
	\centering
	\includegraphics[width=\textwidth]{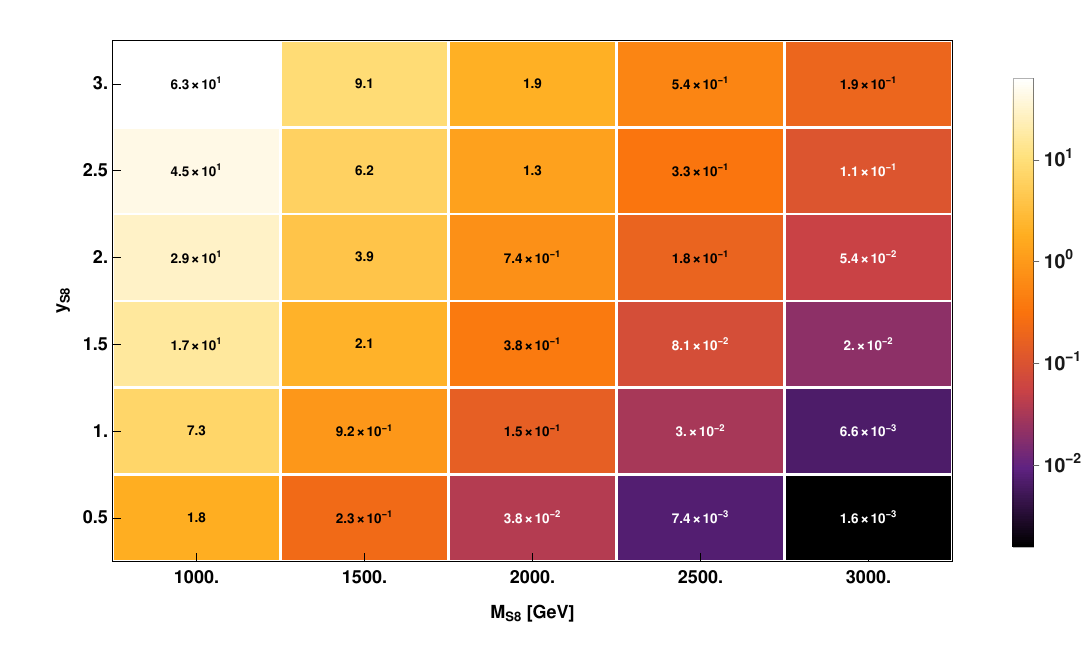}
	\includegraphics[width=\textwidth]{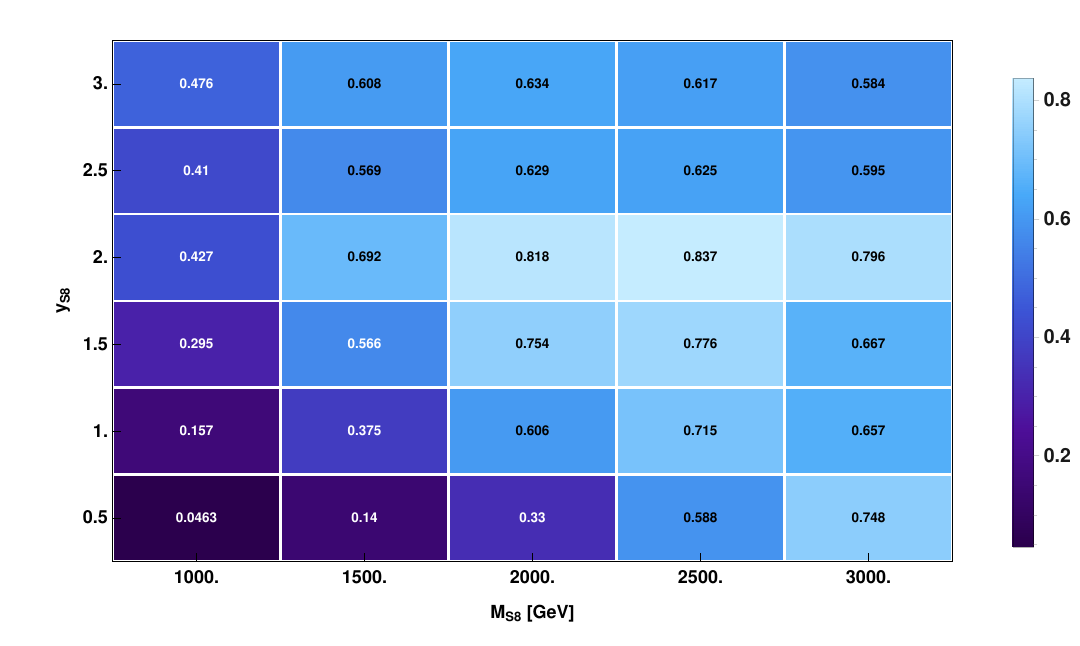}
	\caption{ Two-dimensional representation of the single resonant cross-section of the octet (top) and its ratio with the BSM cross-section (bottom).} 
	\label{fig:Octet1Resonant2D}
\end{figure}

\begin{figure}[p]
\centering
\includegraphics[width=0.85\textwidth]{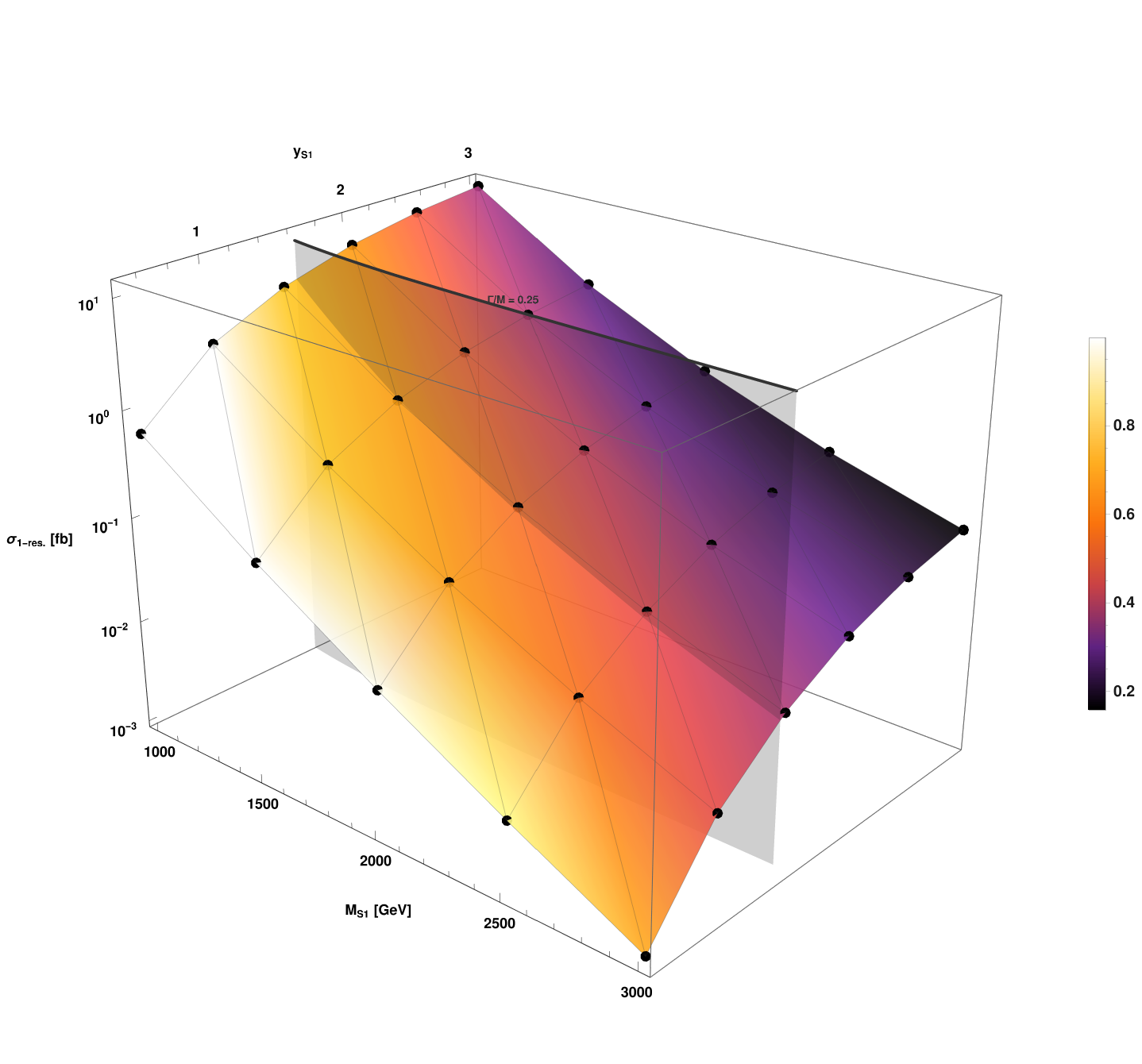}
\includegraphics[width=0.85\textwidth]{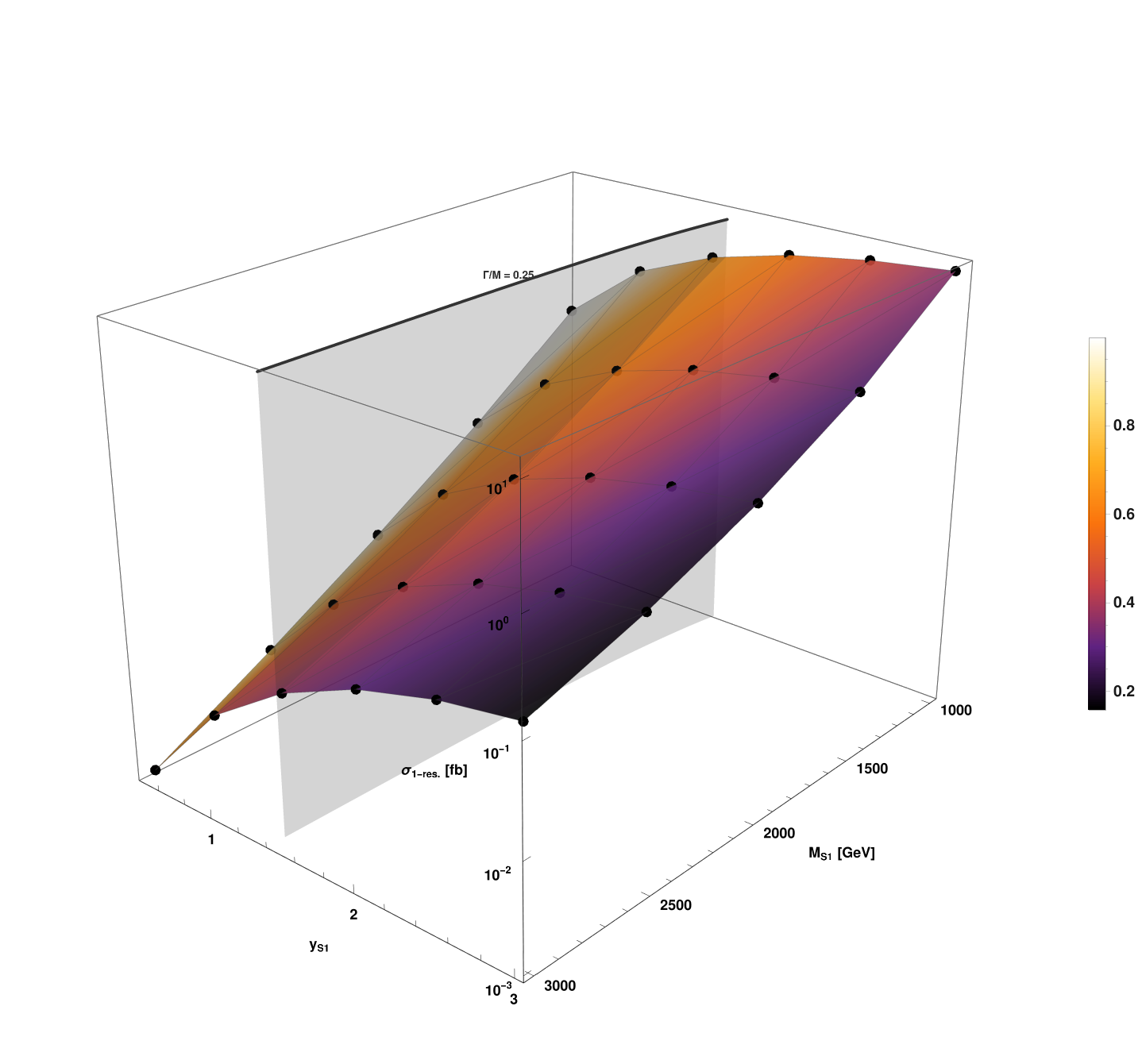}
\caption{ Same as Figure~\ref{fig:Octet1Resonant3D} for the singlet.} 
\label{fig:Singlet1Resonant3D}
\end{figure}

\begin{figure}[p]
\centering
\includegraphics[width=\textwidth]{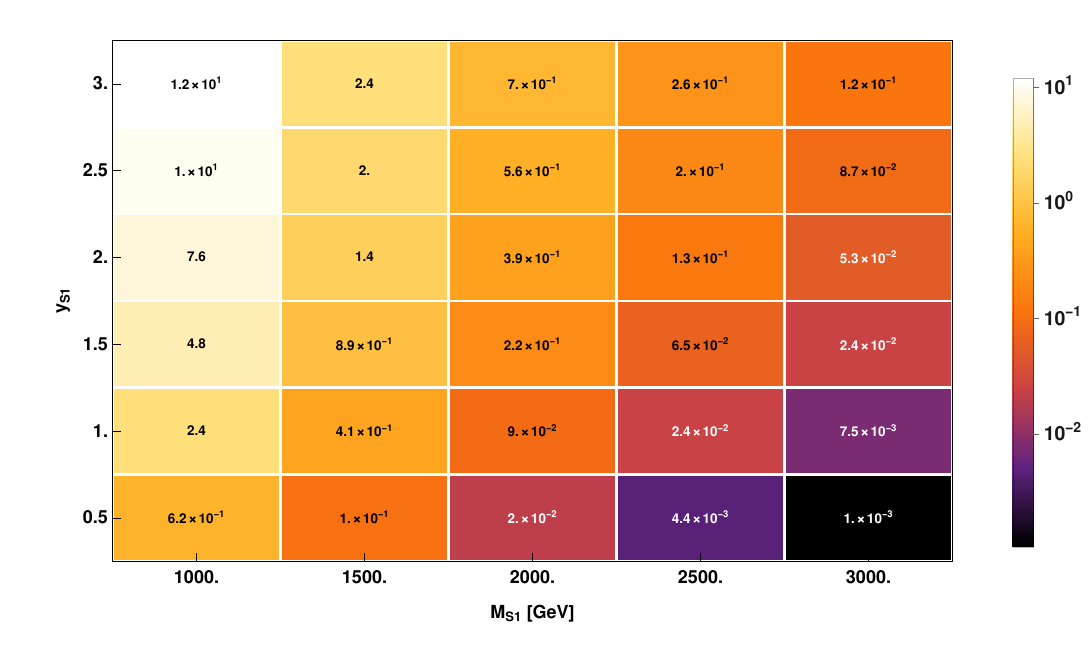}
\includegraphics[width=\textwidth]{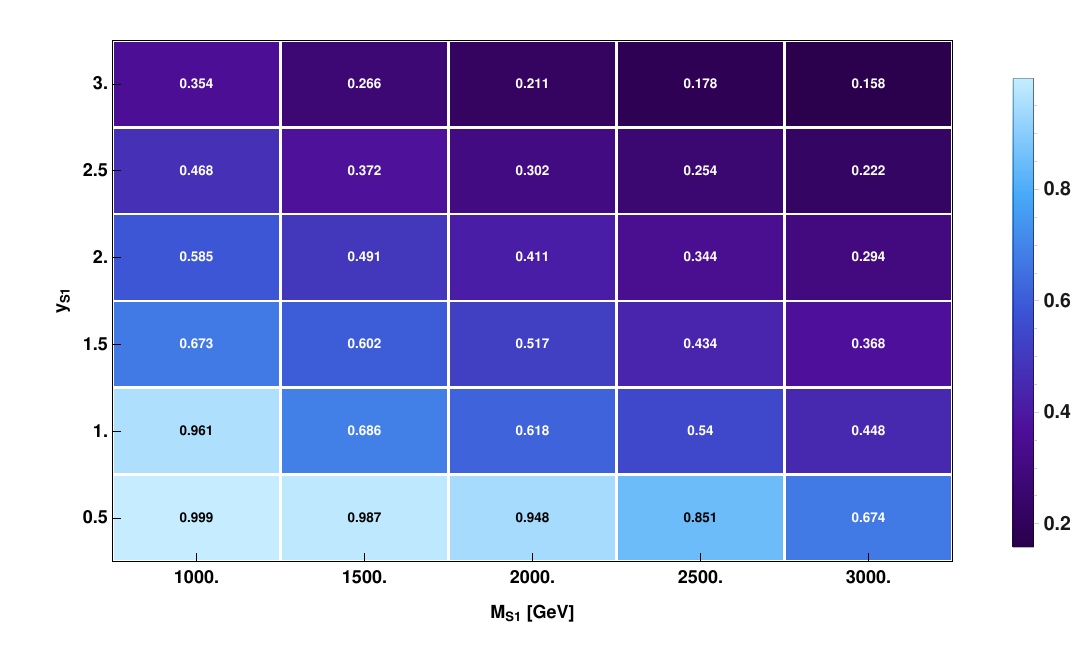}
\caption{ Same as Figure~\ref{fig:Octet1Resonant2D} for the singlet.} 
\label{fig:Singlet1Resonant2D}
\end{figure}

Figures~\ref{fig:Octet1Resonant3D} and \ref{fig:Singlet1Resonant3D} respectively display the front and side views of the three-dimensional representation of the single resonant cross-section of the octet and the singlet. A black plane is added to show the width over mass ratio at 0.1 for the octet and 0.25 for the singlet. The colour represents the ratio with the BSM cross-section. Figures~\ref{fig:Octet1Resonant2D} and \ref{fig:Singlet1Resonant2D} show the single resonant cross-section and its ratio with the BSM cross-section in two dimensions.

%\JT{We see a $y_{\oct}^2$ dependence of the single resonant cross-section for the octet at a given mass as predicted. For instance, when $y_{\oct}$ is multiplied by a factor 3 for $M_{\oct}=1$ TeV, the cross-section increases by a factor 9. At small masses, we saw that the double resonant production dominates and it is only when its value decreases that the single resonant production becomes dominant. This effect is clearly visible at $y_{\oct}=2$ where the single resonant contribution represents 80\% of BSM cross-section for $M_{\oct}>2$ TeV. For $M_{\oct}>2.5$ TeV, it represents more than 50\% of the BSM cross-section for all $y_{\oct}$. It is still the dominant contribution above $y_{\oct}=2$ but its relative contribution is smaller. }

%\JT{For the singlet, the resonant production is the main production mechanism for small $y_{\sing}$ and small $M_{\sing}$. The $y_{\sing}^2$ scaling is visible between $y_{\sing}=0.5$ and $y_{\sing}=1$ at $M_{\sing}=1$ TeV where the cross-section increases by a factor 4. However, the scaling does not hold for the other values of the coupling because finite-width effects give sizeable contributions to the narrow-width approximation.  }

%\newpage

\section{T-channel contributions}

The syntax for the t-channel contributions are similar for both simplified models, only the syntax of the BSM resonance has to be modified accordingly. For the octet, it is
\begin{lstlisting}[language=Python]
generate p p > t t~ t t~ $\textdollar \textdollar$ sig8 QED<=0 QCD<=2 NP<=2 
\end{lstlisting}
and for the singlet,
\begin{lstlisting}[language=Python]
generate p p > t t~ t t~ $\textdollar \textdollar$ sig1 QED<=0 QCD<=2 NP<=2 
\end{lstlisting}
This syntax allows the selection of the middle-left diagrams for the octet and the bottom-right for the singlet.

\begin{figure}[p]
	\centering
	\includegraphics[width=0.85\textwidth]{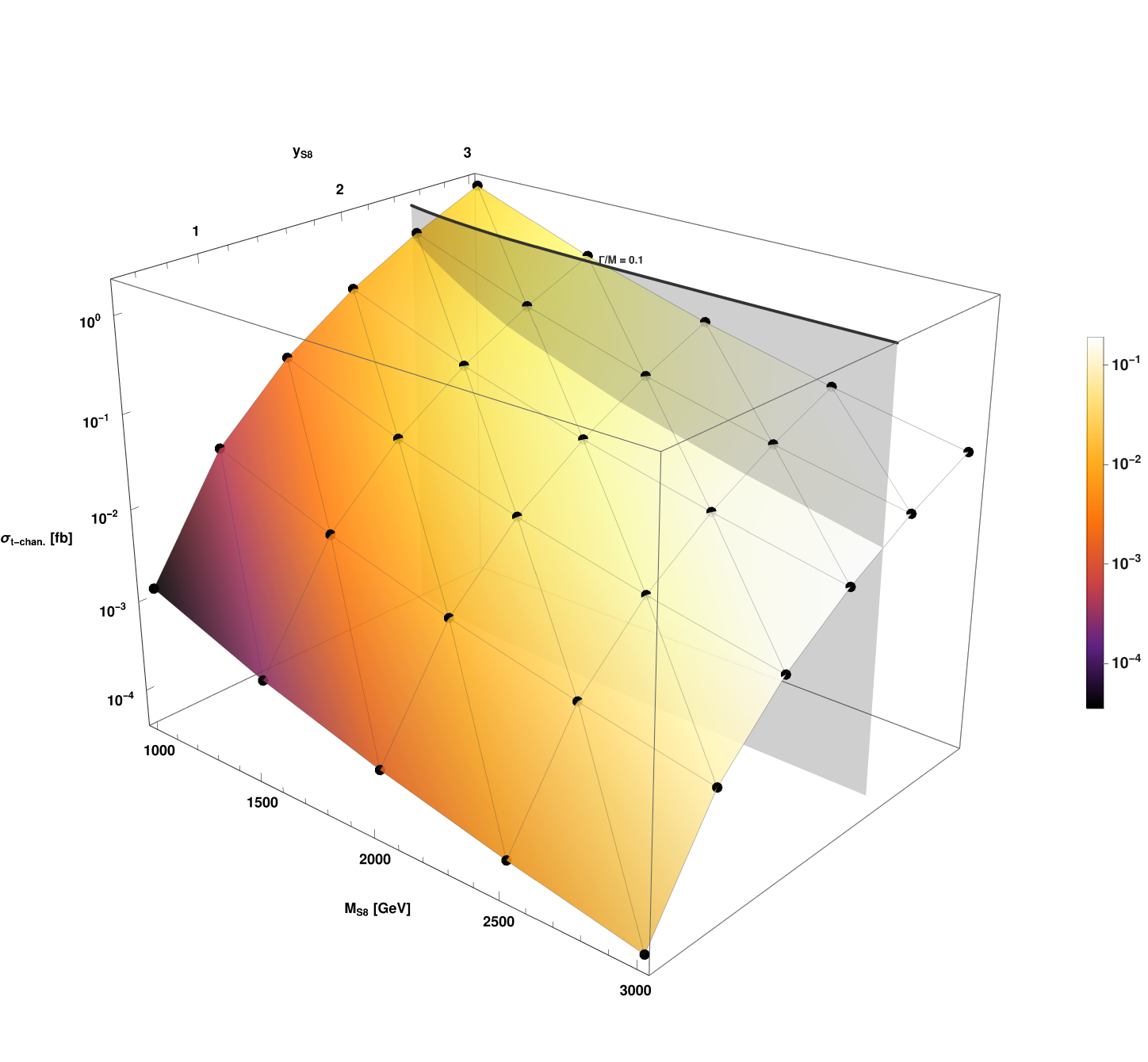}
	\includegraphics[width=0.85\textwidth]{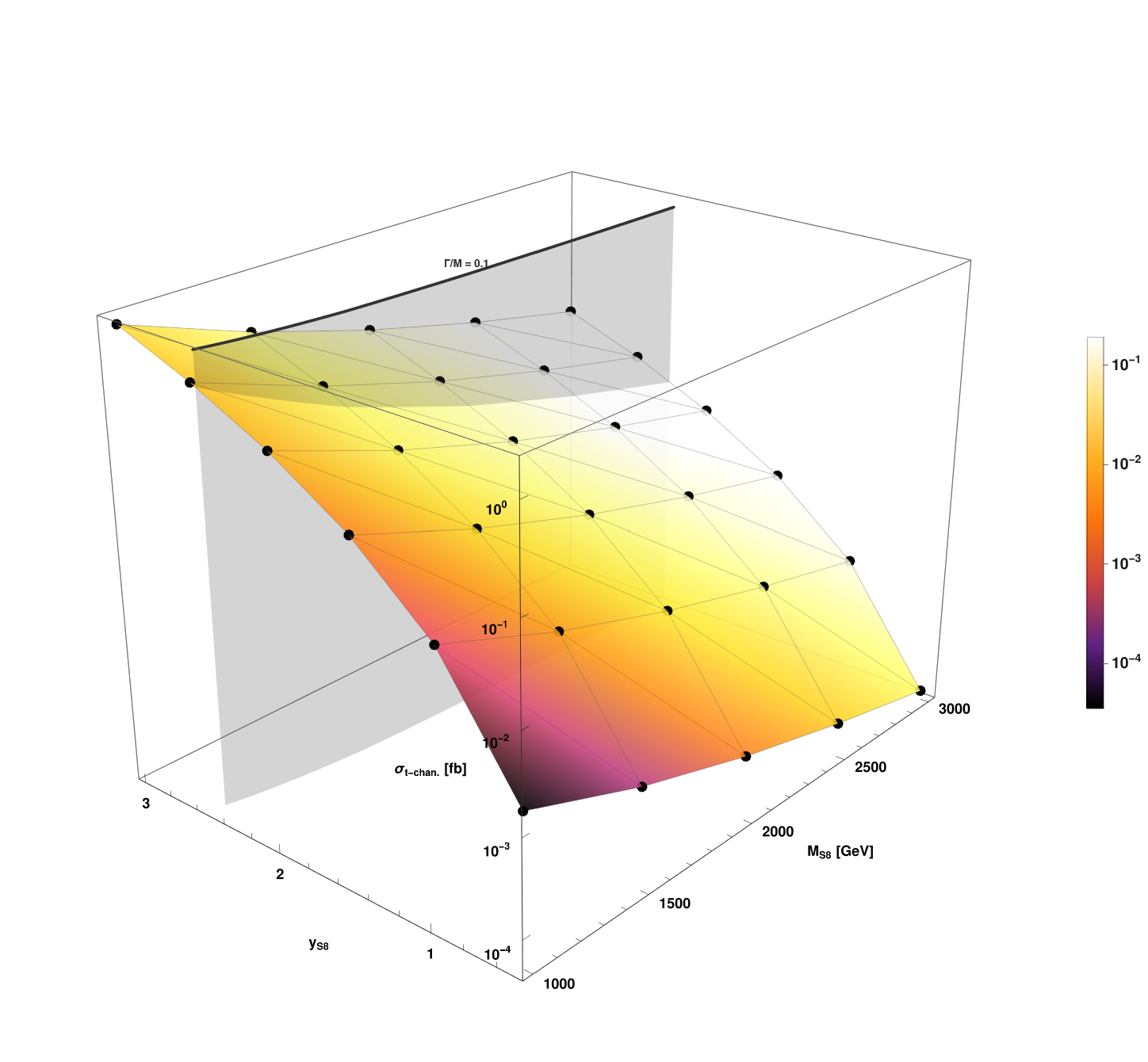}
	\caption{ Three-dimensional representation of the octet t-channel cross-section with a colour legend showing its ratio with the BSM cross-section. A front view of the plot is presented on the top figure and side view on the figure at the bottom. } 
	\label{fig:OctetTChannel3D}
\end{figure}

\begin{figure}[p]
	\centering
	\includegraphics[width=\textwidth]{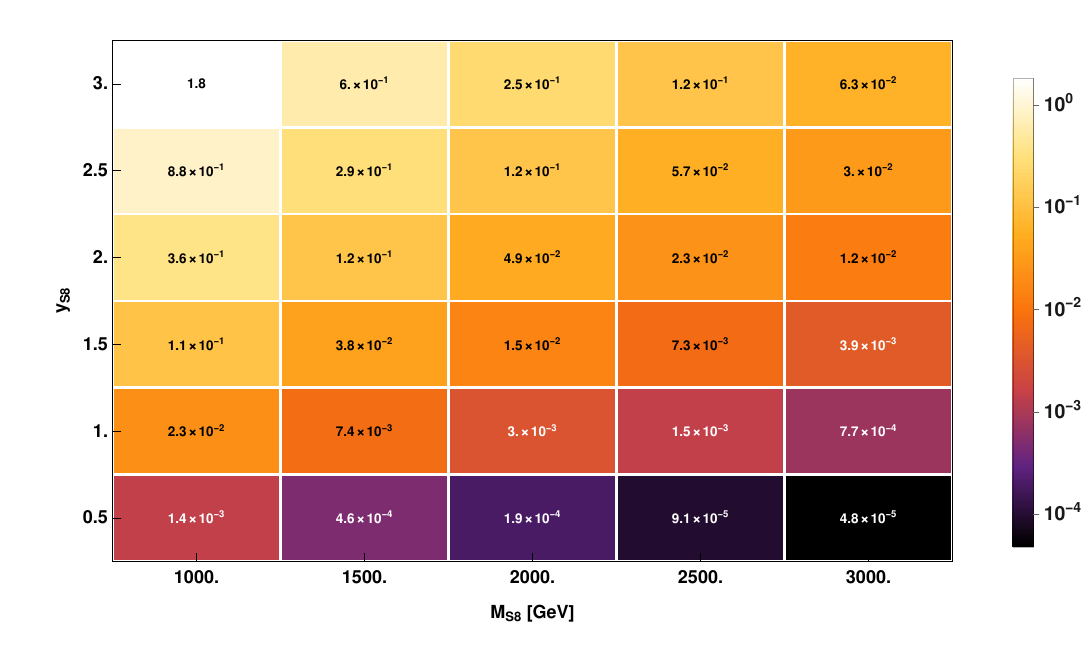}
	\includegraphics[width=\textwidth]{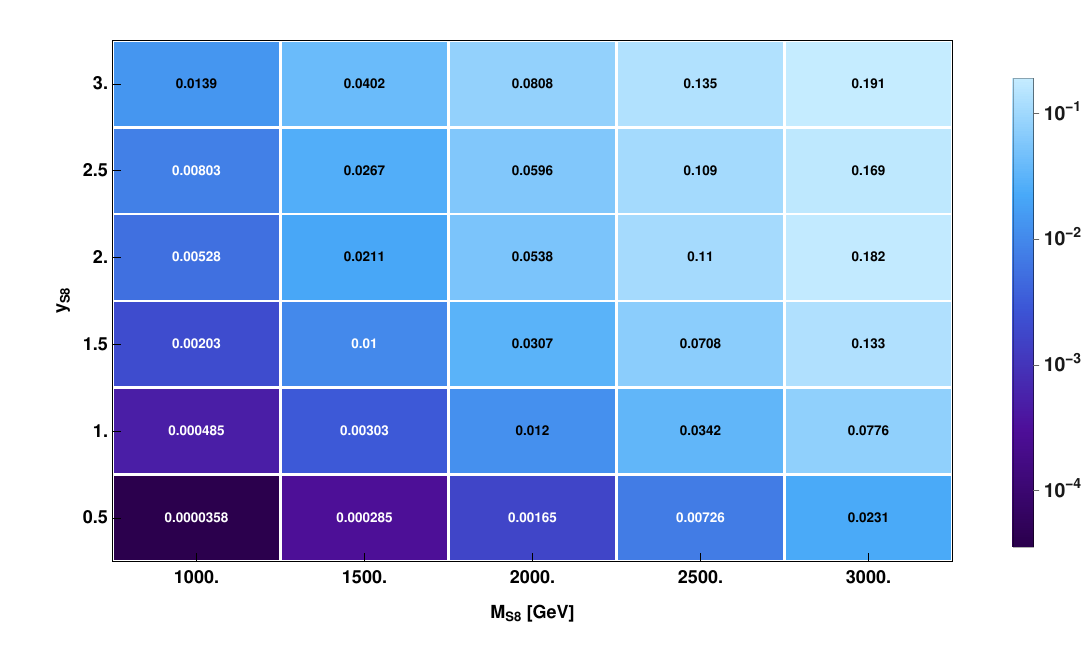}
	\caption{ Two-dimensional representation of the t-channel cross-section of the octet (top) and its ratio with the BSM cross-section (bottom).} 
	\label{fig:OctetTChannel2D}
\end{figure}

\begin{figure}[p]
\centering
\includegraphics[width=0.85\textwidth]{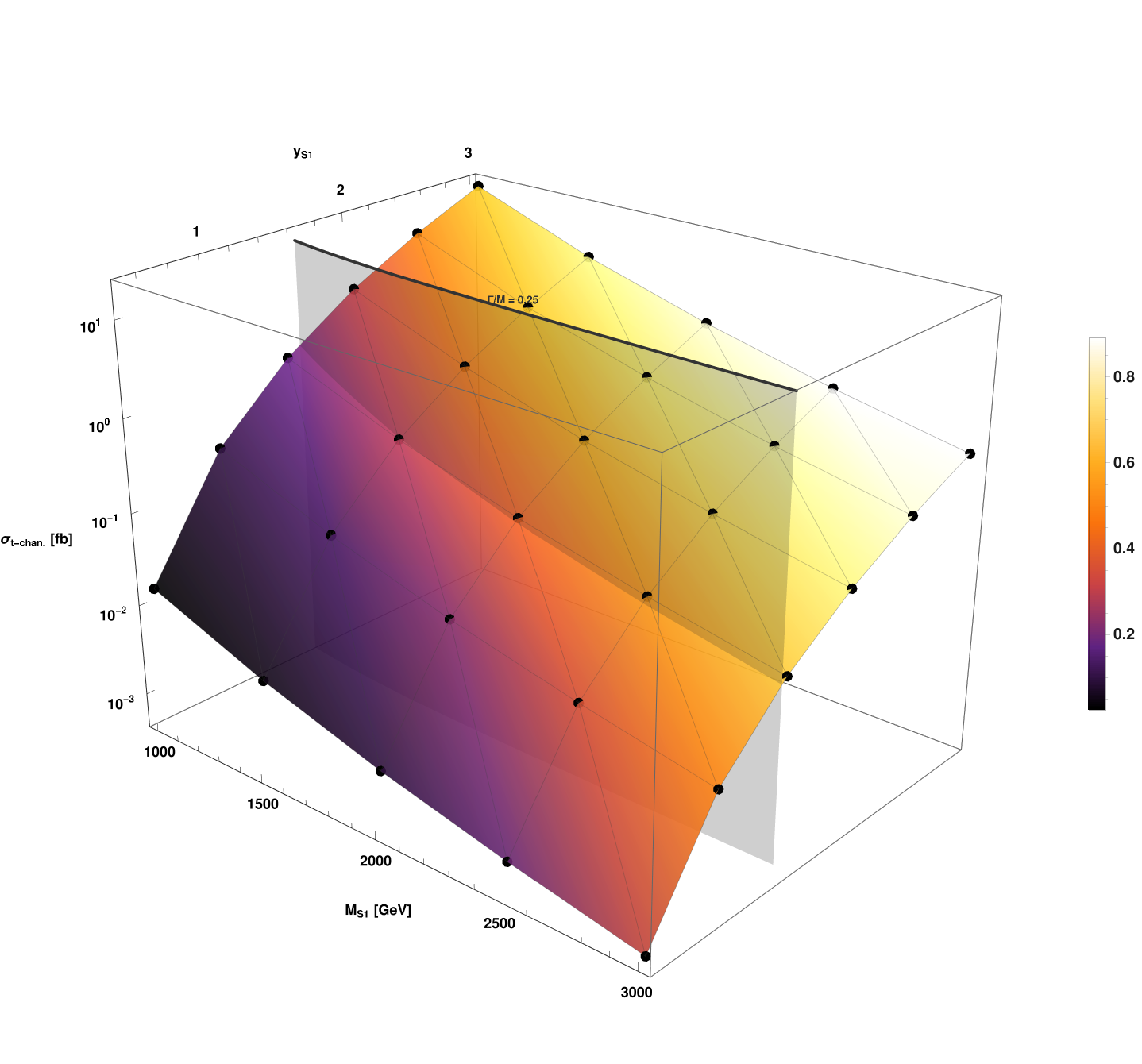}
\includegraphics[width=0.85\textwidth]{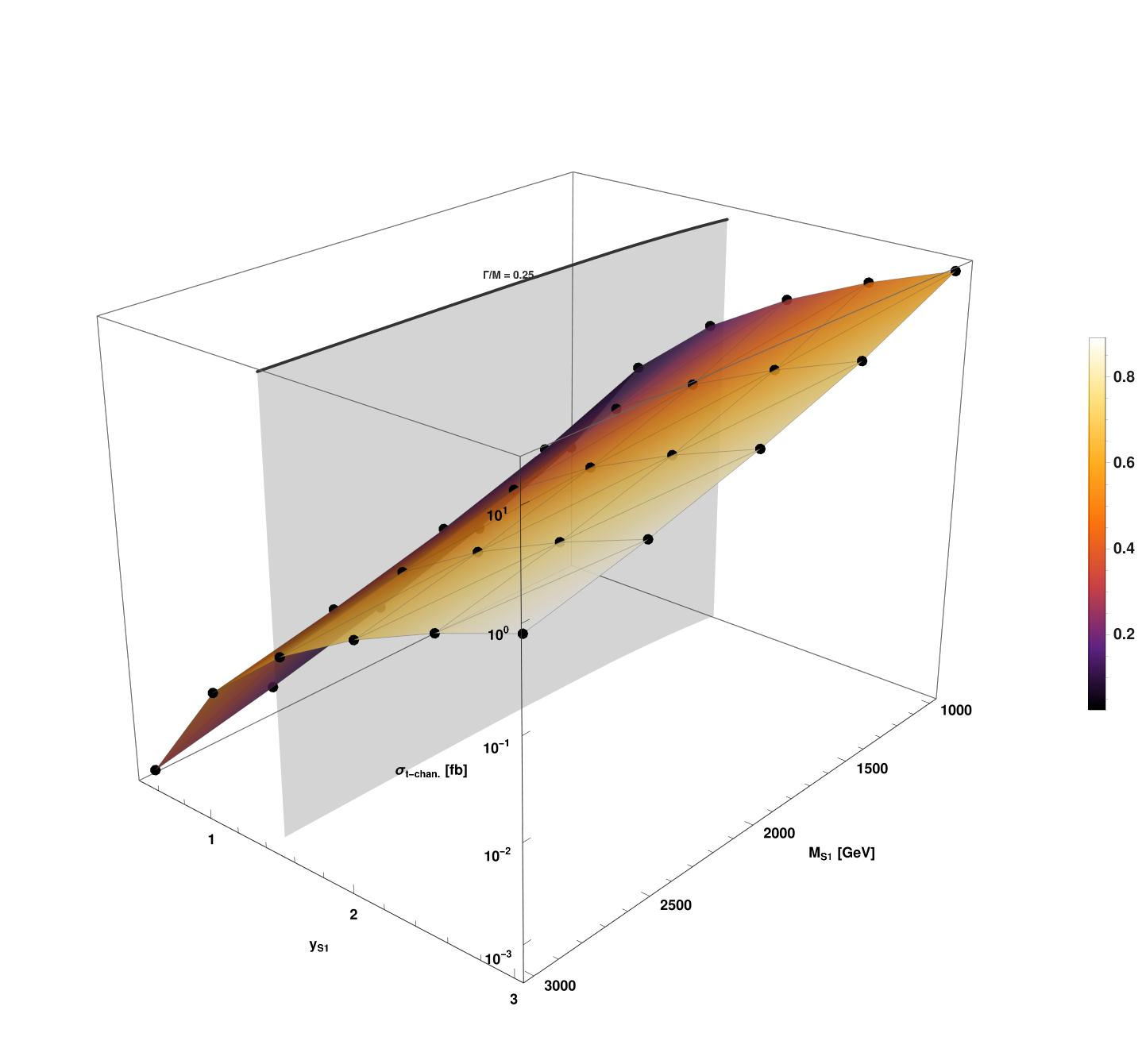}
\caption{ Same as Figure~\ref{fig:OctetTChannel3D} for the singlet.} 
\label{fig:SingletTChannel3D}
\end{figure}

\begin{figure}[p]
\centering
\includegraphics[width=\textwidth]{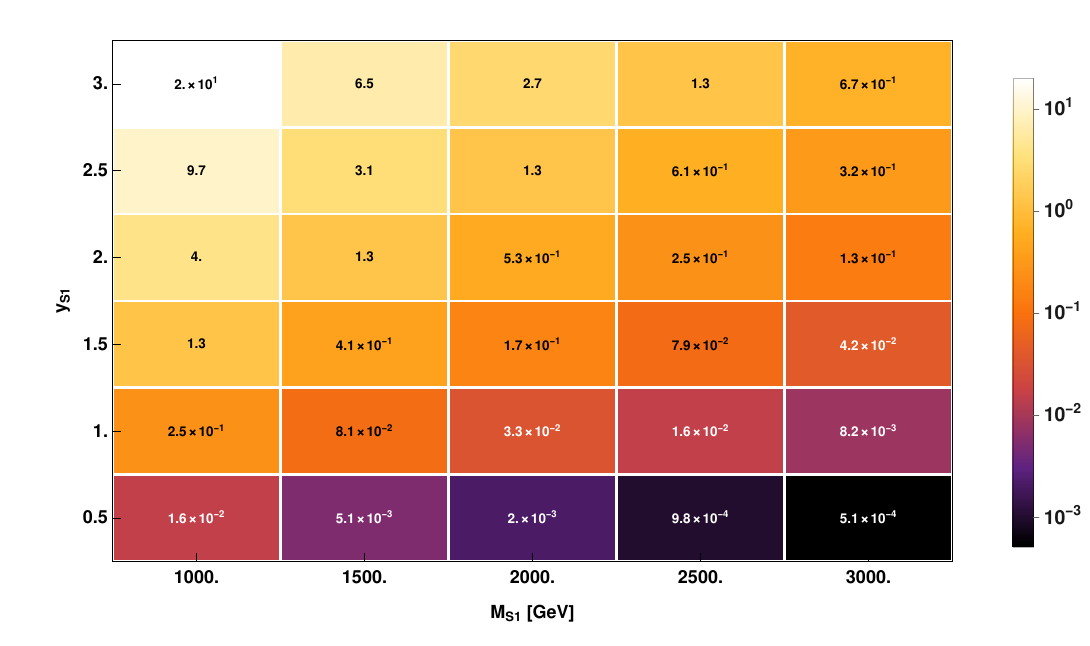}
\includegraphics[width=\textwidth]{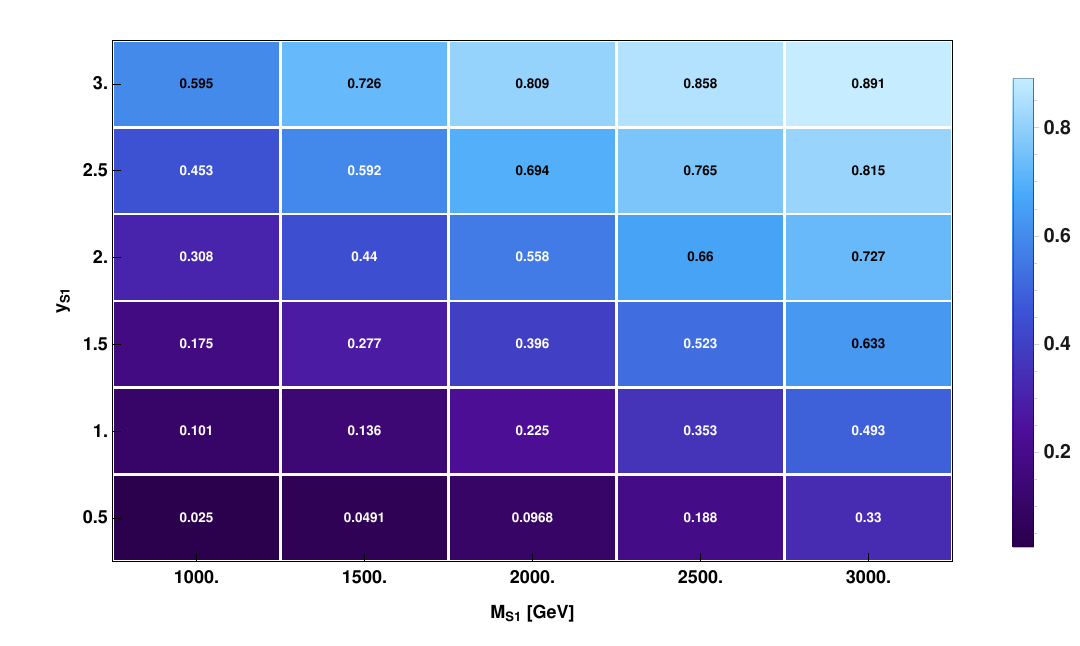}
\caption{ Same as Figure~\ref{fig:OctetTChannel2D} for the singlet.} 
\label{fig:SingletTChannel2D}
\end{figure}

Figures~\ref{fig:OctetTChannel3D} and \ref{fig:SingletTChannel3D} respectively display the front and side views of the three-dimensional representation of the st-channel cross-section of the octet and the singlet. A black plane is added to show the width over mass ratio at 0.1 for the octet and 0.25 for the singlet. The colour represents the ratio with the BSM cross-section. Figures~\ref{fig:OctetTChannel2D} and \ref{fig:SingletTChannel2D} show the t-channel cross-section and its ratio with the BSM cross-section in two dimensions.

%\JT{As expected, the t-channel cross-section grows by a factor $3^4$ when the coupling increases threefold between $y_{\oct}=0.5$ and $y_{\oct}=1.5$ at $M_{\oct}=1$ TeV. This scaling behaviour is valid across the parameter space. The t-channel cross-section is negligible at small masses compared to the BSM cross-section but its contribution increases with larger masses. For $M_{\oct}=1$ TeV, it is at most two orders of magnitude smaller for $y_{\oct}=3$ and five orders of magnitude smaller for $y_{\oct}=0.5$. At $M_{\oct}=3$ TeV, it becomes 20\% for $y_{\oct}=3$ but remains two orders of magnitude smaller for $y_{\oct}=0.5$. }

%\JT{T-channel contributions are much larger for the singlet. For $y_{\sing}=0.5$, their contribution practically double for every 500 GeV step to reach 30\% at $M_{\sing}=3$ TeV. They are already 10\% of the BSM cross-section for $y_{\sing}=1$ with $M_{\sing}=1$ TeV and even 90\% for $y_{\sing}=3$ with $M_{\sing}=3$ TeV. The $y_{\sing}^4$ scaling is valid for small couplings but finite-width effects breaks the scaling rapidly. }

\chapter{Four Top Implementation details}
\label{sec:appendixFourTopNLO}
\pagestyle{fancy}

The counterterms of the simplified models are listed in Sections~\ref{sec:appendixOctetCTs} and \ref{sec:appendixSingletCTs} below. For each counterterm, the relevant 1-loop amplitudes absent from the traditional QCD renormalisation of the SM are presented. They are followed by the complete analytic expressions computed by NLOCT (which include SM contributions even if they are not drawn).

\section{Counterterms of the Octet simplified model}
\label{sec:appendixOctetCTs}

\begin{itemize}
	
\item The gluon wavefunction $\delta Z_{g}$ 
	
%\begin{center}
%\includegraphics[width=0.4\textwidth]{figures/feynman_octet_VV_1.png}
%\includegraphics[width=0.4\textwidth]{figures/feynman_octet_VV_2.png}
%\end{center}
\begin{equation*}
	\begin{split}
		& \includegraphics[width=0.4\textwidth]{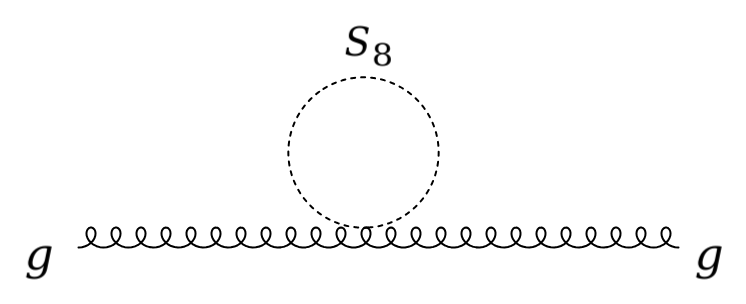}
		\includegraphics[width=0.38\textwidth]{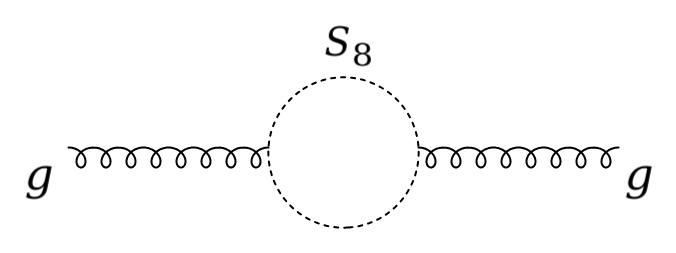} \\
		\delta Z_{g} = & - \frac{5 g_s{}^2 }{24 \pi ^2 \epsilon} + \frac{g_s{}^2}{32 \pi ^2} \log \left(\frac{M_{\oct}^2}{\mu^2}\right) +\frac{g_s{}^2}{24 \pi ^2} \log \left(\frac{m_t^2}{\mu^2} \right).
	\end{split}
\end{equation*}
	
\item The strong coupling $\delta \alpha_s$: 
%\begin{center}
	%\includegraphics[width=0.30\textwidth]{figures/feynman_octet_VVV_1.png}
	%\includegraphics[width=0.35\textwidth]{figures/feynman_octet_VVV_2.png}
%\end{center}
\begin{equation*}
	\begin{split}
		& \includegraphics[width=0.30\textwidth]{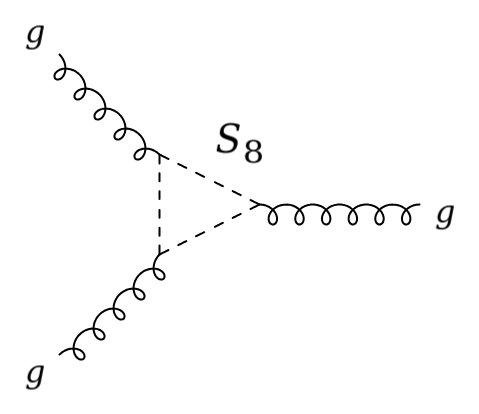}
		  \includegraphics[width=0.35\textwidth]{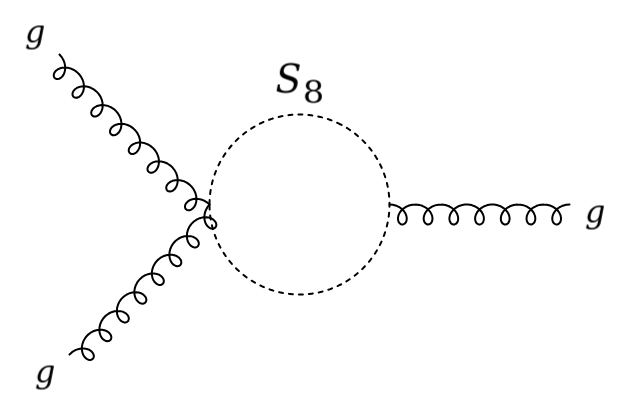} \\
		\delta \alpha_s = & \frac{\alpha_s}{192 \pi^2}  \left(-6 g_s{}^2 \log \left(\frac{M_{\oct}^2}{\mu^2}\right) - 8 g_s{}^2 \log \left(\frac{m_t^2}{\mu^2}\right) \right).
	\end{split}
\end{equation*}

\newpage

\item The top wavefunction $\delta Z_t$  

\begin{center}
	\includegraphics[width=0.4\textwidth]{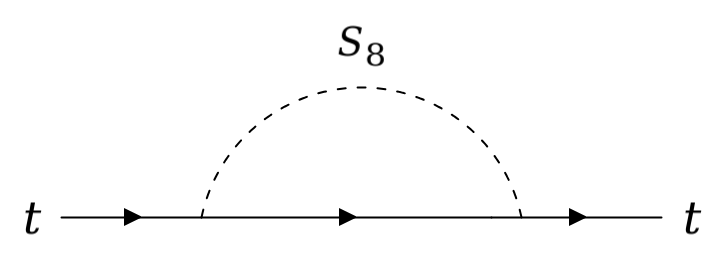}
\end{center}
\begin{equation*}
\begin{split}
	\delta Z_t = & - \frac{g_s{}^2 }{6 \pi ^2 \epsilon} -\frac{g_s{}^2  }{3 \pi ^2} + \frac{7 y_{\oct}{}^2  }{24 \pi ^2} - \frac{y_{\oct}{}^2 M_{\oct}^2 }{8 \pi ^2 m_t^2} \\
	& - \frac{ y_{\oct}{}^2 M_{\oct}^2}{48 \pi ^2 m_t^2} \log \left(\frac{M_{\oct}^2}{\mu^2}\right) \\
	& + \left( \frac{ g_s{}^2}{4 \pi ^2} + \frac{y_{\oct}{}^2}{24 \pi ^2} + \frac{y_{\oct}{}^2 M_{\oct}^2 }{48 \pi ^2 m_t^2} \right) \log \left(\frac{m_t^2}{\mu^2}\right)     \\
	& + \left( \frac{y_{\oct}{}^2}{6 \pi ^2} - \frac{11 y_{\oct}{}^2 M_{\oct}^2 }{48 \pi ^2 m_t^2} + \frac{y_{\oct}{}^2 M_{\oct}^4 }{16 \pi ^2 m_t^4} \right) \log \left(\frac{M_{\oct}^2}{m_t^2}\right) \\
	& -\frac{y_{\oct}{}^2 M_{\oct}^2 }{\pi ^2 \sqrt{M_{\oct}^4-4 M_{\oct}^2 m_t^2}} \log \left(\frac{M_{\oct}+\sqrt{M_{\oct}^2-4 m_t^2}}{2 m_t}\right)  \\
	& +\frac{3 y_{\oct}{}^2 M_{\oct}^4}{4 \pi ^2 m_t^2 \sqrt{M_{\oct}^4-4 M_{\oct}^2 m_t^2}} \log \left(\frac{M_{\oct}+\sqrt{M_{\oct}^2-4 m_t^2}}{2 m_t}\right) \\
	& -\frac{y_{\oct}{}^2 M_{\oct}^6 }{8 \pi ^2 m_t^4 \sqrt{M_{\oct}^4-4 M_{\oct}^2 m_t^2} }  \log \left(\frac{M_{\oct}+\sqrt{M_{\oct}^2-4 m_t^2}}{2 m_t}\right)
\end{split}
\end{equation*}

\newpage

\item The top mass counterterm $\delta m_t$ %receives an additional contribution from a propagating octet to the usual QCD correction from a gluon due to its Yukawa interaction at $\mathcal{O}(y_{\oct}^2)$. }

%\begin{center}
%\includegraphics[width=0.4\textwidth]{figures/feynman_octet_FF.png}
%\end{center}
\begin{equation*}
\begin{split}
	& \includegraphics[width=0.45\textwidth]{figures/feynman_octet_FF.png} \\
	\delta m_t = & -\frac{g_s{}^2 m_t }{3 \pi ^2} + \frac{7 y_{\oct}^2 m_t }{24 \pi ^2} - \frac{y_{\oct}^2 M_{\oct}^2 }{24\pi ^2 m_t} \\
	& - \left( \frac{y_{\oct}^2 m_t }{12 \pi^2} + \frac{y_{\oct}^2 M_{\oct}^2 }{48 \pi ^2 m_t} \right)  \log \left(\frac{M_{\oct}^2}{\mu^2}\right) \\
	& + \left( \frac{g_s{}^2 m_t }{4 \pi ^2} - \frac{y_{\oct}^2 m_t }{24 \pi ^2} + \frac{y_{\oct}^2 M_{\oct}^2 }{48 \pi ^2 m_t} \right) \log \left(\frac{m_t^2}{\mu^2}\right)  \\
	& + \left( \frac{y_{\oct}^2 m_t }{12 \pi ^2} - \frac{5 y_{\oct}^2 M_{\oct}^2 }{48 \pi ^2 m_t} + \frac{y_{\oct}^2 M_{\oct}^4 }{48 \pi ^2 m_t^3} \right)  \log \left(\frac{M_{\oct}^2}{m_t^2}\right)  \\
	& +\frac{y_{\oct}^2 \sqrt{M_{\oct}^4-4 M_{\oct}^2 m_t^2} }{6 \pi ^2 m_t} \log \left(\frac{M_{\oct}+\sqrt{M_{\oct}^2-4 m_t^2}}{2 m_t}\right) \\
	& -\frac{y_{\oct}^2 M_{\oct}^2 \sqrt{M_{\oct}^4-4 M_{\oct}^2 m_t^2} }{24 \pi ^2 m_t^3} \log \left(\frac{M_{\oct}+\sqrt{M_{\oct}^2-4 m_t^2}}{2 m_t}\right)
\end{split}
\end{equation*}

\newpage

\item The octet wavefunction $\delta Z_{\oct}$ 
	
\begin{center}
	\includegraphics[width=0.4\textwidth]{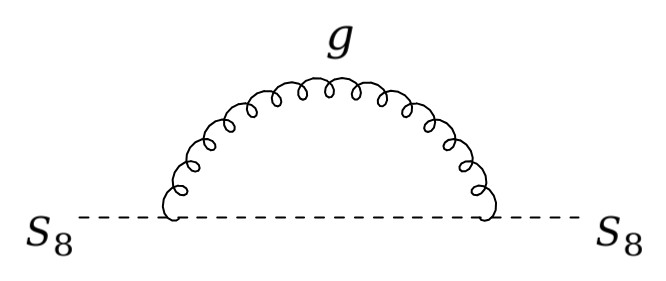}
	\includegraphics[width=0.4\textwidth]{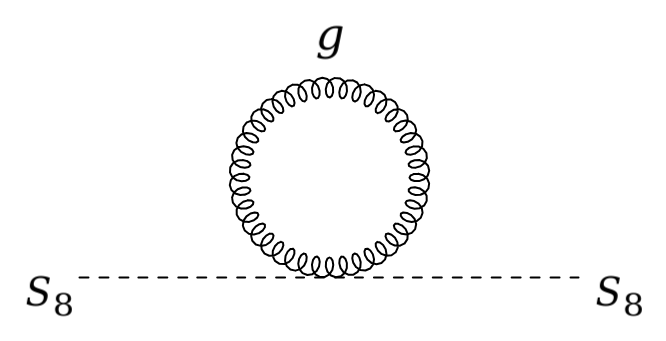}
	\includegraphics[width=0.4\textwidth]{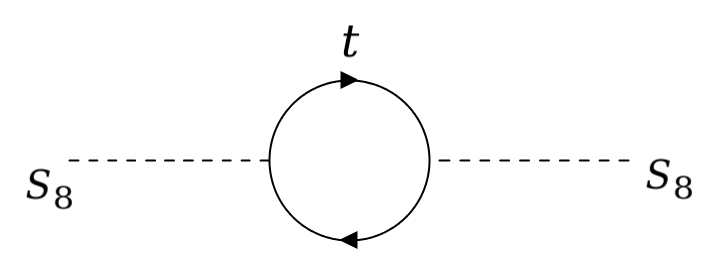}
\end{center}
\begin{equation*}
\begin{split}
	\delta Z_{\oct} = & -\frac{3 g_s{}^2}{8 \pi ^2 \epsilon} - \frac{y_{\oct}{}^2 }{16 \pi ^2} - \frac{y_{\oct}{}^2 m_t^2}{4 \pi ^2 M_{\oct}^2} + \frac{y_{\oct}{}^2}{16 \pi ^2} \log \left(\frac{m_t^2}{\mu^2}\right) \\
	& + \frac{y_{\oct}{}^2 m_t^2 }{8 \pi ^2 M_{\oct} \sqrt{M_{\oct}^2 - 4 m_t^2}} \log \left(\frac{ M_{\oct} \sqrt{M_{\oct}^2 - 4m_t^2} - M_{\oct}^2 + 2 m_t^2}{2 m_t^2}\right) \\
	& + \frac{y_{\oct}{}^2 m_t^4 }{2 \pi ^2 M_{\oct}^3\sqrt{M_{\oct}^2 -4 m_t^2} } \log \left(\frac{ M_{\oct} \sqrt{M_{\oct}^2 - 4m_t^2} - M_{\oct}^2 + 2 m_t^2}{2 m_t^2}\right) \\
	& - \frac{y_{\oct}{}^2 M_{\oct}^2}{16 \pi ^2 M_{\oct} \sqrt{M_{\oct}^2-4 m_t^2}} \log \left(\frac{ M_{\oct} \sqrt{M_{\oct}^2 - 4m_t^2} - M_{\oct}^2 + 2 m_t^2}{2 m_t^2}\right) \\
	%& -\frac{m_t^2 y_{\oct}{}^2 \text{If}\left[M_{\oct}^2=4 m_t^2,1,0\right]}{4 \pi ^2 M_{\oct}^2}+\frac{y_{\oct}{}^2 \text{If}\left[M_{\oct}^2=4m_t^2,1,0\right]}{16 \pi ^2} 
\end{split}
\end{equation*}

\newpage

\item The octet mass counterterm $\delta M_{\oct}$ 

\begin{center}
\includegraphics[width=0.4\textwidth]{figures/feynman_octet_SS_1.png}
\includegraphics[width=0.4\textwidth]{figures/feynman_octet_SS_2.png}
\includegraphics[width=0.4\textwidth]{figures/feynman_octet_SS_3.png}
\end{center}
\begin{equation*}
	\begin{split}
		%& \includegraphics[width=0.4\textwidth]{figures/feynman_octet_SS_1.png} \\
		\delta M_{\oct} = & - \frac{21 g_s{}^2 M_{\oct} }{32 \pi ^2} + \frac{y_{\oct}{}^2 M_{\oct} }{16\pi ^2} - \frac{5 y_{\oct}{}^2 m_t^2 }{16 \pi ^2 M_{\oct}} \\
		&  + \frac{9 g_s{}^2 M_{\oct} }{32 \pi ^2} \log \left(\frac{M_{\oct}^2}{\mu^2}\right) \\
		& + \left( \frac{3 y_{\oct}{}^2 m_t^2 }{16 \pi ^2 M_{\oct}}-\frac{y_{\oct}{}^2 M_{\oct} }{32 \pi ^2} \right)  \log \left(\frac{m_t^2}{\mu^2}\right)  \\
		& -\frac{y_{\oct}{}^2 m_t^2 \sqrt{M_{\oct}^2 - 4 m_t^2} }{8\pi ^2 M_{\oct}^2} \log \left(\frac{ M_{\oct} \sqrt{M_{\oct}^2 - 4 m_t^2} - M_{\oct}^2 + 2 m_t^2 }{2 m_t^2}\right)  \\
		& +\frac{y_{\oct}{}^2 \sqrt{M_{\oct}^2 - 4 m_t^2}}{32 \pi ^2} \log \left(\frac{ M_{\oct} \sqrt{M_{\oct}^2 - 4 m_t^2} - M_{\oct}^2 + 2 m_t^2 }{2 m_t^2}\right) 
	\end{split}
\end{equation*}

\end{itemize}

\newpage

\section{Counterterms of the Singlet simplified model}
\label{sec:appendixSingletCTs}

\begin{itemize}

\item The top wavefunction counterterm $\delta Z_{t}$: 

%\begin{center}
	%\includegraphics[width=0.4\textwidth]{figures/feynman_singlet_FF.png}
%\end{center}
\begin{equation*}
\begin{split}
	& \includegraphics[width=0.45\textwidth]{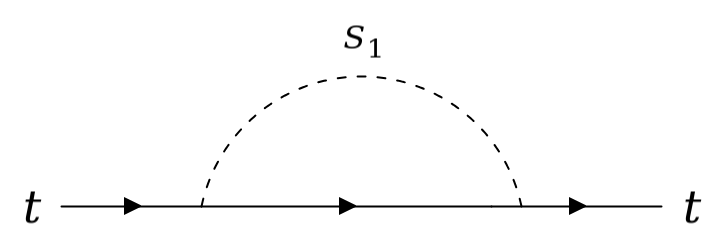} \\
	\delta Z_{t} = & - \frac{g_s{}^2}{6 \pi ^2 \epsilon} - \frac{g_s{}^2 }{3 \pi ^2} + \frac{7 y_{\sing}^2 }{32 \pi ^2} - \frac{3 y_{\sing}^2 M_{\sing}^2 }{32 \pi ^2 m_t^2} \\
	& - \frac{y_{\sing}^2 M_{\sing}^2 }{64 \pi ^2 m_t^2} \log \left(\frac{ M_{\sing}^2}{\mu^2}\right)   \\
	& + \left( \frac{g_s{}^2 }{4 \pi ^2} + \frac{y_{\sing}^2}{32 \pi ^2} + \frac{y_{\sing}^2  M_{\sing}^2 }{64 \pi ^2 m_t^2} \right) \log \left(\frac{ m_t^2}{\mu^2}\right)  \\
	& + \left( \frac{y_{\sing}^2 }{8 \pi ^2} - \frac{5 y_{\sing}^2 M_{\sing}^2 }{32 \pi ^2 m_t^2} + \frac{y_{\sing}^2 M_{\sing}^4 }{32 \pi ^2 m_t^4} \right) \log \left(\frac{ M_{\sing}^2}{m_t^2}\right) \\
	& + \left( - \frac{ y_{\sing}^2 M_{\sing}^2 }{64 \pi ^2 m_t^2} + \frac{y_{\sing}^2 M_{\sing}^4 }{64 \pi ^2 m_t^4} \right) \log \left(\frac{ M_{\sing}^2}{m_t^2}\right)  \\
	& -\frac{3 y_{\sing}^2 M_{\sing}^2}{4 \pi ^2 \sqrt{ M_{\sing}^4-4  M_{\sing}^2 m_t^2}}  \log \left(\frac{ M_{\sing}+\sqrt{ M_{\sing}^2-4  m_t^2}}{2 m_t}\right)  \\
	&  + \frac{9 y_{\sing}^2 M_{\sing}^4 m_t^2 }{16 \pi ^2 m_t^4 \sqrt{ M_{\sing}^4-4  M_{\sing}^2 m_t^2}} \log \left(\frac{ M_{\sing}+\sqrt{ M_{\sing}^2-4  m_t^2}}{2 m_t}\right)  \\
	& -\frac{3 y_{\sing}^2 M_{\sing}^6 }{32 \pi ^2 m_t^4 \sqrt{ M_{\sing}^4-4  M_{\sing}^2 m_t^2}}  \log \left(\frac{ M_{\sing}+\sqrt{ M_{\sing}^2-4  m_t^2}}{2 m_t}\right)
\end{split}
\end{equation*}

\newpage

\item The top mass counterterm $\delta m_t$: %receives an additional contribution from a propagating octet to the usual QCD correction from a gluon due to its Yukawa interaction at $\mathcal{O}(y_{\sing}^2)$. }

%\begin{center}
	%\includegraphics[width=0.4\textwidth]{figures/feynman_octet_FF.png}
%\end{center}
\begin{equation*}
\begin{split}
	& \includegraphics[width=0.45\textwidth]{figures/feynman_singlet_FF.png} \\
	\delta m_t = & - \frac{g_s{}^2 m_t }{3 \pi ^2} + \frac{7 y_{\sing}^2 m_t}{32 \pi ^2} - \frac{ y_{\sing}^2 M_{\sing}^2 }{32 \pi ^2 m_t} \\
	& + \left(- \frac{ y_{\sing}^2 m_t}{16 \pi ^2}  - \frac{ y_{\sing}^2 M_{\sing}^2 }{64 \pi ^2 m_t} \right) \log \left(\frac{ M_{\sing}^2}{\mu^2}\right) \\
	& + \left(  \frac{g_s{}^2 m_t }{4 \pi ^2} - \frac{y_{\sing}^2 m_t }{32 \pi ^2} + \frac{ y_{\sing}^2 M_{\sing}^2 }{64 \pi ^2 m_t}  \right) \log \left(\frac{m_t^2}{\mu^2}\right)  \\
	& + \left(\frac{y_{\sing}^2 m_t }{16 \pi^2} - \frac{5 y_{\sing}^2 M_{\sing}^2}{64 \pi ^2 m_t} + \frac{y_{\sing}^2 M_{\sing}^4 }{64 \pi ^2 m_t^3} \right) \log \left(\frac{ M_{\sing}^2}{m_t^2}\right)  \\
	& +\frac{y_{\sing}^2 }{8 \pi ^2 m_t} \sqrt{ M_{\sing}^4-4  M_{\sing}^2 m_t^2} \log \left(\frac{ M_{\sing}+\sqrt{ M_{\sing}^2-4 m_t^2}}{2 m_t}\right) \\
	& -\frac{ y_{\sing}^2 M_{\sing}^2 }{32 \pi ^2 m_t^3} \sqrt{ M_{\sing}^4-4  M_{\sing}^2 m_t^2} \log \left(\frac{ M_{\sing}+\sqrt{ M_{\sing}^2-4 m_t^2}}{2 m_t}\right)
\end{split}
\end{equation*}

\newpage

\item The scalar singlet wavefunction counterterm $\delta Z_{\sing}$: 

%\begin{center}
	%\includegraphics[width=0.4\textwidth]{figures/feynman_singlet_SS.png}
%\end{center}
\begin{equation*}
	\begin{split}
		& \includegraphics[width=0.4\textwidth]{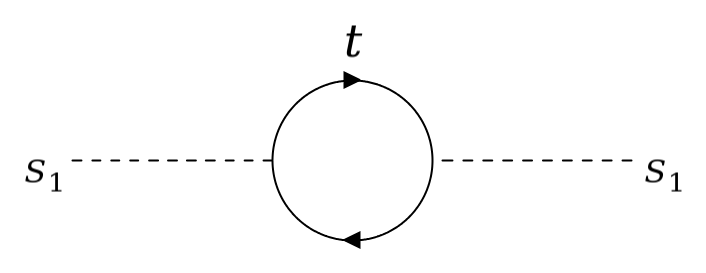} \\
		\delta Z_{\sing} = & - \frac{3 y_{\sing}^2 }{8 \pi ^2} - \frac{3 y_{\sing}^2 m_t^2 }{2 \pi ^2  M_{\sing}^2} \\
		& + \frac{3 y_{\sing}^2 }{8 \pi ^2} \log \left(\frac{m_t^2}{\mu^2}\right) \\
		& + \frac{3 y_{\sing}^2 2 M_{\sing}^2 m_t^2 }{8 \pi ^2  M_{\sing}^3 \sqrt{M_{\sing}^2-4 m_t^2}} \log \left(\frac{ M_{\sing} \sqrt{ M_{\sing}^2-4 m_t^2}- M_{\sing}^2 + 2 m_t^2 }{2 m_t^2}\right) \\
		& + \frac{3 y_{\sing}^2 m_t^4 }{\pi ^2 M_{\sing}^3 \sqrt{ M_{\sing}^2-4 m_t^2}} \log \left(\frac{ M_{\sing} \sqrt{ M_{\sing}^2-4 m_t^2}- M_{\sing}^2 + 2 m_t^2 }{2 m_t^2}\right) \\
		& - \frac{3 y_{\sing}^2 M_{\sing}^4}{8 \pi ^2  M_{\sing}^3 \sqrt{M_{\sing}^2-4m_t^2}} \log \left(\frac{ M_{\sing} \sqrt{ M_{\sing}^2-4 m_t^2}- M_{\sing}^2 + 2 m_t^2 }{2 m_t^2}\right) 
	\end{split}
\end{equation*}

%\newpage

\item The scalar singlet mass counterterm $\delta m_{\sing}$: 

%\begin{center}
	%\includegraphics[width=0.4\textwidth]{figures/feynman_singlet_SS.png}
%\end{center}
\begin{equation*}
	\begin{split}
		& \includegraphics[width=0.4\textwidth]{figures/feynman_singlet_SS.png} \\
		\delta m_{\sing} = & -\frac{15 y_{\sing}^2 m_t^2 }{8 \pi ^2  M_{\sing}}+\frac{3 y_{\sing}^2 M_{\sing} }{8 \pi ^2} \\
		& + \left( \frac{9 y_{\sing}^2 m_t^2}{8 \pi ^2  M_{\sing}} - \frac{3 y_{\sing}^2 M_{\sing}}{16 \pi ^2} \right) \log \left(\frac{m_t^2}{\mu^2}\right)\\
		& +\frac{3 y_{\sing}^2 }{16 \pi^2} \sqrt{M_{\sing}^2-4 m_t^2} \log \left(\frac{  M_{\sing} \sqrt{ M_{\sing}^2-4 m_t^2}- M_{\sing}^2+2 m_t^2}{2 m_t^2}\right) \\
		& -\frac{3 y_{\sing}^2 m_t^2}{4 \pi^2  M_{\sing}^2 } \sqrt{M_{\sing}^2-4 m_t^2} \log \left(\frac{  M_{\sing} \sqrt{ M_{\sing}^2-4 m_t^2}- M_{\sing}^2+2 m_t^2}{2 m_t^2}\right) 
	\end{split}
\end{equation*}

\end{itemize}

\newpage

\section{Code Modifications}
\label{sec:appendixCodeModifs}

To ensure compatibility with the complex mass scheme, some manual adjustments are required before generating the NLO UFO model with \fr. Specifically, we define complex conjugation rules and manipulate the output prior to model export into the UFO format~\cite{Degrande:2011ua, Darme:2023jdn}. We implement first the replacement
\begin{lstlisting}
	CMSConj[X] -> Conjugate[X]
\end{lstlisting}
where \lstinline{X} denotes either the BSM resonance mass, the top quark mass or the strong coupling constant. In addition, we remove complex conjugation from parameters that are assumed real such as the couplings and the renormalisation scale. Generically denoting such a parameter by \lstinline{Y}, this means the replacement
\begin{lstlisting}
	CMSConj[Y] -> Y
\end{lstlisting}
The resulting NLO UFO models are then generated using standard \fr\ functions.

For the colour-singlet scalar simplified model, additional care is required to ensure that NLO calculations with \MG\ include scalar singlets in QCD loop diagrams. This is crucial as these scalars were considered for the computation of UV counterterms by \nloct. To enforce this, we follow the procedure described in Refs.~\cite{Borschensky:2020hot, Borschensky:2021hbo} and modify the \lstinline{is_perturbating()} function in the file \lstinline{base_objects.py}, adding specifically the snippet
\begin{lstlisting}[language=Python]
	if len(int.get('orders')) > 1:
	continue
	## BEGIN ADDITION
	if order in int.get('orders').keys() and \
	abs(self.get('pdg_code')) in [9000001]:
	return True
	## END ADDITION
\end{lstlisting}
As previously indicated, these lines explicitly instruct \MG\ to include the scalar singlet (PDG code 9000001 in our implementation) in QCD loop diagrams. We verified that loops involving at least one scalar singlet were correctly generated and that UV poles cancelled as expected. Then, the four-top process at NLO is generated by \MG with the command:
\begin{lstlisting}[language=Python]
generate p p > t t~ t t~ QED<=0 QCD<=2 NP<=2 [QCD NP]
\end{lstlisting}

As in the full QCD+BSM approach, a custom loop filter must be applied during loop generation in \MG. The precise implementation differs between the scalar octet and scalar singlet cases. In both cases, in the function \lstinline{user_filter()} located in the file named \lstinline{loop_diagram_generation.py}, custom filtering must be activated with:
\begin{lstlisting}[language=Python]
	edit_filter_manually = True
\end{lstlisting}
For the scalar octet model, the following code should then be added to remove triangle loops involving top quarks and scalar octets but no gluons, 
\begin{lstlisting}[language=Python,showstringspaces=false]
	# Apply the custom filter specified if any
	if filter_func:
	try:
	valid_diag = filter_func(diag, structs, model, i)
	except Exception as e:
	raise InvalidCmd("The user-defined filter '%s' did not"%filter+
	" returned the following error:\n       > %s"%str(e))
	## BEGIN ADDITION
	is_incorrect_loop = (9000001 in loop_pdgs) and (6 in loop_pdgs) \
	and (21 not in loop_pdgs)
	if len(diag.get_loop_lines_pdgs())==3 and is_incorrect_loop :
	valid_diag = False
	## END ADDITION
\end{lstlisting}
For the scalar singlet model, the filtering logic is slightly more involved as scalar singlets must be excluded from certain loops based on their topology. This leads to the modification
\begin{lstlisting}[language=Python,showstringspaces=false]
	# Apply the custom filter specified if any
	if filter_func:
	try:
	valid_diag = filter_func(diag, structs, model, i)
	except Exception as e:
	raise InvalidCmd("The user-defined filter '%s' did not"%filter+
	" returned the following error:\n       > %s"%str(e))
	## BEGIN ADDITION
	is_loop_scalar = (9000001 in loop_pdgs)
	is_loop_top = (6 in loop_pdgs)
	is_not_loop_gluon = (21 not in loop_pdgs)
	if len(diag.get_loop_lines_pdgs())<=3 and is_loop_scalar :
	valid_diag = False
	elif len(diag.get_loop_lines_pdgs())==4 and is_loop_scalar \\
	and is_loop_top and is_not_loop_gluon :
	valid_diag = False
	elif len(diag.get_loop_lines_pdgs())==5 and is_loop_scalar \\
	and is_loop_top and is_not_loop_gluon :
	valid_diag = False
	elif len(diag.get_loop_lines_pdgs())==6 and is_loop_scalar \\
	and is_loop_top and is_not_loop_gluon :
	valid_diag = False
	elif len(diag.get_loop_lines_pdgs())>=7 and is_loop_scalar :
	valid_diag = False
	## END ADDITION
\end{lstlisting}

Once the diagram filter was been set up, the four top process can be generated  with the command
\begin{lstlisting}[language=Python]
generate p p > t t~ t t~ QED<=0 QCD<=2 NP<=2 [QCD]
\end{lstlisting}

\chapter{Disentangling the pair and associated production of new scalar resonances}
\label{sec:appendixFourTopObs}
\pagestyle{fancy}

In this section, we list observables that could in principle be used to distinguish between the pair production and associated production of new colour-octet top-philic resonances. These observables exploit the distinct final-state topologies that arise from the two relevant production mechanisms, and were identified from a parton-level study. However, we found that they lose most of their discriminating power after top quark reconstruction from hadron-level events, at least within our framework, due to the smearing of four-momenta and reconstructed masses. For this reason and in order to keep our analysis simple, we have not employed them (except for the first two) to obtain our main results. Instead, we always assumed that pair production was dominating in the considered colour-octet benchmark scenarios. Nevertheless, these observables may still prove useful for characterising selected signal events in real data, especially when used as inputs to a multivariate approach based on Boosted Decision Trees or Neural Networks (which can be readily integrated into our analysis framework as demonstrated in Refs.~\cite{Cornell:2021gut, Cornell:2024dki}).

The goal of these variables is thus to help determine whether selected events are compatible with the production of one or two colour-octet resonances. They probe event shape and are sensitive to the geometry and magnitude of the top quark four-momenta, attempting to assess whether one of the reconstructed top pairs originates from standard QCD interactions.

The following list is not exhaustive but already includes a representative set of key variables.

\begin{itemize}
    \item \textbf{Difference in the top pair masses}: When two resonances are produced, the two top quark pairs in which they decay should have similar invariant masses up to reconstruction effects. This is not necessarily the case if one of the top pairs arises from QCD interactions, and we exploit this variable for the pairing of top quarks in our colour-octet analysis strategy.

    \item \textbf{Largest top transverse momentum}: In the colour-singlet analysis, we pair the two leading-$p_T$ top quarks and assume that they originate from a top-philic resonance decay. This is justified as the tops stemming from the resonance typically carry more transverse momentum than those generated through standard QCD processes.

    \begin{figure}
        \centering
        \includegraphics[width=\linewidth]{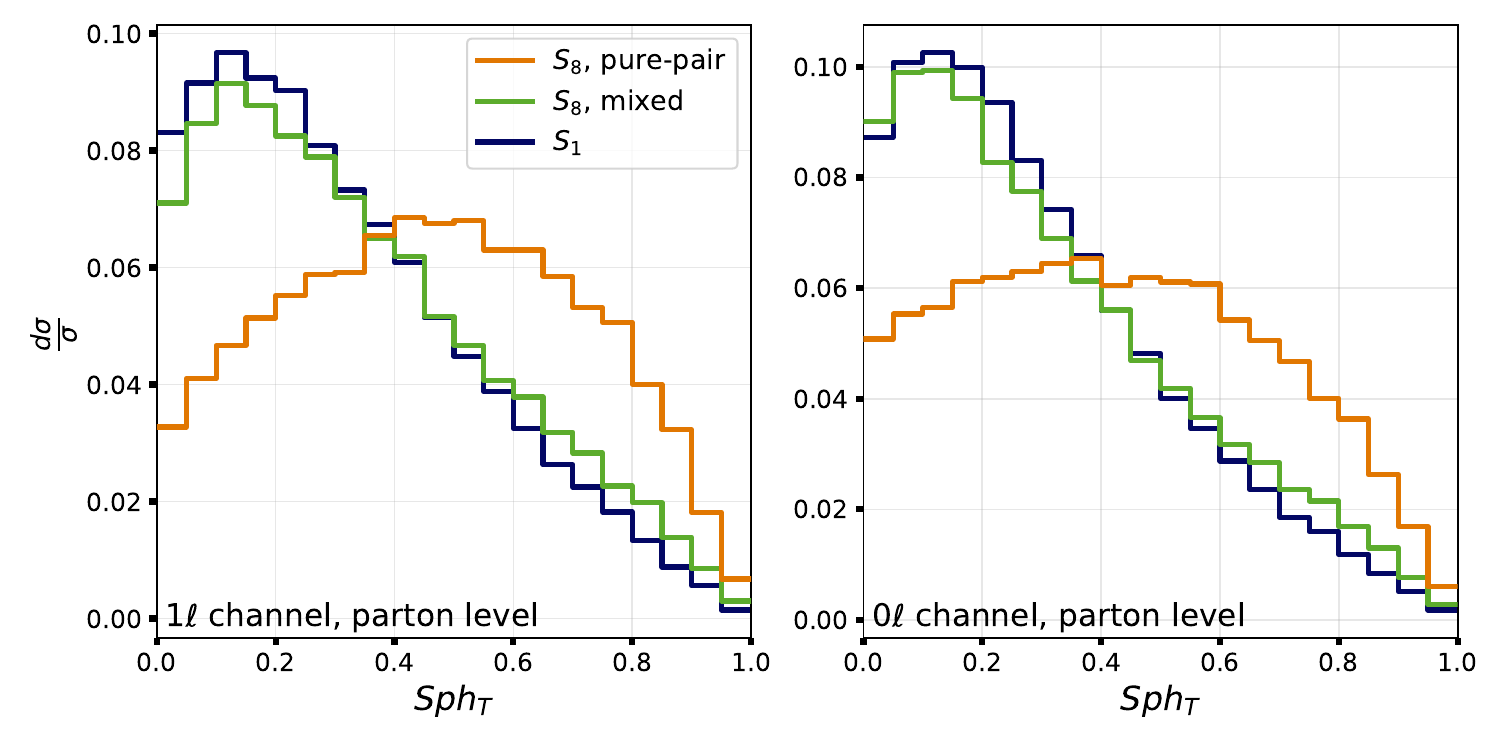}
        \caption{Normalised differential distributions of the transverse sphericity variable in the single-leptonic (left) and fully hadronic (right) channels at parton level. We compare different benchmark points for the colour-octet and colour-singlet models: the curve labelled ``$\oct$, pure-pair'' corresponds to a scenario with $M_{\oct} = 1.3$~TeV and $y_{\oct} = 0.25$ where octet pair production dominates; the ``$\oct$, mixed'' curve refers to the BP1 benchmark of Section~\ref{sec:FourTopTheory} featuring significant contributions from both associated and pair production. Finally, for the singlet case, we consider the BP3 scenario of Section~\ref{sec:FourTopTheory} where only associated production occurs by construction.\label{fig:sphT_lhe}}
    \end{figure}
        
    \item \textbf{Transverse sphericity}: This observable is defined as
    \begin{equation}
        \text{Sph}_T = \frac{2\lambda_2}{\lambda_1 + \lambda_2},
    \end{equation}
    where $\lambda_1 > \lambda_2$ are the eigenvalues of the transverse linearised sphericity tensor
    \begin{equation}
        M_{xy} = \frac{1}{\sum_i |\vec{p}_{T,i}|} \sum_{i=1}^{N_\text{top}} \frac{1}{|\vec{p}_{T,i}|}
        \begin{pmatrix}
            p_{x,i}^2 & p_{x,i}p_{y,i} \\
            p_{y,i}p_{x,i} & p_{y,i}^2
        \end{pmatrix}.
    \end{equation}
    Here, the momenta refer to those of the reconstructed top quarks and the variable ranges from 0 (pencil-like configurations) to 1 (isotropic distributions). Colour-octet pair production typically yields larger $\text{Sph}_T$ values, as all top quarks have four-momenta of comparable magnitude. By contrast, associated production events often feature two top quarks with significantly larger momenta than the others, resulting in a smaller $\text{Sph}_T$ value. This is illustrated in Figure~\ref{fig:sphT_lhe} for the single-leptonic (left) and fully hadronic (right) channels.

    \item \textbf{Transverse thrust}: This observable is defined as
    \begin{equation}
        \text{Thr}_T = 1 - \max_{\hat{n}_T} \frac{\sum_i |\vec{p}_{T,i} \cdot \hat{n}_T|}{\sum_i |\vec{p}_{T,i}|},
    \end{equation}
    where the sum runs over the transverse momenta of the different top quarks and $\hat{n}_T$ corresponds to the unit vector in the transverse plane that maximises the projection of all different momenta. This variable behaves similarly to transverse sphericity: values close to zero correspond to back-to-back top pairs, while values closer to $1 - 2/\pi$ signal more isotropic events.

    \item \textbf{Opening angle}: Since resonances are usually produced nearly at rest, the opening angle $\theta$ between the top quarks in which they decay tends to be close to $\pi$. This is not always the case for top quarks produced via QCD interactions. The relevant angle is defined as usual through $\cos\theta = \vec{p}_1 \cdot \vec{p}_2 / (|\vec{p}_1|\, |\vec{p}_2|)$, where $\vec{p}_{1,2}$ are the momenta of two top quarks.

    \item \textbf{Scalar triple product}: This variable is defined as
    \begin{equation}
        \frac{(\vec{p}_1 \times \vec{p}_2) \cdot \vec{p}_3}{|\vec{p}_1 \times \vec{p}_2|\, |\vec{p}_3|},
    \end{equation}
    where $\vec{p}_{1,2,3}$ are the momenta of three top quarks. This variable vanishes if the momenta are coplanar and reaches $\pm1$ when they are orthogonal. Events from associated production tend to yield values close to zero, whereas pair production favours more spread-out configurations with values closer to $\pm1$, although this behaviour also depends on the number of leptons in the final state.
\end{itemize}

\end{appendices}

\clearpage
\addtocontents{toc}{\protect\setcounter{tocdepth}{0}}
\addcontentsline{toc}{chapter}{Bibliography}
\bibliography{bibliography}
\bibliographystyle{ieeepes}

%%%%%%%%%%%%%%%%%%%%%%%%%%%%%%%%%%%%%%
\end{document}